\documentclass{aa}

\usepackage{natbib}
\usepackage[varg]{txfonts}
\usepackage{booktabs}
\usepackage{bm}
\usepackage{comment}
\usepackage{graphicx}  
\usepackage{subfig}
\usepackage{upgreek}
\usepackage{multirow}
\bibpunct{(}{)}{;}{a}{}{,} 
\usepackage{color, colortbl}
 
\usepackage{framed, color}
\definecolor{shadecolor}{gray}{.85}

\definecolor{LightCyan}{rgb}{0.88,1,1}
 
\begin{document}

\title{Revisiting the BE99 method for the study of outflowing gas in protostellar jets}
 
\author{T.\,Sperling\inst{1}
\and J.\,Eislöffel\inst{1}}
\institute{Thüringer Landessternwarte, Sternwarte 5, D-07778, Tautenburg, Germany}
\date{Received:  12-08-2024  / Accepted: 21-11-2024}

\abstract {An established method measuring the hydrogen ionisation fraction in shock excited gas is the BE99 method, which utilises six bright forbidden emission lines of [S\,II]$\lambda\lambda$6716, 6731, [N\,II]$\lambda\lambda$6548, 6583, and [O\,I]$\lambda\lambda$6300, 6363. The main assumptions of this technique are that the gas is in a low-excitation state ($x_\text{e}< 0.3$) and the equilibrium of the underlying ionisation network is reached.} {We aim to extent the BE99 method by including more emission lines in the blue and near-infrared part of the spectrum ($\lambda = 3\,500-11\,000\,\AA$), and considering higher hydrogen ionisation fractions ($x_\text{e} > 0.3$). In addition, we investigate how a non-equilibrium state of the gas and the presence of extinction influence the BE99 technique.} {We numerically solve a network of ionisation reactions in the time-domain, which lead to the BE99 equilibrium. We apply the BE99 method on synthetic spectra also in the non-equilibrium state. We extent the BE99 technique to higher ionisation fractions by considering additional reactions, which involve higher ionisation states of oxygen, nitrogen, and sulphur. We test our concepts on the low excitation outflow of Par Lup 3-4 and the high excitation 244-440 Proplyd in Orion.} {Plenty additional emission line ratios can in principle be exploited as extended curves (or stripes) in the ($x_\text{e}, T_\text{e}$)-diagram. If the BE99 equilibrium is reached and extinction is corrected for, all stripes overlap in one location in the ($x_\text{e}, T_\text{e}$)-diagram indicating the existing gas parameters. We find that the BE99 equilibrium is reached faster than the hydrogen recombination time. The application to the Par Lup 3-4 outflow shows that the classical BE99 lines together with the [N\,I]$\lambda\lambda$5198+5200 lines do not meet in one location in the ($x_\text{e}, T_\text{e}$)-diagram. This indicates that the gas parameters derived from the classical BE99 method are not fully consistent with other observed line ratios. A multi-line approach is necessary to determine the gas parameters. From our analysis we derive $n_e \sim 45\,000\,\text{cm}^{-3}-53\,000\,\text{cm}^{-3}$, $T_e = 7\,600\,\text{K}-8\,000\,\text{K}$, and $x_e \sim 0.027-0.036$ for the Par Lup 3-4 outflow.  For the 244-440 Proplyd we were able to use the line ratios of [S\,II]$\lambda\lambda$6716+6731, [O\,I]$\lambda\lambda$6300+6363, and [OII]$\lambda\lambda$7320, 7330 in the BE99 diagram to estimate the ionisation fraction at knot E3 ($x_e = 0.58\pm 0.05$). } {The BE99 method can be extended utilising more emission line ratios in the ($x_\text{e}, T_\text{e}$)-diagram and considering higher ionisation states. Exploiting new line ratios reveals more insights on the state of the gas. Our analysis indicates, however, that a multi-line approach is more robust in deriving gas parameters, especially for high density gas.}   
\maketitle

\section{Introduction}

Protostellar outflows are essential consequences when young stars are forming \citep[e.g.][]{frank_2014, bally_2016, pascucci_2023}. They transport away a substantial amount of excess angular momentum from the protostellar system and thus prevent the spinning up of the protostar. In order to fully understand the nature of their gas excitation and the launching mechanism it is important to accurately measure the physical conditions of these outflows, that is the electron density $n_e$, electron temperature $T_e$, and ionisation fraction $x_e$. These key parameters can be constrained observationally by analysing the emitted forbidden emission lines (FELs), which are attributed to the presence of shocks. Protostellar outflows are typically traced by FELs such as, e.g.,  [S\,II]$\lambda\lambda 4068, 4076$, [O\,I]$\lambda\lambda 6300, 6363$, [S\,II]$\lambda\lambda 6716, 6731$, and [N\,II]$\lambda\lambda 6548, 6583$  indicating low-excitation gas conditions ($x_e \lesssim 0.3$, $n_e \sim 10^2-10^6\,\text{cm}^{-3}, T_e \sim 5\,000-20\,000\,\text{K}$).
 
In this context different methods are being used to derive the gas parameters: 1. Shock models can be used for comparing observed FELs with predictions \citep[e.g.][]{hollenbach_1989, hartigan_1994}. Although this approach is the natural choice as shocks cause the observed emission, this approach introduces a strong model-dependency and degeneracy. The geometry of the shock, the shock type (C-shock, J-shock, radiative shock) and other parameters (shock velocities, preshock densities, magnetic fields) substantially influence the resulting FEL spectra; 2. Individual line ratios that are sensitive to one or more gas parameters can be used as diagnostic tools \citep[e.g.][]{osterbrock_book, natta_2014, fang_2018, nisini_2024}. In this framework the extinction and the gas parameters are typically constrained separately and/or consecutively -- depending on the available line ratios. For example, electron densities and temperatures are typically inferred from the [S\,II]$\lambda$6716/[S\,II]$\lambda$6731 or [O\,I]$\lambda\lambda$ (6300+6363)/[O\,I]$\lambda$5577 line ratios, respectively. The ionisation fraction is then derived from further line ratios (e.g. [N\,II]$\lambda$6583/[O\,I]$\lambda$6300) under the assumption of a certain ionisation model with $n_e$ and $T_e$ as previously determined parameters. This method is in essence a multi-step approach. Often only a range for $n_e$, $T_e$, and $x_e$ can be constrained and some available lines remain unused. 3. Another approach is to find the best excitation model, whereby again no detailed shock physics is involved -- that describes simultaneously all observed line ratios that are attributed to the same emitting gas \citep[e.g.][]{hartigan_2007, giannini_2015, giannini_2019}. Mathematically, this represents a nonlinear minimisation problem. Model grid calculations are usually preferred over model fitting, which in turn can be computationally very extensive already for a handful of parameters and a grid with moderate resolution. As a caveat, not all appearing lines can be blindly combined, as they can come from different gas phases.  
 4. A compromise between computationally expensive grid calculations and an in-depth analysis of a few diagnostic diagrams is the BE99-method \citep{bacciotti_1999}, where only one diagnostic diagram is utilised to derive $n_e$, $T_e$, and $x_e$ \citep[e.g.][]{Nisini_2005, podio_2006, ray_2007, coffey_2008, podio_2009, mauri_2014}. The BE99-method utilises six forbidden emission lines ([O\,I]$\lambda\lambda 6300, 6363$, [S\,II]$\lambda\lambda 6716, 6731$, and [N\,II]$\lambda\lambda 6548, 6583$) in the visual part of the electromagnetic spectrum and works in the framework of collisional ionisation equilibrium, where the atoms and ions are assumed to exist mostly in their ground state. In a nutshell, the BE99 diagnostic diagram is constructed as follows: From the observed six FELs three line ratios with their errors  are calculated: O\,I/N\,II, O\,I/S\,II, and N\,II/S\,II (O\,I := [O\,I]$\lambda\lambda$6300+6363, N\,II := [N\,II]$\lambda\lambda$6548+6583, and S\,II := [S\,II]$\lambda\lambda$6716+6731). These three ratios are then compared with theoretical predictions in the $(x_e, T_e)$-parameter space. If the observed line ratio with error margins is consistent with a given set of values in the parameter space -- say $\{x_e^i, T_e^j\}$ --  this point is accepted as potential representation of the data. For all three line ratios these points are collected and effectively appear as three extended curves, that is stripes, in the $(x_e, T_e)$-parameter space. The location, where all stripes overlap, gives the determined gas parameters. \\
All the above mentioned methods are based on time-independent models, that is they assume that the emitting gas is in equilibrium. The hydrogen recombination time $\tau_\text{rec}$ or dynamical time scale $\tau_\text{dyn}$ are usually given as justification for their usage. Since protostellar outflows are not stationary -- as proper motions studies show -- these and potentially other time-scales play a crucial role.  \\
Furthermore, with new spectroscopic instruments such as XShooter, UVES, or MUSE not only the visual part of the spectrum can be studied but also other valuable emission lines that are connected to the BE99 lines in the blue or near-infrared are accessible. In turn, the BE99 method can be extended in a natural way, i.e. by including more emission line ratios as stripes in the ($x_\text{e}, T_\text{e}$)-parameter space potentially providing more information about the present gas conditions. \\
In this study we explore the idea of extending the BE99 method by utilising additional line ratios in the spectral range of $3\,500-11\,000\,\AA$ and considering higher ionisation fractions ($x_\text{e} > 0.3$). Furthermore, we numerically solve a network of ionisation reactions, on which the BE99 ionisation balance is based on, in the time-domain to investigate how the non-equilibrium situation affects the BE99 method. We study the time scales of reaching the equilibrium to inspect how well the equilibrium assumption is justified when applied to protostellar outflows. We test our propositions on the low excitation outflow of Par Lup 3-4 and the high excitation 244-440 Proplyd in Orion.  

\section{The BE99 method in the time domain}\label{BE99}
 
 One of the main assumption of the BE99 method is that the reaction equilibrium in the involved atoms and ions is reached. In this Section we investigate the situation, when the BE99 ionisation balance is not yet reached. We use a simple network of ionisation reactions that lead to the equilibrium ionisation balance of the BE99 method. 
 
\subsection{Deriving the time-dependent reaction rates}\label{sec:time_dependent_reactions}
 
Four elements  are involved in the BE99 technique. These are nitrogen (N), oxygen (O), sulphur (S), and hydrogen (H). Their relevant  ionisation states are denoted  as
$Y^k \in \{ \text{N}^0, \text{N}^+, \text{O}^0, \text{O}^+, \text{S}^+, \text{H}^0, \text{H}^+  \}$. Their number densities are denoted as $n(Y^k)$. The electron density is $n_\text{e}$. The total number densities of an element $Y$, here introduced as $n_\text{tot}(Y)$, are given by the sum of all relevant ionisation states. In a low-excitation medium we expect sulphur to be mostly singly ionised. Nitrogen, oxygen, and hydrogen coexist in their neutral and singly ionised form simultaneously. Thus, their total number densities and ionisation fractions are given by
\begin{equation}\label{eq:sulfur}
n_\text{tot}( Y)  = n( Y^0) + n( Y^{+}),
\end{equation}
and
\begin{equation}
x_Y  = n(Y^{+})/\left[ n(Y^0) + n(Y^{+})\right]. \label{eq:ionisation_fraction_N} 
\end{equation}
\noindent
Three atomic processes are considered in the BE99 technique: 1. direct and inverse charge exchange, 2. collisional ionisation, 3. radiative/dielectronic recombination. Photo-ionisation is neglected. The BE99 method assumes ionisation equilibrium in the involved reactions. Without this assumption a time dependent version of the BE99 method can be formulated from the underlying set of reactions, that is
\begin{equation}\label{eq:reactions_A}
 \text{N}^0 +\text{H}^+   \underset{\delta_\text{N}'}{\stackrel{\delta_\text{N}}{\rightleftharpoons}} \text{N}^+ +\text{H}^0   \hspace{1cm}   (\text{charge exchange}) ,
\end{equation}
\begin{equation}
\text{N}^0  +\text{e}^-   \overset{C_\text{N}}{\longrightarrow}  \text{N}^+  +2\text{e}^-  \hspace{1cm}  (\text{coll. ionisation}),
\end{equation}
\begin{equation}
 \text{N}^+  +\text{e}^-   \overset{\alpha_\text{N}}{\longrightarrow}  \text{N}^0 \hspace{2cm}  (\text{recombination}), 
\end{equation}
\begin{equation}
\text{O}^0 +\text{H}^+  \underset{\delta_\text{O}'}{\stackrel{\delta_\text{O}}{\rightleftharpoons}} \text{O}^+ +\text{H}^0   \hspace{1cm}   (\text{charge exchange}) ,
\end{equation}
\begin{equation}
\text{O}^0  +\text{e}^-    \overset{C_\text{O}}{\longrightarrow}  \text{O}^+  +2\text{e}^-  \hspace{1cm}   (\text{coll. ionisation}),
\end{equation}
\begin{equation}
 \text{O}^+  +\text{e}^-   \overset{\alpha_\text{O}}{\longrightarrow}  \text{O}^0  \hspace{2cm}   (\text{recombination}), 
\end{equation}
\begin{equation}
 \text{H}^0  +\text{e}^-   \overset{C_\text{H}}{\longrightarrow}  \text{H}^+  +2\text{e}^-   \hspace{1cm}   (\text{coll. ionisation}),
\end{equation}
\begin{equation}\label{eq:reactions_J}
\text{H}^+  +\text{e}^-   \overset{\alpha_\text{H}}{\longrightarrow}  \text{H}^0  \hspace{2cm}  (\text{recombination})   .
\end{equation}
The respective reaction rates are indicated above or below the arrow of each reaction. They are functions of the $T_\text{e}$ (Appendix\,A).  \\
From the above reactions (Eqs.\,\ref{eq:reactions_A}--\ref{eq:reactions_J}) the full time-dependent system consisting of seven coupled ordinary differential equations -- one for each involved particle N$^\text{o}$, N$^+$, O$^\text{o}$, O$^+$, H$^\text{o}$, H$^+$, e$^-$ -- can be derived straight forwardly (Appendix\,A). 
 In the BE99 method, no assumptions for the hydrogen reactions are involved. In turn, without invoking any knowledge about additional reactions with hydrogen, the hydrogen ionisation fraction $x_\text{e}$ remains a free parameter in the BE99 method. In the excitation model, however, the ionisation fraction in the equilibrium will be a function of the electron temperature $T_\text{e}$ and not a free parameter. This model then provides the opportunity to study the  BE99 method in the time domain.

\subsection{The initial conditions}
 
\begin{figure}  
\resizebox{\hsize}{!}{\includegraphics[trim=0 0 0 0, clip, width=0.9\textwidth]{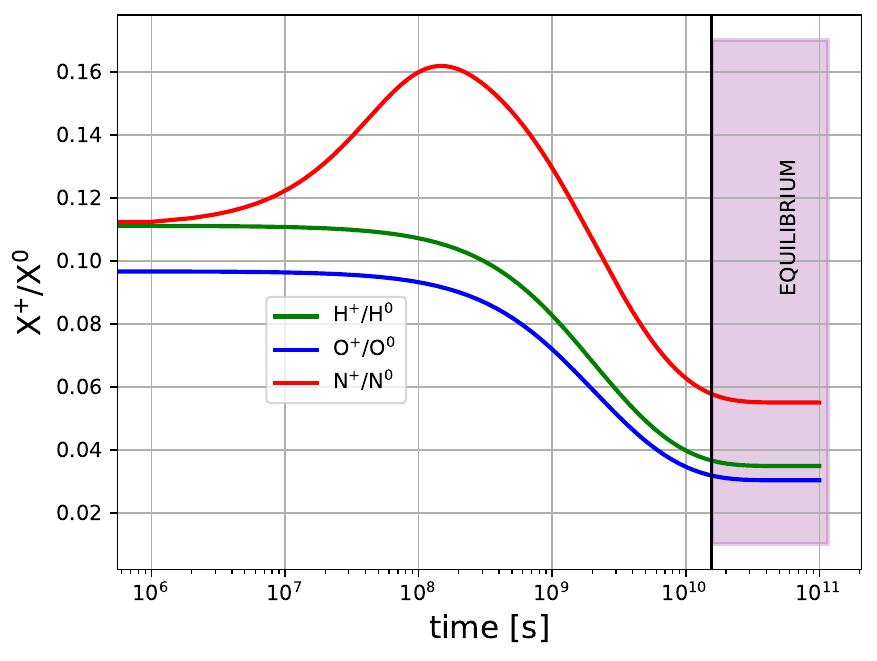}}
\caption{\small{Time resolved numerical solution of the reaction rates for the ionisation fractions of hydrogen, oxygen, and nitrogen (Eqs.\,\ref{eq:reactions_A}-\ref{eq:reactions_J}). The initial conditions at $t_0=0$ were $n_\text{e}^{init} = 2\,000\,\text{cm}^{-3}$, $x_\text{e}^{init} = 0.1$, $x_\text{O}^{init} = 0.1$, and $x_\text{N}^{init}  = 0.1$. They have already evolved to the values at the starting point of this plot. The reaction equilibrium is reached after $\tau_\text{eq} \approx 2\times 10^{10}\,\text{s}$ as indicated as pink shaded area. We fix $T_\text{e} = \text{const} = 10\,500\,\text{K}$ during the simulation.}}\label{fig:ionisation_ratios} 
\end{figure}

 The stated system of ordinary differential equations can be integrated numerically by specifying eight initial number densities  $n_\text{init}(\text{N}^0)$, $n_\text{init}(\text{N}^+)$, $n_\text{init}(\text{O}^0)$, $n_\text{init}(\text{O}^+)$, $n_\text{init}(\text{H}^0)$, $n_\text{init}(\text{H}^+)$, $n_\text{init}(\text{S}^+)$, and $n_\text{e}^\text{init}$.  To keep our model reasonably simple, we assume the gas in isothermal condition. Since we assume constant element abundances and global charge neutrality over the whole integration, only four initial values instead of eight fully determine the reactions. In order to invoke physically relevant initial conditions we chose $n_\text{e}^{init}$, $x_\text{e}^{init}$, $x_\text{N}^{init}$, and $x_\text{O}^{init}$ as a set of independent initial parameters.  Element abundances are taken from \citet{asplund_2009}. Their stated values are $\text{N}/\text{H} = 6.76\times 10^{-5}$, 
  $\text{O}/\text{H} = 4.90 \times 10^{-4}$, and $\text{S}/\text{H} =  1.32 \times 10^{-5}$. We note that these values differ from the values used in the BE99 paper ($\text{N}/\text{H} = 1.1\times 10^{-4}$, 
  $\text{O}/\text{H} = 6.0 \times 10^{-4}$, and $\text{S}/\text{H} =  1.6 \times 10^{-5}$). Different assumed element abundances and updated atomic data affect the results of any diagnostic method. \\
 The transformations between the two sets of initial conditions can be derived straight forwardly. We find \begin{equation}\label{eq:01}
 n(\text{H}^{+})   =  n(\text{H}^{0})\cdot x_\text{e}/(1-x_\text{e}).
 \end{equation}
The element abundances are invoked by  
 \begin{equation}\label{eq:02}
 n(\text{N}^0) + n(\text{N}^{+}) =  \text{N}/\text{H}  \cdot \left( n(\text{H}^0) + n(\text{H}^{+}) \right), 
 \end{equation}
 \begin{equation}\label{eq:03}
 n(\text{O}^0) + n(\text{O}^{+}) = \text{O}/\text{H}  \cdot \left( n(\text{H}^0) + n(\text{H}^{+}) \right),
 \end{equation}
 \begin{equation}\label{eq:04}
 n(\text{S}^{+}) = \text{S}/\text{H}  \cdot \left( n(\text{H}^0) + n(\text{H}^{+}) \right).
 \end{equation}
Furthermore, the electron density is fixed by global charge neutrality   
\begin{equation}
n(\text{e}^-) = n(\text{H}^+) + n(\text{O}^+) + n(\text{N}^+) + n(\text{S}^+).
\end{equation}
With these equations we can derive 
\begin{equation}\label{eq:002}
n(\text{H}^0)  =   \frac{(1-x_\text{e})\,n_\text{e}}{x_\text{e} + \frac{\text{S}}{\text{H}} + \left(\frac{\text{N}}{\text{H}}\right) x_\text{N}   + \left(\frac{\text{O}}{\text{H}}\right) x_\text{O}  }.
 \end{equation}
Thus, from Eqs.\,\ref{eq:01}-\ref{eq:002} the initial conditions can be transformed one-to-one by specifying $n_\text{e}^{init}$, $x_\text{e}^{init}$, $x_\text{N}^{init}$, and $x_\text{O}^{init}$. The ionisation parameters are constrained to $x_\text{e}^{init}$, $x_\text{N}^{init}$, $x_\text{O}^{init}$ $\in [0,1]$, with 0 representing a neutral medium and 1 being fully ionised (in that element). We integrate the full system of the coupled ordinary differential equations numerically until the reaction equilibrium is reached using the Runge-Kutta-solver implemented in Python \citep{python_scipy_2020}. An example of a numerical solution with $T_e = \text{const}$ = 10\,500\,K is depicted in Fig.\,\ref{fig:ionisation_ratios}. Figure\,1 shows how the three ratios $\text{H}^+/\text{H}^\text{o}$, $\text{O}^+/\text{O}^\text{o}$, and $\text{N}^+/\text{N}^\text{o}$ change with time until the reaction equilibrium is reached after about $2\times 10^{10}\,\text{s} \approx 634$\,years. The ratios $\text{H}^+/\text{H}^\text{o}$ and $\text{O}^+/\text{O}^\text{o}$ show a very similar evolution over the shown integration time, whereas the $\text{N}^+/\text{N}^\text{o}$ ratio behaves differently (an extra maximum at about $10^8$\,s). This behaviour is a consequence of the dominant charge exchange reactions between oxygen and hydrogen in the reaction network (see following Section\,\ref{sec:equi}). Both ratios $\text{H}^+/\text{H}^\text{o}$ and $\text{O}^+/\text{O}^\text{o}$ are locked in -- with $\text{O}^+/\text{O}^\text{o} \lesssim \text{H}^+/\text{H}^\text{o}$ -- shortly after the start of the reactions.

\begin{figure}  
\resizebox{\hsize}{!}{\includegraphics[trim=0 0 0 0, clip, width=0.9\textwidth]{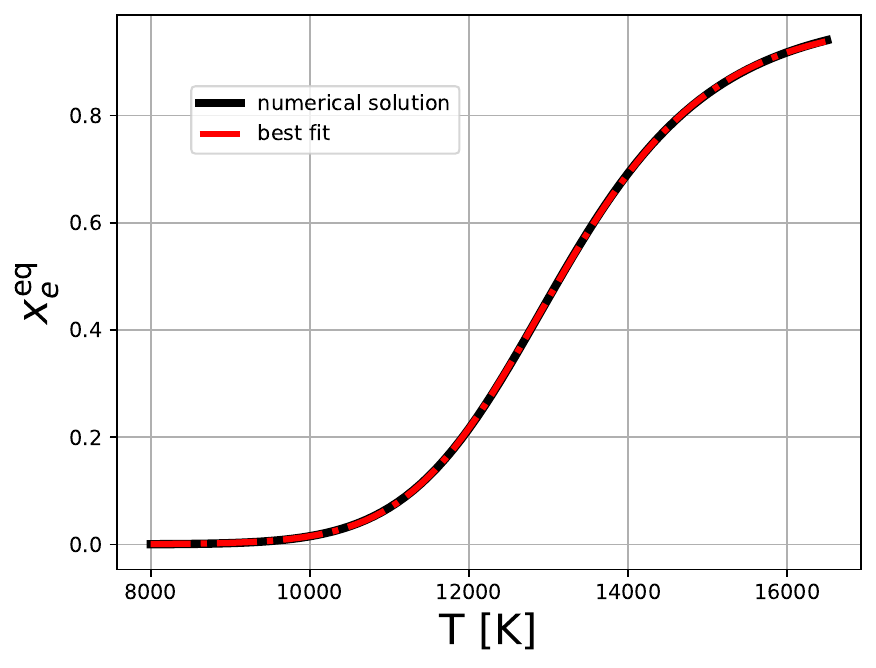}}
\caption{\small{The hydrogen ionisation fraction of the BE99 reactions in the equilibrium $x_\text{e}^\text{eq}$ as a function of the electron temperature $T_\text{e}$.}}\label{fig:equilibrium_ionisation_fraction} 
\end{figure}

  \subsection{The reaction equilibrium}\label{sec:equi} 
   
An interesting situation occurs when a stable equilibrium is reached after some time $t$. For the equilibrium we require
  \begin{equation}\label{equ:equilibrium_ode}
  d[Y^k]/dt  = 0 ,\quad \quad \text{for all } Y^k \text{ and for } t> \tau_\text{eq}.
  \end{equation}
By invoking these constraints to the full system of ODEs, we can derive the ionisation balance in the equilibrium. They are
  \begin{align}
\frac{\text{N}^+}{\text{N}^0}\bigg\rvert_\text{eq} &= \frac{ \delta_\text{N}  \cdot [\text{H}^{+}]_\text{eq} + C_\text{N} \cdot    [\text{e}^{-}]_\text{eq} }{\delta^{'}_\text{N}  \cdot [\text{H}^0]_\text{eq} + \alpha_\text{N}  \cdot  [\text{e}^{-}]_\text{eq}}, \label{eq:nitrogen_equ}     \\
\frac{\text{O}^+}{\text{O}^0}\bigg\rvert_\text{eq} &= \frac{ \delta_\text{O}  \cdot [\text{H}^{+}]_\text{eq} + C_\text{O} \cdot    [\text{e}^{-}]_\text{eq} }{\delta^{'}_\text{O}  \cdot [\text{H}^0]_\text{eq} + \alpha_\text{O}  \cdot  [\text{e}^{-}]_\text{eq}} ,\label{eq:oxygen_equ}  \\
 \frac{\text{H}^+}{\text{H}^0}\bigg\rvert_\text{eq} &= \frac{\delta^{'}_\text{O} \cdot [\text{O}^{+}]_\text{eq} + \delta^{'}_\text{N} \cdot [\text{N}^{+}]_\text{eq} + C_\text{H} \cdot [\text{e}^{-}]_\text{eq}}{\delta_\text{O}  \cdot [\text{O}^0]_\text{eq} + \delta_\text{N}  \cdot [\text{N}^0]_\text{eq} + \alpha_\text{H}  \cdot [\text{e}^-]_\text{eq}} .\label{eq:hydrogen_equ}    
      \end{align}
We note that only  Eqs.\,\ref{eq:nitrogen_equ} and \ref{eq:oxygen_equ} appear in the BE99 method, since the reactions with hydrogen are not constrained by any assumption. 
    Charge  exchange with oxygen dominates for $T_\text{e} \sim 5\,000-15\,000\,\text{K}$, since $\delta_\text{O}, \delta^{'}_\text{O} \sim 10^{-9}\,\text{cm}^{3}\,\text{s}^{-1}$ and $ C_\text{O}, \alpha_\text{O}\sim 10^{-12}-10^{-13}\,\text{cm}^{3}\,\text{s}^{-1}$. Thus, the ionisation states of oxygen and hydrogen are tightly locked via  
    \begin{equation}
    \frac{\text{O}^+}{\text{O}^0}\bigg\rvert_\text{eq}   \approx \left( \frac{\delta_\text{O}}{\delta_\text{O}'}\right) \cdot \left(\frac{\text{H}^+}{\text{H}^0}\right)\bigg\rvert_\text{eq} \approx  \left(\frac{8}{9}\right)\cdot  \left(\frac{\text{H}^+}{\text{H}^0}\right)\bigg\rvert_\text{eq} \quad \text{for} \quad  T_\text{e} \sim 10^4\,\text{K}
    \end{equation}  
    as shown in \citet{osterbrock_book}. Furthermore, we can assume $n_\text{e} \approx n(\text{H}^+)$ (neutral plasma condition).  Thus we have 
\begin{equation}
 n(\text{H}^0) \approx (1-x_\text{e})\cdot  n_\text{e}/x_\text{e}\label{eq:hydrogen_ionisation}
\end{equation}    
and the ionisation equilibrium for nitrogen (Eq.\,\ref{eq:nitrogen_equ}) and oxygen (Eq.\,\ref{eq:oxygen_equ})  can be approximated by   
    \begin{align} 
\frac{\text{N}^+}{\text{N}^0}\bigg\rvert_\text{eq} &\approx  \frac{\left( \delta_\text{N}    +  C_\text{N}  \right)\cdot x_e^\text{eq}   }{ \left(\alpha_\text{N}      -  \delta^{'}_\text{N}  \right)  x_e^\text{eq}     + \delta^{'}_\text{N}  }   , \label{eq:BE99_EQUILIBRIUM_N}  \\
\frac{\text{O}^+}{\text{O}^0}\bigg\rvert_\text{eq} &\approx \frac{\left( \delta_\text{O}    +  C_\text{O}  \right)\cdot x_e^\text{eq}   }{ \left(\alpha_\text{O}      -  \delta^{'}_\text{O}  \right)  x_e^\text{eq}     + \delta^{'}_\text{O}  } \approx \left( \frac{\delta_\text{O}}{\delta_\text{O}'}\right) \cdot \frac{x_\text{e}^\text{eq}}{1-x_\text{e}^\text{eq}}      \label{eq:BE99_EQUILIBRIUM_O}
      \end{align} 
 as demonstrated by \citet{bacciotti_1999}. The above Eqs.\,\ref{eq:BE99_EQUILIBRIUM_N} and \ref{eq:BE99_EQUILIBRIUM_O} are in full agreement with our time-dependent simulations displayed in Figure\,\ref{fig:ionisation_ratios}.

\subsection{The ionisation fraction in the equilibrium}\label{sec:ionisation_fraction_equ}

Extensive numerical experiments confirm that the ionisation fraction in the equilibrium ($x_\text{e}^{\text{eq}}$) is a function of only $T_\text{e}$. In detail, we integrated the time-dependent reaction model shown in Fig.\,\ref{fig:ionisation_ratios} until the equilibrium state for a fine temperature range ($T_e = 8\,000-16\,500\,\text{K}$) and grid of initial conditions.  We plot $x_\text{e}^{\text{eq}}$ in Fig.\,\ref{fig:equilibrium_ionisation_fraction} and fit a four parametric function to the curve 
\begin{equation}\label{eq:f(T)}
x_\text{e}^{\text{eq}}(T_\text{e}) = A \cdot \left(1 + \text{exp}\left[-B\cdot(T_\text{e}-T_0)\right] \right)^{C}   
\end{equation}
with  $A=0.9785$, $B=8.778\times 10^{-4}$, $T_0=12\,000\,\text{K}$, $C=-2.1762$. The function $x_\text{e}^{\text{eq}}(T_\text{e})$ looks qualitatively similar to the hydrogen ionisation fraction derived from the Saha-Equation. However, both functions are quantitatively very different (see Appendix\,B). In turn, the Saha-Equation does not properly describe the BE99 gas conditions in the equilibrium. \\
In principle, Eq.\,\ref{eq:f(T)} offers the opportunity to measure the ionisation fraction directly from the electron temperature, i.e. without the need to construct a BE99 diagram. The line ratio [O\,I]$\lambda\lambda$(6300+6363)/[O\,I]$\lambda$5577 would be sufficient for that purpose as it is sensitive to $T_\text{e}$. However, this way of determining the ionisation fraction requires that the proposed ionisation reactions of our model are the dominant ones and the temperature stays constant with time.  Using Eq.\,\ref{eq:f(T)} for determining the ionisation fraction would therefore introduce a model dependency, which the BE99 technique deliberately avoids.

 \subsection{Time scales:  $\tau_\text{rec}$, $\tau_\text{eq}$, $\tau_\text{dyn}$}
 
The BE99 method is based on the assumption that the gas under investigation is in equilibrium. The  hydrogen  recombination time  
\begin{equation}\label{eq:recombination_time}
\tau_\text{rec}\sim 1/(n_\text{e}\alpha_\text{H}(T_\text{e})) \approx 10^{5}\,\text{yr}/n_\text{e} [\text{cm}^{-3}]
\end{equation} 
with $\alpha_\text{H}$ as hydrogen recombination rate coefficient is usually used to justify this assumption \citep{bacciotti_1995, osterbrock_book, hartigan_2007}. The line of reasoning is that the ionisation fractions of O$^+$/O$^0$ and  H$^+$/H$^0$  are coupled through charge exchange and therefore ultimately with $x_\text{e}^{\text{eq}}$. In this regard \citet{osterbrock_book} noted that \textit{deviations from ionisation equilibrium are ordinarily damped out in times of this [$\tau_\text{rec}$] order of magnitude}.  However, oxygen and nitrogen react on different time scales as compared to hydrogen and in addition the hydrogen  recombination time is derived for a reaction system involving only hydrogen, i.e. no coupling with other reactants. Thus, the hydrogen recombination time may not provide an appropriate justification, if the equilibrium is reached or not.  
In the following, we therefore solve the ODEs numerically to study the how fast the equilibrium is reached.\\
We define that the equilibrium is reached ($\tau_\text{eq}$), when the ionisation fractions of hydrogen, nitrogen, and oxygen stay constant with time and deviate only by  5\% from the ionisation equilibrium (Eqs.\,\ref{eq:BE99_EQUILIBRIUM_N} and \ref{eq:BE99_EQUILIBRIUM_O}).
With this definition we numerically found a crude estimate:
\begin{equation}\label{eq:equilibrium_time_scale}
\tau_\text{eq}  \approx  10\,\text{yr} \,\frac{  x_\text{e}^{init}}{n_\text{e}^{init}\,[\text{cm}^{-3}]}  \, \text{exp}\left(\frac{145528}{T_\text{e}\,[\text{K}]}\right).
\end{equation}
\noindent 
We can compare the equilibrium times scale (Eq.\,\ref{eq:equilibrium_time_scale}) with the hydrogen  recombination time  (Eq.\,\ref{eq:recombination_time}) by plugging in realistic numbers for $T_\text{e}$, $n_\text{e}$, and $x_\text{e}$. We base our illustration on the \citet{bacciotti_1999} measurements of the HH\,34 jet. 
 The brightest knot J of the HH\,34 jet beam is located at about $\theta = 20\arcsec$ away from the source (Fig.\,6a--f therein). An estimate of $\tau_\text{eq}$ can be done by taking the values of $T_\text{e}$, $n_\text{e}$, and $x_\text{e}$ at the apex of knot J, for which we read $T_\text{e} \approx 8\,000\,\text{K}$, $n_\text{e}\approx 500\,\text{cm}^{-3}$, and $x_\text{e} \approx 0.05$. With Eq.\,\ref{eq:equilibrium_time_scale} we therefore derive $\tau_\text{eq}\approx  80\,000 \,\text{yrs}$.
 The  hydrogen  recombination time  $\tau_\text{rec}$ derived from Eq.\,\ref{eq:recombination_time} gives a value of about 200\,yr, which is more than two orders of magnitude smaller than the equilibrium time scale. From that comparison the medium before and behind knot J of the HH\,34 jet has not enough time to reach the reaction equilibrium and in turn the BE99 technique would not be applicable here. In comparison, the dynamical time scale (dynamical shock age) of the jet can be estimated via
\begin{equation}
\tau_\text{dyn} = \theta \cdot D/v_t \approx 4.74\times \frac{\theta [''] \cdot D [\text{pc}]}{v_t [\text{km}\,\text{s}^{-1}]} \,\text{yrs}
\end{equation}
with $D$ as distance to the target, $v_t$ as transverse velocity, and $\theta$ as projected angle on the plane of sky. For the HH34 knot J the dynamical time scale is on the order of the hydrogen  recombination time, i.e. $\tau_\text{dyn}\approx 200\,\text{yrs}$ for $D = 480\,\text{pc}$ and $v_t = 220\,\text{km}\,\text{s}^{-1}$. In order to resolve the stated \textit{time-scale-problem} ($\tau_\text{eq}>> \tau_\text{rec}, \tau_\text{dyn}$), we introduce a new time scale in the following Section \ref{sec:time_resolved_BE99} that is more suitable for describing the state of the system.

\subsection{Resolving the BE99 method in the time domain}\label{sec:time_resolved_BE99}

\begin{figure} 
\resizebox{\hsize}{!}{\includegraphics[trim=0 0 0 0, clip, width=0.9\textwidth]{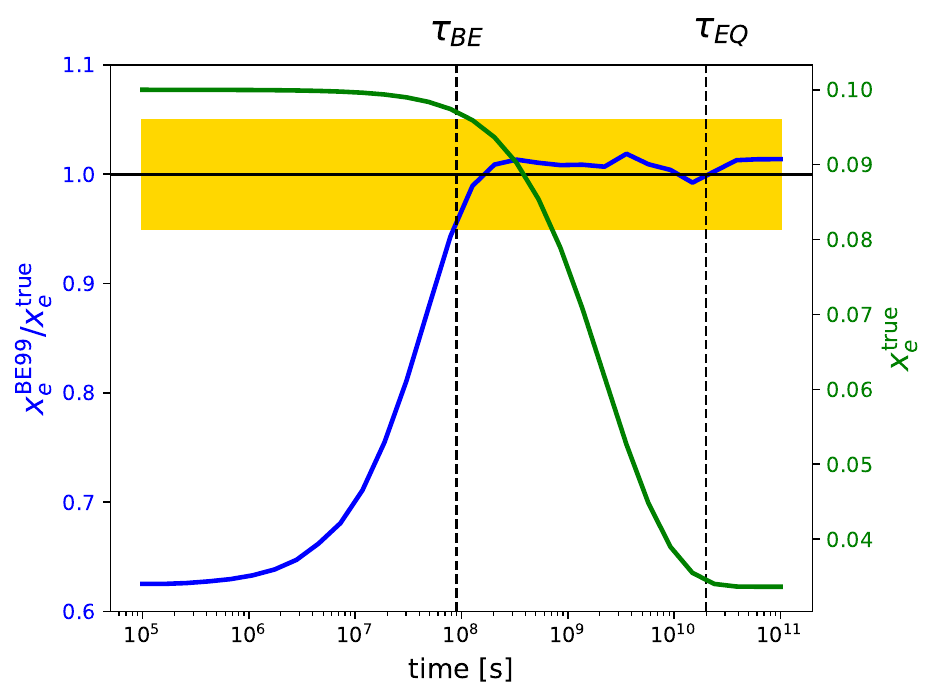}}
\caption{\small{Time resolved BE99 method for the model of Fig.\,\ref{fig:ionisation_ratios}. The yellow horizontal stripe indicates the region, where the BE99 method measures an ionisation fraction  to $5\,\%$ accuracy with respect to the ionisation fraction. Within $\tau_\text{BE} \approx 10^8\,\text{s}$ the BE99 method converges -- long before the reaction equilibrium ($x_e^\text{eq} = 0.0337$)  is reached ($\tau_{\text{BE}}<<\tau_\text{eq}$).}}\label{fig:BE_time_scale} 
\end{figure}

\begin{figure}  
\resizebox{\hsize}{!}{\includegraphics[trim=0 0 0 0, clip, width=0.9\textwidth]{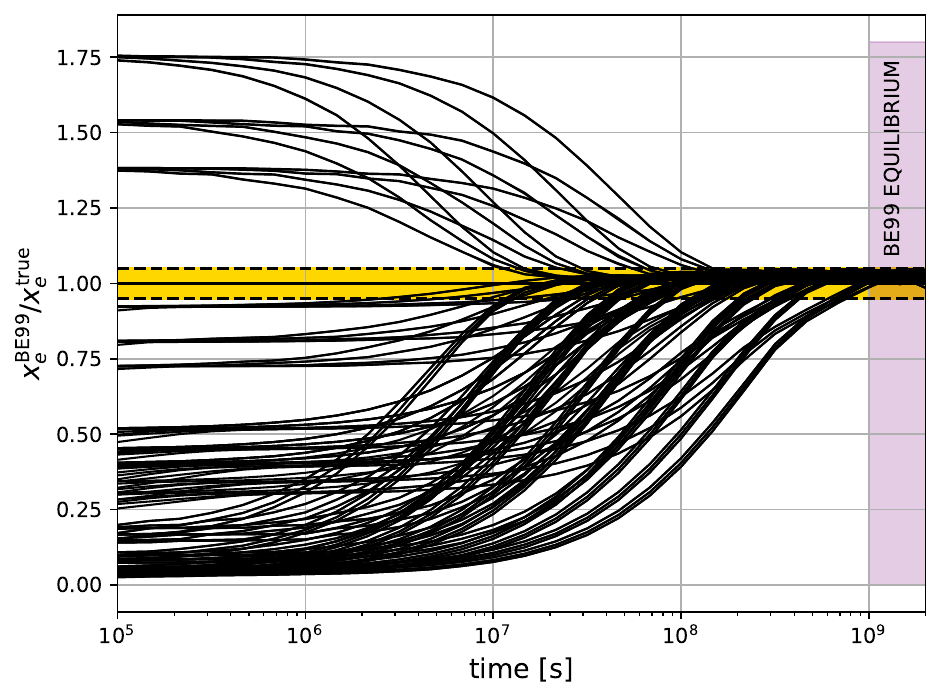}}
\caption{\small{Convergence of the BE99 method in the time domain. In total 192 models with different initial values have been integrated. Parameters: $T_\text{e} [\text{K}]= 8000, 9000, 10000$; $n_\text{e}^\text{init}  [\text{cm}^{-3}]= 1000, 2000, 5000, 10000$; $x_\text{e}^\text{init} = 0.05, 0.1, 0.2, 0.3$; $x_\text{N}^\text{init} = 0.01, 0.1$;   $x_\text{O}^\text{init} = 0.01, 0.1$.  Each black curve represents one model, where the BE99 method  (assuming equilibrium) has been applied following the gas through the non-equilibrium state. For all these models the BE99 method measures the true gas conditions up to 5\,$\%$ accuracy (indicated as yellow horizontal stripe) after $\sim 10^9\,\text{s}$. }}\label{fig:be99_time} 
\end{figure}

The critical question for the applicability of the BE99 method may actually not be if the gas is in reaction equilibrium - this may never be the case in any outflow region - but instead how far off the gas is from the equilibrium.  \\
In order to answer this question we follow the numerical solution of our model in the time domain. At each point in time we have full knowledge of the true gas conditions and can therefore calculate synthetic spectra in the non-equilibrium and equilibrium state. These (model-dependent) synthetic spectra can be used as input for the (model-independent) BE99 method. This situation corresponds to the observers perspective who does not know if the  equilibrium is reached.  
 Applying the BE99 technique out of equilibrium will give measurements of the electron density, temperature and ionisation fraction ($n_e^\text{BE99}$, $T_\text{e}^\text{BE99}$, $x_\text{e}^\text{BE99}$) that do not represent the true gas conditions ($n_\text{e}^\text{true}$, $T_\text{e}^\text{true}$, $x_\text{e}^\text{true}$). As the system evolves in time from a non-equilibrium to the equilibrium state the BE99 method will provide higher accuracy in these three gas parameters, i.e. it converges.  In order to quantify this  transition phase we introduce the time scale $\tau_\text{BE}$. We define $\tau_\text{BE}$ as the time at which the BE99 method provides a measurement of the  ionisation fraction within 5\% of the true ionisation fraction, which is comparable to a typical measurement error of excellent observational data:
\begin{equation}\label{condition_new_BE}
\left\| 1 -  x_\text{e}^\text{BE99}/x_\text{e}^\text{true} \right\| < 0.05 \quad \text{for} \quad t > \tau_\text{BE} .
\end{equation}
\noindent 
We visualise the convergence of the BE99 method in the time domain in Figure\,\ref{fig:BE_time_scale}: already within $\tau_\text{BE} \approx 10^{8}\,\text{s}\,(3\,\text{yrs})$ the BE99 method measures the true ionisation within the  5\% error margin.  In comparison, the  hydrogen recombination time derived from Eq.\,\ref{eq:recombination_time} is about $\tau_\text{rec} \approx 50\,\text{yrs}$, implying $\tau_\text{BE} < \tau_\text{rec} << \tau_\text{eq}$. In conclusion, neither $ \tau_\text{eq}$ nor $\tau_\text{rec}$ are indicating the relevant time scale for the applicability of the BE99 method -- it is rather  $\tau_\text{BE}$. We tested this hypothesis by integrating 192 models with different input parameters (Fig.\,\ref{fig:be99_time}). All the tested models have a BE99 equilibrium time smaller than $10^{9}\,\text{s}\approx 32\,\text{yrs}$. This result supports the notion that the BE99 equilibrium is reached fast enough -- even faster that the hydrogen  recombination time.

\begin{figure*} 
\centering
\subfloat{\includegraphics[trim=0 0 0 0, clip, width=0.2 \textwidth]{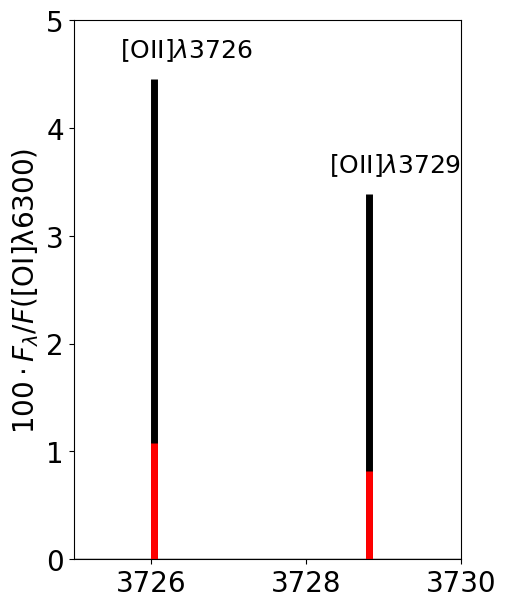}}
\hfill
\subfloat{\includegraphics[trim=0 0 0 0, clip, width=0.185 \textwidth]{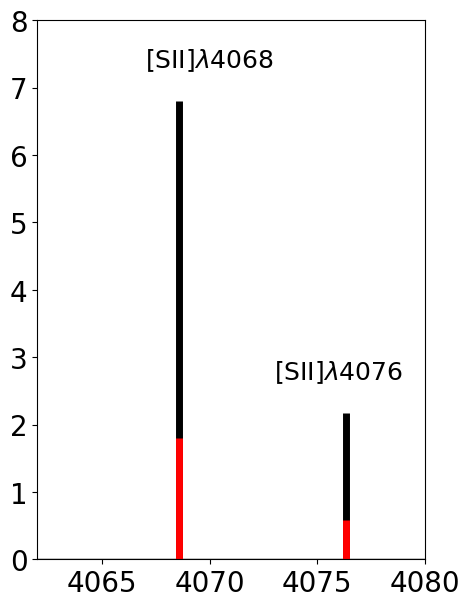}}
\hfill
\subfloat{\includegraphics[trim=0 0 0 0, clip, width=0.192 \textwidth]{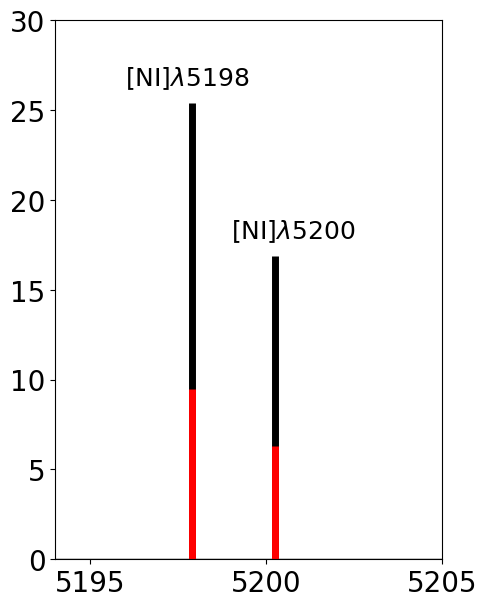}}
\hfill 
\subfloat{\includegraphics[trim=0 0 0 0, clip, width=0.19 \textwidth]{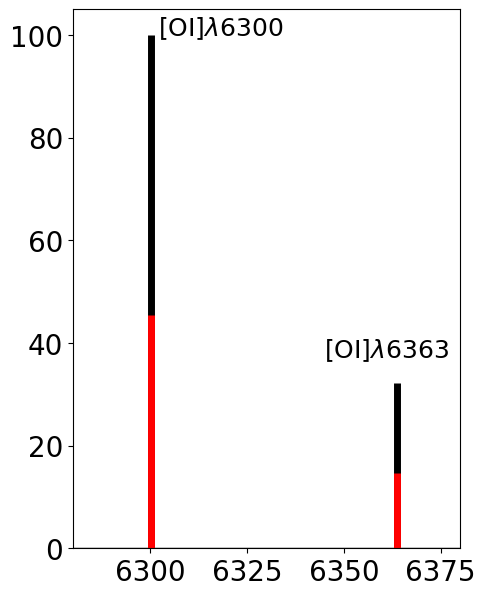}}
\hfill
\subfloat{\includegraphics[trim=0 0 0 0, clip, width=0.173 \textwidth]{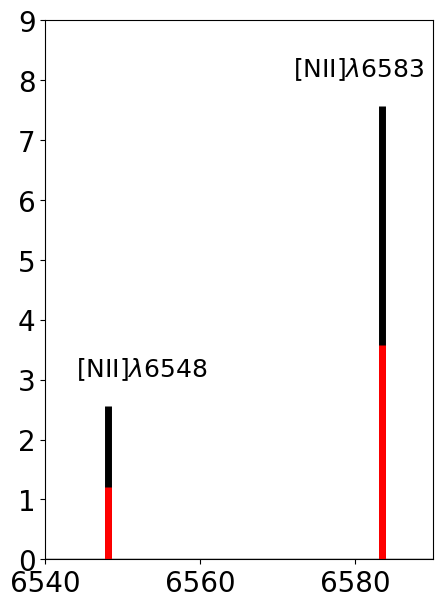}}
\hfill  \\
\subfloat{\includegraphics[trim=0 0 0 0, clip, width=0.215 \textwidth]{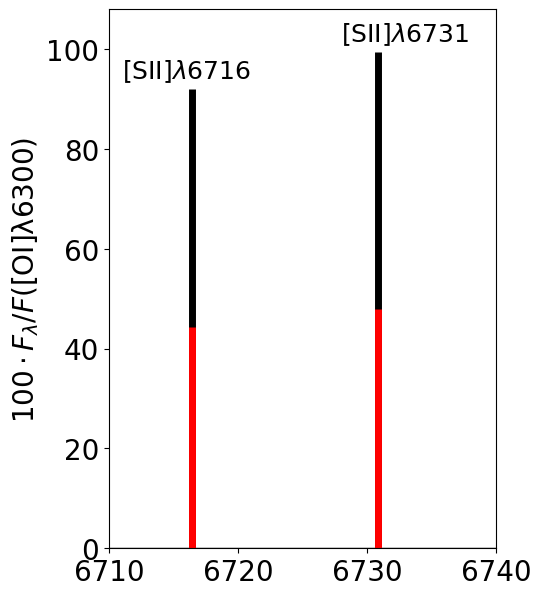}}
\hfill
\subfloat{\includegraphics[trim=0 0 0 0, clip, width=0.2 \textwidth]{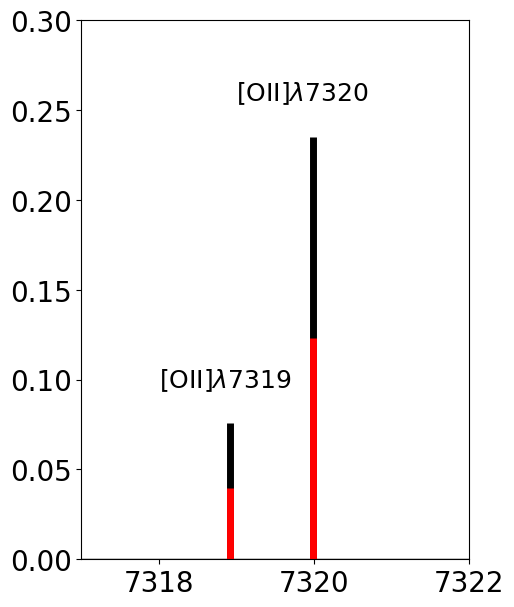}}
\hfill
\subfloat{\includegraphics[trim=0 0 0 0, clip, width=0.2 \textwidth]{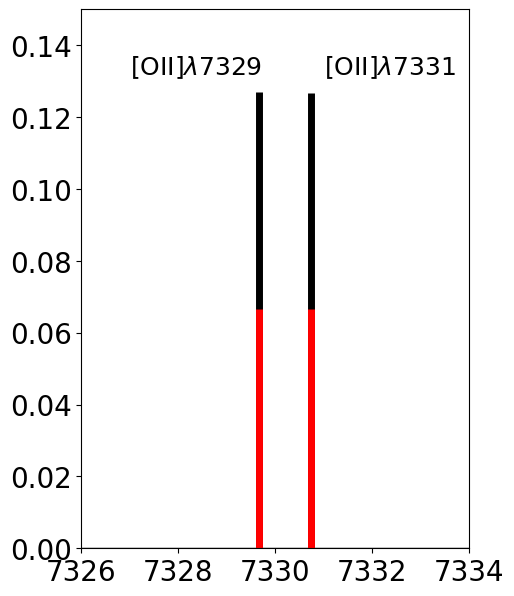}}
\hfill 
\subfloat{\includegraphics[trim=0 0 0 0, clip, width=0.17 \textwidth]{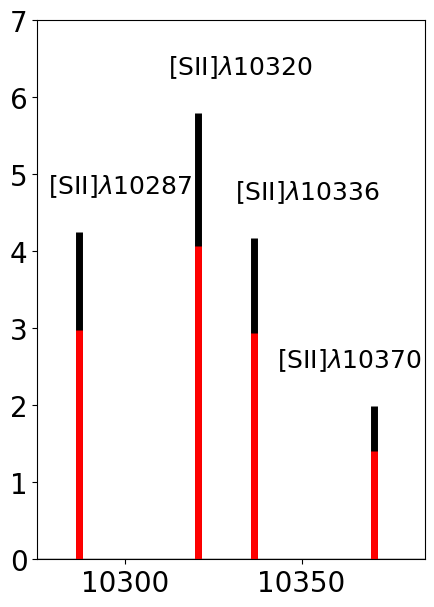}}
\hfill
\subfloat{\includegraphics[trim=0 0 0 0, clip, width=0.2 \textwidth]{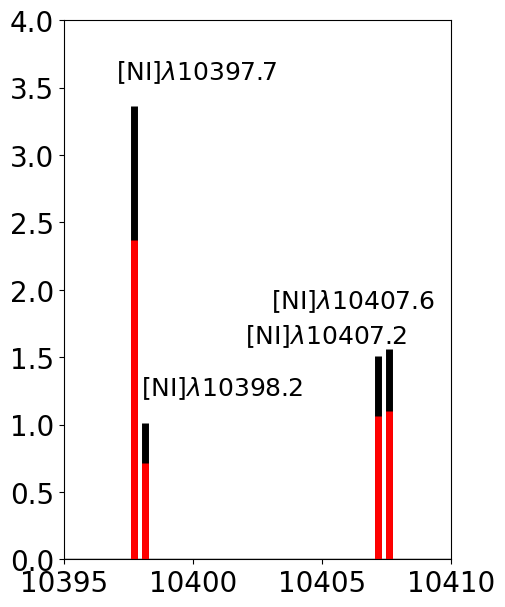}}
\hfill  \\
\subfloat{\includegraphics[trim=0 96 0 165, clip, width=0.99 \textwidth]{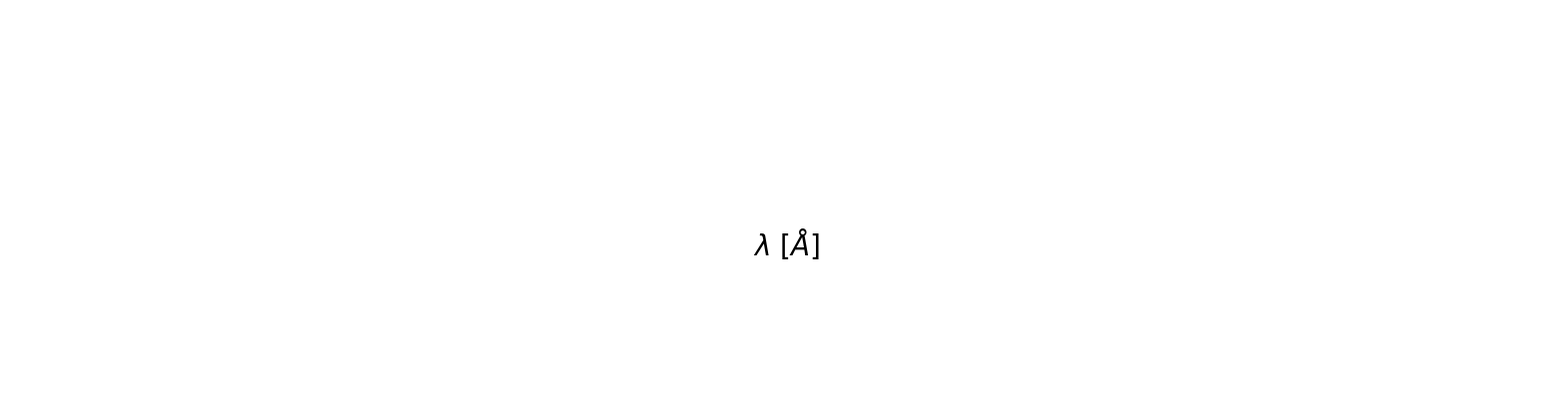}}
\hfill  
\caption{\small{Predicted BE99e  emission lines. Black vertical lines are the synthetic line emissivities normalised to the [O\,I]$\lambda$6300 line. Red vertical lines are the reddened line emissivities assuming the extinction law of \citet{cardelli_1989} with $A_V = 1.0$ and $R_V =3.1$. The equilibrium conditions are $T_\text{e} = 10\,500\,\text{K}$, $n_\text{e}^\text{eq} = 672.65\,\text{cm}^{-3}$, $x_\text{e}^\text{eq} = 0.0336$.}}\label{fig:be_extension}
\end{figure*} 

 \begin{figure}[h!] 
\resizebox{\hsize}{!}{\includegraphics[trim=65 540 300 50, clip, width=0.9\textwidth]{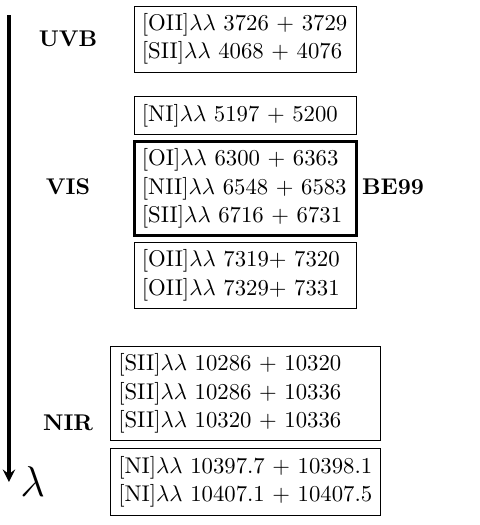}}
\caption{\small{Possible combinations of lines in the BE99e method.  \textbf{Note:} In the $(T_\text{e}, x_\text{e})$-plane the choice of the numerator/denominator for a given line ratio is irrelevant, e.g. OI/NII and NII/OI represent the same stripe.}}\label{fig:be_lines_boxes} 
\end{figure}

\section{The BE99 method with more line ratios (BE99e)}

In the classical BE99 method three line ratios are involved: OI/NII, SII/NII, and OI/SII. All six associated emission lines are located close to each other in the visible part of the spectrum and thus extinction correction is not crucial. In addition, the three stripes of line ratios most probably always meet in one common region in the $(x_\text{e}, T_\text{e})$-space when applied to observations of protostellar outflows. From the BE99 diagram it is not possible to see if the equilibrium is reached or if the extinction determination is correct or the BE99 assumptions are sufficiently fulfilled. However, other line ratios of the involved species in the BE99 ionisation equilibrium in the UVB and NIR part of the spectrum can in principle be used to extend the BE99 diagram in a natural way. 
Figure\,\ref{fig:be_extension} shows a synthetic emission line spectrum of the model in the equilibrium with the strongest emission lines of [O\,I], [N\,II], [S\,II], [N\,I], and [O\,II] in the spectral range of $  3\,500-11\,000\,\AA$. These lines are also comparably bright in protostellar outflows and thus can be included in the diagnostics. Other transitions such as the [N\,II]$\lambda$5754 line are typically too faint.  From the inspection of this spectrum many extensions of the BE99 diagram, hereby labelled as BE99e, are possible. We note that no additional assumption or equation has been invoked by transitioning from BE99 to BE99e. A selection of line ratios for constructing  BE99e  diagrams are depicted in Figure\,\ref{fig:be_lines_boxes}. Including any combination of the depicted line ratios in the BE99e diagram is straight forward and can give more information about the state of the gas.  Observers can in principle choose according to the dataset available which stripes to use for determining the ionisation fraction.  They do not necessarily have to rely on the six classic BE99 lines. \\
In order to demonstrate how the new line ratios can be exploited we present predictions of how the various new stripes in the $(x_\text{e}, T_\text{e})$ diagram look like in the time domain together with the influence of extinction for the model (Appendix C).  As examples we present four BE99e diagrams at five time steps ($10^4\,\text{s}$, $10^6\,\text{s}$, $10^8\,\text{s}$, $10^{10}\,\text{s}$, $10^{12}\,\text{s}$) and three extinction values ($A_V = 0.0, 0.3, 0.6$) in Figs.\,\ref{fig:all_BE_diagrams}-\ref{fig:all_BE_diagrams_D}. We note, that emission lines towards shorter wavelengths are more affected by extinction than emission lines in the near-infrared. In order to illustrate this effect we plot  in Figure\,\ref{fig:be_extension} the  extinction correction factor for $A_V=1.0$ and $R_V=3.1$. The intrinsic and observed line fluxes at wavelength $\lambda$ are denoted by $F_{\text{int}}(\lambda)$ and $F_{\text{obs}}(\lambda)$. We assume that  
 \begin{equation}
 F_{\text{obs}}(\lambda) = F_{\text{int}}(\lambda) \cdot 10^{-0.4 \, A_V\, c(\lambda, R_V)}, 
 \end{equation}\label{eq:extinction_law}
   with $c(\lambda, R_V)$ as normalised reddening curve and $A_V$ as visual extinction  \citep[e.g.][]{osterbrock_book}. From \cite{cardelli_1989} we adopt the expression for $c(\lambda, R_V)$.

\subsection{Determining the extinction using the  BE99e-diagram}

Applying the extended  BE99e method on observational data requires that the line emissivities are first corrected for extinction via e.g. the Balmer decrement, Paschen decrement, Millers method, or line ratios of iron (e.g. [Fe\,II]1.32\,$\upmu$m/1.63\,$\upmu$m). All the mentioned methods come with flaws and uncertainties, which the observer tries to minimise. If the gas is in equilibrium and the extinction is corrected for properly all stripes overlap in one region in the $(x_\text{e}, T_\text{e})$-space. 
However, the critical case occurs, when not all stripes are crossing in one region.  In such a situation one should try to find a visual extinction $A_{V}^\text{{BE}}$ in Eq.\,\ref{eq:extinction_law}, for which the stripes overlap.
If such an extinction value exists the gas is close enough to the equilibrium. This procedure should be repeated until a satisfactory $A_V^\text{true} = A_V + A_{V}^\text{{BE}}$ is found. Also the electron density and electron temperature can be treated in the same way. By changing them one can in principle check if the stripes will cross in one location.

\subsection{Insights from the extra stripes in the $(x_\text{e}, T_\text{e})$ diagram}

We briefly discuss the global behaviour of the new stripes in the $(x_\text{e}, T_\text{e})$ diagram depending on the non-equilibrium/equilibrium state and the extinction. Our main findings are:

\begin{itemize}
\item In the classic BE99 diagram all three stripes always meet in one location in the $(x_\text{e}, T_\text{e})$-space. From the diagram alone it cannot be seen if the equilibrium condition is fulfilled or if extinction is correctly determined. In our models, the classic BE99 method is robust in measuring the electron density and electron temperature, i.e. the influence of extinction and non-equilibrium conditions are marginal. 
\item A plethora of BE99e  diagrams can be constructed with the same underlying assumptions of the original BE99 paper. In contrast, these new diagrams are sensitive to the equilibrium state and the extinction. In the equilibrium state all stripes cross in one location in the $(x_\text{e}, T_\text{e})$-space measuring the true gas conditions, if the spectra are correctly dereddened. If the equilibrium is not reached or the spectra have not been properly dereddened the stripes spread over the $(x_\text{e}, T_\text{e})$-space not crossing in one location. Instead they potentially cross in many locations which are not representing the true gas conditions. Some stripes of line ratios are more sensitive to non-equilibrium or extinction influence than others. For example the ratios of OI/[S\,II]$\lambda\lambda$4068+4076, NII/[OII]$\lambda\lambda$3726+3729, OI/[N\,I]$\lambda\lambda$5198+5200 are very sensitive to both influences. Less sensitive line ratios are e.g. SII/[S\,II]$\lambda\lambda$4068+4076, SII/[N\,I]$\lambda\lambda$5198+5200, SII/[N\,I]$\lambda\lambda$10397+10398. 
\item With the additional  BE99e  diagrams a determination of  $x_\text{e}$ and $T_\text{e}$ is possible even if not all classical BE99 lines are available. Line ratios can be exchanged according to the lines available.
\item  Some emission line ratios represent almost the same stripes in the $(x_\text{e}, T_\text{e})$-space. For example OI/[N\,I]$\lambda\lambda$10397+10398 overlaps almost perfectly with OI/[N\,I]$\lambda\lambda$10407+10408.  
\end{itemize}

\section{Extension to higher ionisation fractions (BE99e+)}\label{sec:ionisation}

The BE99  and BE99e methods work for a low-ionisation  medium ($x_\text{e} < 0.3$). We can, however, extend it to higher ionisation degrees.  We indicate this extension by BE99e+.  In a high-excitation medium we expect to have several ionisation states of oxygen, nitrogen, sulphur, and hydrogen. We therefore include the following species  
 \begin{equation}
 Y^k =\left(\text{O}^0, \text{O}^+, \text{O}^{2+}, \text{N}^0, \text{N}^+,  \text{N}^{2+}, \text{S}^+, \text{S}^{2+},  \text{S}^{3+},  \text{H}^0, \text{H}^+ \right) .
 \end{equation}
\noindent
Additional to the   reactions  rates Eqs.\,\ref{eq:reactions_A}--\ref{eq:reactions_J}  there are  
\begin{align}
 \text{O}^+  +\text{e}^-  & \overset{C_{\text{O}^+}}{\longrightarrow}  \text{O}^{2+}  +2\text{e}^-   \hspace{1cm}   (\text{coll. ionisation}), \\
 \text{O}^{2+}  +\text{e}^-  & \overset{\alpha_{\text{O}^+}}{\longrightarrow}  \text{O}^+   \hspace{2cm}   (\text{recombination}),\\
\text{O}^+ +\text{H}^+ &  \underset{\delta_{\text{O}^+}'}{\stackrel{\delta_{\text{O}^+}}{\rightleftharpoons}}  \text{O}^{2+} +\text{H}^0  \hspace{1cm}   (\text{charge exchange})  ,\\
 \text{N}^+  +\text{e}^-  & \overset{C_{\text{N}^+}}{\longrightarrow}  \text{N}^{2+}  +2\text{e}^-   \hspace{1cm}   (\text{coll. ionisation}), \\
 \text{N}^{2+}  +\text{e}^-  & \overset{\alpha_{\text{N}^+}}{\longrightarrow}  \text{N}^+   \hspace{2cm}   (\text{recombination}),\\
\text{N}^+ +\text{H}^+ &  \underset{\delta_{\text{N}^+}'}{\stackrel{\delta_{\text{N}^+}}{\rightleftharpoons}}  \text{N}^{2+} +\text{H}^0 \hspace{1cm}   (\text{charge exchange}) , \\
 \text{S}^+  +\text{e}^-  & \overset{C_{\text{S}^+}}{\longrightarrow}  \text{S}^{2+}  +2\text{e}^-   \hspace{1cm}   (\text{coll. ionisation}), \\
 \text{S}^{2+}  +\text{e}^-  & \overset{\alpha_{\text{S}^+}}{\longrightarrow}  \text{S}^+  \hspace{2cm}   (\text{recombination}) ,\\
\text{S}^{2+}  +\text{e}^-  & \overset{C_{\text{S}^{2+}}}{\longrightarrow}  \text{S}^{3+}  +2\text{e}^-  \hspace{1cm}   (\text{coll. ionisation}) , \\
 \text{S}^{3+}  +\text{e}^-  & \overset{\alpha_{\text{S}^{2+}}}{\longrightarrow}  \text{S}^{2+} \hspace{2cm}   (\text{recombination})  ,\\
 \text{S}^+ +\text{H}^+ &  \underset{\delta_{\text{S}^+}'}{\stackrel{\delta_{\text{S}^+}}{\rightleftharpoons}}  \text{S}^{2+} +\text{H}^0 \hspace{1cm}   (\text{charge exchange}),\\
\text{S}^{2+} +\text{H}^+  &  \underset{\delta_{\text{S}^{2+}}'}{\stackrel{\delta_{\text{S}^{2+}}}{\rightleftharpoons}} \text{S}^{3+} +\text{H}^0 \hspace{1cm}   (\text{charge exchange}).
\end{align}
From these equations it is straight forward to derive a set of ordinary differential equations describing the time dependent reaction network (Appendix\,A).  \\
By invoking the equilibrium condition (Eqs.\,\ref{equ:equilibrium_ode}) to the full system of ODEs, we can derive the new ionisation balance in the equilibrium for BE99e+. They are
\begin{align}
 \frac{\text{O}^{+}}{ \text{O}^0}  &  = \frac{\left( \delta_\text{O}    +  C_\text{O}  \right)\cdot x_e   }{ \left(\alpha_\text{O}      -  \delta^{'}_\text{O}  \right)  x_e     + \delta^{'}_\text{O}  } ,  \\
 \frac{\text{O}^{2+}}{ \text{O}^+}  &  = \frac{\left( \delta_{\text{O}^+}    +  C_{\text{O}^+}  \right)\cdot x_e   }{ \left(\alpha_{\text{O}^+}      -  \delta^{'}_{\text{O}^+}  \right)  x_e     + \delta^{'}_{\text{O}^+}  }  , \\
 \frac{\text{N}^{+} }{\text{N}^0}  & = \frac{\left( \delta_\text{N}    +  C_\text{N}  \right)\cdot x_e   }{ \left(\alpha_\text{N}      -  \delta^{'}_\text{N}  \right)  x_e     + \delta^{'}_\text{N}  }   ,\\
  \frac{\text{N}^{2+}}{ \text{N}^+}  &  = \frac{\left( \delta_{\text{N}^+}    +  C_{\text{N}^+}  \right)\cdot x_e   }{ \left(\alpha_{\text{N}^+}      -  \delta^{'}_{\text{N}^+}  \right)  x_e     + \delta^{'}_{\text{N}^+}  }  , \\
  \frac{\text{S}^{2+}}{ \text{S}^+}  &  = \frac{\left( \delta_{\text{S}^+}    +  C_{\text{S}^+}  \right)\cdot x_e   }{ \left(\alpha_{\text{S}^+}      -  \delta^{'}_{\text{S}^+}  \right)  x_e     + \delta^{'}_{\text{S}^+}  }  , \\
 \frac{\text{S}^{3+}}{ \text{S}^+}  &  = \frac{\left( \delta_{\text{S}^{2+}}    +  C_{\text{S}^{2+}}  \right)\cdot x_e   }{ \left(\alpha_{\text{S}^{2+}}      -  \delta^{'}_{\text{S}^{2+}}  \right)  x_e     + \delta^{'}_{\text{S}^{2+}}  }   .
\end{align} 
With the additional species of O$^{2+}$ and S$^{2+}$ we can include several new line ratios such as $\text{SIII} = \text{[S\,III]}\lambda9068 + \text{[S\,III]}\lambda9530$ or $\text{OIII} = \text{[O\,III]}\lambda4958+ \text{[O\,III]}\lambda5007$.
With that plenty new line ratios in the BE99 diagram can be added straight forwardly including line ratios such as OI/SIII or NII/OIII.  
The presented equations for the reaction network in the equilibrium represent an extended excitation model such as the one described by \citet[e.g.][]{giannini_2015}.

\section{Extension including other species (BE99e++)}\label{sec:other_species}
 
 In principle, the described method may be extended to include species other than oxygen, nitrogen, and sulphur. In order to determine the ionisation fraction at least two ionisation states of the same species have to be included, that is for example Fe$^+$/Fe$^{2+}$ or Ne$^+$/Ne$^{2+}$. The ionisation balance $Y^{+i}/Y^{+i+1}$ can be found by writing down all relevant physical reactions as described in Section\,\ref{sec:ionisation} and solve the reaction network for the equilibrium. The main challenge here may be to identify  the relevant reactions.  The magnitude of the reaction rates, however, indicate the dominant reactions. The ionisation balance $Y^{+i}/Y^{+i+1}$ is ideally described as a function of $x_\text{e}$ and $T_\text{e}$, i.e.
 \begin{equation}
Y^{+i}/Y^{+i+1}\bigg\rvert_\text{eq}  = f(x_\text{e},T_\text{e} )  .  
\end{equation}
Line ratios of $Y^{+i}$ and $Y^{+i+1}$ (BE99e++ method)  can be included as stripes in the $(x_\text{e}, T_\text{e})$-diagram straightforwardly.    

\subsection{Non-thermal contributions}

The outlined theory of extending the BE99 method in various directions is limited by the assumption that all lines are thermally excited. Other effects such as a) photodissociation of OH molecules in the disk surface layers \citep[e.g.][]{gorti_2011, rigliaco_2013}, b) photoionisation \citep[e.g.][]{bally_2001, riaz_2015, mendez_delgado_2021}, c) fluorescent excitation \citep{bautista_1999}, or d) pumping \citep{ferland_2012, nemer_2020} can contribute to the observed line emission and preclude the applicability of the BE99 method. Line ratios such as [S\,II]$\lambda$4068/[O\,I]$\lambda$6300 or [O\,I]$\lambda$6300/[O\,I]$\lambda$5577 can give further clues on the excitation mechanism. For example for the low-velocity component of protostellar outflows of T Tauri stars  \cite{natta_2014} and \citet{fang_2018} found that the observed line ratios are consistent with thermal excitation. For the high-velocity component typically shock models are used to explain the observed line ratios. However, each source is different and thus the assumption of thermal excitation must be justified as well.

\subsection{Other caveats and limitations}

Element abundances can substantially differ from solar abundances in protostellar outflows and jets due to, for example, grain sputtering or different evolutionary pathways of the observed material. Deviant element abundances change the position of the stripes in any BE99 diagram and therefore the determined gas parameters. Furthermore, collisions with other species such as hydrogen atoms are not yet included in the excitation model. The lack of atomic data prevents this so far. Line emissivities from the excitation model are expected to be affected to an unknown amount.\\
 Another limiting aspect is the critical density ($n_{\text{crit}}$) of the utilised FELs. Although its definition is not consistent in the literature, its meaning can be described as follows: For a given energy level and temperature the critical density refers to the density of the collision partners, at which the de-excitation rate is equal to the radiative decay rate from that level. In the high-density limit (here: $n_e \gg n_{\text{crit}}$) that implies that the line luminosity per atom or ion coming from that level does not increase with the density of the collision partners anymore, that is it is not sensitive to density variations. The frequent collisional excitation and de-excitation essentially set up a Boltzmann population and the emerging emission lines are quenched. A common misconception is that these quenched lines disappear, which they in fact do not. However, line ratios, where quenched lines are involved, may not be used as diagnostics. 
The utilised emission lines have distinct critical densities $n_\text{crit}$ in the range of $n_\text{crit} = 10^3-10^8\,\text{cm}^{-3}$ \citep[see Table\,3 in][]{giannini_2019}. In turn, line ratios can be affected by quenching or are just not sensitive to certain gas conditions. For instance, the optical sulphur lines, which are used for determining the electron densities, are only sensitive around $n_e\sim 10^{4}\,\text{cm}^{-3}$. In the high-density limit these lines are quenched and not sensitive to the density anymore. In the BE99 method, this line ratio can in principle be substituted by the [O\,II]$\lambda\lambda$3726/3729 line ratio. However, also the blue [O\,II] lines have similar critical densities as compared to the optical sulphur lines. In turn, regions with high electron densities, for example close to the driving source, may not be accessible able to  any BE99 method.

\section{Applications of the extended BE99 methods}

In the following Section we apply the excitation model described in Section\,\ref{sec:ionisation} to observational data of two targets with very distinct excitation conditions. 


\subsection{Application to the (high excitation) Proplyd 244-440}

\begin{figure}[h!] 
\resizebox{\hsize}{!}{\includegraphics[trim=0 0 0 0, clip, width=0.9\textwidth]{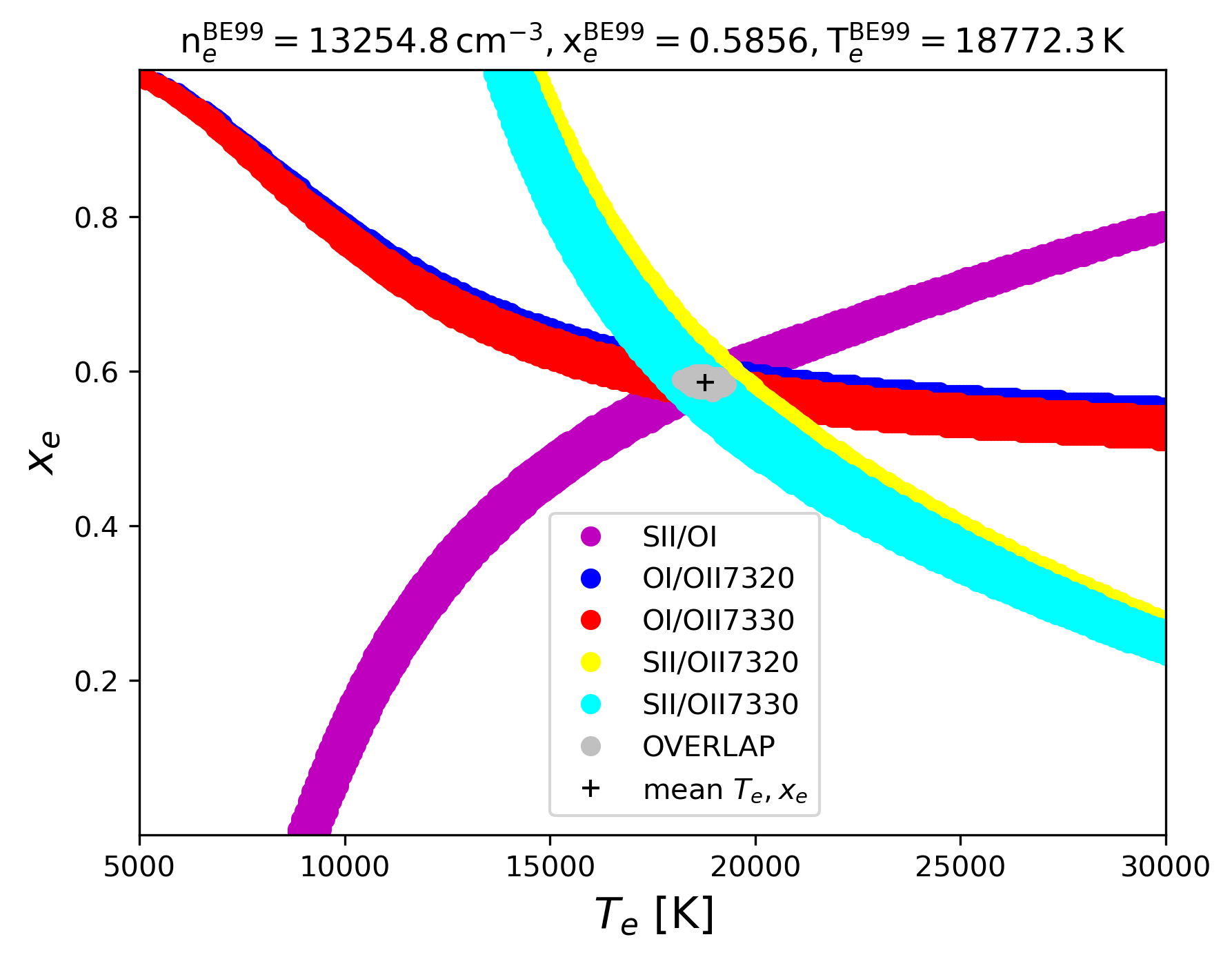}}
\caption{\small{An application of the additional line ratios (BE99e+) to determine the gas parameters of the outflow region associated with knot E3 of 244-440 Orion proplyd \citep{kirwan_2023}. }}\label{fig:be_application_kirwan} 
\end{figure}

In order to demonstrate the applicability of additional stripes in the $(x_\text{e}, T_\text{e})$-diagram we take a look at recently published MUSE data of the outflow associated with the 244-440 Orion proplyd in \citet{kirwan_2023}. The determination of the ionisation fraction via the BE99 method was not possible as stated therein. No [N\,II] line fluxes are specified in the paper (see Table\,1 therein) indicating that they are either not detected or other unspecified problems prevent their usage. However, other line fluxes of oxygen and sulphur have been measured making a determination of the ionisation fraction possible with the outlined extension of the  BE99e+  diagrams. For example, the extracted dereddened line fluxes of [O\,I]$\lambda\lambda$6300+6363, [S\,II]$\lambda\lambda$6716+6731, [O\,II]$\lambda$7320, and [O\,II]$\lambda$7330 at knot E3 of Proplyd 244-440 are stated in Table\,1 therein. Since no line errors are specified we assumed that the corresponding errors are $2\,\%$. Higher/lower errors would thicken/thin down the stripes in the BE99e+  diagram. Even without the [N\,II] lines at hand it is now possible to generate a BE99e+ diagram based on the line ratios of OI/SII, OI/OII, SII/OII. We present the BE99e+  diagram based on the stated line ratios in Fig.\,\ref{fig:be_application_kirwan}. From Figure\,\ref{fig:be_application_kirwan} we see that: The line ratios do cross in a common region in the $(x_\text{e}, T_\text{e})$-space as predicted by our model. We measure a electron temperature of $T_\text{e} = 18\,800\,\text{K}\pm 1\,000\,\text{K}$ and an ionisation fraction of about $x_\text{e} = 0.58\pm 0.05$. This indicates that this region is in a high-excitation state, which is consistent with the expected gas conditions.

\subsection{Application to the (low excitation) Par Lup 3-4 outflow}

Par Lup 3-4 ($16^\text{h}$\,$08^\text{m}$\,$51.44^\text{s}$, $-39^\text{d}$\,$05^\text{m}$\,$30.5^\text{s}$) is a well studied protostellar outflow of the Lupus\,III cloud. A detailed source description can be found in  \citet{bacciotti_2011}. Its molecular outflow was investigated in \citet{miranda_2020}. Observations with the medium resolution slit spectrograph X-Shooter revealed that the protostellar outflow associated with Par Lup 3-4 (HH\,600) displays low excitation \citep{whelan_2014}. 
 
\subsubsection{Observations and data reduction}

In our analysis we make use of the archival X-Shooter observations of Par Lup 3-4 (programme ID: 085.C-0238(A), date of observations: April 7 2010). X-Shooter has three arms   covering a spectral range of about $3\,000-5\,590\,\AA$ (UVB), $5\,590-10\,240\,\AA$ (VIS), and $10\,240-24\,800\,\AA$ (NIR). In the case of Par Lup 3-4 the slit was aligned with the outflows P.A. of about $130^\text{o}$. The slit widths were $1\farcs 0$, $0\farcs 9$, and $0\farcs 9$ implying a resolving power 5\,100 ($58\,\text{km}\,\text{s}^{-1}$), 8\,800 ($34\,\text{km}\,\text{s}^{-1}$), and 5\,600 ($53\,\text{km}\,\text{s}^{-1}$) for the UVB, VIS, and NIR arm, respectively. The seeing during the observations was about $0\farcs 9-1\farcs 2$. Exposure times were $4\times 900\,\text{s}$ in each arm.  The X-Shooter spectra of Par Lup 3-4 were analysed  in \citet{bacciotti_2011}, \citet{giannini_2013}, and \citet{whelan_2014} -- hereafter B11, G13, and W14, respectively. \\
We re-reduced  the archival X-Shooter data for Par Lup 3-4 using the up-to-date Esoreflex environment \citep[v.\,3.6.1, ][]{esoreflex_2013} to obtain the 1D and 2D spectra. The data reduction includes bias subtraction, flat
fielding, wavelength calibration, and optimal spectrum extraction. We checked the 1D and 2D spectra for consistent relative flux calibration in the UVB, VIS, and NIR arms -- no relative flux adjustment was necessary as in the data reduction of W14. The pixel scales in the 2D data are $0\farcs 16\,\text{pixel}^{-1}$ in the UVB and VIS arms, and $0\farcs 21\,\text{pixel}^{-1}$ in the NIR arm. The spectral samplings are $0.2\,\AA$ in the UVB and VIS arms, and $0.6\,\AA$ in the NIR arm. Cosmics were removed with the Lacosmic routine implemented in Python \citep{van_dokkum_2001}. No photospheric correction was needed as the photospheric features are marginally present.  We adopt the systemic velocity of $v_{\text{LSR}} = 5.5\,\text{km}\,\text{s}^{-1}$ stated in B11. 
We extracted the position-velocity (PV) diagrams for the relevant emission lines by fitting and removing a 2D Gaussian around the emission line region. We present the PV diagrams in Figures\,\ref{fig:all_pv_diagrams_boxes}. In total $26$ line regions were inspected for our analysis. These include the six optical BE99 lines plus 20 further line regions lying in the spectral range XShooter.

\subsubsection{Extracting line fluxes}\label{sec:par_lup_line_fluxes}

\begin{figure}[h!] 
\resizebox{\hsize}{!}{\includegraphics[trim=0 0 0 0, clip, width=0.7\textwidth]{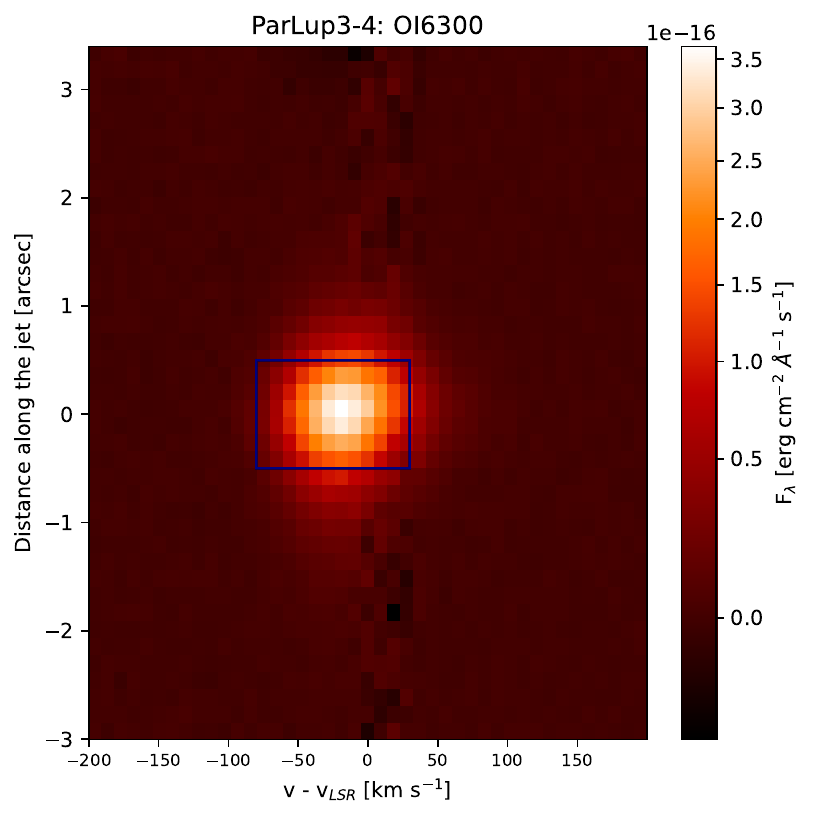}}
\caption{\small{The position-velocity diagrams for the [O\,I]$\lambda$6300 line of Par Lup 3-4.Line fluxes have been extracted within the sketched box.}}\label{fig:Par-lup_OI6300_example} 
\end{figure}

We are interested in the physical conditions of the outflow close to the driving source, that is the innermost one arcsec region. This region displays a clear detection in most of the BE99 lines. Following W14 and G13 we restrict our analysis to the region of $-0\farcs 5$ to  $+0\farcs 5$ with respect to the position of the driving source.  We decided to extract flux values in this region confined spectrally by $-80\,\text{km}\,\text{s}^{-1}$ and $ +30\,\text{km}\,\text{s}^{-1}$. Initially, we extracted fluxes in five consecutive pixel rows in spatial direction at the driving source. However, we noticed no substantial change in the relative line fluxes. \\
 Figure\,\ref{fig:Par-lup_OI6300_example} displays the flux box for the [O\,I]$\lambda$6300 line of Par Lup 3-4 as an example for the flux extraction. Since we only need relative line fluxes for our analysis we normalise all extracted fluxes  to the line flux of the bright [O\,I]$\lambda$6300 line, i.e. we set its flux to $F(\text{[O\,I]}\lambda 6300) = 100$. The extracted normalised flux values with their 1$\sigma$-errors or upper limits  are stated in Table\,\ref{table:line_fluxes}.  \\
For our line diagnostics we adopt $A_V = 0.0$, which was derived from the analysis of the near-infrared iron lines in the Par Lup 3-4 outflow (G13). We did not utilise line ratios of e.g. [S\,II]$\lambda$4068/[S\,II]$\lambda$10320, [S\,II]$\lambda$4076/[S\,II]$\lambda$10336 to derive the extinction \citep{miller_1968}, since no extinction value is consistent with the observed Miller line ratios. This is puzzling and implies that either some of the [S\,II] emission in the UVB arm is overluminous or the [S\,II] emission in the NIR arm is underluminous.

\begin{figure*} 
\centering
\subfloat[]{\includegraphics[trim=0 0 0 0, clip, width=0.3 \textwidth]{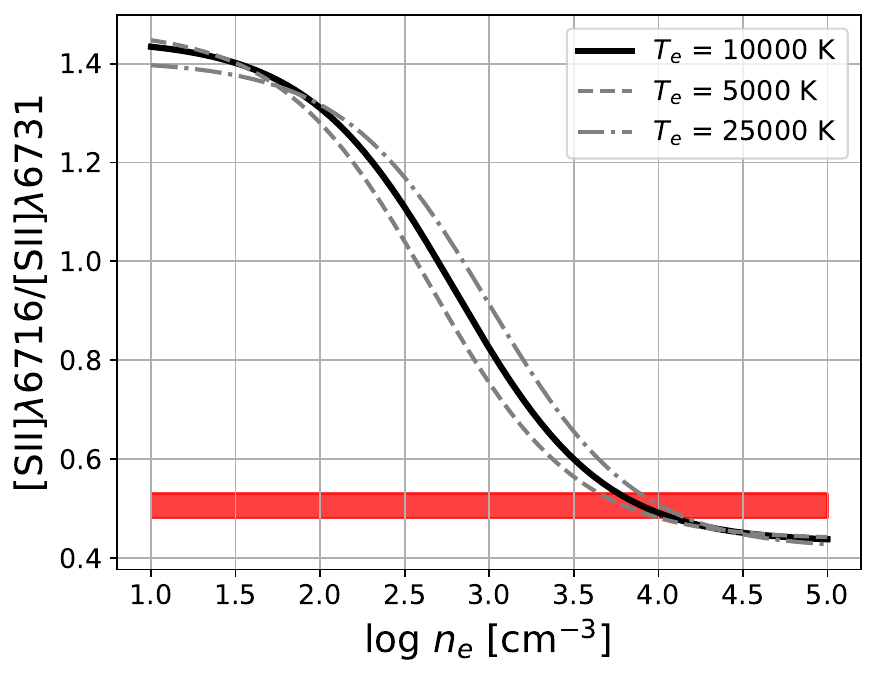}}
\hfill 
\subfloat[]{\includegraphics[trim=0 0 0 0, clip, width=0.3 \textwidth]{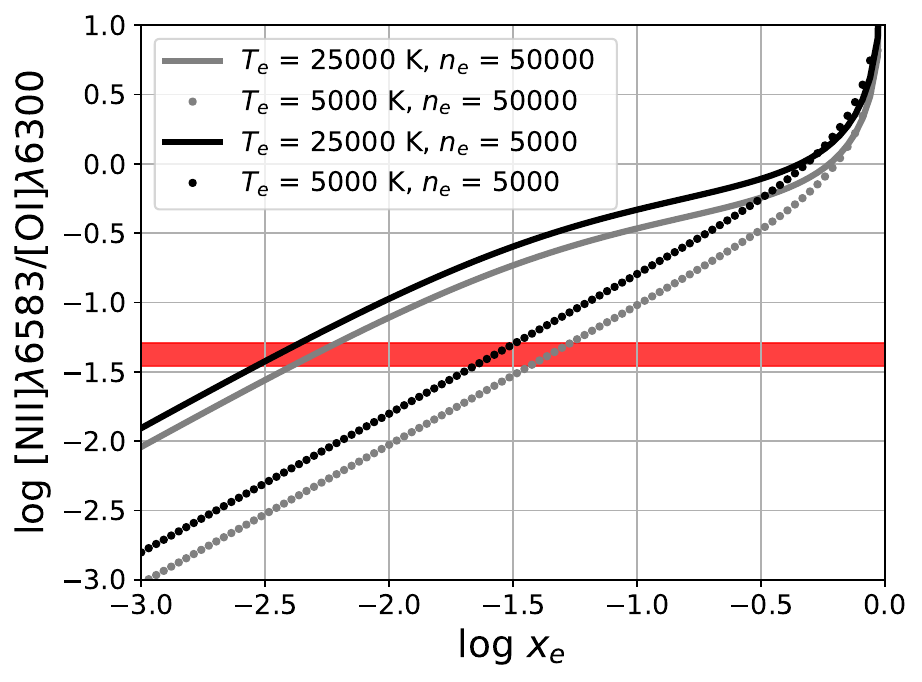}}
\hfill  
\subfloat[]{\includegraphics[trim=0 0 0 0, clip, width=0.3 \textwidth]{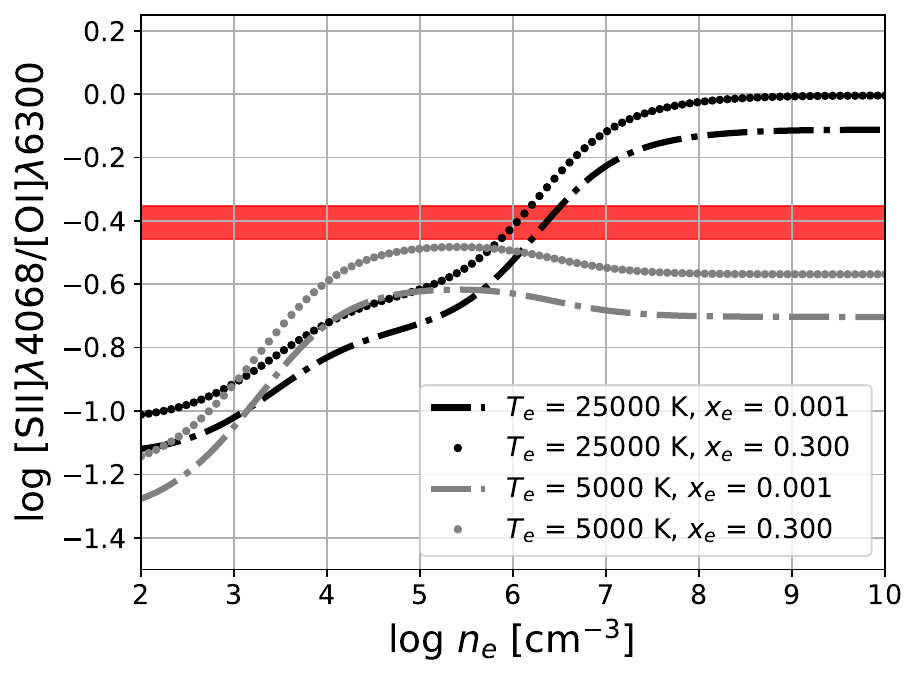}}
\hfill 
\caption{\small{Diagnostic diagrams for the observed line fluxes of Par Lup 3-4. The red horizontal stripes indicate the measured line flux ratios including their $1\sigma$ error margins. The black and grey curves are theoretical line ratios derived from our underlying excitation model. They represent reproductions of Fig.\,A.2. in \citet{giannini_2019}.}}\label{fig:all_diagnostics}
\end{figure*} 

\begin{figure*} 
\centering
\subfloat{\includegraphics[trim=0 0 0 0, clip, width=0.9 \textwidth]{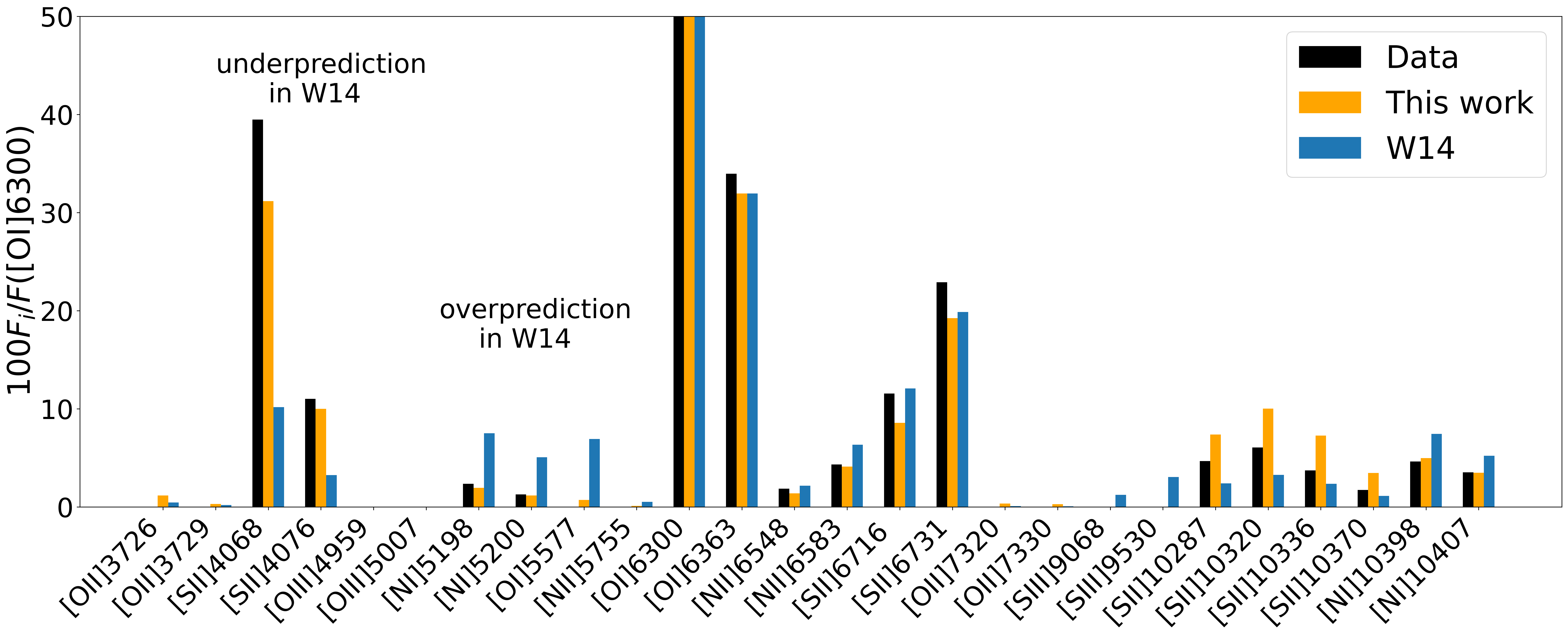}}
\hfill  
\caption{\small{A comparison of the observed emission line ratios (black bars) with the best model (yellow bars) found by the multi-line analysis and the W14 model found via the classical BE99 method (blue bars). [O\,I]$\lambda$6300 is reproduced by both models, the plot therefore zooms in on the weaker lines. We see an overall agreement of the observed data with the two models -- especially with the optical BE99 lines. The [O\,I]$\lambda$5577, [N\,I]$\lambda$5198, and [N\,I]$\lambda$5200 emission lines, however, are substantially overpredicted and the [S\,II]$\lambda$4068 and[S\,II]$\lambda$4076 lines are heavily underpredicted in W14. The near-infrared lines of sulphur and nitrogen are well reproduced.}}\label{fig:all_best_model_fits_compar}
\end{figure*} 

\subsubsection{NIR line fluxes in W14}\label{sec:nir_line_fluxes}

For a consistency check we also list the W14 flux values in Table\,\ref{table:line_fluxes}. Comparing our extracted relative flux values with the ones presented in W14 we see a very good agreement within the errorbars for the VIS and UVB arms. However, the NIR flux values in W14 are systematically too high by a factor of four as compared to our values.  W14 investigated two outflows, namely Par Lup 3-4 and ESO H$\alpha$ 574. In their data reduction for ESO H$\alpha$ 574 W14 found that the NIR arm fluxes are too low at about $25\,\%$ when comparing the continuum in the 1D spectra of the VIS and UVB arms. For Par Lup 3-4 this was not the case, which is in full agreement with our analysis. We suspect that W14 nonetheless applied the correction factor found for ESO H$\alpha$ 574 also on the NIR fluxes of Par\,Lup\,3-4, which fully explains the stated inconsistency. The usage of different pipeline versions of Esoreflex can explain small differences in the relative line fluxes as displayed in Table\,\ref{fig:all_pv_diagrams_boxes}, but not the systematically too high NIR fluxes in W14.  

\subsubsection{Time scale of the outflow}\label{sec:time_scale_parlup}
 
We can estimate the dynamical scale and hydrogen recombination time  of one of the two outflow lobes of Par Lup 3-4. (We do not know the equilibrium time scale for the outflow, but as our model calculations show this time scale is not decisive.) The distance to Par Lup 3-4 is $D\sim 155\,\text{pc}$ \citep{miranda_2020}. We assume a jet velocity of $v_t = 170\,\text{km}\,\text{s}^{-1}$ and a jet length of $0\farcs 5$ (W14). With that we get a dynamical time scale of about 2\,yrs. The hydrogen recombination time (Eq.\,\ref{eq:recombination_time}) is about $\tau_\text{rec} \sim 1-10\,\text{yrs}$ assuming $n_e \sim 10^4-10^5\,\text{cm}^{-3}$.
 Thus the gas of the Par Lup 3-4 outflows (both lobes assuming symmetry) should have had enough time to  stay close enough to the equilibrium. 

\subsubsection{Physical conditions of Par Lup 3-4}

In order to investigate the physical conditions in the Par Lup 3-4 outflow we first check  if thermal excitation can explain the observed line ratios \citep[e.g.][]{osterbrock_book, giannini_2015, giannini_2019, nisini_2024}. Figures\,\ref{fig:all_diagnostics}a-c show three exemplary diagnostic diagrams as a function of $T_e$, $n_e$, and $x_e$ for the observed line ratios of [S\,II]$\lambda$6716/[S\,II]$\lambda$6731, [S\,II]$\lambda$4068/[O\,I]$\lambda$6300, and [N\,II]$\lambda$6583/[O\,I]$\lambda$6300. 
The observed line ratio of [S\,II]$\lambda$6716/[S\,II]$\lambda$6731 $\sim 0.5$ is close to the high density limit indicating high electron densities, that is $n_e \gtrsim 5\times 10^3\,\text{cm}^{-3}$. Such high electron densities are indeed expected to be measured close to the driving source of the outflow \citep{fang_2018}. The ionisation fraction can be constrained from the [N\,II]$\lambda$6583/[O\,I]$\lambda$6300 line ratio. For high electron densities the observed [N\,II]$\lambda$6583/[O\,I]$\lambda$6300 line ratio indicates $x_e \lesssim 0.03$ for $T_e = 5\,000-25\,000\,\text{K}$. The observed [S\,II]$\lambda$4068/[O\,I]$\lambda$6300 line ratio is consistent with these gas parameters. In conclusion, all observed line ratios are consistent with the thermal excitation. \\
W14 determined the gas parameters $n_e$, $T_e$, and $x_e$ via the classical BE99 method. The stated values in Table\,1 therein are: $n_e=4\,300\pm 200\,\text{cm}^{-3}$, $T_e=31\,000\pm 500\,\text{K}$, and $x_e=0.003\,\pm 0.0002$ for the blue lobe and $n_e=4\,600\pm 300\,\text{cm}^{-3}$, $T_e=24\,000\pm 600\,\text{K}$, and $x_e=0.0063\,\pm 0.0002$ for the red lobe. These values are consistent with the values that we derive for the full flux box when applying the classical BE99 method. When adding additional line ratios (BE99e or BE99e+), such as OI/[N\,I]$\lambda\lambda$5198+5200, SII/[N\,I]$\lambda\lambda$5198+5200, and NII/[N\,I]$\lambda\lambda$5198+5200, however, not all stripes cross in one location in the $(x_e, T_e)$-diagram. We extensively tried to bring the stripes together by applying different extinction values, but they never cross in one location. Thus, the gas parameters derived from the classical BE99 method are not in good agreement with the observations and any BE99 method fails.

\subsubsection{Fitting line ratios to excitation model}\label{sec:excitation_model_par_lup34}

Apart from the diagnostic diagrams and the BE99e and BE99e+ methods we could take the whole spectrum and fit an excitation model to it, that is we utilise all observed lines at once and not separately. For such a multi-line approach with $n$ observed line fluxes ($F^\text{obs}_i$ with $i = 1, \dots , n$) following \citet{hartigan_2007} we calculated an extensive grid of excitation models with the three parameters $n_e$, $T_e$, and $x_e$. Our model grid spans:  $n_e = 2\,000\,\text{cm}^{-3} - 200\,000\,\text{cm}^{-3}$ spaced linearly in $2\,000$ grid points ($\Delta n_e = 100\,\text{cm}^{-3}$), $T_e = 5\,000\,\text{K}-30\,000\,\text{K}$ spaced linearly in $250$ grid points ($\Delta T_e = 100\,\text{K}$), and $x_e = 0.001-0.200$ spaced linearly in $200$ grid points ($\Delta x_e = 0.001$). In total $2\,000\times 250\times 200 = 10^8$ models with all relevant synthetic line fluxes ($F^\text{model}_i$) are calculated to find the best fit to the observed line fluxes. The best model in our model grid is then found by minimising the residuals defined by 
\begin{equation}
C = \sum_{k=1}^{n(n-1)/2} \left(\frac{r_k - m_k}{\sigma_k}\right)^2,
\end{equation}
with $r_k = \log \left(F^\text{obs}_i/F^\text{obs}_j\right)$, $m_k = \log \left(F^\text{model}_i/F^\text{model}_j\right)$, and $\sigma_k$ as propagated line ratio errors ($i,j = 1,\dots , n$ with $i < j$).  We also tested  non-zero extinction values to check for better fits, but the residuals were always better for $A_V = 0.0$.\\
We compare the observed line fluxes to the modelled line fluxes in Figure\,\ref{fig:all_best_model_fits_compar}. It is remarkable how good the excitation model fits to the Par Lup 3-4 data. In fact, all observed lines are consistent with the best excitation model within 10\,$\%$, and all the non-detections are consistent with the model as well. The multi-line approach indicates the following excitation condition in the Par Lup 3-4 outflow: $n_e \sim 45\,000\,\text{cm}^{-3}-53\,000\,\text{cm}^{-3}$, $T_e = 7\,600\,\text{K}-8\,000\,\text{K}$, and $x_e \sim 0.027-0.036$. These values substantially differ from the ones derived from the classical BE99 method. The electron density is about 6 times higher than the one derived from BE99. The electron temperature is about three times lower than from BE99 and the ionisation fraction is about four times higher than from BE99. We think that the multi-line approach by \citet{hartigan_2007} best describes the observations of Par Lup 3-4 due to the following reasons: 1. The [O\,I]$\lambda$5577 line is not detected towards Par Lup 3-4 but should be, if the classical BE99 conditions prevail, 2. The [S\,II]$\lambda$4068 are substantially underpredicted by the BE99 model. The BE99 method may not give proper results for the Par Lup 3-4 outflows because of the high electron densities at the driving source (quenching effects).  

 \section{Conclusions}

We have numerically integrated a set of ionisation reactions in the time domain which represent the BE99 reaction network.   We used this model to generate synthetic emission line spectra of the atomic/ionic species that are involved in the BE99 equilibrium (NI, NII, SII, OI, OII). From these spectra we were able to study how the system evolves from a non-equilibrium to the equilibrium state. Our analysis demonstrates that the equilibrium of the BE99 reactions is reached on time-scales shorter than the hydrogen recombination time. This, however, is not critical for the application of the original BE99 method, since the BE99 method converges faster than the hydrogen recombination time.  \\
We propose a natural extension of the BE99 method, that is based on the same assumptions as in the original BE99 paper (BE99e), by including many more line ratios as stripes in the $(x_\text{e}, T_\text{e})$-diagram in the spectral range of about $3\,700-10\,400\,\AA$. If the BE99 conditions are fulfilled all the new stripes in the $(x_\text{e}, T_\text{e})$-space meet in one region indicating the true gas parameters. If, however, the system is out of the equilibrium, or extinction has not been corrected properly, or the determined gas parameters are not correct the stripes do not overlap in one but in many regions. Furthermore we put forward an extension of the BE99 method towards higher ionisation fractions ($x_\text{e}>0.3$, BE99e+). Additional line ratios including lines of for instance [S\,III] and [O\,III] can be exploited. Other species such as Fe$^+$/Fe$^{2+}$ or Ne$^+$/Ne$^{2+}$ can in principle be included straightforwardly.   \\
We tested the outlined ideas on two examples of distinct targets, that is the high-excitation Proplyd 244-440 and the low-excitation outflow Par Lup 3-4. For Proplyd 244-440 we were able to derive the ionisation fraction even without the [N\,II]$\lambda\lambda$6548+6583 lines. Line ratios of [S\,II]$\lambda\lambda$6716+6731 and [O\,I]$\lambda\lambda$6300+6363 with [O\,II]$\lambda$7320 or [O\,II]$\lambda$7330 can be used as substitutes. \\
We extensively analysed the emission lines of the protostellar outflow Par Lup 3-4. When including the NI = [N\,I]$\lambda$5198+[N\,I]$\lambda$5200 line ratio in the BE99e+  diagram we see that not all stripes overlap in one region in the $(x_\text{e}, T_\text{e})$. Our analysis shows that neither extinction nor non-equilibrium effects are likely responsible for that notion. A comparison with a multi-line approach revealed that the most probable explanation is that any  BE99 method underestimates the electron density due to quenching of the optical sulphur lines  in this case. The application of the BE99 method close to the driving source, where densities are likely very high, that is on the order of the critical density of the sulphur lines, has to be checked carefully by e.g. inspecting other line ratios. 

\begin{acknowledgements}  
The authors thank the anonymous referee for helping to improve the quality of the paper. This work has
been supported by the Deutsche Forschungsgemeinschaft (DFG) in the framework of the YTTHACA Project (Young stars at Tübingen, Tautenburg, Hamburg
\& ESO – A Coordinated Analysis) under the programme ID EI 409/20-1.
\end{acknowledgements}

\bibliographystyle{aa} 
\bibliography{papers}  

\appendix

\setcounter{table}{0}
\renewcommand{\thetable}{A.\arabic{table}}
  
\setcounter{figure}{0}
\renewcommand\thefigure{\thesection A.\arabic{figure}}  

\setcounter{equation}{0}
\renewcommand{\theequation}{A.\arabic{equation}}  

\section*{Appendix A - Reaction rates and atomic data}\label{appendix:A}
   
\subsection*{Charge transfer rates}\label{appendix:Charge transfer rates}

We use the charge transfer rates $\delta_\text{N}'$, $\delta_{\text{N}^+}'$, $\delta_{\text{S}^+}'$, and $\delta_{\text{S}^{2+}}'$ stated in \citet{kingdon_1996}. 
Since the stated rate for oxygen in therein  is only  valid for temperatures of $T_\text{e}= 10-10\,000\,\text{K}$ we use a simple power law fit to the data of \citet{stancil_1999}. For $T_\text{e} = 500-50\,000\,\text{K}$ we find the best fit to be
\begin{equation}
\delta_\text{O}'(T_\text{e})  = 5.262\times 10^{-11}\,T_\text{e}[\text{K}]^{0.3986} \quad [\text{cm}^{3}\,\text{s}^{-1}].
\end{equation}
The charge transfer rates $\delta_{Y^k}'$ and $\delta_{Y^k}$     
are connected by the Milne relation \citep[e.g.][]{osterbrock_book}. For example for oxygen the charge transfer reaction is 
 \begin{equation}
 \text{O}^0(^3\text{P}_{2,1,0}) +\text{H}^+    \underset{\delta_\text{O}'}{\stackrel{\delta_\text{O}}{\rightleftharpoons}}  \text{O}^+(^4\text{S}_{3/2}^0) +\text{H}^0(^2\text{S}_{1/2}) .
\end{equation}
Here, we assume that only ground state transitions as indicated in brackets in the above Eq.\,A.2 are relevant. The function $\delta_\text{O}(T_\text{e})$ can then be calculated via  
\begin{equation}
\frac{\delta_\text{O}(T_\text{e})}{\delta_\text{O}'(T_\text{e})} = \frac{g(^4\text{S}^\text{o}_{3/2})\times g(^4\text{S}^\text{o}_{1/2})}{g(^3\text{P}_0) + g(^3\text{P}_1) + g(^3\text{P}_2)}  \,  \text{exp}\left(-\frac{\Delta E}{ k_\text{B}T_\text{e}}\right),
\end{equation} 
where $\Delta E=\chi (\text{O}^0) - \chi (\text{H}^0)$ is the energy difference of the two involved ionisation potentials. In the same way we calculate $\delta_\text{N}$, $\delta_{\text{O}^+}$, $\delta_{\text{N}^+}$, $\delta_{\text{S}^+}$, and $\delta_{\text{S}^{2+}}$ from the below reactions 
\begin{align}
 \text{O}^0(^3\text{P}_{2,1,0}) +\text{H}^+ &  \underset{\delta_\text{O}'}{\stackrel{\delta_\text{O}}{\rightleftharpoons}}  \text{O}^+(^4\text{S}_{3/2}^0) +\text{H}^0(^2\text{S}_{1/2}) ,  \\
  \text{N}^0(^4\text{S}_{3/2}^0) +\text{H}^+ & \underset{\delta_\text{N}'}{\stackrel{\delta_\text{N}}{\rightleftharpoons}}  \text{N}^+(^3\text{P}_{0,1,2}) +\text{H}^0(^2\text{S}_{1/2}) , \\
 \text{O}^+(^4\text{S}_{3/2}^0) +\text{H}^+ &  \underset{\delta_{\text{O}^+}'}{\stackrel{\delta_{\text{O}^+}}{\rightleftharpoons}}  \text{O}^{2+}(^3\text{P}_{2,1,0}) +\text{H}^0(^2\text{S}_{1/2}) ,  \\
  \text{N}^+(^3\text{P}_{0,1,2}) +\text{H}^+ & \underset{\delta_{\text{N}^+}'}{\stackrel{\delta_{\text{N}^+}}{\rightleftharpoons}}  \text{N}^{2+}(^2\text{P}_{1/2, 3/2}^0) +\text{H}^0(^2\text{S}_{1/2})   , \\
 \text{S}^+(^4\text{S}_{3/2}^0) +\text{H}^+ &  \underset{\delta_{\text{S}^+}'}{\stackrel{\delta_{\text{S}^+}}{\rightleftharpoons}}  \text{S}^{2+}(^3\text{P}_{2,1,0}) +\text{H}^0(^2\text{S}_{1/2}) ,  \\
  \text{S}^{2+}(^3\text{P}_{0,1,2}) +\text{H}^+ & \underset{\delta_{\text{S}^{2+}}'}{\stackrel{\delta_{\text{S}^{2+}}}{\rightleftharpoons}}  \text{S}^{3+}(^2\text{P}_{1/2, 3/2}^0) +\text{H}^0(^2\text{S}_{1/2})   .
 \end{align} 
We take the ionisation potentials from Appendix D of \citet{draine_book_2011}

\subsection*{Collisional ionisation rates}\label{appendix:Collisional ionisation rates} 

The reaction rates for collisional ionisation are taken from \citet{landini_1990}. In the range of $T_\text{e} = 10^4 - 10^8\,\text{K}$ (hot plasma) the analytic expression is used
\begin{equation}
\begin{split}
C(T_\text{e}) = A_\text{col} \,T_\text{e}[\text{K}]^{1/2} \left(1 + 0.1  \frac{T_\text{e}[\text{K}]}{T_\text{col}} \right)^{-1}  \\
\text{exp}\left(-\frac{T_\text{col}}{T_\text{e}[\text{K}]}\right) \quad [\text{cm}^{3}\,\text{s}^{-1}].
\end{split}
\end{equation} 
The exponential factor $\text{exp}\left(-T_\text{col}/T_\text{e}[\text{K}]\right)$ in the above equation may be read as activation energy barrier for the collisional ionisation with electrons, i.e. roughly the ionisation potential. \\
 For the hydrogen collisional ionisation rates (units:  $[\text{cm}^{3}\,\text{s}^{-1}]$) we take the formula from \citet{dopita_book}
 \begin{equation}
C_\text{H}(T_\text{e}) = 2.5 \times 10^{-10}   T_\text{e}[\text{K}]^{1/2}  \left( 1 +  \frac{T_\text{e}[\text{K}]}{78945} \right) \text{exp}\left(-\frac{157890}{T_\text{e}[\text{K}]}\right)  .
\end{equation} 

\subsection*{Recombination rates}\label{appendix:Recombination rates}

Two processes contribute to the total recombination rate $\alpha_\text{tot}$ \citep{landini_1990}. These are the radiative $\alpha_\text{R}$ and the dielectronic recombination rates $\alpha_\text{D}$  
\begin{equation}
\alpha_\text{tot} = \alpha_\text{R} + \alpha_\text{D}.
\end{equation}
In the range of $T_\text{e} = 10^4 - 10^8\,\text{K}$  the analytic expressions for $\alpha_\text{R}$ are given by \citet{landini_1990}
\begin{equation}
\alpha_\text{R}(T_\text{e})  = A_\text{rad} \, T_4^{-\text{X}_\text{rad}}  [\text{cm}^{3}\,\text{s}^{-1}].
\end{equation} 
We note that in \citet{landini_1990} the sign of $\text{X}_\text{rad}$ is not correct. The radiative recombination rates stated therein are in agreement with  \citet{Pequignot_1991}.  We take $\alpha_\text{D}$ from \citet{nussbaumer_storey_1983}, which are valid for $T_\text{e}\sim 10^3-6\times 10^4\,\text{K}$.  Explicitly they are given by (units: $\text{cm}^{3}\,\text{s}^{-1}$)
\begin{equation}
\alpha_\text{D}(T_\text{e})  = 10^{-12}\left(\frac{a}{T_4} + b + c  T_4 + d T_4^2 \right)\, T_4^{-3/2} \,\text{exp}\left(-\frac{f}{T_4}\right).
\end{equation}
 For the hydrogen recombination rate (case B) we take the formula presented in \citet{osterbrock_book}
\begin{equation}
\alpha_\text{H}(T_\text{e})  =  3.69\times 10^{-10} \,T_\text{e}[\text{K}]^{-0.79}  [\text{cm}^{3}\,\text{s}^{-1}].
\end{equation}
 For sulphur we take the dielectronic recombination rates from \citet{landini_1990}. The analytic expression is 
\begin{equation}
\begin{split}
\alpha_\text{D}(T_\text{e})   = A_\text{di} \, T_\text{e}[\text{K}]^{-3/2} \,\text{exp}\left(-\frac{T_0}{T_\text{e}[\text{K}]}\right)\\
 \left(1 + B_\text{di} \,\text{exp}\left(-\frac{T_i}{T_\text{e}[\text{K}]}\right)\right)  [\text{cm}^{3}\,\text{s}^{-1}] .
\end{split}
\end{equation}

\subsection*{Five level system}\label{Five level system}

\begin{figure} 
\centering
\subfloat{\includegraphics[trim=0 0 0 0, clip, width=0.49 \textwidth]{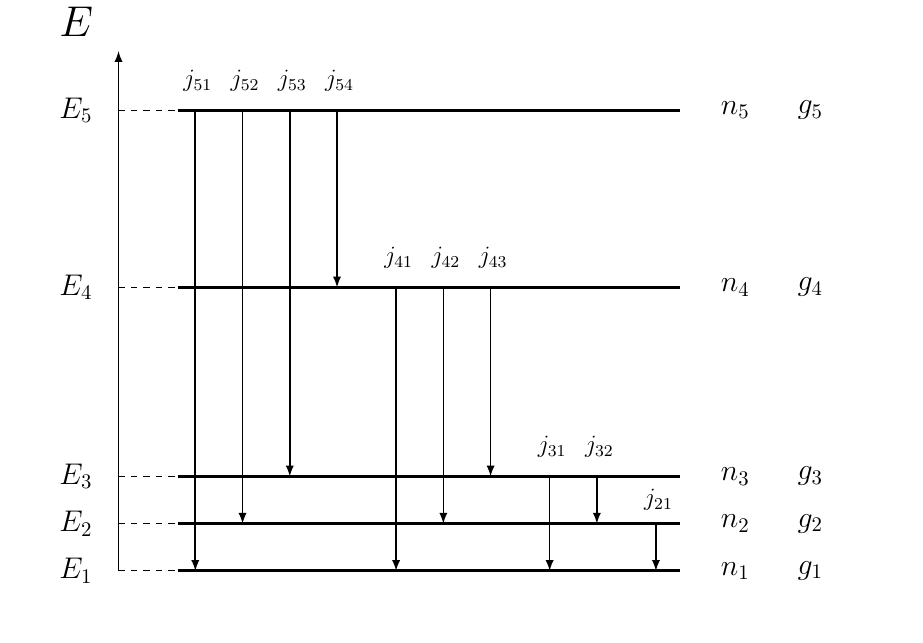}}
\hfill
\caption{\small{Grotrian diagrams for a five level system.}}\label{fig:synthetic_spectrum}
\end{figure}

In this paragraph we describe the numerical solution of a five level atomic/ionic system of species $Y^k$ (see Fig.\,\ref{fig:synthetic_spectrum}). 
The energy levels, (internal) number densities, and statistical weights are labelled as $E_i(Y^k)$, $n_i(Y^k)$, and $g_i(Y^k)$ with $i=1,\dots 5$, respectively. 
In a five level system the internal number densities $n_i(Y^k)$ are connected to the total number density $n(Y^k)$ via
 \begin{align}\label{Eq:internal_levels}
 n(Y^k) & = \sum_{i=1}^5 n_i(Y^k)  .
 \end{align}
 We can assume that  $n(Y^k)$ is known from the ionisation balance and the internal level population are the quantities we want to determine. We consider three processes (rate coefficients in brackets): (a) spontaneous decay ($A_{ul}$), (b) collisional de-excitation ($C_{ul}$), and (c) collisional excitation ($C_{lu}$). Here, the indices $u$ and $l$ indicate the upper level and the lower level, respectively. In an equilibrium ($\dot{n}_i(Y^k)=0$ for all $i = 1\dots 5$) the rate equation for a fixed energy level $i$ is given by \citep[e.g.][]{osterbrock_book}
 \begin{align}
 0    = & \sum_{j>i}\,\left( A_{ji}+ C_{ji}\right)  n_j  + \sum_{j<i}\,C_{ji}  n_j\nonumber\\
   & -  \sum_{j<i}\,\left( A_{ij} + C_{ij}\right)  n_i  - \sum_{j>i}\,C_{ij}  n_i .
 \end{align}
We note that the internal level populations reach the equilibrium on very small time scales so that the equilibrium assumption here is justified. 
In a five level system only four rate equations are  linearly independent. Therefore, we can replace the last rate equation ($i=5$) with Eq.\,\ref{Eq:internal_levels} to formulate the matrix equation 
 \begin{equation}\label{Eq:matrix_equation}
        \begin{pmatrix} 
         M_{11} &  M_{12} & M_{13} &  M_{14} & M_{15} \\
         M_{21} &  M_{22} & M_{23} &  M_{24} & M_{25} \\
         M_{31} &  M_{32} & M_{33} &  M_{34} & M_{35} \\
         M_{41} &  M_{42} & M_{43} &  M_{44} & M_{45} \\
         1 &  1 & 1 &  1 & 1 \\ \end{pmatrix} \cdot \begin{pmatrix} 
         n_1(Y^\text{k}) \\
         n_2(Y^\text{k}) \\
         n_3(Y^\text{k}) \\
         n_4(Y^\text{k}) \\
         n_5(Y^\text{k}) \\ \end{pmatrix} = \begin{pmatrix} 
        0 \\
         0 \\
         0 \\
         0 \\
        n(Y^\text{k}) \\ \end{pmatrix}.
      \end{equation}
 \noindent     
 The coefficients of the matrix $\hat{M}$ are functions of  $A_{ul}, C_{lu}, C_{ul}$. Explicitly, they are    
       \begin{align}
         M_{11} & = -\left[C_{15} + C_{14} + C_{13} + C_{12}\right] , \\
         M_{12} & = C_{21} +A_{21} ,  \\
         M_{13} & = C_{31} +A_{31},  \\
         M_{14} & = C_{41} +A_{41},  \\
         M_{15} & = C_{51} +A_{51} ,  
            \end{align}
             \begin{align}
         M_{21} & = C_{12} \\
         M_{22} & = - \left[C_{25} + C_{24}+C_{23} +  C_{21} + A_{21} \right] ,\\
         M_{23} & = C_{32} +A_{32}, \\
         M_{24} & = C_{42} +A_{42},\\
         M_{25} & = C_{52} +A_{52}, 
            \end{align}
          \begin{align}
         M_{31} & = C_{13} ,  \\
         M_{32} & = C_{23} , \\
         M_{33} & = - \left[A_{31} + A_{32} + C_{31} + C_{32} + C_{35} + C_{34}\right] , \\
         M_{34} & =  A_{43} + C_{43} , \\
         M_{35} & =  A_{53} + C_{53} , 
            \end{align}
          \begin{align}
         M_{41} & = C_{14},\\
         M_{42} & = C_{24},\\
         M_{43} & = C_{34}, \\
         M_{44} & = - \left[A_{41} + A_{42} + A_{43}  +C_{41} + C_{42} + C_{43} + C_{45} \right], \\
         M_{45} & =A_{54} + C_{54} . 
            \end{align}       
\noindent            
Equation \ref{Eq:matrix_equation} can be solved numerically for the internal number densities $n_i(Y^k)$. In the case of dominant electron collisions, the collision rates $C_{ul}(n_\text{e}, T_\text{e}, Y^k)$ and $C_{lu}(n_\text{e}, T_\text{e}, Y^k)$ are given by
\begin{align}  
C_{ul} & =    \frac{8.629\times 10^{-6} n_\text{e}}{\sqrt{T_\text{e}[\text{K}]}}\cdot \frac{\Upsilon_{lu}(T_\text{e}, Y^k)}{g_u}, \\ 
C_{lu} & =    \frac{8.629\times 10^{-6} n_\text{e}}{\sqrt{T_\text{e}[\text{K}]}}\cdot \frac{\Upsilon_{lu}(T_\text{e}, Y^k)}{g_l}\cdot \text{exp}\left(-\frac{E_u-E_l}{k_\text{B} T_\text{e}}\right).
\end{align}
 The collisional strengths ($\Upsilon_{lu} = \Upsilon_{ul}$) can be expressed as analytical functions in terms of the electron temperature.      
 We adopt the expressions from \citet{draine_book_2011} as presented in Tables \ref{Tab:OII}--\ref{Tab:SIII}. The Einstein coefficients are adopted from the NIST atomic database. Synthetic emissivities are calculated via 
\begin{equation}
j_{ul}(Y^k) = 1/4\pi\cdot n_{u}(Y^k)\cdot A_{ul}(Y^k)\cdot E_{ul}(Y^k).
\end{equation}

 \subsection*{The reaction network of the BE-method}
 
  For simplicity, we indicate number densities with square brackets, i.e. [N$^+$] is short for $n(\text{N}^+)$. The seven reaction rates  described in Section\,\ref{sec:time_dependent_reactions} are:
  \begin{align}
     \frac{d[\text{N}^0]}{dt}  = & - \delta_\text{N}  \cdot [\text{N}^0] \cdot [\text{H}^{+}] + \delta^{'}_\text{N} \cdot [\text{N}^{+}]\cdot [\text{H}^0] \nonumber\\
      &  - C_\text{N} \cdot  [\text{N}^0] \cdot [\text{e}^{-}] + \alpha_\text{N}\cdot [\text{N}^{+}]  \cdot  [\text{e}^{-}],\\
     \frac{d[\text{N}^+]}{dt}   = & +\delta_\text{N}  \cdot [\text{N}^0] \cdot [\text{H}^{+}] - \delta^{'}_\text{N} \cdot [\text{N}^{+}]\cdot [\text{H}^0] \nonumber\\
     & + C_\text{N} \cdot  [\text{N}^0] \cdot [\text{e}^{-}] - \alpha_\text{N}\cdot [\text{N}^{+}]  \cdot  [\text{e}^{-}],\\
      \frac{d[\text{O}^0]}{dt}  = & - \delta_\text{O}  \cdot [\text{O}^0] \cdot [\text{H}^{+}] + \delta^{'}_\text{O} \cdot [\text{O}^{+}]\cdot [\text{H}^0]\nonumber \\
      & - C_\text{O} \cdot  [\text{O}^0] \cdot [\text{e}^{-}] + \alpha_\text{O}\cdot [\text{O}^{+}]  \cdot  [\text{e}^{-}],\\
     \frac{d[\text{O}^+]}{dt}   = & +\delta_\text{O}  \cdot [\text{O}^0] \cdot [\text{H}^{+}] - \delta^{'}_\text{O} \cdot [\text{O}^{+}]\cdot [\text{H}^0] \nonumber\\
     & + C_\text{O} \cdot  [\text{O}^0] \cdot [\text{e}^{-}] - \alpha_\text{O}\cdot [\text{O}^{+}]  \cdot  [\text{e}^{-}],\\
     \frac{d[\text{H}^0]}{dt}  = &  + \delta_\text{N}  \cdot [\text{N}^0] \cdot [\text{H}^{+}] - \delta^{'}_\text{N} \cdot [\text{N}^{+}]\cdot [\text{H}^0]\nonumber\\
     & + \delta_\text{O}  \cdot [\text{O}^0] \cdot [\text{H}^{+}] - \delta^{'}_\text{O} \cdot [\text{O}^{+}]\cdot [\text{H}^0]  \nonumber \\
     & -C_\text{H}\cdot [\text{H}^0]\cdot [\text{e}^-] + \alpha_\text{H}\cdot [\text{H}^+]\cdot [\text{e}^-],\\
   \frac{d[\text{H}^+]}{dt}   = & - \delta_\text{N}  \cdot [\text{N}^0] \cdot [\text{H}^{+}] + \delta^{'}_\text{N} \cdot [\text{N}^{+}]\cdot [\text{H}^0]\nonumber \\
   & - \delta_\text{O}  \cdot [\text{O}^0] \cdot [\text{H}^{+}] + \delta^{'}_\text{O} \cdot [\text{O}^{+}]\cdot [\text{H}^0]\nonumber\\
    & +C_\text{H}\cdot [\text{H}^0]\cdot [\text{e}^-] - \alpha_\text{H}\cdot [\text{H}^+]\cdot [\text{e}^-], \\
     \frac{d[\text{e}^-]}{dt}   = & + C_\text{N} \cdot  [\text{N}^0] \cdot [\text{e}^{-}] - \alpha_\text{N}\cdot [\text{N}^{+}]  \cdot  [\text{e}^{-}]\nonumber \\
     & + C_\text{O} \cdot  [\text{O}^0] \cdot [\text{e}^{-}] - \alpha_\text{O}\cdot [\text{O}^{+}] \cdot [\text{e}^{-}]\nonumber\\
     & +C_\text{H}\cdot [\text{H}^0]\cdot [\text{e}^-] - \alpha_\text{H}\cdot [\text{H}^+]\cdot [\text{e}^-] .   
   \end{align}
We note, that element conservation and charge neutrality are ensured:  
\begin{equation}
\sum_{k, \text{ same element}} \frac{d[Y^k]}{dt}= 0
\end{equation}
and
\begin{equation}
d[\text{e}^-]/dt  = d[\text{N}^+]/dt +  d[\text{O}^+]/dt +  d[\text{H}^+]/dt.
\end{equation}

 \subsection*{The reaction network of the extended BE-method}

From the stated reactions in Section\,\ref{sec:ionisation}  we derive the following set of differential equations: 
   \begin{align}
  \frac{d[\text{O}^0]}{dt}  = & - \delta_\text{O}  \cdot [\text{O}^0] \cdot [\text{H}^{+}] + \delta^{'}_\text{O} \cdot [\text{O}^{+}]\cdot [\text{H}^0]\nonumber \\
      & - C_\text{O} \cdot  [\text{O}^0] \cdot [\text{e}^{-}] + \alpha_\text{O}\cdot [\text{O}^{+}]  \cdot  [\text{e}^{-}],\\
    \frac{d[\text{O}^+]}{dt}   = & +\delta_\text{O}  \cdot [\text{O}^0] \cdot [\text{H}^{+}] - \delta^{'}_\text{O} \cdot [\text{O}^{+}]\cdot [\text{H}^0] \nonumber\\
     & + C_\text{O} \cdot  [\text{O}^0] \cdot [\text{e}^{-}] - \alpha_\text{O}\cdot [\text{O}^{+}]  \cdot  [\text{e}^{-}]\nonumber \\
 & -C_{\text{O}^+} \cdot  [\text{O}^+] \cdot [\text{e}^{-}] + \alpha_{\text{O}^+}\cdot [\text{O}^{2+}]  \cdot  [\text{e}^{-}]\nonumber \\
 & -\delta_{\text{O}^+}  \cdot [\text{O}^+] \cdot [\text{H}^{+}] + \delta^{'}_{\text{O}^+} \cdot [\text{O}^{2+}]\cdot [\text{H}^0] ,\\
  \frac{d[\text{O}^{2+}]}{dt}  = &+ C_{\text{O}^+} \cdot  [\text{O}^+] \cdot [\text{e}^{+}] - \alpha_{\text{O}^+}\cdot [\text{O}^{2+}]  \cdot  [\text{e}^{-}] \nonumber \\
      &   + \delta_{\text{O}^+}  \cdot [\text{O}^+] \cdot [\text{H}^{+}] - \delta^{'}_{\text{O}^+} \cdot [\text{O}^{2+}]\cdot [\text{H}^0],
   \end{align}
and
   \begin{align}
     \frac{d[\text{N}^0]}{dt}  = & - \delta_\text{N}  \cdot [\text{N}^0] \cdot [\text{H}^{+}] + \delta^{'}_\text{N} \cdot [\text{N}^{+}]\cdot [\text{H}^0] \nonumber\\
      &  - C_\text{N} \cdot  [\text{N}^0] \cdot [\text{e}^{-}] + \alpha_\text{N}\cdot [\text{N}^{+}]  \cdot  [\text{e}^{-}],\\
     \frac{d[\text{N}^+]}{dt}   = & +\delta_\text{N}  \cdot [\text{N}^0] \cdot [\text{H}^{+}] - \delta^{'}_\text{N} \cdot [\text{N}^{+}]\cdot [\text{H}^0] \nonumber\\
     & + C_\text{N} \cdot  [\text{N}^0] \cdot [\text{e}^{-}] - \alpha_\text{N}\cdot [\text{N}^{+}]  \cdot  [\text{e}^{-}]\\ 
 & -C_{\text{N}^+} \cdot  [\text{N}^+] \cdot [\text{e}^{-}] + \alpha_{\text{N}^+}\cdot [\text{N}^{2+}]  \cdot  [\text{e}^{-}]\nonumber \\
 & -\delta_{\text{N}^+}  \cdot [\text{N}^+] \cdot [\text{H}^{+}] + \delta^{'}_{\text{N}^+} \cdot [\text{N}^{2+}]\cdot [\text{H}^0] ,\\
  \frac{d[\text{N}^{2+}]}{dt}  = &+ C_{\text{N}^+} \cdot  [\text{N}^+] \cdot [\text{e}^{+}] - \alpha_{\text{N}^+}\cdot [\text{N}^{2+}]  \cdot  [\text{e}^{-}] \nonumber \\
      &   + \delta_{\text{N}^+}  \cdot [\text{N}^+] \cdot [\text{H}^{+}] - \delta^{'}_{\text{N}^+} \cdot [\text{N}^{2+}]\cdot [\text{H}^0], 
   \end{align}
and
   \begin{align}
     \frac{d[\text{S}^+]}{dt}  &=  - C_{\text{S}^{+}} \cdot  [\text{S}^+] \cdot [\text{e}^{-}] + \alpha_{\text{S}^+}\cdot [\text{S}^{2+}]  \cdot  [\text{e}^{-}] \nonumber\\
        & - \delta_{\text{S}^+}  \cdot [\text{S}^+] \cdot [\text{H}^{+}] + \delta^{'}_{\text{S}^+} \cdot [\text{S}^{2+}]\cdot [\text{H}^0]  ,\\
     \frac{d[\text{S}^{2+}]}{dt}   = & + C_{\text{S}^{+}} \cdot  [\text{S}^+] \cdot [\text{e}^{-}] - \alpha_{\text{S}^{+}}\cdot [\text{S}^{2+}]  \cdot  [\text{e}^{-}] \nonumber\\
      & -C_{\text{S}^{2+}} \cdot  [\text{S}^{2+}] \cdot [\text{e}^{-}] + \alpha_{\text{S}^{2+}}\cdot [\text{S}^{3+}]  \cdot  [\text{e}^{-}]\nonumber \\
     & +\delta_{\text{S}^+}  \cdot [\text{S}^+] \cdot [\text{H}^{+}] - \delta^{'}_{\text{S}^+} \cdot [\text{S}^{2+}]\cdot [\text{H}^0] \nonumber\\
 & -\delta_{\text{S}^{2+}}  \cdot [\text{S}^{2+}] \cdot [\text{H}^{+}] + \delta^{'}_{\text{S}^{2+}} \cdot [\text{S}^{3+}]\cdot [\text{H}^0] ,\\
  \frac{d[\text{S}^{3+}]}{dt}  = &+ C_{\text{S}^{2+}} \cdot  [\text{S}^{2+}] \cdot [\text{e}^{+}] - \alpha_{\text{S}^{2+}}\cdot [\text{S}^{3+}]  \cdot  [\text{e}^{-}] \nonumber \\
      &   + \delta_{\text{S}^{2+}}  \cdot [\text{S}^{2+}] \cdot [\text{H}^{+}] - \delta^{'}_{\text{S}^{2+}} \cdot [\text{S}^{3+}]\cdot [\text{H}^0]. 
   \end{align}
For hydrogen we have
   \begin{align} 
     \frac{d[\text{H}^0]}{dt}  =& + \delta_\text{O}  \cdot [\text{O}^0] \cdot [\text{H}^{+}] - \delta^{'}_\text{O} \cdot [\text{O}^{+}]\cdot [\text{H}^0]  \nonumber \\
 &  + \delta_\text{N}  \cdot [\text{N}^0] \cdot [\text{H}^{+}] - \delta^{'}_\text{N} \cdot [\text{N}^{+}]\cdot [\text{H}^0]\nonumber\\
     & -C_\text{H}\cdot [\text{H}^0]\cdot [\text{e}^-] + \alpha_\text{H}\cdot [\text{H}^+]\cdot [\text{e}^-] \nonumber \\
& + \delta_{\text{O}^+}  \cdot [\text{O}^+] \cdot [\text{H}^{+}] - \delta^{'}_{\text{O}^+} \cdot [\text{O}^{2+}]\cdot [\text{H}^0]  \nonumber \\
& + \delta_{\text{N}^+}  \cdot [\text{N}^+] \cdot [\text{H}^{+}] - \delta^{'}_{\text{N}^+} \cdot [\text{N}^{2+}]\cdot [\text{H}^0]  \nonumber \\
& + \delta_{\text{S}^+}  \cdot [\text{S}^+] \cdot [\text{H}^{+}] - \delta^{'}_{\text{S}^+} \cdot [\text{S}^{2+}]\cdot [\text{H}^0]  \nonumber \\
& + \delta_{\text{S}^{2+}}  \cdot [\text{S}^{2+}] \cdot [\text{H}^{+}] - \delta^{'}_{\text{S}^{2+}} \cdot [\text{S}^{3+}]\cdot [\text{H}^0]  ,\\
   \frac{d[\text{H}^+]}{dt}   = & -  \frac{d[\text{H}^0]}{dt}.
   \end{align}
 The equation for the electron density can be obtained from the intrinsic  charge neutrality
 \begin{align}
 \frac{d[\text{e}^-]}{dt} &=   \frac{d[\text{H}^+]}{dt} +  \frac{d[\text{O}^+]}{dt}  + \frac{d[\text{N}^+]}{dt}  \nonumber \\
  & + 2   \frac{d[\text{O}^{2+}]}{dt} + 2\frac{d[\text{N}^{2+}]}{dt} +  2\frac{d[\text{S}^{2+}]}{dt} \nonumber \\ 
 & +  3\frac{d[\text{S}^{3+}]}{dt}
\end{align}    
  Element conservation is ensured, since  
  \begin{equation}
\sum_{k, \text{ same element}} \frac{d[Y^k]}{dt}= 0
\end{equation}
We present the Einstein coefficients, line transitions, and collisional strengths with electrons for N, N$^+$, O, O$^+$, O$^{2+}$, S$^+$, and S$^{2+}$ in Tables\,\ref{Tab:SII}-\ref{Tab:OIII}.

\section*{Appendix B - The Saha equation}\label{appendix:B}

 \begin{figure}[h!] 
\resizebox{\hsize}{!}{\includegraphics[trim=0 0 0 0, clip, width=0.9\textwidth]{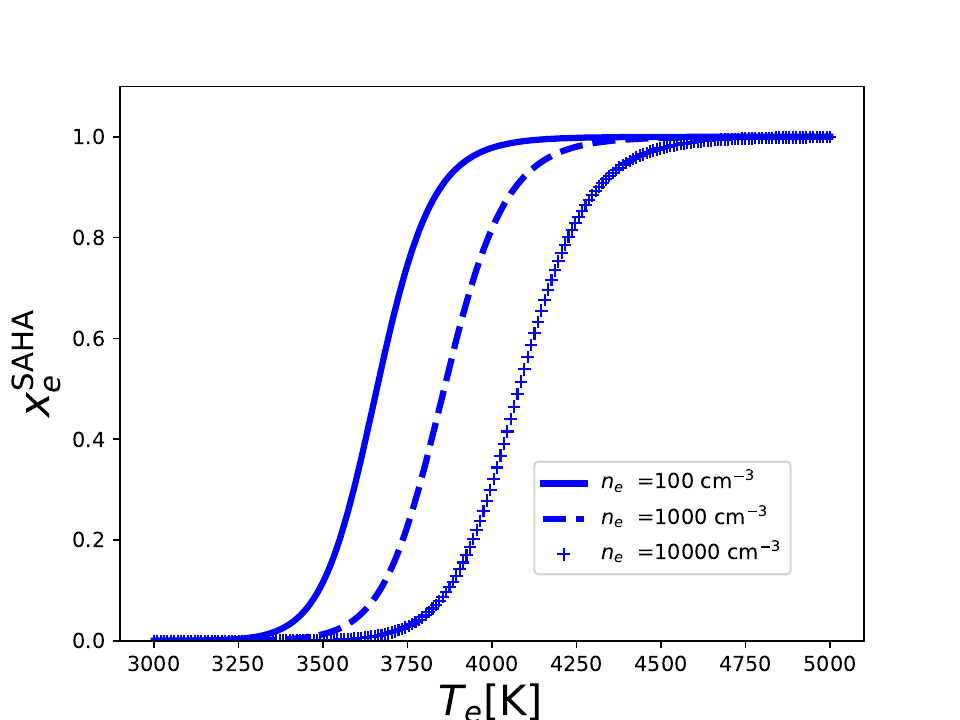}}
\caption{\small{The hydrogen ionisation fraction from the Saha equation. A comparison with Fig.\,\ref{fig:equilibrium_ionisation_fraction} shows the differences of the ionisation fraction of the BE99 equilibrium.}}\label{fig:saha_equation} 
\end{figure}

We can compare the function $x_e^\text{eq}(T_e)$ that has been numerically derived from solving the network of ODEs with the Saha equation \citep[e.g.][]{osterbrock_book}. The Saha equation is given by
\begin{equation}
\frac{n(Y^{+i+1})  \,n_\text{e}}{n(Y^{+i})} = \left(\frac{2g_{i+1}}{g_i}\right) \left( \frac{2\pi m_\text{e} k_\text{B} T_\text{e}}{h^2} \right)^{3/2}\,\text{exp}\left(- \frac{\chi}{k_\text{B} T_\text{e}} \right) .
\end{equation}
Here, $m_\text{e}$ is  the electron mass, $k_\text{B}$ is the Boltzmann constant, $h$  is the Planck constant, and $\chi$ is the energy needed to ionise $Y^i$ from the ground state to $Y^{i+1}$.
In the case of hydrogen we have $n(Y^{+i+1}) = n(\text{H}^+)$, $n(Y^{+i}) = n(\text{H}^0)$, $g_{i+1} = 1$, $g_{i} = g(^2\text{S}_{1/2})=2$, and the ionisation potential of hydrogen $\chi(\text{H}^0) =  13.59844\,\text{eV}$ leading to 
\begin{equation}
\frac{n(\text{H}^+)}{n(\text{H}^0)} = \left(\frac{1}{n_\text{e}}\right) \left( \frac{2\pi\, m_\text{e} \,k_\text{B}\, T_\text{e}}{h^2} \right)^{3/2}\,\text{exp}\left(- \frac{\chi(\text{H}^0)}{k_\text{B} T_\text{e}} \right) .
\end{equation}
We can assume that $n(\text{H}^+) \approx n_\text{e}$ and introduce the Saha hydrogen  ionisation fraction via Eq.\,\ref{eq:01}: 
\begin{equation}
\frac{x_\text{e}^\text{Saha}}{1-x_\text{e}^\text{Saha}} =:f 
\end{equation}
with 
\begin{equation}
 f =  \left(\frac{1}{n_\text{e}}\right) \left( \frac{2\pi \,m_\text{e}\, k_\text{B} T_\text{e}}{h^2} \right)^{3/2}\,\text{exp}\left(- \frac{\chi(\text{H}^0)}{k_\text{B} T_\text{e}} \right) .
\end{equation}
The above Equation can be rearranged to 
 \begin{equation}
 x_\text{e}^\text{Saha}(T_e, n_e)=  f/(1 + f) .
\end{equation} 
 We plot the Saha hydrogen ionisation fraction for three electron densities in Figure\,\ref{fig:saha_equation}.

\begin{table}[htb]  
\centering \scalebox{0.77}{
\begin{tabular}{||c | c c c c  |}
\hline\hline 
  $j_{ul}$ & Transition ($u$--$l$)  & $\lambda_\text{air}$ [$\AA$] & $A_{ul}$ [s$^{-1}$] & $\Upsilon_{ul}(T_\text{e}, \text{O}^+)$, $T_4 = T_\text{e}/10^4\,\text{K}$    \\  
\hline    
   &  &   & &   \\
 $j_{51}$ &  $^2$P$_{1/2}^\text{o}$--$^4$S$_{3/2}^\text{o}$  & 2470.2 &  $2.12\times 10^{-2}$    &  $0.1395\,T_4^{0.02642 + 0.01077\,\text{ln}\,T_4}$   \\            
 $j_{52}$ &  $^2$P$_{1/2}^\text{o}$--$^2$D$_{5/2}^\text{o}$  & 7318.9 &  $5.19\times 10^{-2}$ & $0.3541\,T_4^{0.06713 + 0.02157\,\text{ln}\,T_4}$     \\
 $j_{53}$ &  $^2$P$_{1/2}^\text{o}$--$^2$D$_{3/2}^\text{o}$ &  7329.7 &   $8.672\times 10^{-2}$   &  $0.3296\,T_4^{0.0775 + 0.0291\,\text{ln}\,T_4}$   \\
 $j_{54}$ &  $^2$P$_{1/2}^\text{o}$--$^2$P$_{3/2}^\text{o}$ &  5.03\,\text{mm} &  $1.41\times 10^{-10}$    & $0.3228\,T_4^{0.04077 + 0.02892\,\text{ln}\,T_4}$    \\ 
          &                      &          & &     \\
 $j_{41}$ &  $^2$P$_{3/2}^\text{o}$--$^4$S$_{3/2}^\text{o}$ & 2470.3 &  $5.22\times 10^{-2}$  & $0.2827\,T_4^{0.02589 + 0.009331\,\text{ln}\,T_4}$    \\           
 $j_{42}$ &  $^2$P$_{3/2}^\text{o}$--$^2$D$_{5/2}^\text{o}$ & 7320.0 & $9.907\times 10^{-2}$   &  $0.8437\,T_4^{0.08544 + 0.02737\,\text{ln}\,T_4}$    \\ 
 $j_{43}$ &  $^2$P$_{3/2}^\text{o}$--$^2$D$_{3/2}^\text{o}$ & 7330.7 & $5.34\times 10^{-2}$   & $0.4923\,T_4^{0.06732 + 0.02241\,\text{ln}\,T_4}$      \\ 
          &                      &          & &    \\
 $j_{31}$ &  $^2$D$_{3/2}^\text{o}$--$^4$S$_{3/2}^\text{o}$  & 3726.0  &  $1.776\times 10^{-4}$   & $0.5488\,T_4^{0.01254 + 0.005683\,\text{ln}\,T_4}$     \\      
 $j_{32}$ &  $^2$D$_{3/2}^\text{o}$--$^2$D$_{5/2}^\text{o}$ &  499.5\,$\upmu$m  &  $1.30\times 10^{-7}$   & $1.432\,T_4^{-0.1505 + 0.03903\,\text{ln}\,T_4}$   \\ 
          &                      &          &  &       \\
 $j_{21}$ &  $^2$D$_{5/2}^\text{o}$--$^4$S$_{3/2}^\text{o}$ & 3728.8  &  $3.058\times 10^{-5}$  & $0.802\,T_4^{0.01821 + 0.007941\,\text{ln}\,T_4}$    \\
          &                      &          & &       \\
\hline\hline
\end{tabular} }
 \caption{\small{Atomic data for O\,II. Formulae for $\Upsilon_{ul}$ are taken from fit to \citet{tayal_2007_OII}.  We note -- as already pointed out by \citet{kisielius_2009} -- that a few rows in Table\,3 of \citet{tayal_2007_OII} are mixed-up, that is the transitions are not labelled correctly. In Table\,3 therein the entries for the transition $1$ to $4$ must be $1$ to $5$ and vice versa. The entries for the transition $2$ to $4$ must be $2$ to $5$ and vice versa. The entries for the transition $3$ to $4$ must be $3$ to $5$ and vice versa. The validity range for the stated $\Upsilon_{ul}$ is $T_\text{e} = 4\,000-100\,000\,\text{K}$. }}\label{Tab:OII}  
\end{table}

\begin{table}   
\centering \scalebox{0.77}{
\begin{tabular}{||c| c c c c |}
\hline\hline
 $j_{ul}$ & Transition  ($u$--$l$)  & $\lambda_\text{air}$ [$\AA$] & $A_{ul}$ [s$^{-1}$]  & $\Upsilon_{ul}(T_\text{e}, \text{N}^0)$, $T_4 = T_\text{e}/10^4\,\text{K}$ \\  
\hline    
   &  &   & &    \\
 $j_{51}$ & $^2$P$_{3/2}^\text{o}$--$^4$S$_{3/2}^\text{o}$   & 3466.5  &  $ 6.5\times 10^{-3}$    & $  0.1095\,T_4^{0.7545 - 0.1418\,\text{ln}\,T_4}$   \\            
 $j_{52}$ & $^2$P$_{3/2}^\text{o}$--$^2$D$_{5/2}^\text{o}$   & 10397.7 &  $ 6.12\times 10^{-2}$    & $  0.1968\,T_4^{0.5257 - 0.007914\,\text{ln}\,T_4}$  \\
 $j_{53}$ & $^2$P$_{3/2}^\text{o}$--$^2$D$_{3/2}^\text{o}$   &  10407.2 &  $ 2.741\times 10^{-2}$    &  $  0.3684\,T_4^{0.486 + 0.007193\,\text{ln}\,T_4}$   \\
 $j_{54}$ & $^2$P$_{3/2}^\text{o}$--$^2$P$_{1/2}^\text{o}$   &  $25.9\,\text{mm}$ &  $ 5.17\times 10^{-13}$    &  $  0.1248\,T_4^{0.693 - 0.06741\,\text{ln}\,T_4}$  \\ 
          &                      &          & &     \\
 $j_{41}$ & $^2$P$_{1/2}^\text{o}$--$^4$S$_{3/2}^\text{o}$   &  3466.5 &  $ 2.6\times 10^{-3}$    &   $  0.05486\,T_4^{0.7599 - 0.1449\,\text{ln}\,T_4}$  \\           
 $j_{42}$ & $^2$P$_{1/2}^\text{o}$--$^2$D$_{5/2}^\text{o}$  & 10398.1  & $ 3.45\times 10^{-2}$      &  $  0.1424\,T_4^{0.4652 + 0.01418\,\text{ln}\,T_4}$  \\ 
 $j_{43}$ & $^2$P$_{1/2}^\text{o}$--$^2$D$_{3/2}^\text{o}$  & 10407.6 &  $ 5.313\times 10^{-2}$   &  $  0.1401\,T_4^{0.5374 - 0.01256\,\text{ln}\,T_4}$  \\ 
          &                      &          & &      \\
 $j_{31}$ & $^2$D$_{3/2}^\text{o}$--$^4$S$_{3/2}^\text{o}$ & 5197.9 &  $ 2.034\times 10^{-5}$    &   $  0.3365\,T_4^{0.7217 - 0.1246\,\text{ln}\,T_4}$  \\      
 $j_{32}$ & $^2$D$_{3/2}^\text{o}$--$^2$D$_{5/2}^\text{o}$ & $1.15\,\text{mm}$ & $1.07\times 10^{-8}$   &  $  0.2557\,T_4^{0.9563 - 0.1536\,\text{ln}\,T_4}$  \\ 
          &                      &          &  &      \\
 $j_{21}$ & $^2$D$_{5/2}^\text{o}$--$^4$S$_{3/2}^\text{o}$  & 5200.3  & $7.561\times 10^{-6}$   &  $  0.2244\,T_4^{0.7219 - 0.1247\,\text{ln}\,T_4}$   \\
          &                      &          & &      \\
\hline\hline
\end{tabular} }
 \caption{\small{Atomic data for N\,I. Formulae for $\Upsilon_{ul}$ are taken from fit to data from \citet{tayal_2006_NI}. The validity range for the stated $\Upsilon_{ul}$ is $T_\text{e} = 500-50\,000\,\text{K}$.}}\label{Tab:NI}
\end{table}

\begin{table}[ht]   
\centering \scalebox{0.77}{
\begin{tabular}{||c| c c c c |}
\hline\hline 
  $j_{ul}$ & Transition ($u$--$l$)  & $\lambda_\text{air}$ [$\AA$] & $A_{ul}$ [s$^{-1}$]   & $\Upsilon_{ul}(T_\text{e}, \text{N}^+)$, $T_4 = T_\text{e}/10^4\,\text{K}$ \\  
\hline    
   &  &   & &   \\  
 $j_{51}$ & $^1$S$_0$--$^3$P$_0$ & --$^a$ & 0.0$^a$ & $0.0352\,T_4^{0.066 + 0.018\,\text{ln}\,T_4}$  \\            
 $j_{52}$ & $^1$S$_0$--$^3$P$_1$ & $3062.8$ & $3.18\times 10^{-2}$ & $0.105\,T_4^{0.070 + 0.021\,\text{ln}\,T_4}$   \\
 $j_{53}$ & $^1$S$_0$--$^3$P$_2$ & $3070.5$ &  $1.55\times 10^{-4}$   & $0.176\,T_4^{0.065 + 0.017\,\text{ln}\,T_4}$  \\
 $j_{54}$ & $^1$S$_0$--$^1$D$_2$ & $5754.6$ & $1.14 $  & $0.806\,T_4^{-0.175 - 0.014\,\text{ln}\,T_4}$  \\
          & &  &   &  \\
 $j_{41}$ &  $^1$D$_2$--$^3$P$_0$ & $6527.2$ & $5.25\times 10^{-7}$  & $0.303\,T_4^{0.053 + 0.009\,\text{ln}\,T_4}$ \\            
 $j_{42}$ &  $^1$D$_2$--$^3$P$_1$ & $6548.0$ & $9.84922\times 10^{-4}$   & $0.909\,T_4^{0.053 + 0.010\,\text{ln}\,T_4}$  \\
 $j_{43}$ &  $^1$D$_2$--$^3$P$_2$ & $6583.5$ & $2.91865\times 10^{-3}$  & $1.51\,T_4^{0.054 + 0.011\,\text{ln}\,T_4}$   \\
          & &  &   &   \\
 $j_{31}$ &   $^3$P$_2$--$^3$P$_0$ &  $76.4\,\upmu\text{m}$ & $1.12\times 10^{-12}$    & $0.273\,T_4^{0.166 + 0.030\,\text{ln}\,T_4}$  \\            
 $j_{32}$ &  $^3$P$_2$--$^3$P$_1$ & $121.8\,\upmu\text{m}$ & $7.46\times 10^{-6}$   &  $1.15\,T_4^{0.137 + 0.024\,\text{ln}\,T_4}$  \\
          & &  &  &      \\
 $j_{21}$ &  $^3$P$_1$--$^3$P$_0$ & $205.3\,\upmu\text{m}$ &  $2.08\times 10^{-6}$   & $0.431\,T_4^{0.099 + 0.014\,\text{ln}\,T_4}$  \\
    &  &   &  &    \\ 
\hline\hline
\end{tabular} }
\caption{\small{Atomic data for N\,II. Formulae for $\Upsilon_{ul}$ are taken from fit to data from \citet{hudson_2005_NII}.  $^a$ Transition is strictly forbidden ($J = 0 \longrightarrow 0$). The validity range for the stated $\Upsilon_{ul}$ is $T_\text{e} = 5\,000-30\,000\,\text{K}$.}}\label{Tab:NII}   
\end{table}

\begin{table}[htb]   
\centering \scalebox{0.77}{
\begin{tabular}{||c| c c c c  |}
\hline\hline 
  $j_{ul}$ & Transition  ($u$--$l$)  & $\lambda_\text{air}$ [A] & $A_{ul}$ [s$^{-1}$]  & $\Upsilon_{ul}(T_e, \text{O}^{2+})$ , $T_4 = T_e/10^4\,\text{K}$ \\  
\hline    
   &  &   & &   \\  
 $j_{51}$ & $^1$S$_0$--$^3$P$_0$ & --$^a$ & 0.0$^a$  &  $0.03223\,T_4^{0.1055+0.04667\,\text{ln}\,T_4}$   \\            
 $j_{52}$ & $^1$S$_0$--$^3$P$_1$ & 2320.9  & $2.15\times 10^{-1}$  & $0.09669\,T_4^{0.1055+0.04667\,\text{ln}\,T_4}$    \\
 $j_{53}$ & $^1$S$_0$--$^3$P$_2$ & 2331.4  &  $6.34\times 10^{-4}$    & $0.16115 \,T_4^{0.1055+0.04667\,\text{ln}\,T_4}$  \\
 $j_{54}$ & $^1$S$_0$--$^1$D$_2$ & 4363.2  &  1.71  & $0.521\,T_4^{0.1683 - 0.04775\,\text{ln}\,T_4}$  \\
          & &  &   &     \\
 $j_{41}$ &  $^1$D$_2$--$^3$P$_0$ & 4931.2  & $2.41\times 10^{-6}$   & $0.2441\,T_4^{0.1042+0.02924\,\text{ln}\,T_4}$   \\            
 $j_{42}$ &  $^1$D$_2$--$^3$P$_1$ & 4958.9  & $6.21457\times 10^{-3}$    & $0.7323 \,T_4^{0.1042+0.02924\,\text{ln}\,T_4}$   \\
 $j_{43}$ &  $^1$D$_2$--$^3$P$_2$ & 5006.8  &  $1.81352\times 10^{-2}$ & $1.2205\,T_4^{0.1042+0.02924\,\text{ln}\,T_4}$  \\
          & &  &   &    \\
 $j_{31}$ &   $^3$P$_2$--$^3$P$_0$ & $32.7\,\mu\text{m}$   & $3.17\times 10^{-11}$     & $0.2572\,T_4^{0.07632+ 0.01861\,\text{ln}\,T_4}$   \\            
 $j_{32}$ &  $^3$P$_2$--$^3$P$_1$ &  $51.8\,\mu\text{m}$  &  $9.76\times 10^{-5}$    &  $1.231\,T_4^{0.05185 + 0.008704\,\text{ln}\,T_4}$  \\
          & &  &  &     \\
 $j_{21}$ &  $^3$P$_1$--$^3$P$_0$ &  $88.3\,\mu\text{m}$  &  $2.61\times 10^{-5}$    & $0.5217\,T_4^{0.02985 - 0.0007505\,\text{ln}\,T_4}$   \\
    &  &   &  &   \\ 
\hline\hline
\end{tabular} }
\caption{\small{Atomic data for O\,III. Formulae for $\Upsilon_{ul}$ are taken from fit to data from \citet{aggarwal_1999_OIII}.  $^a$ Transition is strictly forbidden ($J = 0 \longrightarrow 0$). The validity range for the stated $\Upsilon_{ul}$ is $T_\text{e} = 5\,000-30\,000\,\text{K}$. }}\label{Tab:OIII}   
\end{table}

 \clearpage
 
\begin{table*}[ht]   
\centering \scalebox{0.75}{
\begin{tabular}{||c|c c c|c c|}
\hline\hline  
 & &  \textbf{S\,II} & &   &   \\  
 $j_{ul}$ & Transition  ($u$--$l$) & $\lambda_\text{air}$ [$\AA$] & $A_{ul}$ [s$^{-1}$] & $\Upsilon_{ul}(T_\text{e}, \text{S}^+)$, $T_4 = T_\text{e}/10^4\,\text{K}$ & $T_\text{range}/T_4$   \\  
\hline    
   &  &   &  & &   \\  
 $j_{51}$ &  $^2$P$_{3/2}^\text{o}$--$^4$S$_{3/2}^\text{o}$ & $4068.6$ & $1.92\times 10^{-1}$  &   $1.42\,T_4^{0.041 - 0.001\,\text{ln}\,T_4}$ & $0.5-3$ \\            
 $j_{52}$ &  $^2$P$_{3/2}^\text{o}$--$^2$D$_{3/2}^\text{o}$ &  $10286.7$  &  $ 1.148\times 10^{-1}$   & $2.39\,T_4^{-0.006}$ & $0.5-3$\\
 $j_{53}$ &  $^2$P$_{3/2}^\text{o}$--$^2$D$_{5/2}^\text{o}$ & $10320.5$   & $1.567\times 10^{-1}$   & $4.06\,T_4^{0.005\,\text{ln}\,T_4}$ & $0.5-3$ \\
 $j_{54}$ &  $^2$P$_{3/2}^\text{o}$--$^2$P$_{1/2}^\text{o}$ & $214.1\,\upmu\text{m}$   &  $9.14\times 10^{-7}$  & $1.80\,T_4^{0.032 - 0.001\,\text{ln}\,T_4}$ & $0.5-3$ \\
          & &  &  &  &   \\
 $j_{41}$ &  $^2$P$_{1/2}^\text{o}$--$^4$S$_{3/2}^\text{o}$ &  $4076.3$  &  $7.72\times 10^{-2}$ & $0.704\,T_4^{0.042 + 0.006\,\text{ln}\,T_4}$ & $0.5-3$\\            
 $j_{42}$ &  $^2$P$_{1/2}^\text{o}$--$^2$D$_{3/2}^\text{o}$ &   $10336.4$ &  $1.424\times 10^{-1}$   & $1.47\,T_4^{0.014  \,\text{ln}\,T_4}$  & $0.5-3$ \\
 $j_{43}$ &  $^2$P$_{1/2}^\text{o}$--$^2$D$_{5/2}^\text{o}$ &  $10370.5$  & $6.81\times 10^{-2}$  & $1.78\,T_4^{-0.012- 0.006\,\text{ln}\,T_4}$ & $0.5-4$ \\
          & &  &  &  &   \\
 $j_{31}$ &  $^2$D$_{5/2}^\text{o}$--$^4$S$_{3/2}^\text{o}$ & $6716.4$   & $2.019\times 10^{-4}$   & $3.83\,T_4^{-0.070 - 0.022\,\text{ln}\,T_4}$ & $0.5-4$\\            
 $j_{32}$ &  $^2$D$_{5/2}^\text{o}$--$^2$D$_{3/2}^\text{o}$ & $314.5\,\upmu\text{m}$    & $3.46\times 10^{-7}$   & $6.89\,T_4^{-0.103 - 0.022\,\text{ln}\,T_4}$  &  $0.5-5$\\
          & &  &  &  &    \\
 $j_{21}$ &  $^2$D$_{3/2}^\text{o}$--$^4$S$_{3/2}^\text{o}$ & $6730.8$   &  $6.84\times 10^{-4}$  & $2.56\,T_4^{-0.071 - 0.023\,\text{ln}\,T_4}$  & $0.5-4$\\
    &  &   &  &  &    \\ 
\hline\hline
\end{tabular}  }
\caption{\small{Atomic data for S\,II. Formulas for $\Upsilon_{ul}$ are taken from the fit to the data in \citet{tayal_2010_SII}.}  }\label{Tab:SII}  
\end{table*}

\begin{table*}[ht]    
\centering \scalebox{0.75}{
\begin{tabular}{||c | c c c| c c|} 
\hline\hline
  &  & \textbf{O\,I} &  &   &    \\  
 $j_{ul}$ & Transition  ($u$--$l$)  & $\lambda_\text{air}$ [$\AA$] & $A_{ul}$ [s$^{-1}$] & $\Upsilon_{ul}(T_\text{e}, \text{O}^0)$, $T_4 = T_\text{e}/10^4\,\text{K}$ & $T_\text{range}/T_4$ \\  
\hline    
   &  &   & & &  \\  
 $j_{51}$ & $^1$S$_0$--$^3$P$_2$ & $2958.4$ & $2.42\times 10^{-4}$  & $\left(0.032666\,T_4^{1.50}\right)/\left(1 + 0.80\,T_4^{1.125} \right)$ & $0.5-10$ \\            
 $j_{52}$ & $^1$S$_0$--$^3$P$_1$ & $2972.3$ & $7.54\times 10^{-2}$   & $\left(0.019600\,T_4^{1.50}\right)/\left(1 + 0.80\,T_4^{1.125} \right)$ &  $0.5-10$ \\
 $j_{53}$ & $^1$S$_0$--$^3$P$_0$ &  --$^a$  & 0.0$^a$  & $\left(0.0065333\,T_4^{1.50}\right)/\left(1 + 0.80\,T_4^{1.125} \right)$ &  $0.5-10$\\
 $j_{54}$ & $^1$S$_0$--$^1$D$_2$ & $5577.3$ & $1.26$ & $\left(0.116\,T_4^{0.53}\right)/\left(1 + 0.111\,T_4^{0.160} \right)$  &  $0.5-10$ \\
          & &  & & &   \\
 $j_{41}$ & $^1$D$_2$--$^3$P$_2$ & $6300.3$ & $5.6511\times 10^{-3}$   & $\left(0.237777\,T_4^{1.43}\right)/\left(1 + 0.605\,T_4^{1.105} \right)$ &  $0.5-10$ \\            
 $j_{42}$ & $^1$D$_2$--$^3$P$_1$ & $6363.8$ & $1.82339\times 10^{-3}$  & $\left(0.1426666\,T_4^{1.43}\right)/\left(1 + 0.605\,T_4^{1.105} \right)$ &  $0.5-10$ \\
 $j_{43}$ & $^1$D$_2$--$^3$P$_0$ & $6391.7$ & $8.60\times 10^{-7}$   & $\left(0.047555\,T_4^{1.43}\right)/\left(1 + 0.605\,T_4^{1.105} \right)$  &  $0.5-10$\\
          & &  &  & &  \\
 $j_{31}$ & $^3$P$_0$--$^3$P$_2$ &   $44.1\,\upmu\text{m}$ &  $1.34\times 10^{-10}$  & $0.0136\,T_4^{0.61} + 0.0186\,T_4^{1.49}$ & $ 0.05-1 $ \\    
  &   &     &   &  $0.0321\,T_4^{1.0}$ & $ > 1  $ \\            
 $j_{32}$ & $^3$P$_0$--$^3$P$_1$ &   $145.5\,\upmu\text{m}$ &  $1.75\times 10^{-5}$   & $0.00166\,T_4^{0.71} + 0.0288\,T_4^{1.97}$  & $ 0.05-1 $   \\
   &  &     &    &   $0.0283\,T_4^{1.5}  $  & $  > 1  $  \\
          & &  &   & &   \\
 $j_{21}$ & $^3$P$_1$--$^3$P$_2$ &   $63.2\,\upmu\text{m}$ &  $8.91\times 10^{-5}$   & $0.041\,T_4^{0.69} + 0.064\,T_4^{1.72}$ & $ 0.05-1 $  \\
  &   &    &    &     $0.106\,T_4^{1.1}  $ &  $ >1  $ \\
    &  &   & & &    \\ 
\hline\hline
\end{tabular} }
\caption{\small{Atomic data for O\,I. Formulae for $\Upsilon_{ul}$ are taken from fit to data from \citet{Pequignot_1990_OI}.  $^a$ Transition is strictly forbidden ($J = 0 \longrightarrow 0$ ).}}\label{Tab:OI}  
\end{table*}

\begin{table*}[ht]   
\centering \scalebox{0.75}{
\begin{tabular}{||c|c c c| c c |}
\hline\hline
  & &  \textbf{S\,III} & &   &  \\  
  $j_{ul}$ & Transition ($u$--$l$)  & $\lambda_\text{air}$ [A] & $A_{ul}$ [s$^{-1}$]  & $\Upsilon_{ul}(T_e, \text{S}^{2+})$, $T_4 = T_e/10^4\,\text{K}$ & $T_\text{range}/T_4$ \\  
\hline    
   &  &   & & &     \\  
 $j_{51}$ & $^1$S$_0$--$^3$P$_0$ & --$^a$ & 0.0$^a$   &   $0.131\,T_4^{0.043 + 0.031\,\text{ln}\,T_4}$ & $0.5-4$ \\            
 $j_{52}$ & $^1$S$_0$--$^3$P$_1$ & 3721.6   & $6.61\times 10^{-1}$  &  $0.3932\,T_4^{0.04237 + 0.03498\,\text{ln}\,T_4}$ & $0.5-4$  \\
 $j_{53}$ & $^1$S$_0$--$^3$P$_2$ & 3797.2   &  $8.82\times 10^{-3}$  & $0.6548\,T_4^{0.04257 + 0.03514\,\text{ln}\,T_4}$ & $0.5-4$ \\
 $j_{54}$ & $^1$S$_0$--$^1$D$_2$ & 6312.1   &  2.08  & $1.399\,T_4^{0.1418 + 0.03214\,\text{ln}\,T_4}$  & $0.5-4$  \\
          & &  &   & &   $0.5-4$ \\
 $j_{41}$ &  $^1$D$_2$--$^3$P$_0$ & 8829.4  & $5.25\times 10^{-6}$  & $0.7799\,T_4^{-0.01658 + 0.0203\,\text{ln}\,T_4}$  & $0.5-4$ \\            
 $j_{42}$ &  $^1$D$_2$--$^3$P$_1$ & 9068.6  & $1.85394\times 10^{-2}$   & $2.341\,T_4^{-0.01686 + 0.02001\,\text{ln}\,T_4}$ &  $0.5-4$ \\
 $j_{43}$ &  $^1$D$_2$--$^3$P$_2$ & 9530.6  & $4.8009\times 10^{-2}$   &  $3.898\,T_4^{-0.01676 + 0.02021\,\text{ln}\,T_4}$  & $0.5-4$  \\
          & &  &   & &   \\
 $j_{31}$ &   $^3$P$_2$--$^3$P$_0$ & $12.0\,\mu\text{m}$   & $4.11\times 10^{-8}$    &  $1.310\,T_4^{-0.065 + 0.052\,\text{ln}\,T_4}$ & $0.5-4$ \\            
 $j_{32}$ &  $^3$P$_2$--$^3$P$_1$ &  $18.7\,\mu\text{m}$  &  $2.06\times 10^{-3}$    & $7.845\,T_4^{-0.1633 - 0.007318\,\text{ln}\,T_4}$ & $0.5-10$ \\
          & &  &  & &     \\
 $j_{21}$ &  $^3$P$_1$--$^3$P$_0$ &  $33.5\,\mu\text{m}$  &  $4.79\times 10^{-4}$   & $3.916\,T_4^{-0.2281 -0.04135\,\text{ln}\,T_4}$ & $0.5-10$ \\
    &  &   &  & &  \\ 
\hline\hline
\end{tabular} }
\caption{\small{Atomic data for S\,III. Formulae for $\Upsilon_{ul}$ are taken from fit to data from \citet{tayal_1999_SIII}.  $^a$ Transition is strictly forbidden ($J = 0 \longrightarrow 0$).}}\label{Tab:SIII}   
\end{table*}      

\clearpage
\onecolumn

\section*{Appendix C - Extended BE99 diagrams}\label{appendix:D}

\nopagebreak
\begin{figure*}[h] 
\centering
\subfloat[\small{$A_V =0.0$, $t=10^6\,\text{s}$}]{\includegraphics[trim=0 0 0 0, clip, width=0.32  \textwidth]{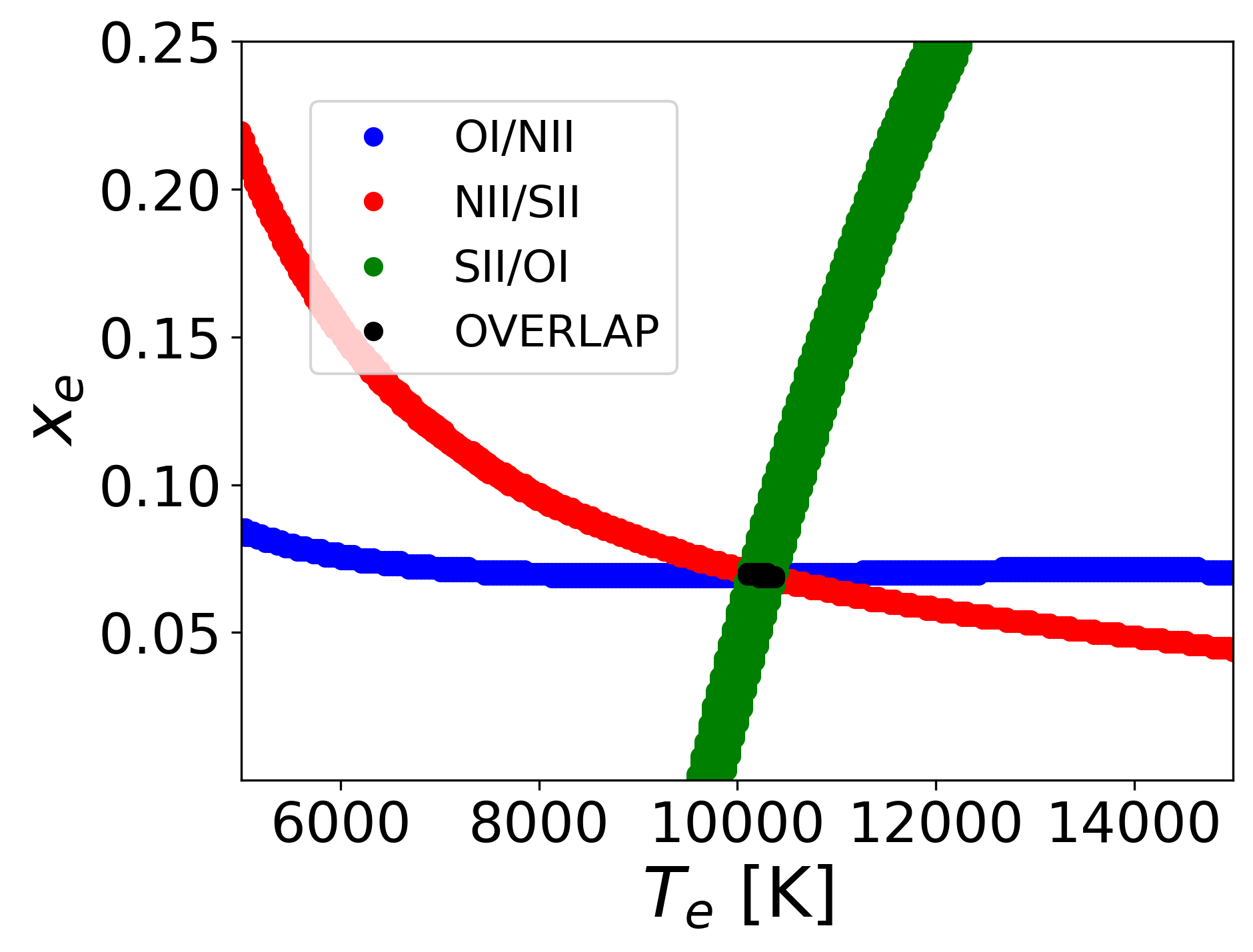}} 
\hspace{1em}   
\subfloat[\small{$A_V =0.3$, $t=10^6\,\text{s}$}]{\includegraphics[trim=0 0 0 0, clip, width=0.31  \textwidth]{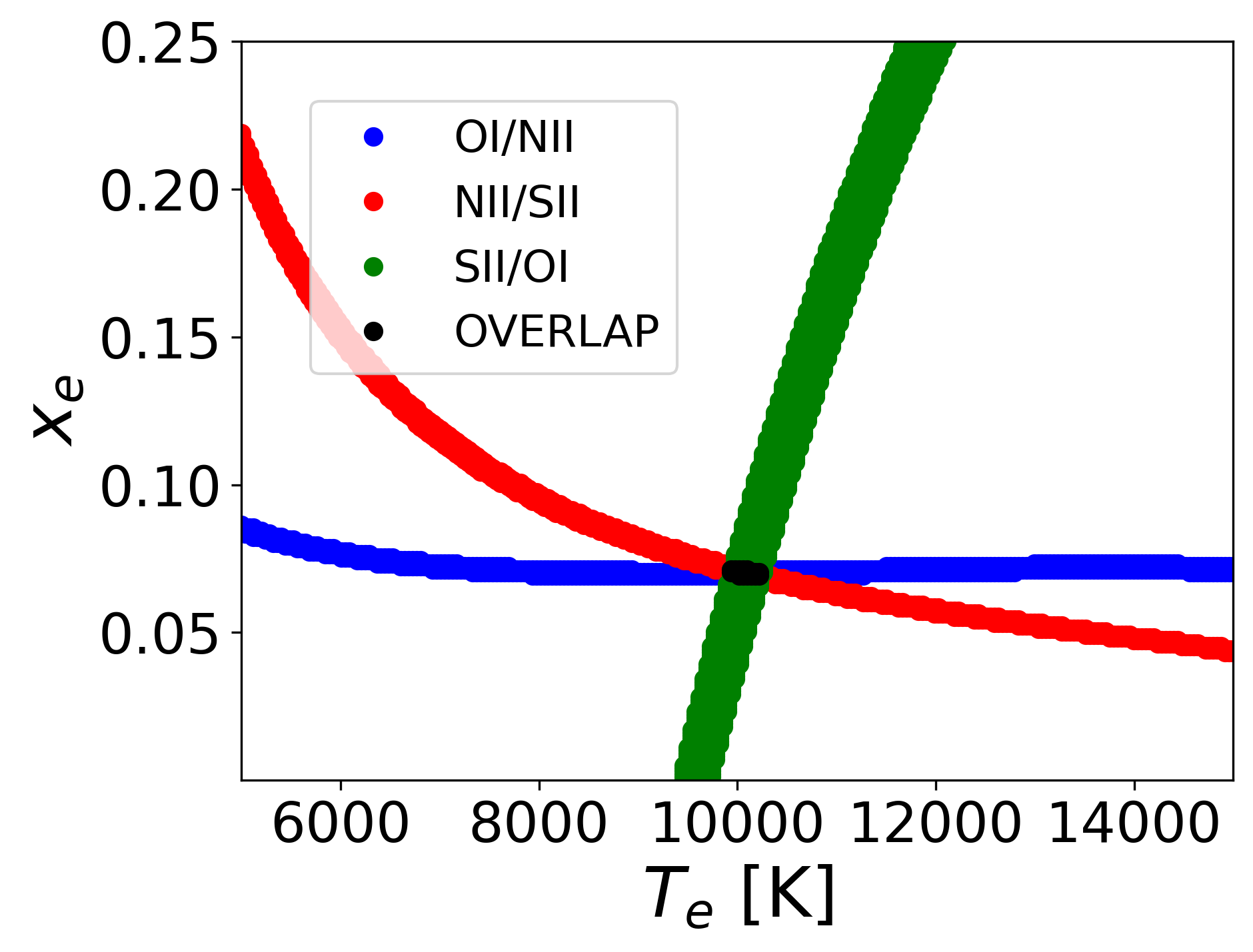}}
\hspace{1em} 
\subfloat[\small{$A_V =0.6$, $t=10^6\,\text{s}$}]{\includegraphics[trim=0 0 0 0, clip, width=0.31\textwidth]{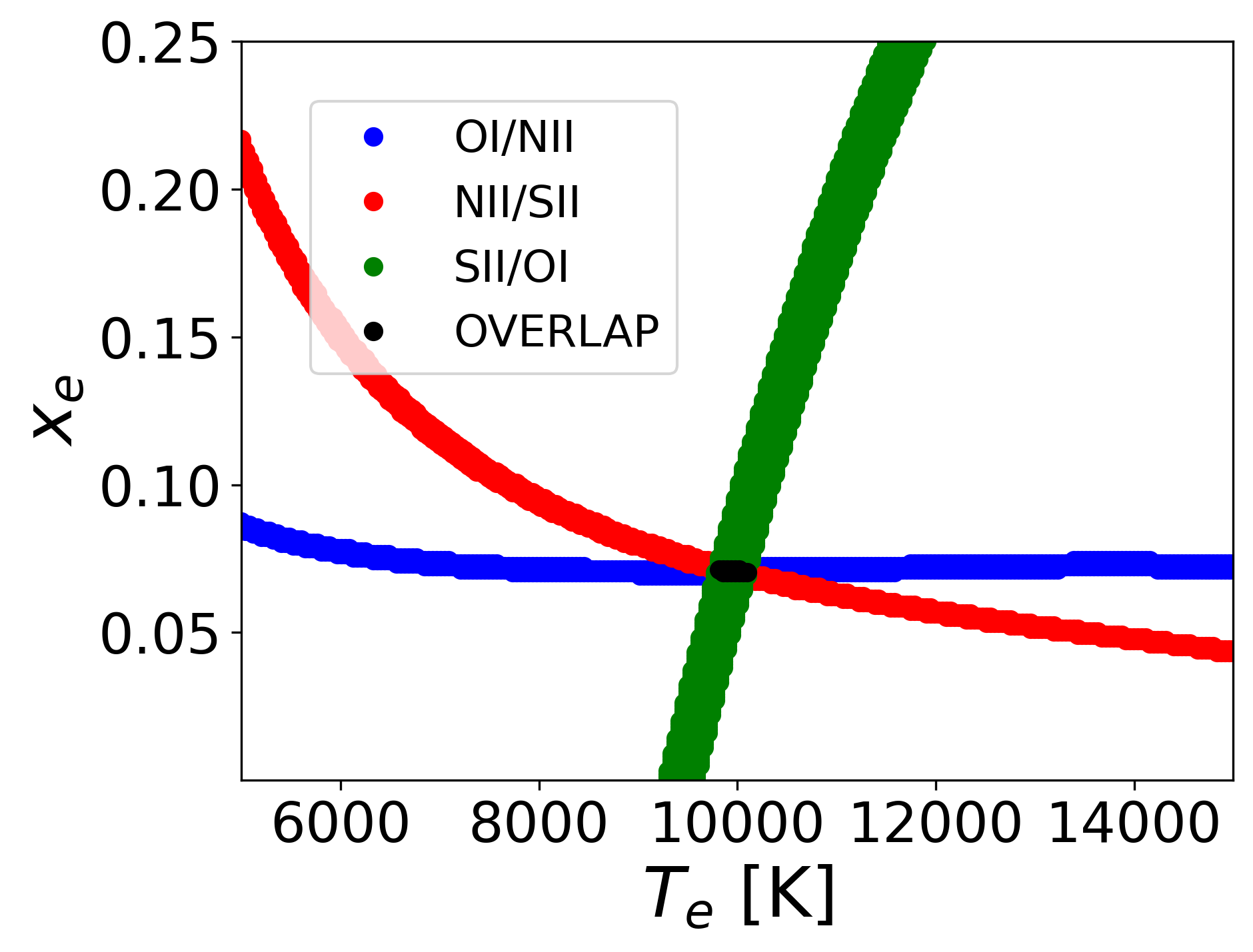}} 
\hspace{1em} \\
\subfloat[\small{$A_V =0.0$, $t=10^8\,\text{s}$}]{\includegraphics[trim=0 0 0 0, clip, width=0.31 \textwidth]{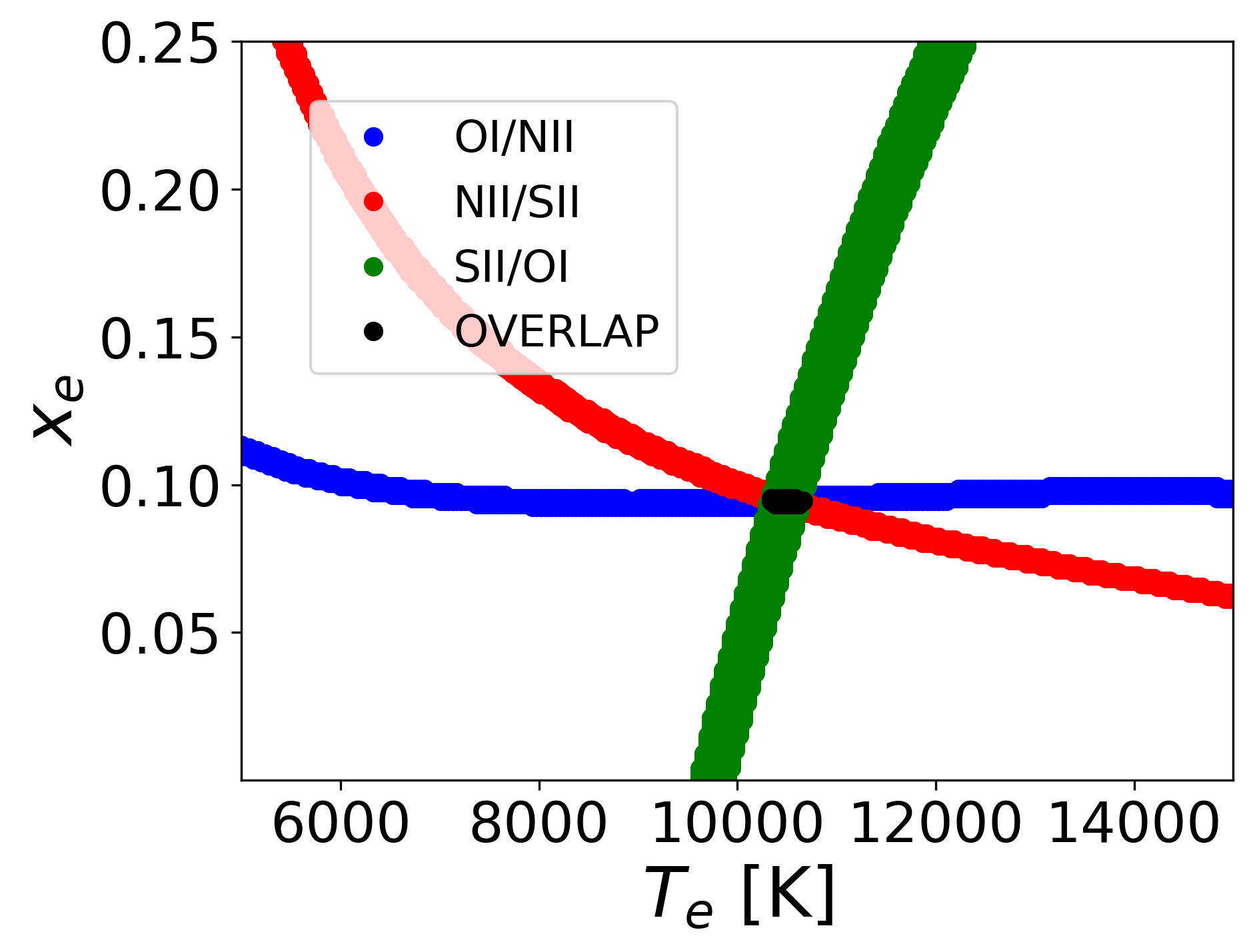}} 
\hspace{1em}   
\subfloat[\small{$A_V =0.3$, $t=10^8\,\text{s}$}]{\includegraphics[trim=0 0 0 0, clip, width=0.31 \textwidth]{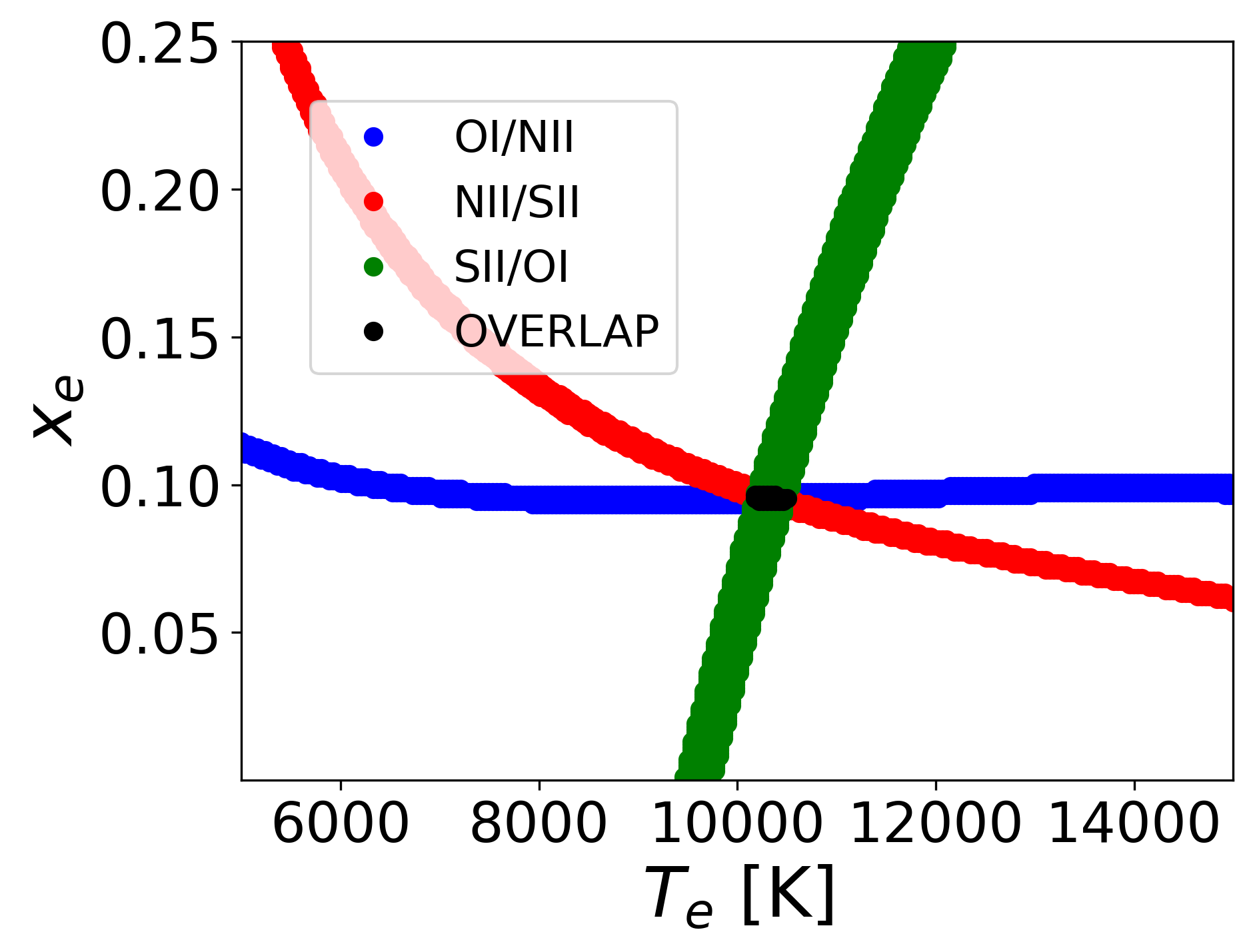}}
\hspace{1em} 
\subfloat[\small{$A_V =0.6$, $t=10^8\,\text{s}$}]{\includegraphics[trim=0 0 0 0, clip, width=0.31 \textwidth]{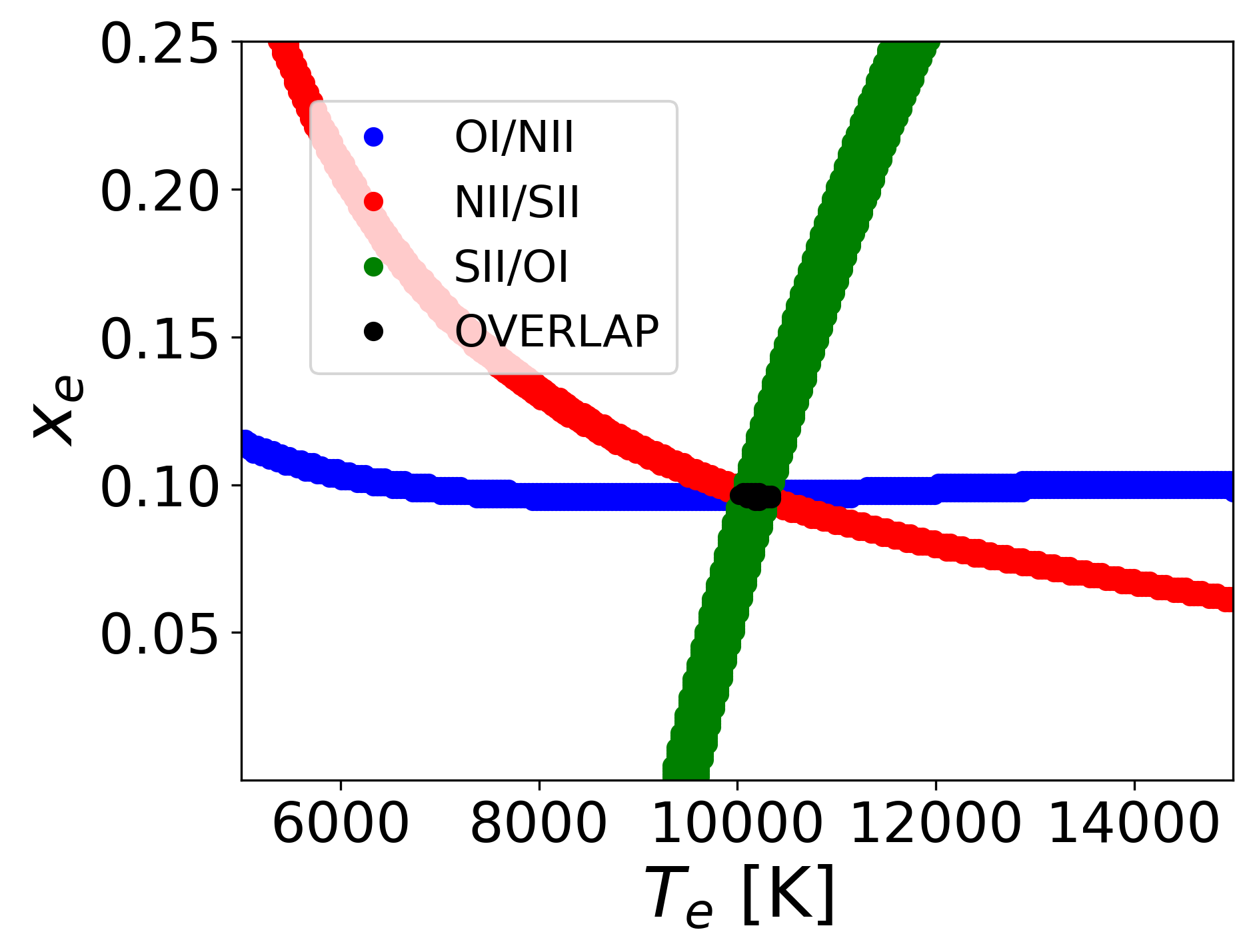}} 
\hspace{1em}   \\
\subfloat[\small{$A_V =0.0$, $t=10^{10}\,\text{s}$}]{\includegraphics[trim=0 0 0 0, clip, width=0.31 \textwidth]{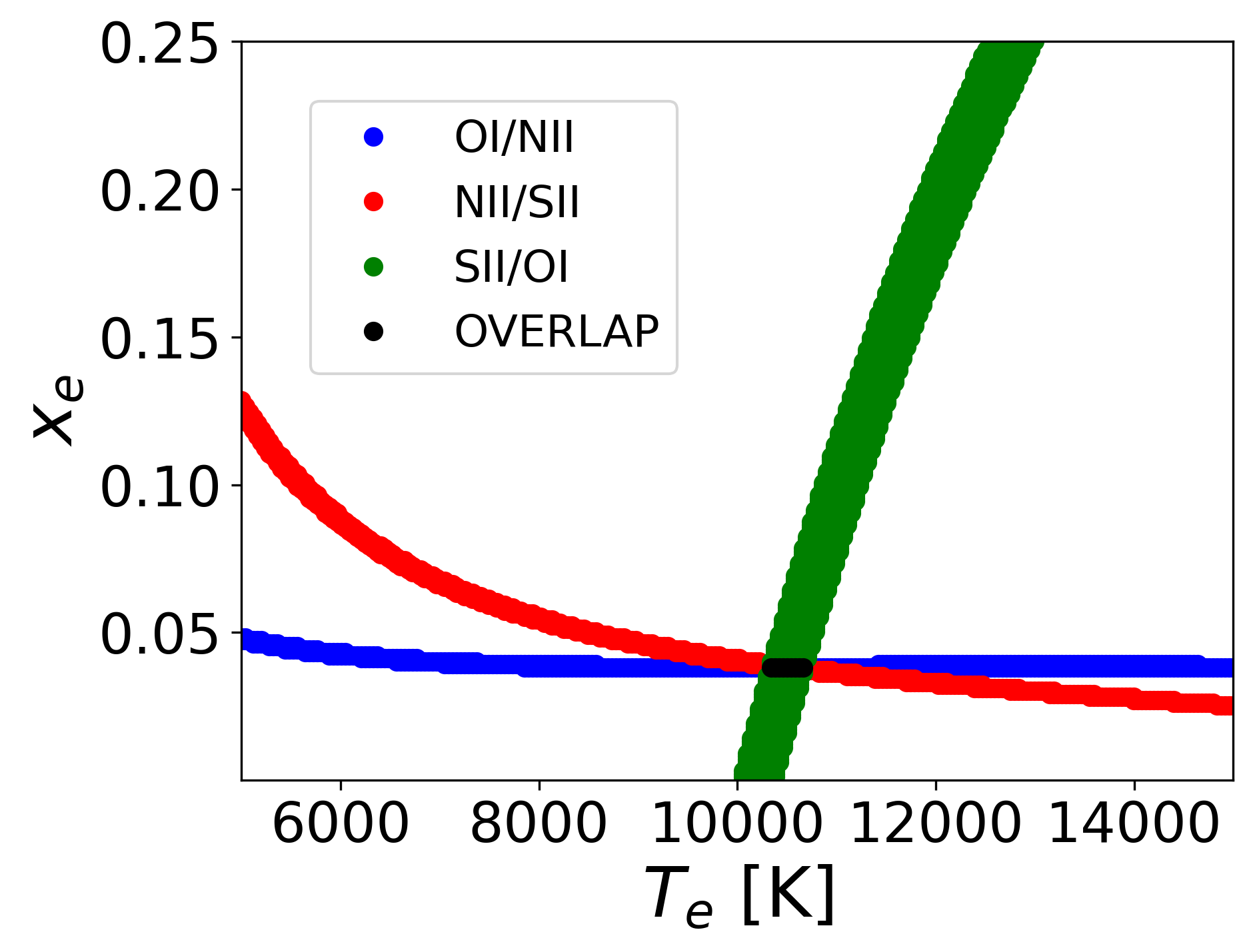}} 
\hspace{1em}   
\subfloat[\small{$A_V =0.3$, $t=10^{10}\,\text{s}$}]{\includegraphics[trim=0 0 0 0, clip, width=0.31  \textwidth]{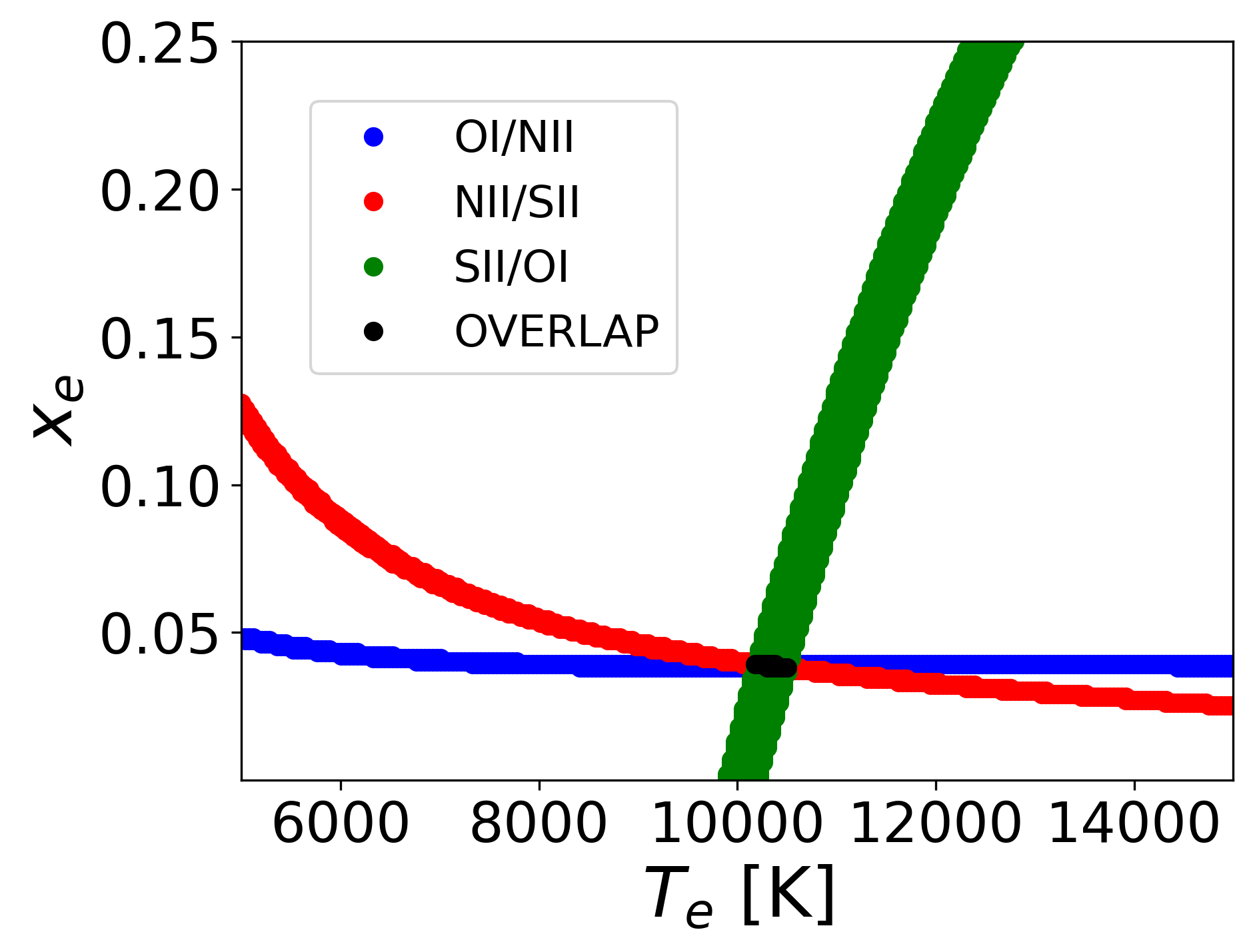}} 
\hspace{1em} 
\subfloat[\small{$A_V =0.6$, $t=10^{10}\,\text{s}$}]{\includegraphics[trim=0 0 0 0, clip, width=0.31  \textwidth]{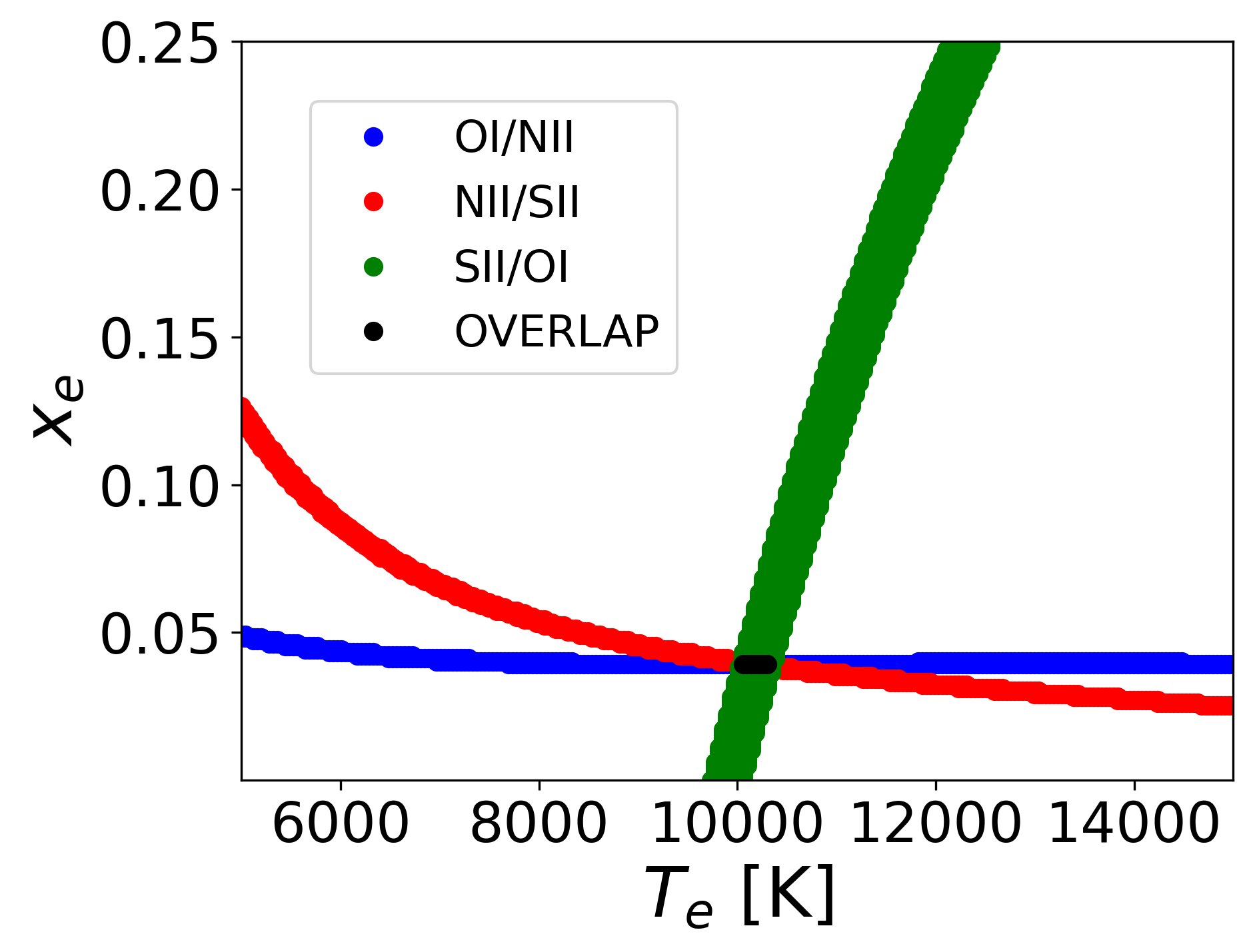}} 
\hspace{1em}   \\
\subfloat[\small{$A_V =0.0$, $t=10^{12}\,\text{s}$}]{\includegraphics[trim=0 0 0 0, clip, width=0.31  \textwidth]{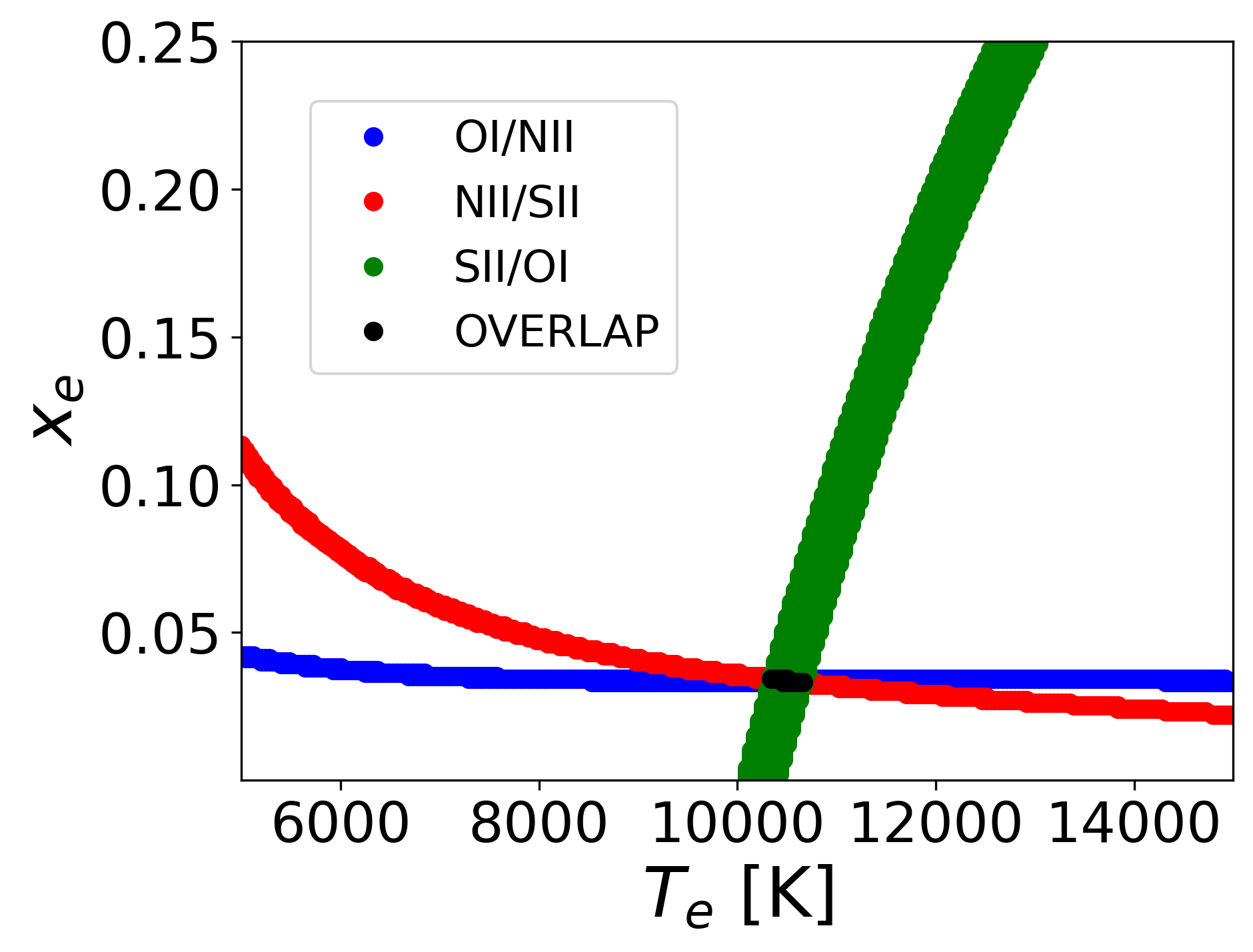}}
\hspace{1em} 
\subfloat[\small{$A_V =0.3$, $t=10^{12}\,\text{s}$}]{\includegraphics[trim=0 0 0 0, clip, width=0.31\textwidth]{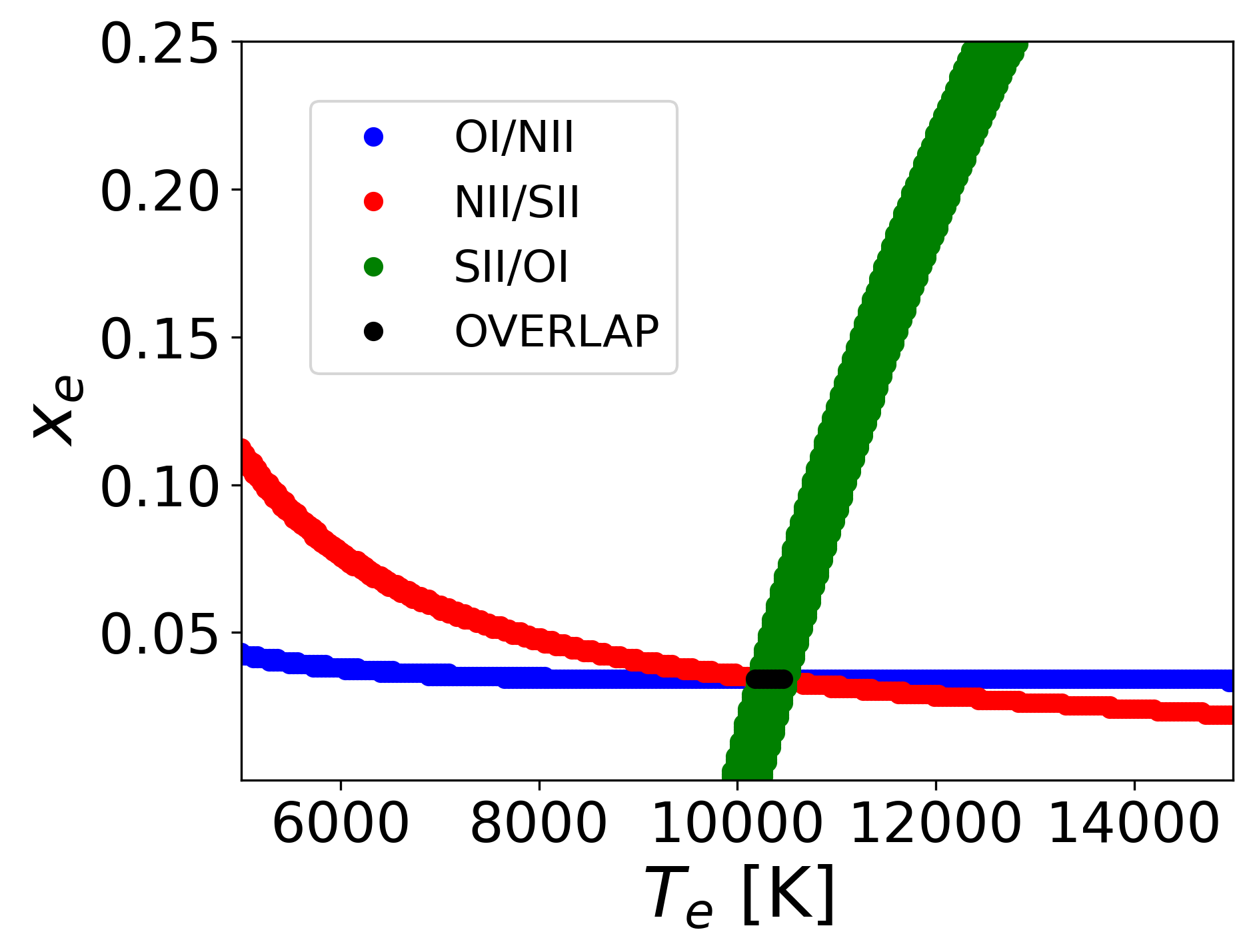}} 
\hspace{1em}   
\subfloat[\small{$A_V =0.6$, $t=10^{12}\,\text{s}$}]{\includegraphics[trim=0 0 0 0, clip, width=0.31 \textwidth]{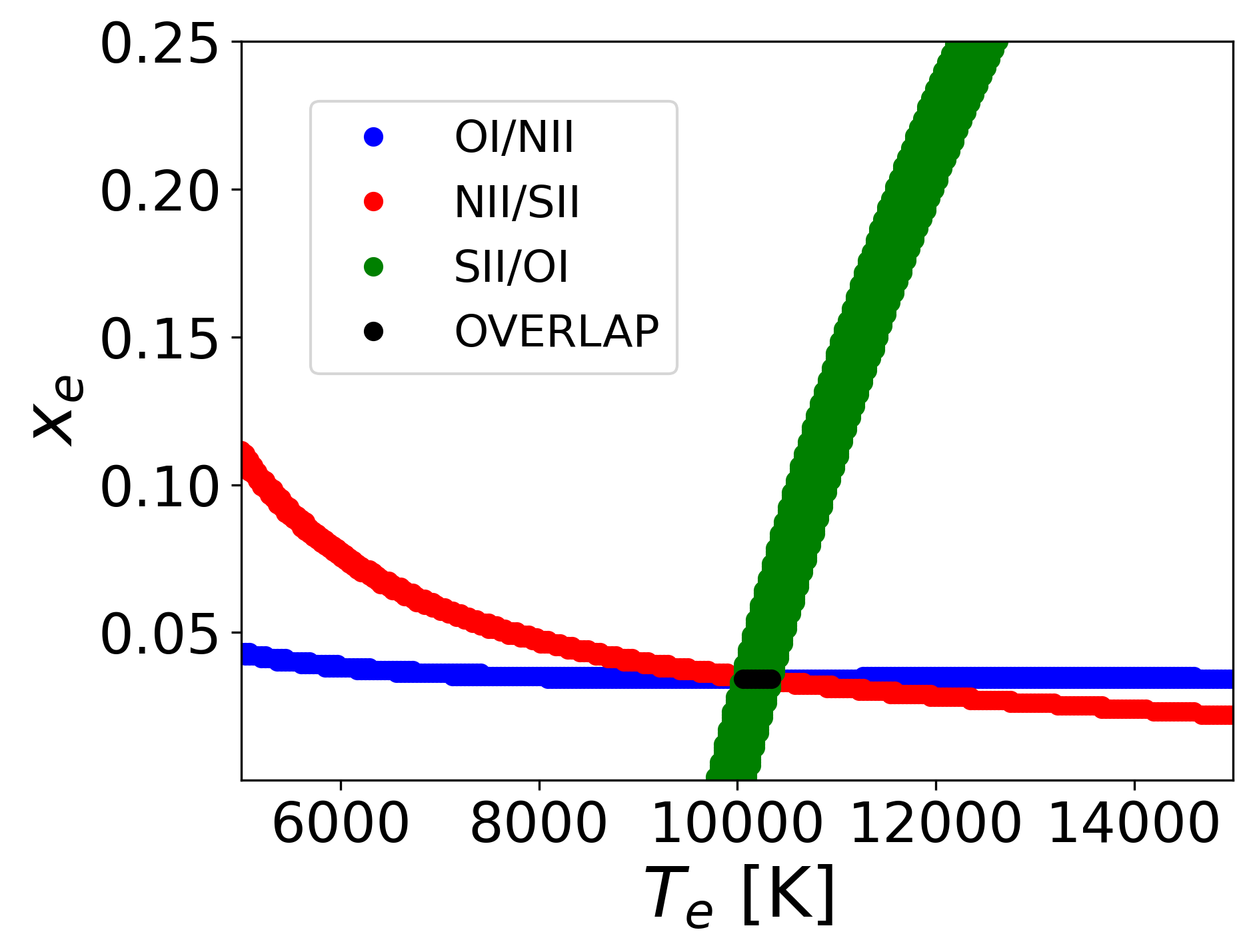}} 
\hspace{1em}  
\caption{\small{The BE99 diagrams in the time domain for the model in Fig.\,\ref{fig:ionisation_ratios}. We integrated the model until  $t =$ $10^6\,\text{s}$, $10^8\,\text{s}$, $10^{10}\,\text{s}$, and $10^{12}\,\text{s}$ and reddened the synthetic line fluxes with three extinction values ($A_V= 0.0, 0.3, 0.6\,\text{mag}$). These line fluxes are then used as input for constructing the shown BE99 diagrams. At each time and for each assumed extinction value the three stripes of line ratios (red, blue, green) meet in one location in the ($x_e$-$T_e$) diagram (black region). Even before reaching the reaction equilibrium ($\tau_\text{EQ} \approx 2\times 10^{10}\,\text{s}$) the BE99 method converges (see Fig.\,\ref{fig:BE_time_scale}).}}\label{fig:all_BE_diagrams}
\end{figure*} 
 
\clearpage
 
\begin{figure*}[h] 
\centering
\subfloat[\small{$A_V =0.0$, $t=10^6\,\text{s}$}]{\includegraphics[trim=0 0 0 0, clip, width=0.31  \textwidth]{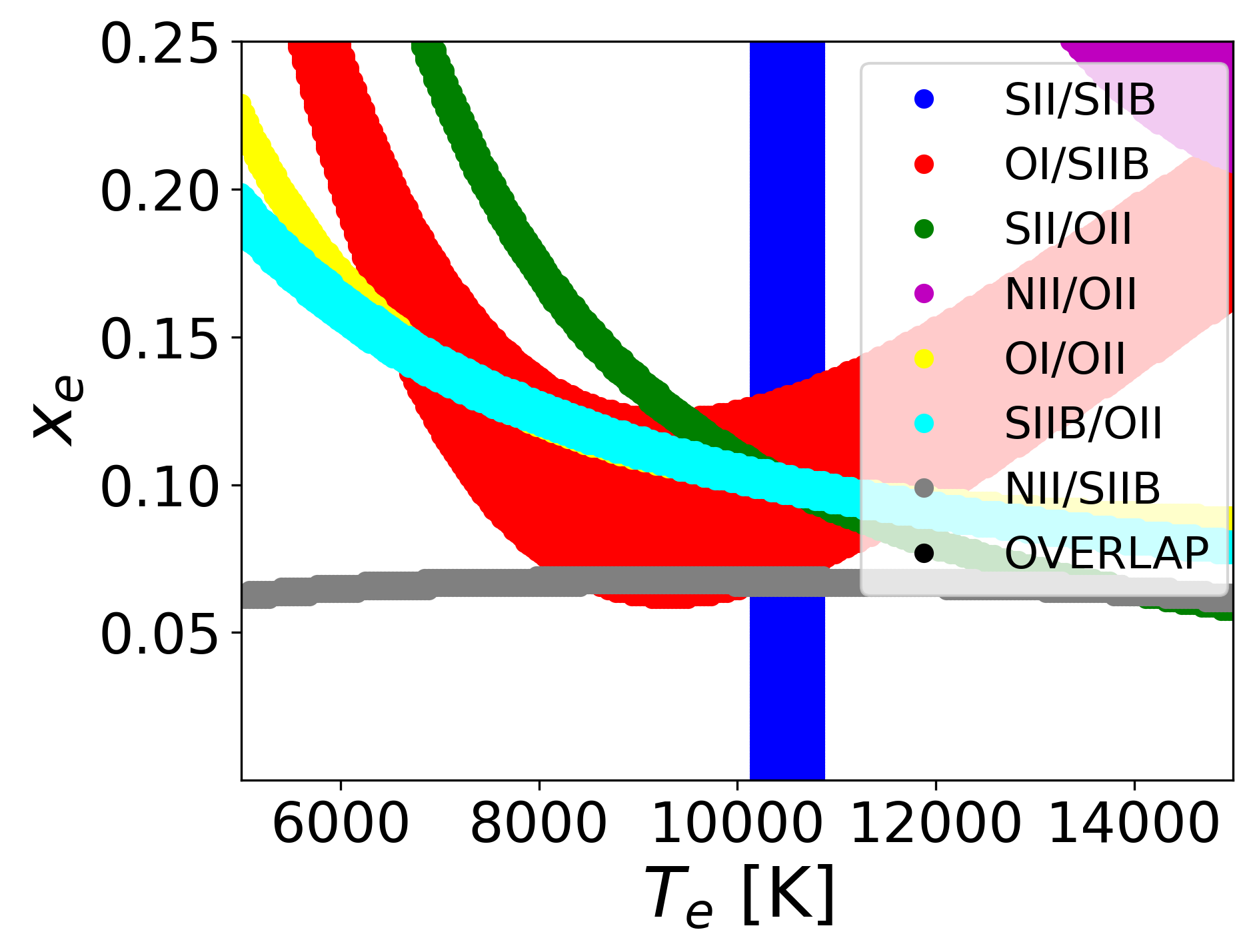}} 
\hspace{1em}   
\subfloat[\small{$A_V =0.3$, $t=10^6\,\text{s}$}]{\includegraphics[trim=0 0 0 0, clip, width=0.31  \textwidth]{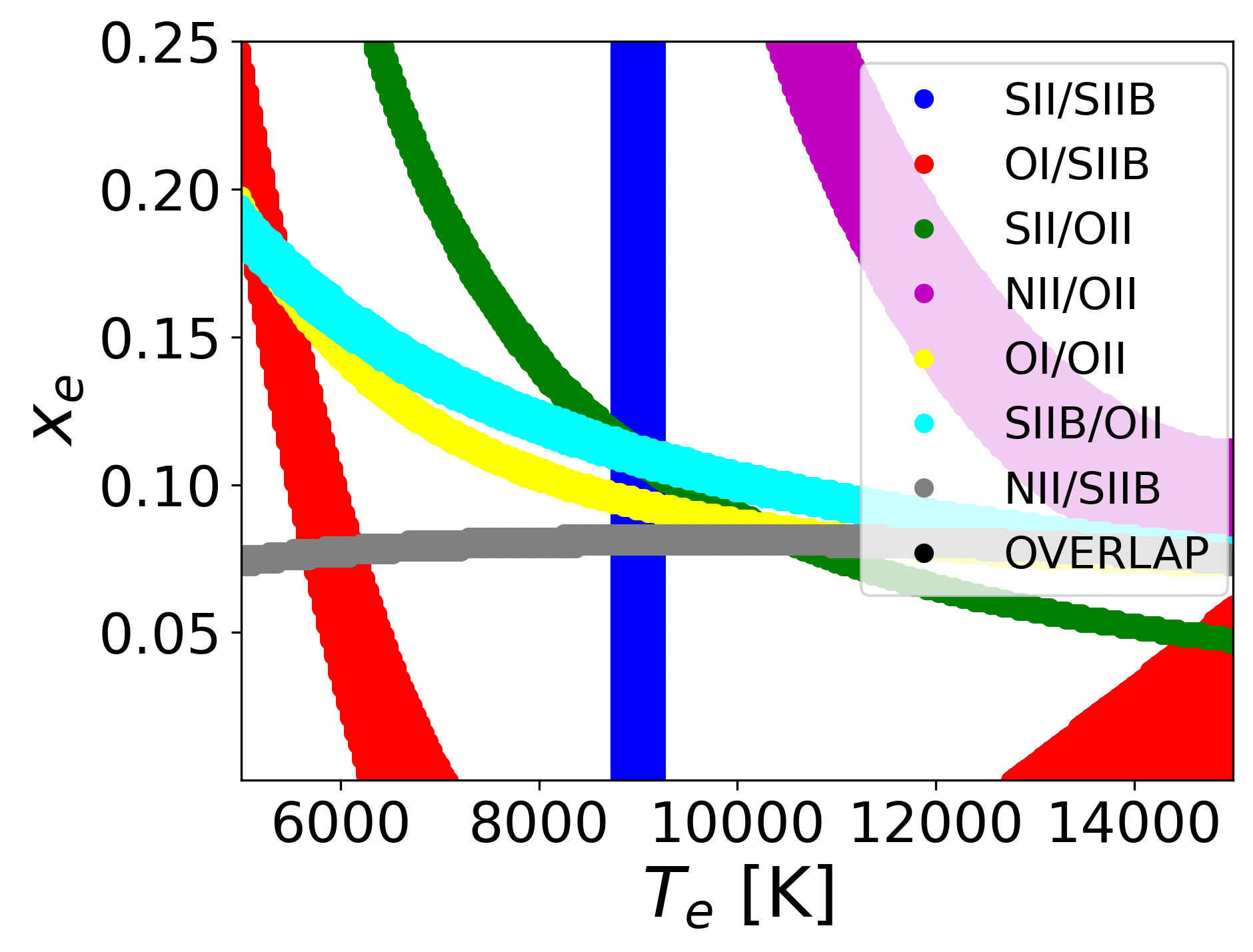}}
\hspace{1em} 
\subfloat[\small{$A_V =0.6$, $t=10^6\,\text{s}$}]{\includegraphics[trim=0 0 0 0, clip, width=0.31 \textwidth]{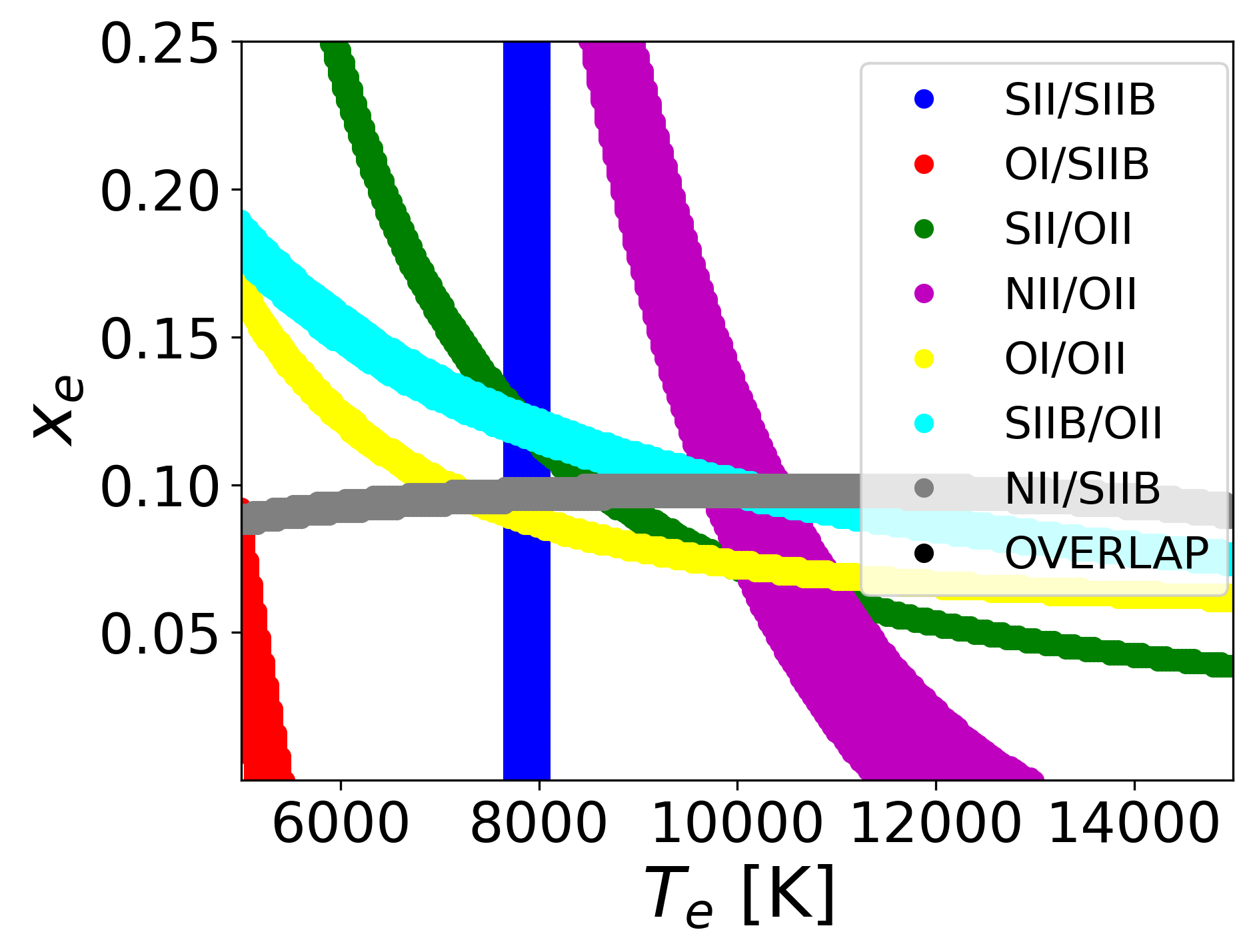}} 
\hspace{1em} \\
\subfloat[\small{$A_V =0.0$, $t=10^8\,\text{s}$}]{\includegraphics[trim=0 0 0 0, clip, width=0.31 \textwidth]{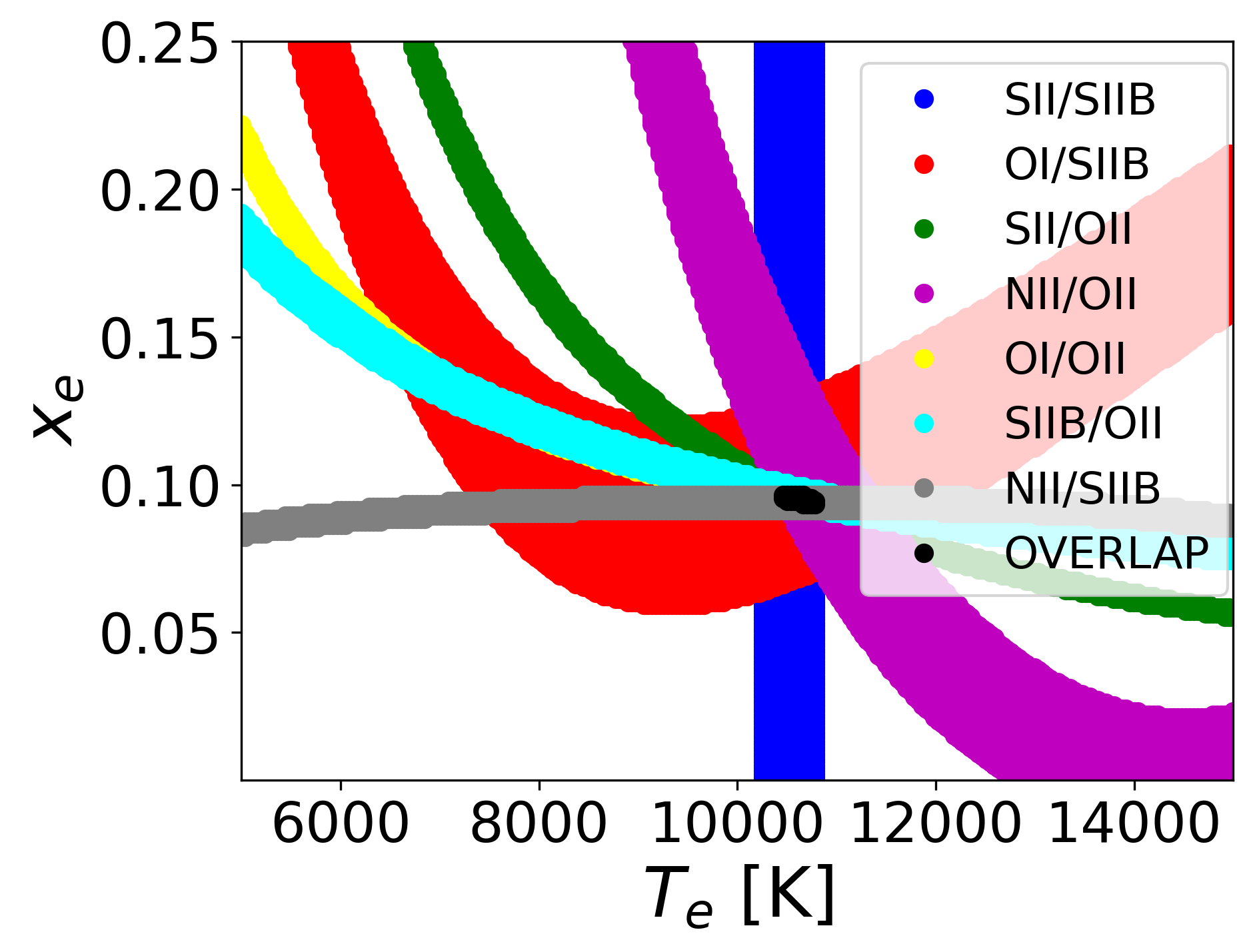}} 
\hspace{1em}   
\subfloat[\small{$A_V =0.3$, $t=10^8\,\text{s}$}]{\includegraphics[trim=0 0 0 0, clip, width=0.31 \textwidth]{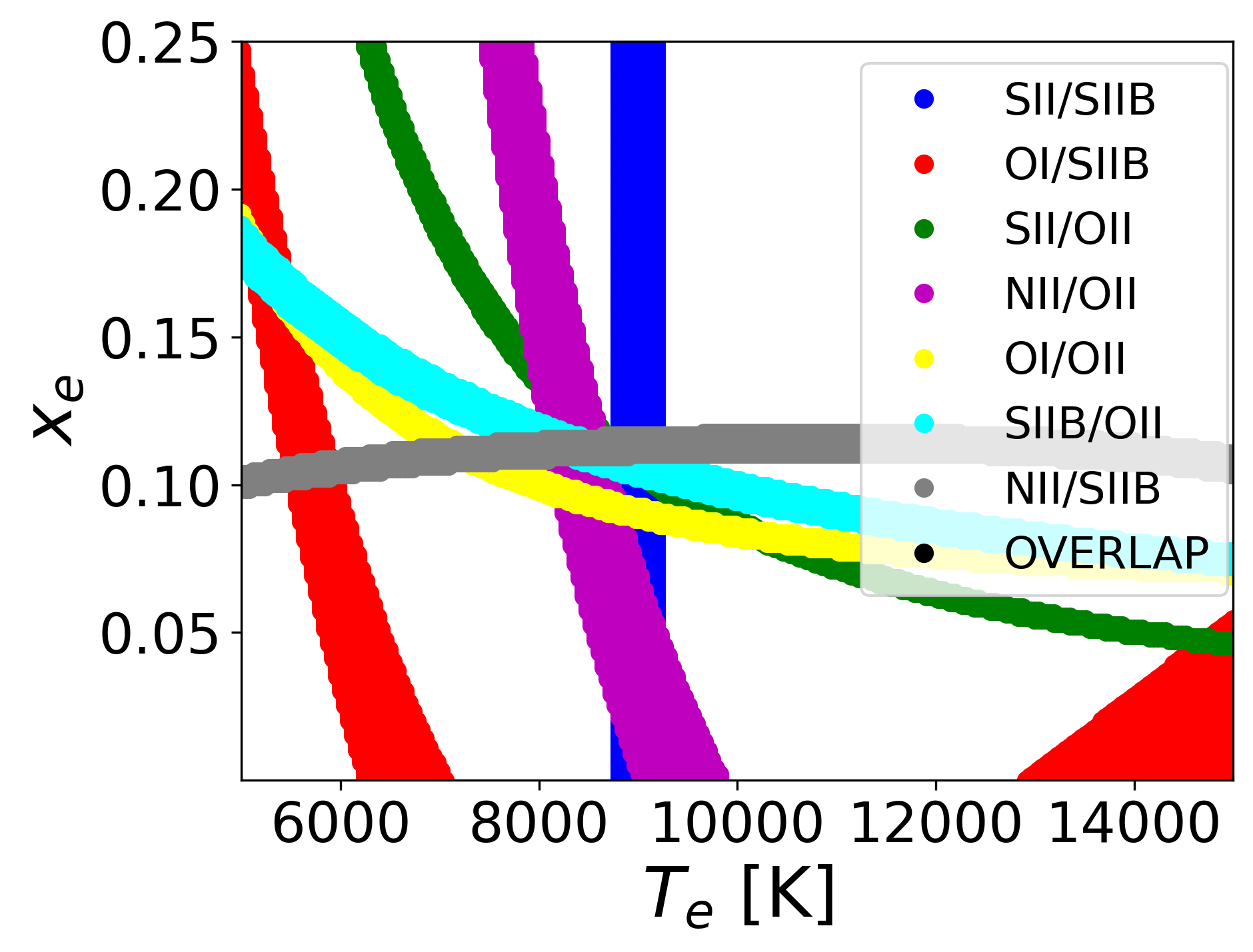}}
\hspace{1em} 
\subfloat[\small{$A_V =0.6$, $t=10^8\,\text{s}$}]{\includegraphics[trim=0 0 0 0, clip, width=0.31 \textwidth]{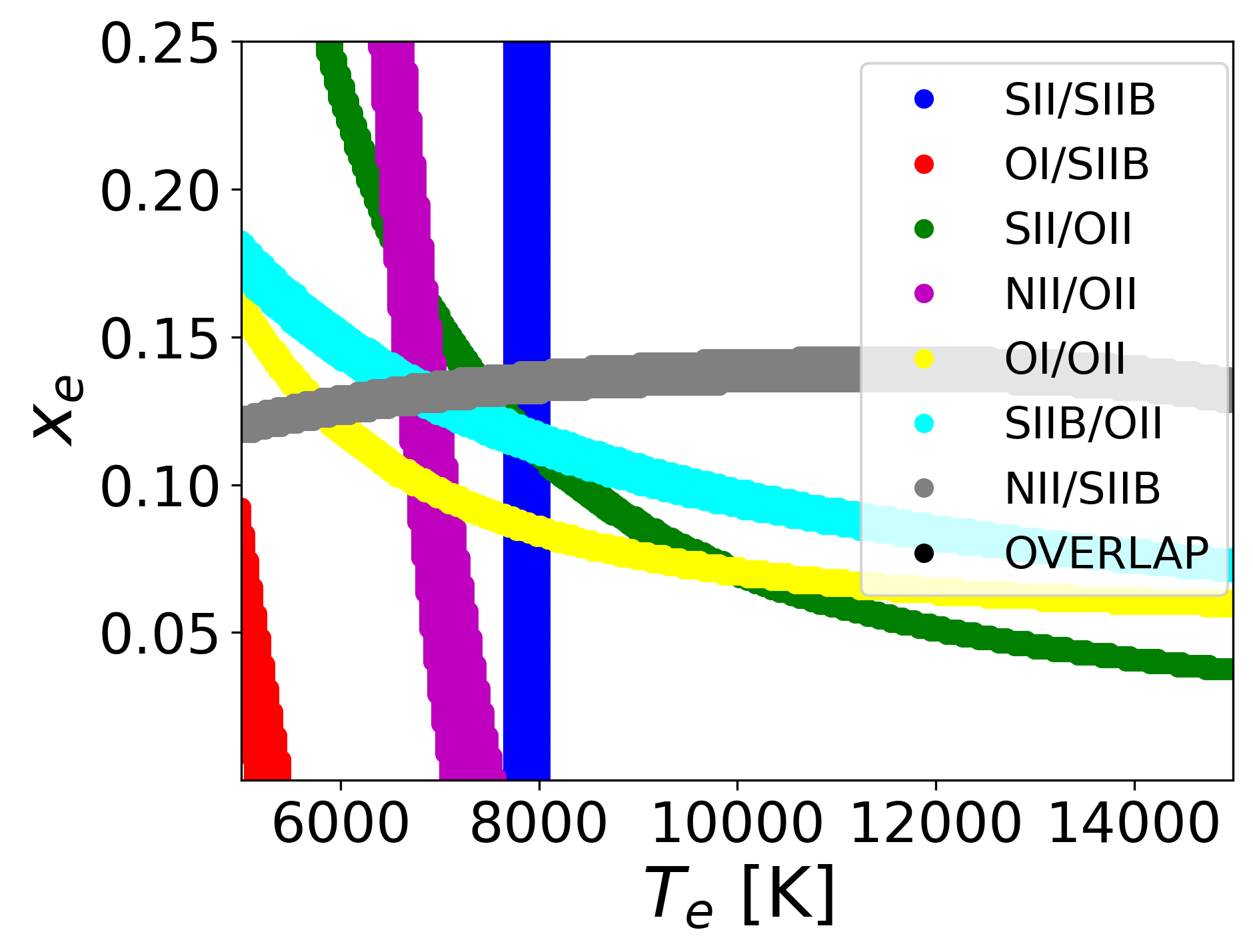}} 
\hspace{1em} \\
\subfloat[\small{$A_V =0.0$, $t=10^{10}\,\text{s}$}]{\includegraphics[trim=0 0 0 0, clip, width=0.31 \textwidth]{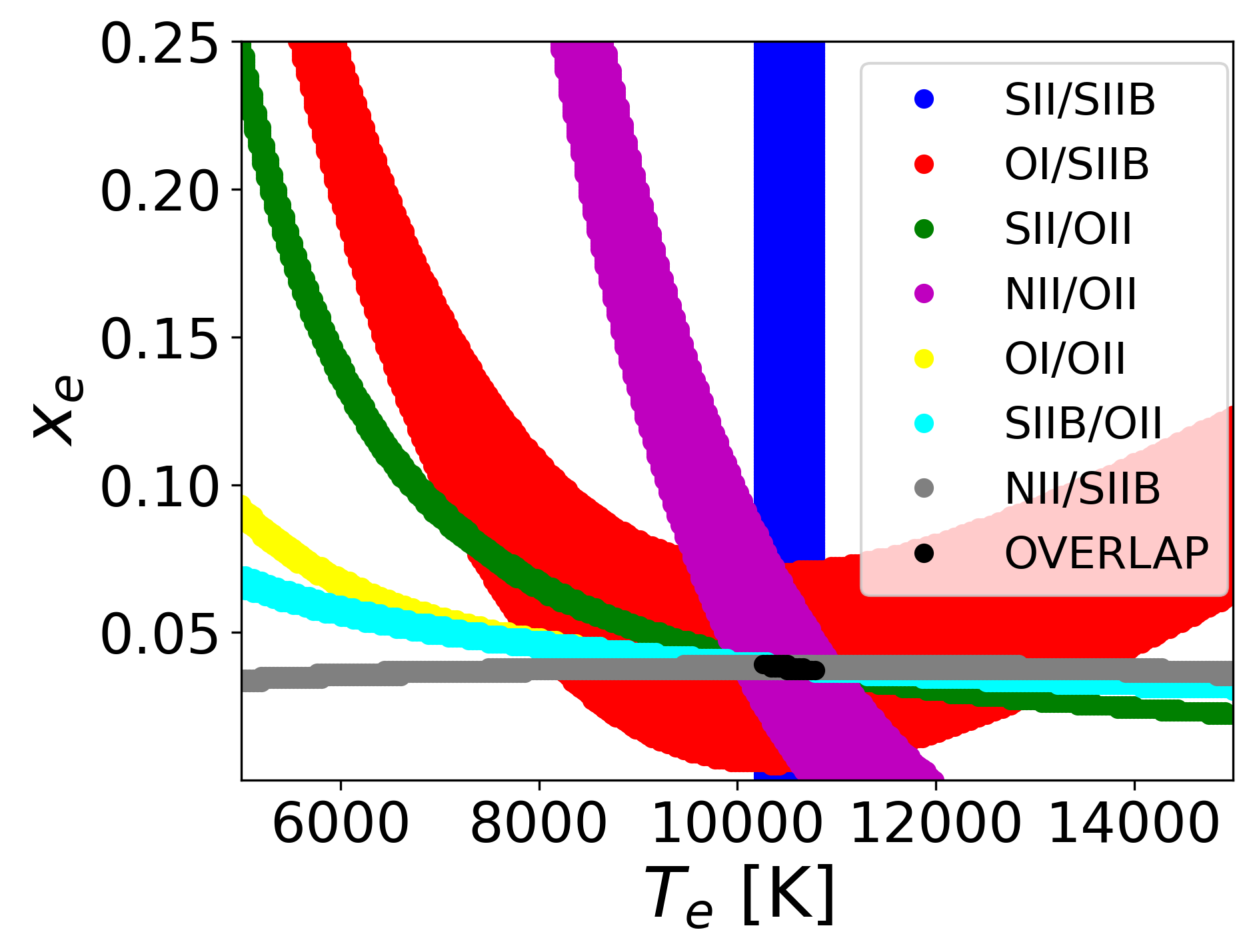}} 
\hspace{1em}   
\subfloat[\small{$A_V =0.3$, $t=10^{10}\,\text{s}$}]{\includegraphics[trim=0 0 0 0, clip, width=0.31  \textwidth]{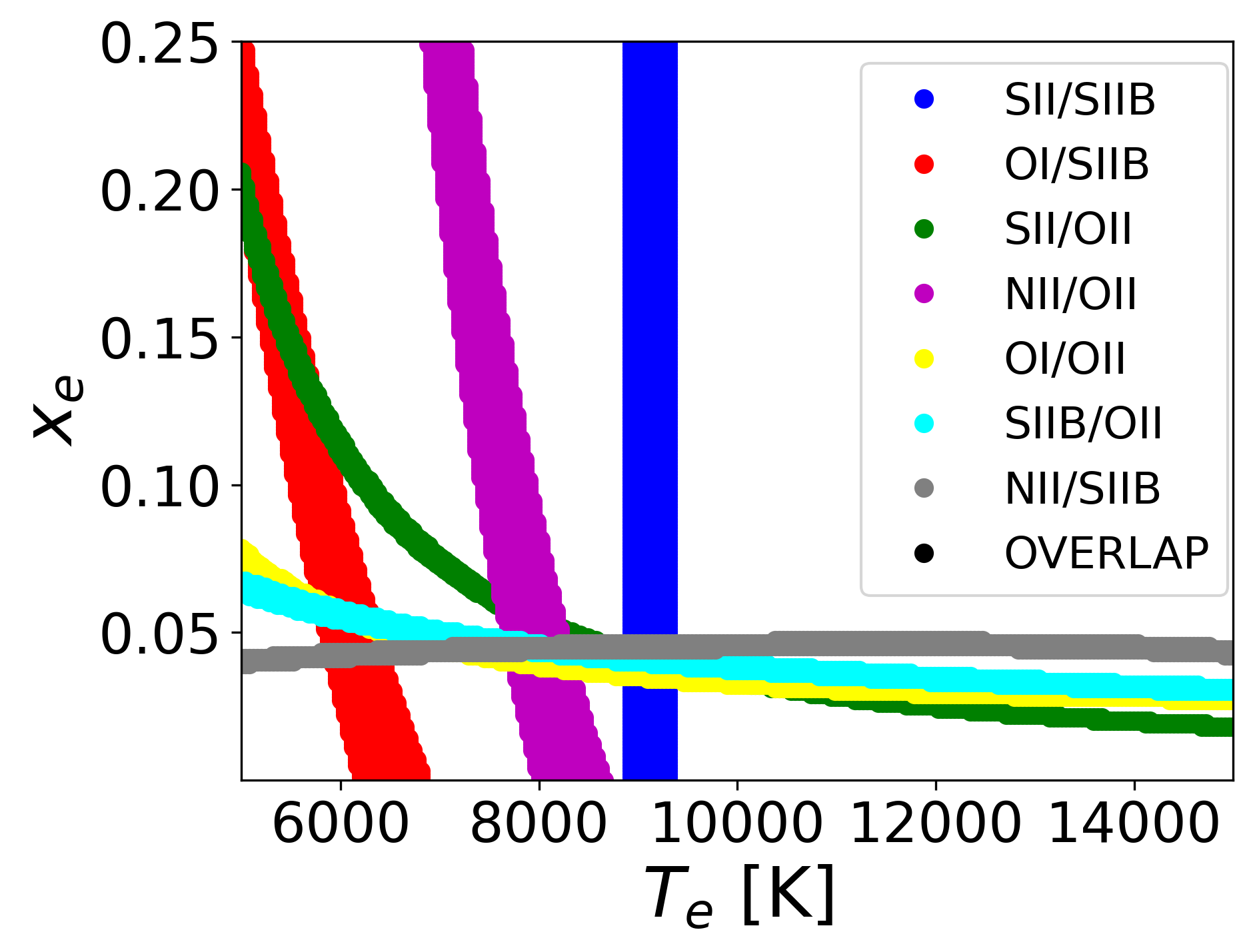}} 
\hspace{1em}   
\subfloat[\small{$A_V =0.6$, $t=10^{10}\,\text{s}$}]{\includegraphics[trim=0 0 0 0, clip, width=0.31  \textwidth]{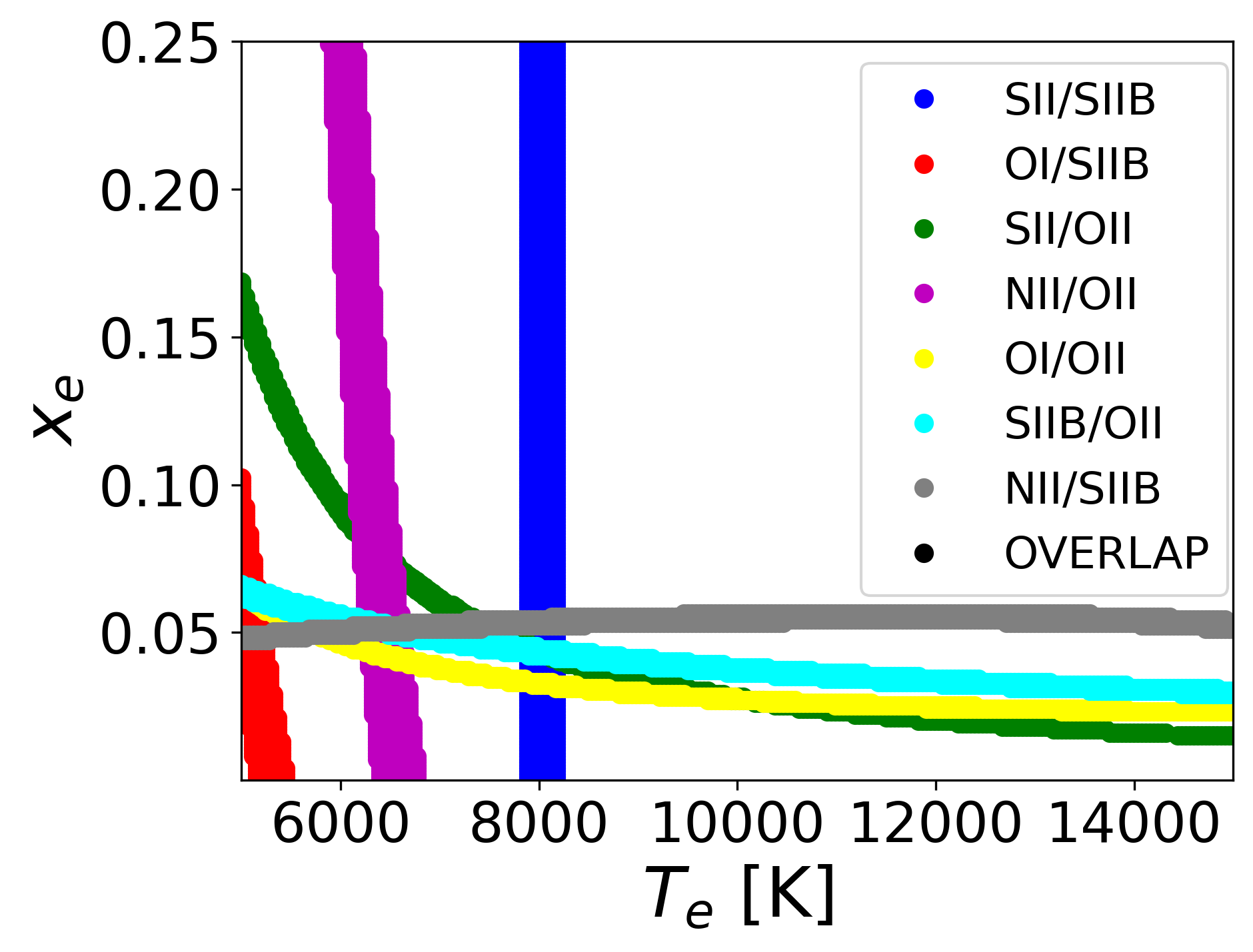}} 
\hspace{1em}  \\
\subfloat[\small{$A_V =0.0$, $t=10^{12}\,\text{s}$}]{\includegraphics[trim=0 0 0 0, clip, width=0.31  \textwidth]{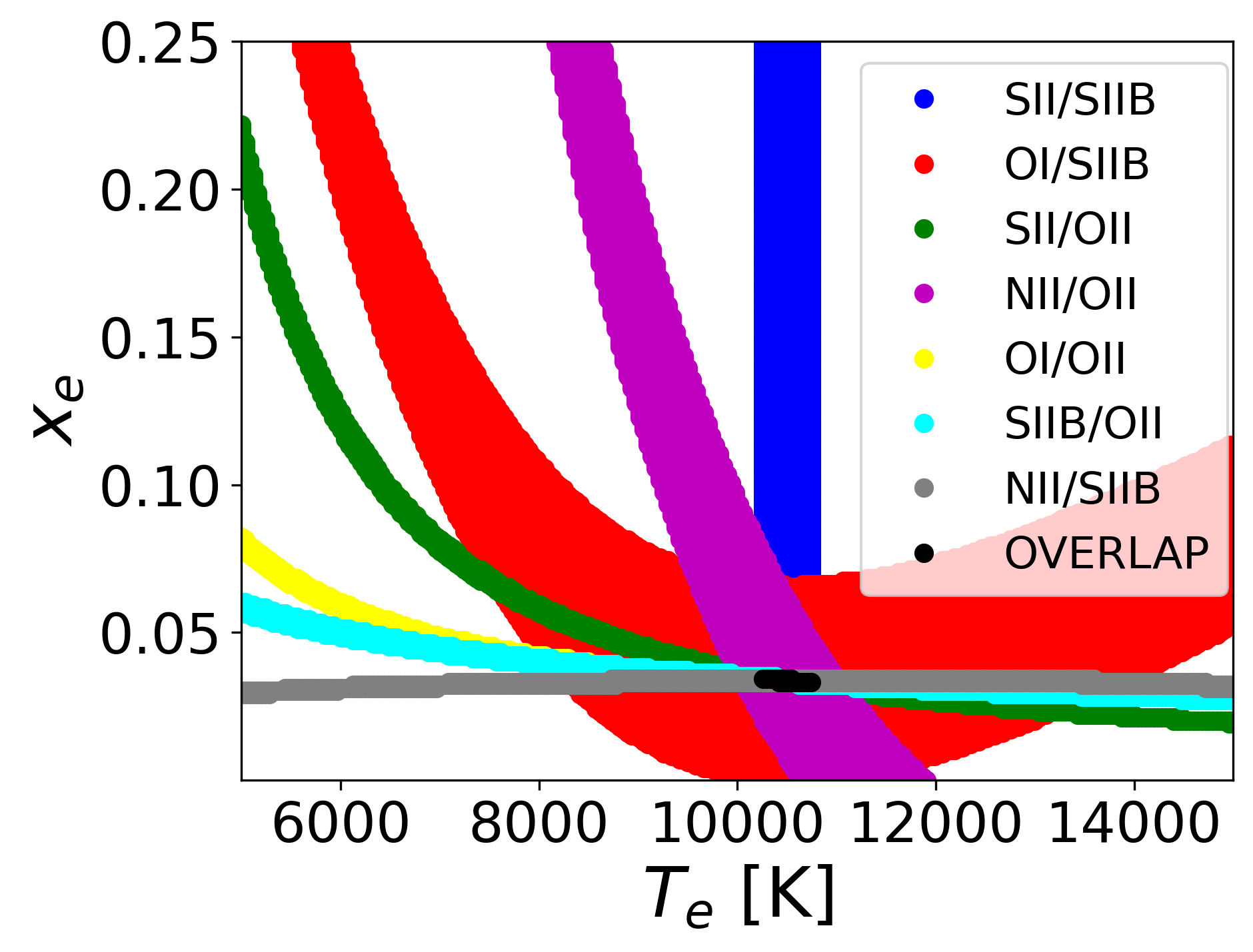}}
\hspace{1em} 
\subfloat[\small{$A_V =0.3$, $t=10^{12}\,\text{s}$}]{\includegraphics[trim=0 0 0 0, clip, width=0.31 \textwidth]{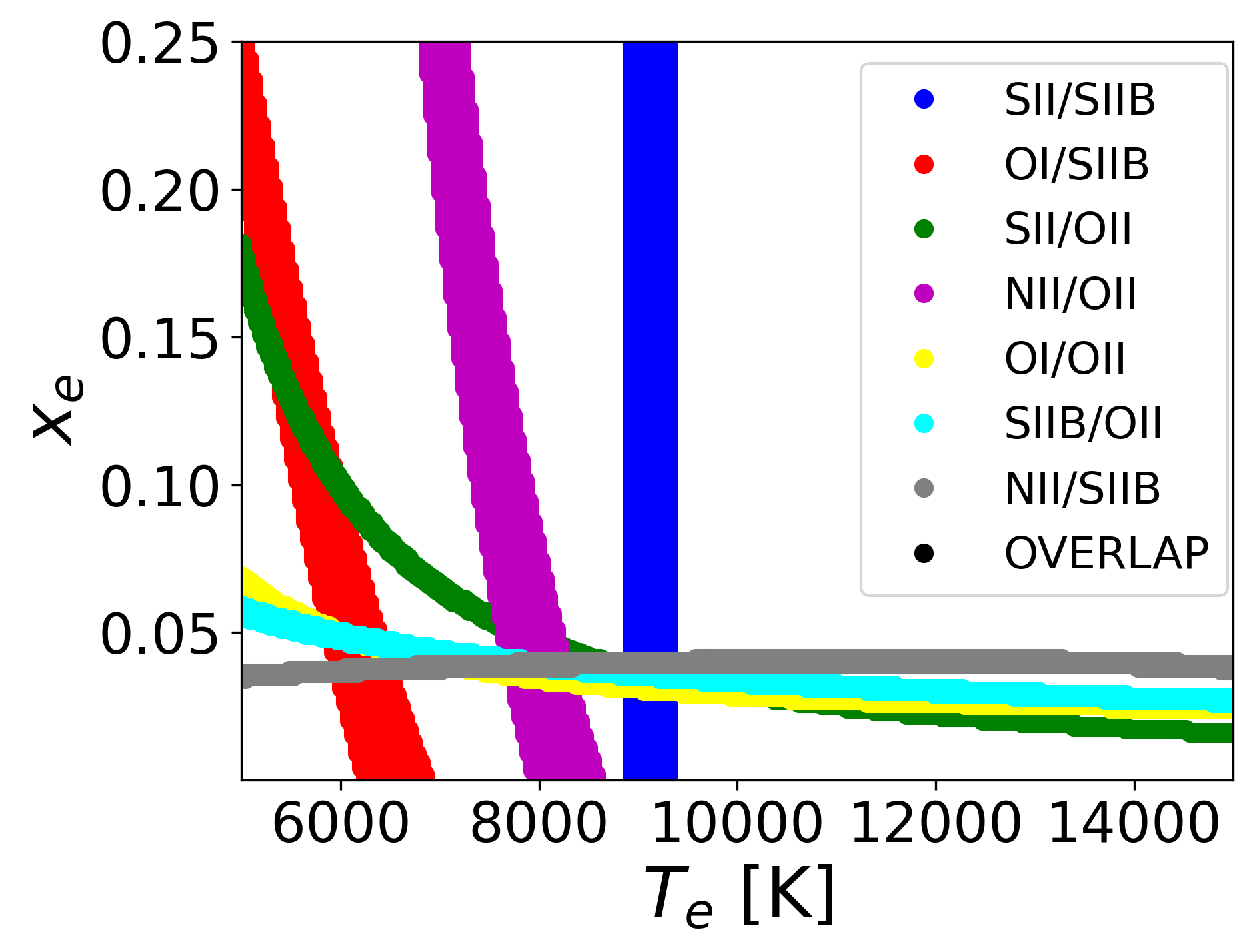}} 
\hspace{1em}   
\subfloat[\small{$A_V =0.6$, $t=10^{12}\,\text{s}$}]{\includegraphics[trim=0 0 0 0, clip, width=0.31  \textwidth]{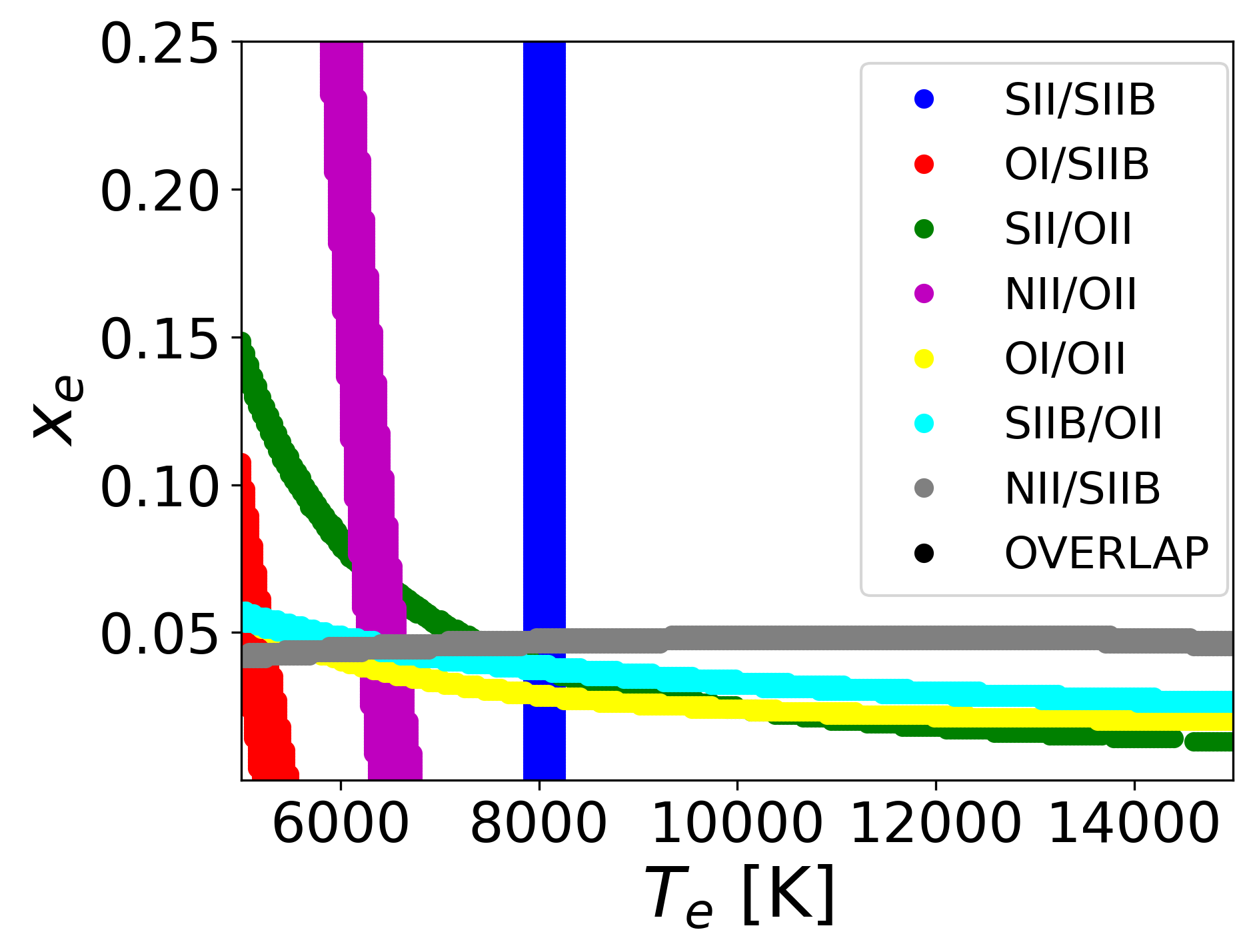}}
\hspace{1em} 
\caption{\small{An extended version of the BE99 diagrams (BE99e) as depicted in Fig.\,\ref{fig:all_BE_diagrams}. When including further line ratios in the diagnostics (indicated as coloured stripes), the presence of extinction or an out-of-equilibrium situation can cause the stripes to not overlap in one location on the ($x_e$-$T_e$) diagram. Abbreviation: SIIB = [S\,II]$\lambda\lambda$4068+4076, OII=[O\,II]$\lambda\lambda$3726+3729.}}\label{fig:all_BE_diagrams_A}
\end{figure*} 

 \clearpage
   
\begin{figure*}[h] 
\centering
\subfloat[\small{$A_V =0.0$, $t=10^6\,\text{s}$}]{\includegraphics[trim=0 0 0 0, clip, width=0.31  \textwidth]{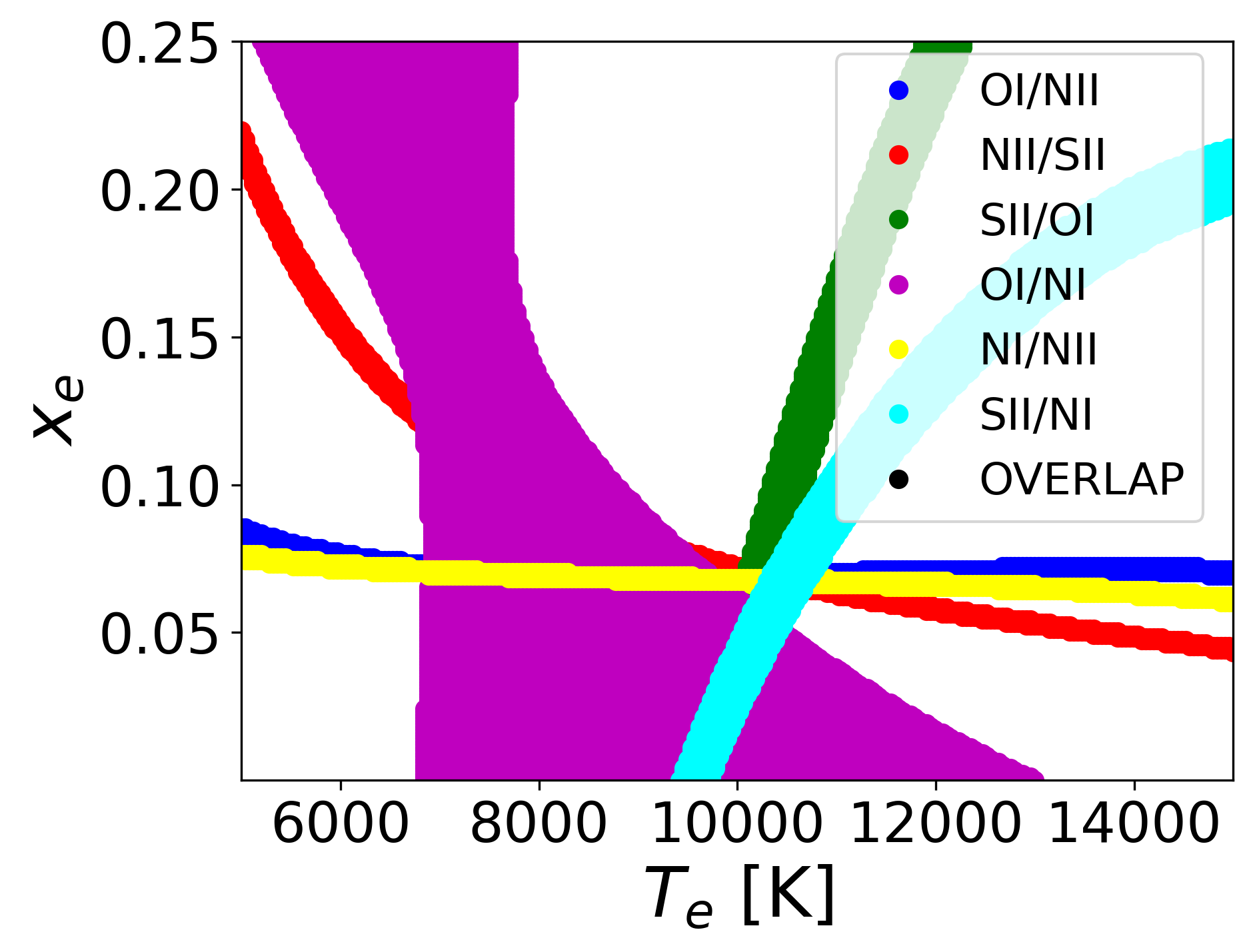}} 
\hspace{1em}   
\subfloat[\small{$A_V =0.3$, $t=10^6\,\text{s}$}]{\includegraphics[trim=0 0 0 0, clip, width=0.31  \textwidth]{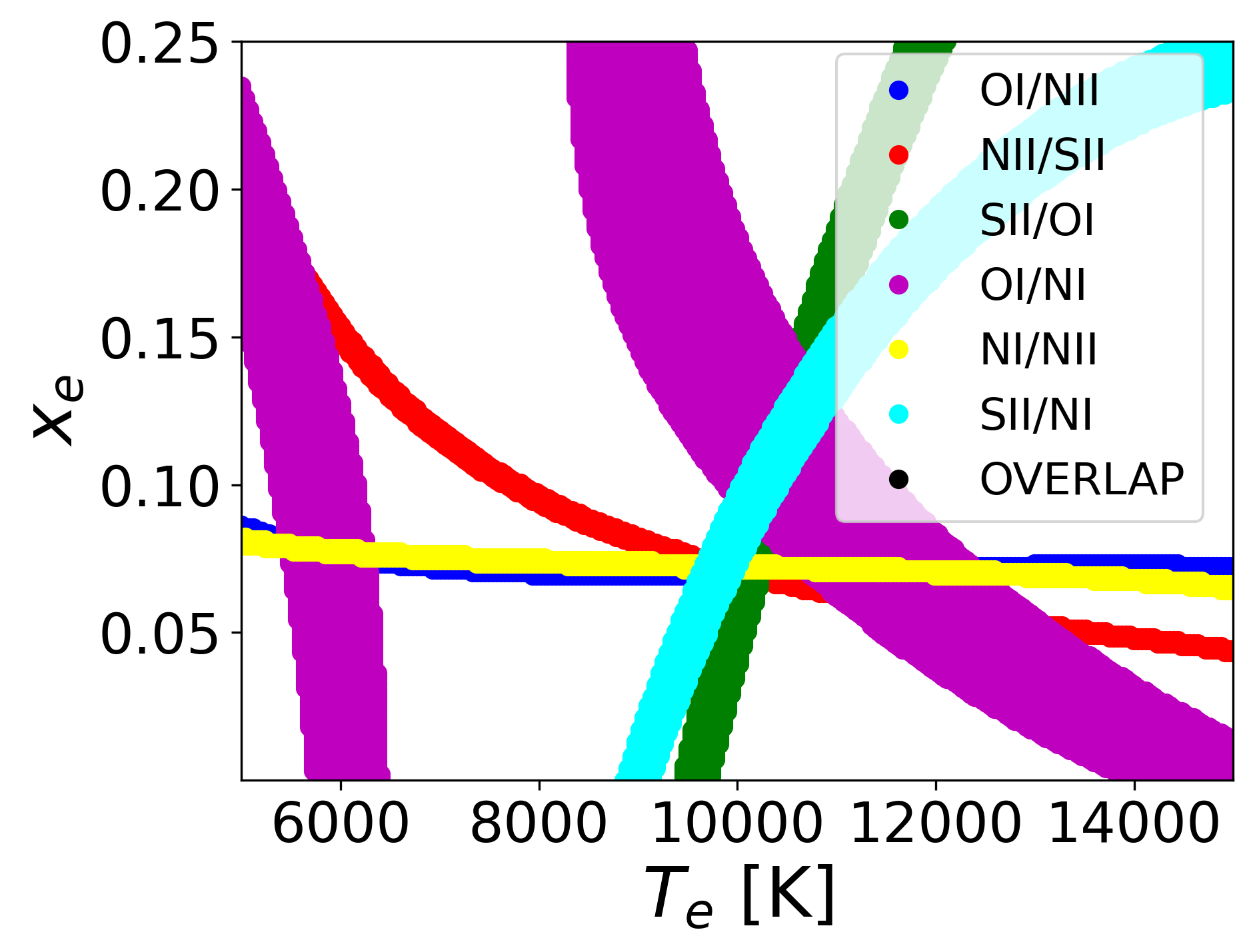}}
\hspace{1em} 
\subfloat[\small{$A_V =0.6$, $t=10^6\,\text{s}$}]{\includegraphics[trim=0 0 0 0, clip, width=0.31  \textwidth]{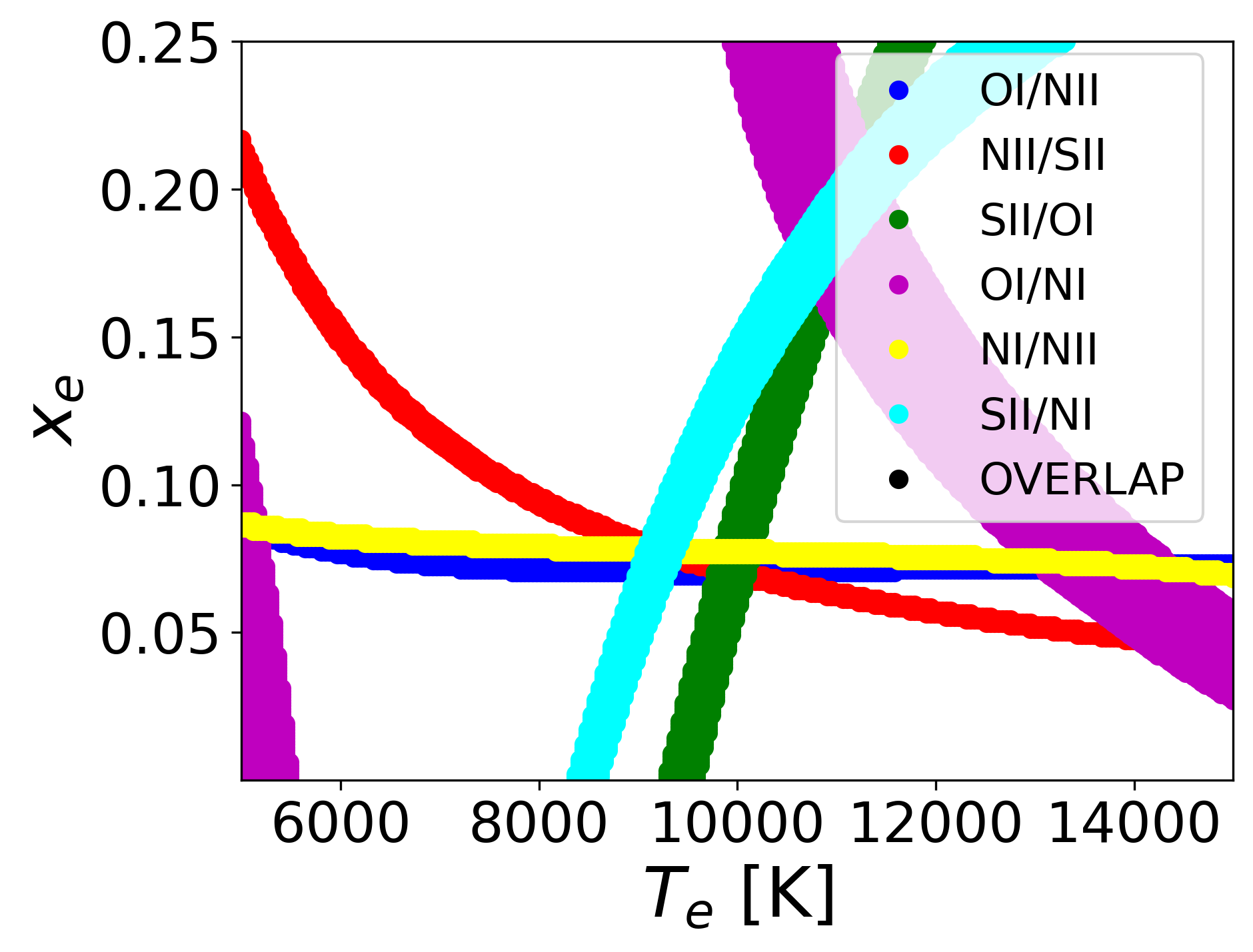}}
\hspace{1em} \\
\subfloat[\small{$A_V =0.0$, $t=10^8\,\text{s}$}]{\includegraphics[trim=0 0 0 0, clip, width=0.31 \textwidth]{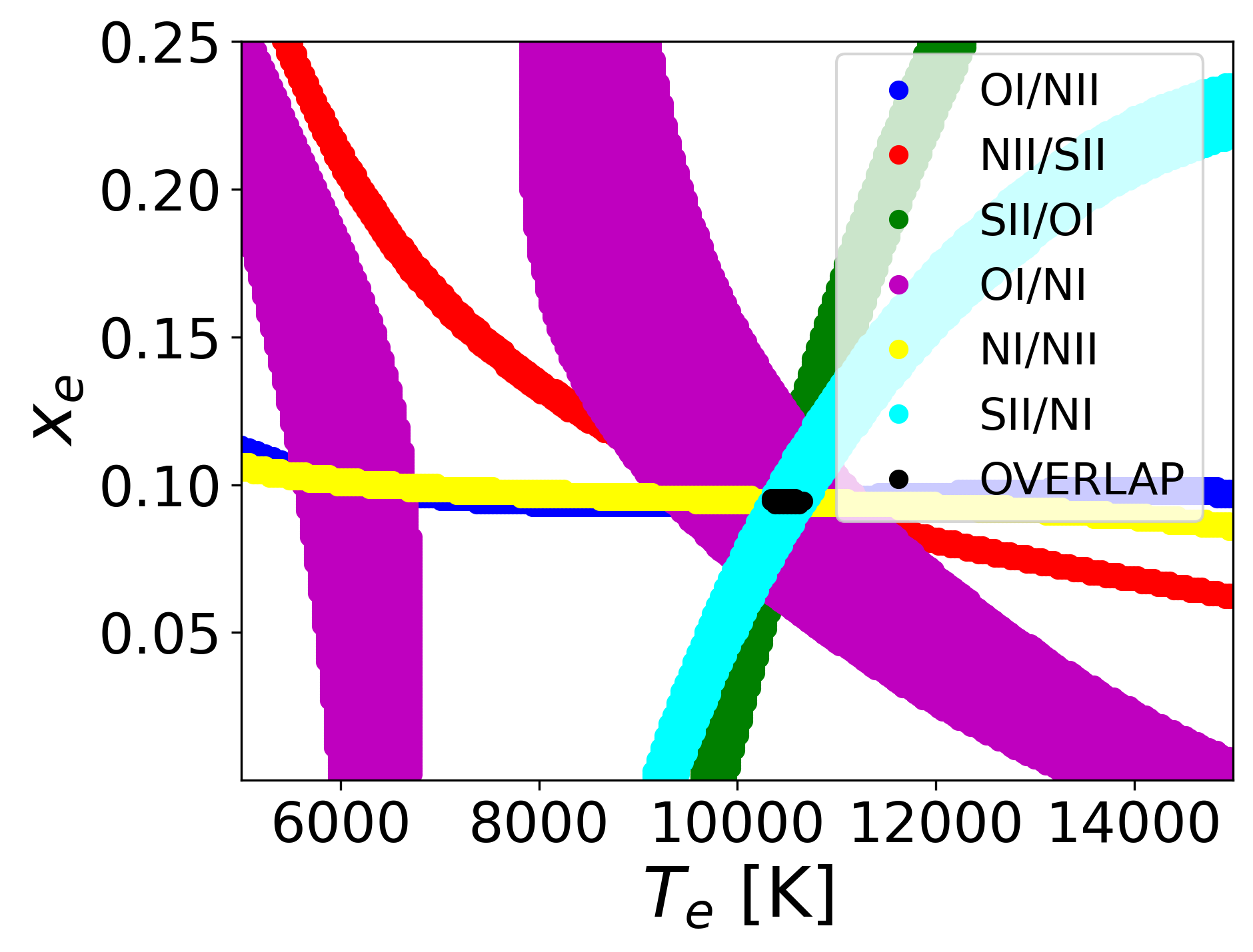}} 
\hspace{1em}   
\subfloat[\small{$A_V =0.3$, $t=10^8\,\text{s}$}]{\includegraphics[trim=0 0 0 0, clip, width=0.31 \textwidth]{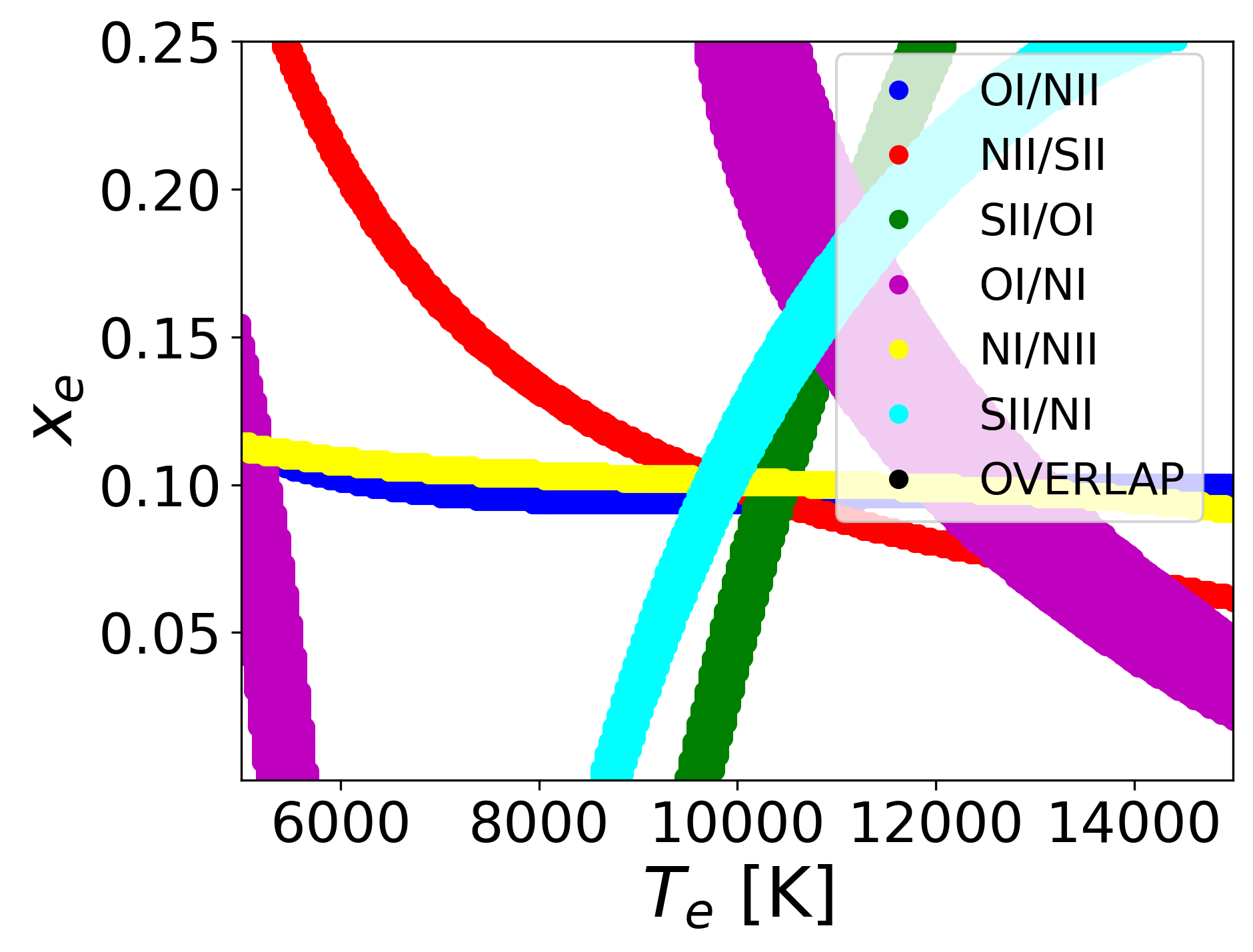}}
\hspace{1em}
\subfloat[\small{$A_V =0.6$, $t=10^8\,\text{s}$}]{\includegraphics[trim=0 0 0 0, clip, width=0.31 \textwidth]{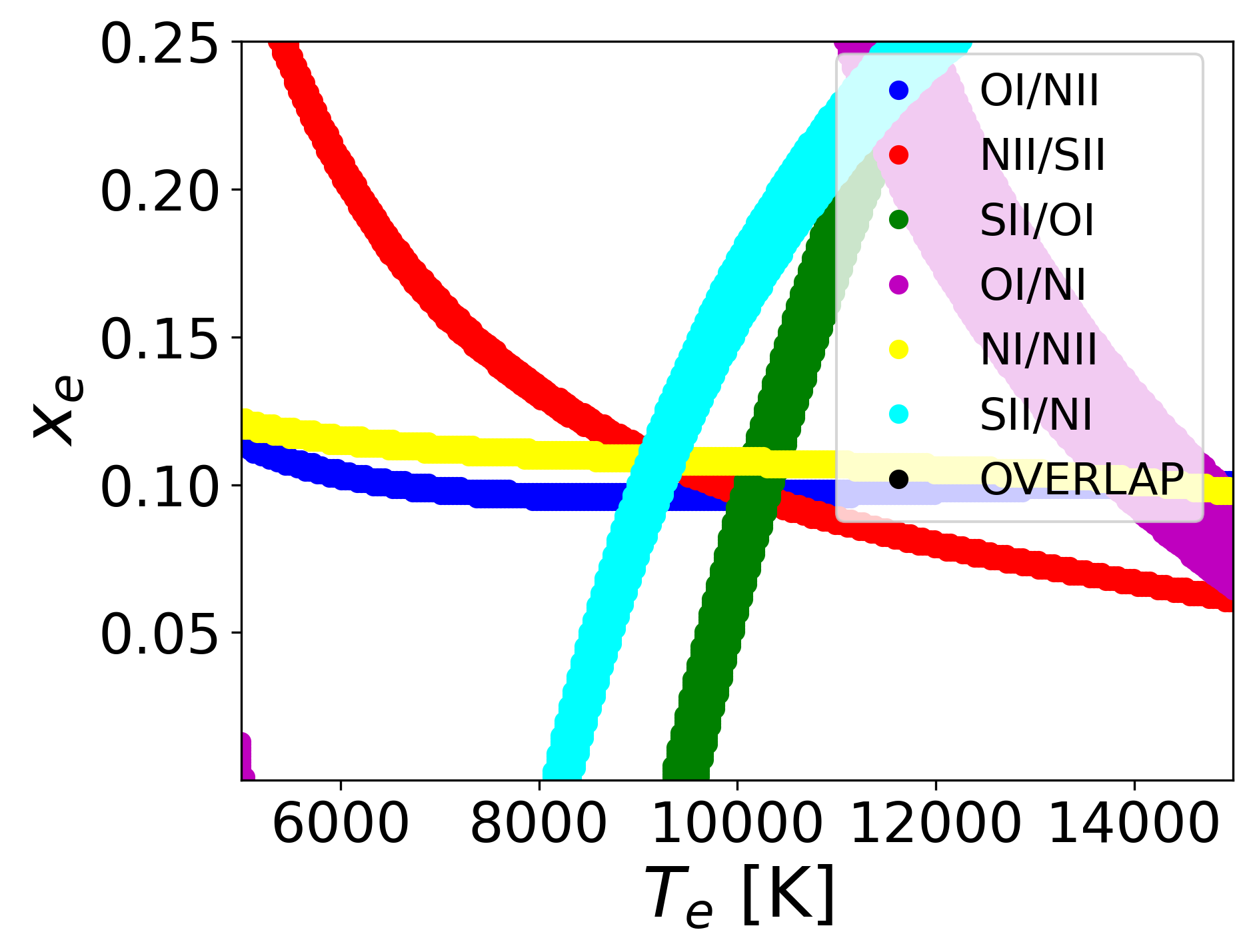}}
\hspace{1em}  \\
\subfloat[\small{$A_V =0.0$, $t=10^{10}\,\text{s}$}]{\includegraphics[trim=0 0 0 0, clip, width=0.31  \textwidth]{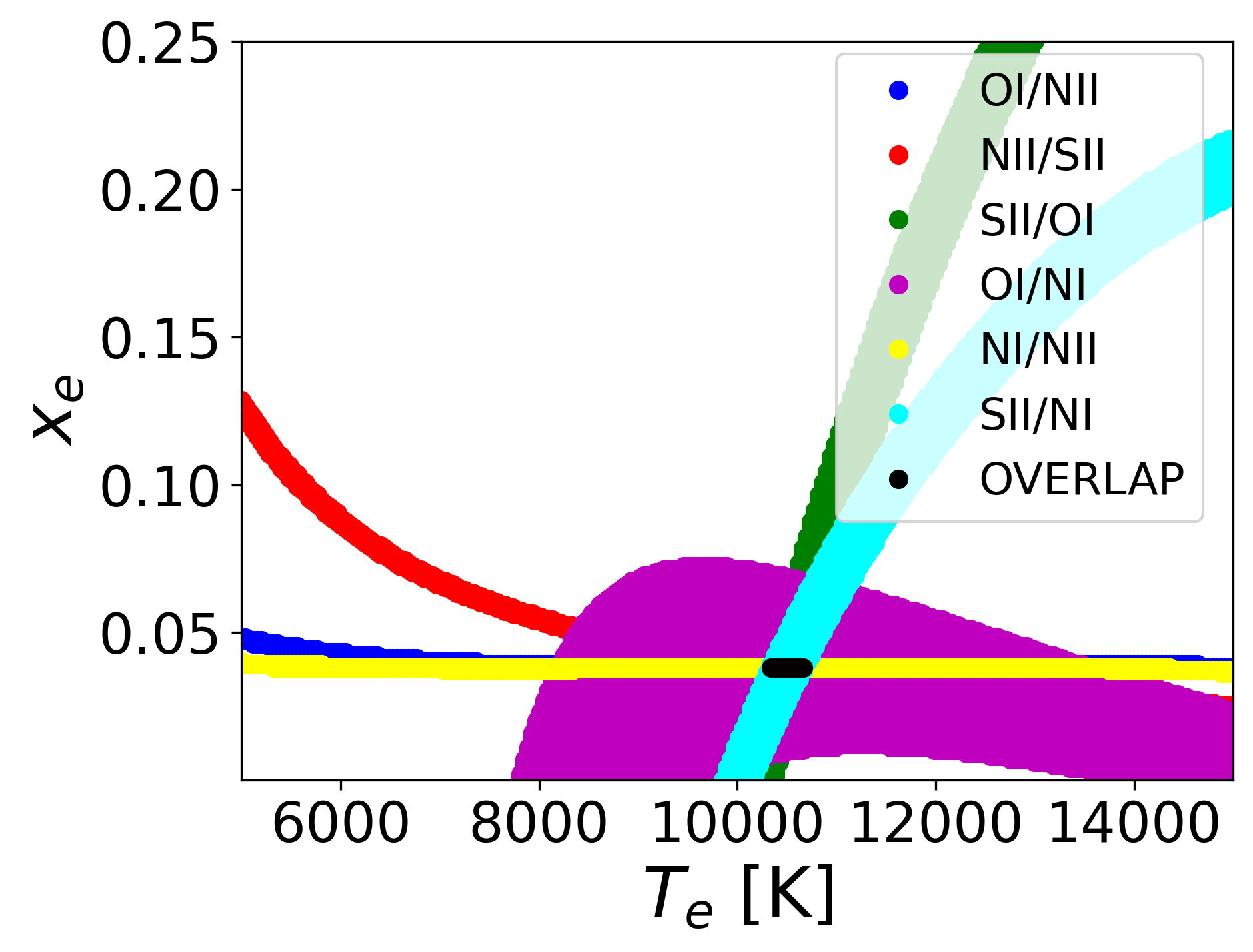}} 
\hspace{1em}   
\subfloat[\small{$A_V =0.3$, $t=10^{10}\,\text{s}$}]{\includegraphics[trim=0 0 0 0, clip, width=0.31  \textwidth]{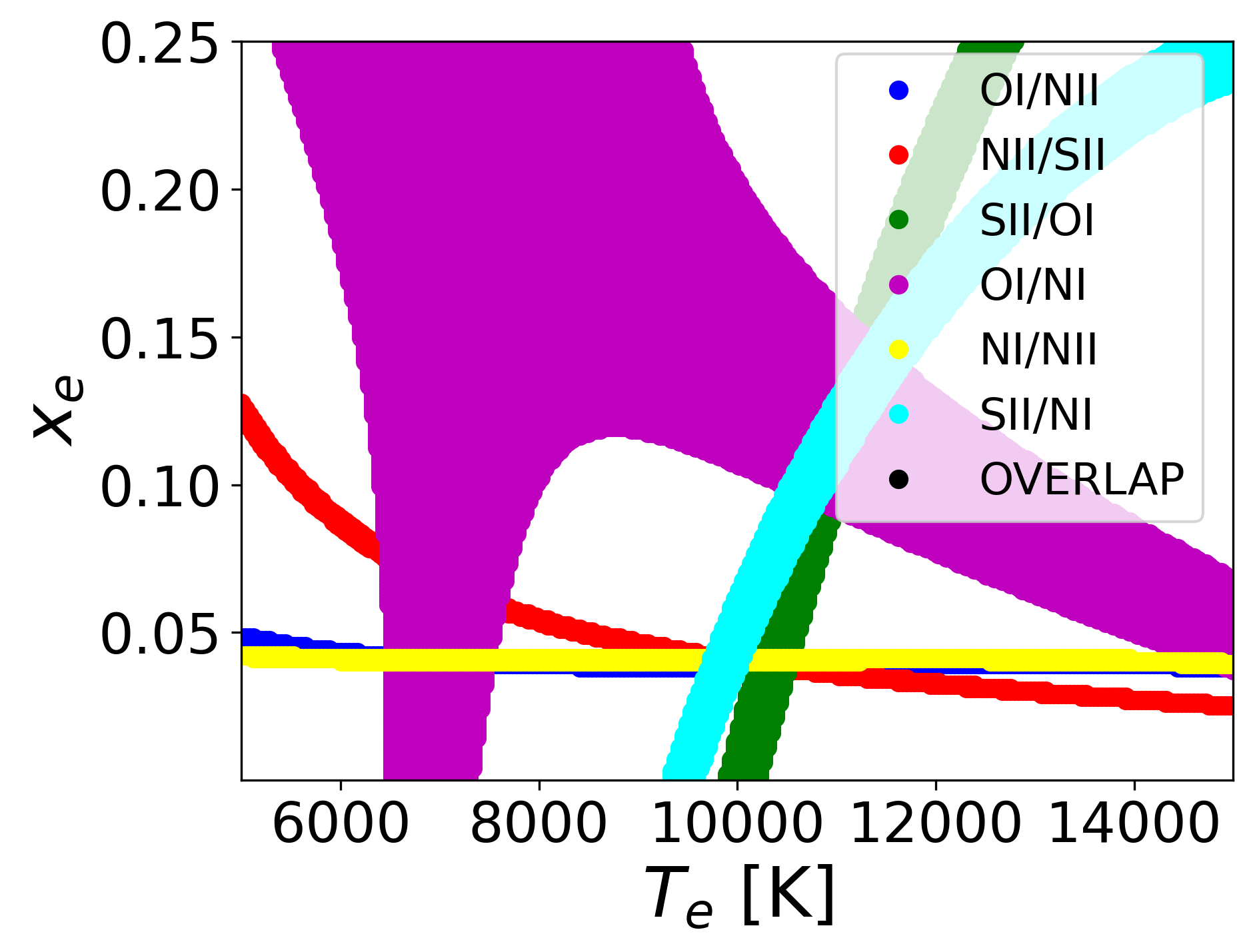}} 
\hspace{1em} 
\subfloat[\small{$A_V =0.6$, $t=10^{10}\,\text{s}$}]{\includegraphics[trim=0 0 0 0, clip, width=0.31 \textwidth]{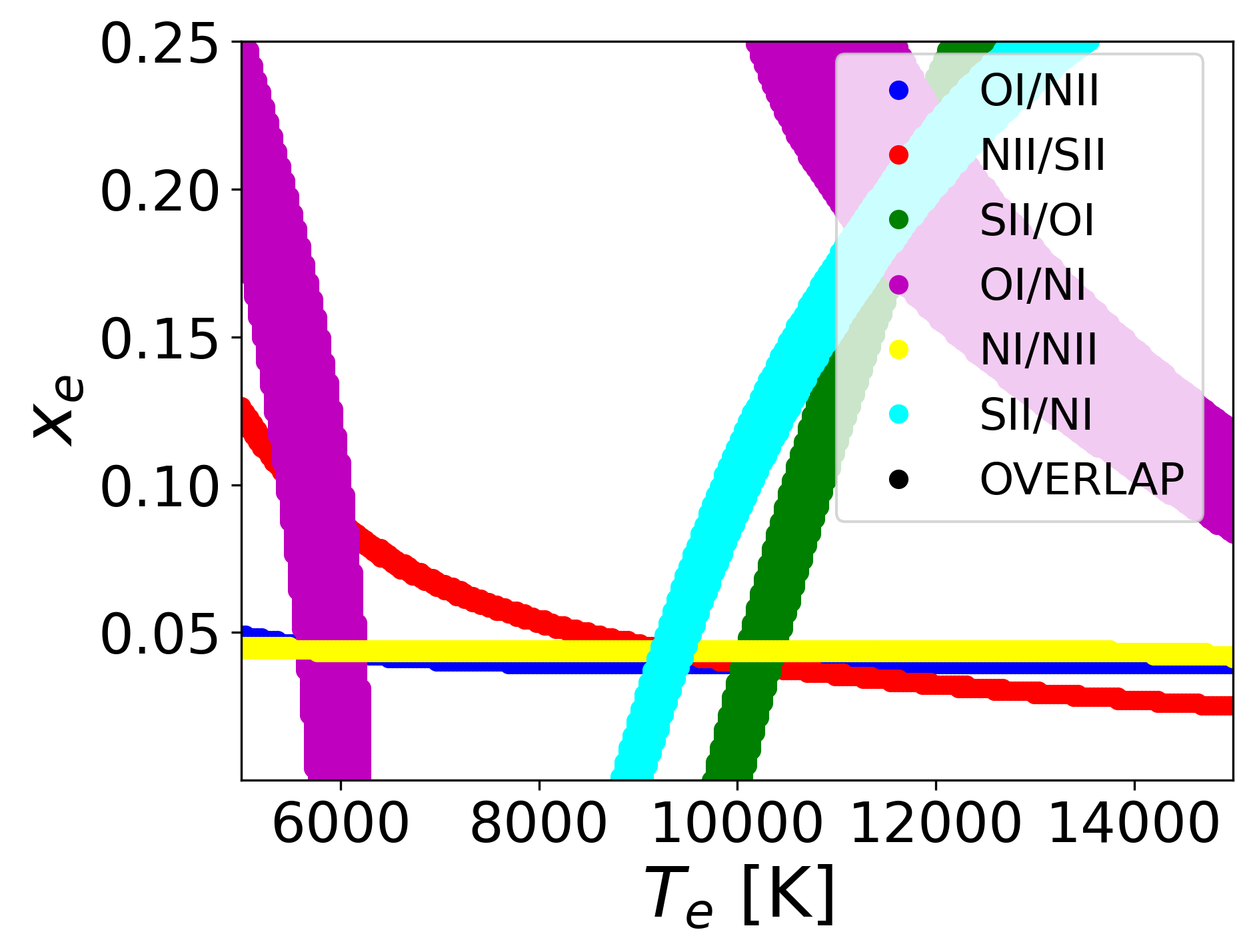}} 
\hspace{1em}   \\
\subfloat[\small{$A_V =0.0$, $t=10^{12}\,\text{s}$}]{\includegraphics[trim=0 0 0 0, clip, width=0.31  \textwidth]{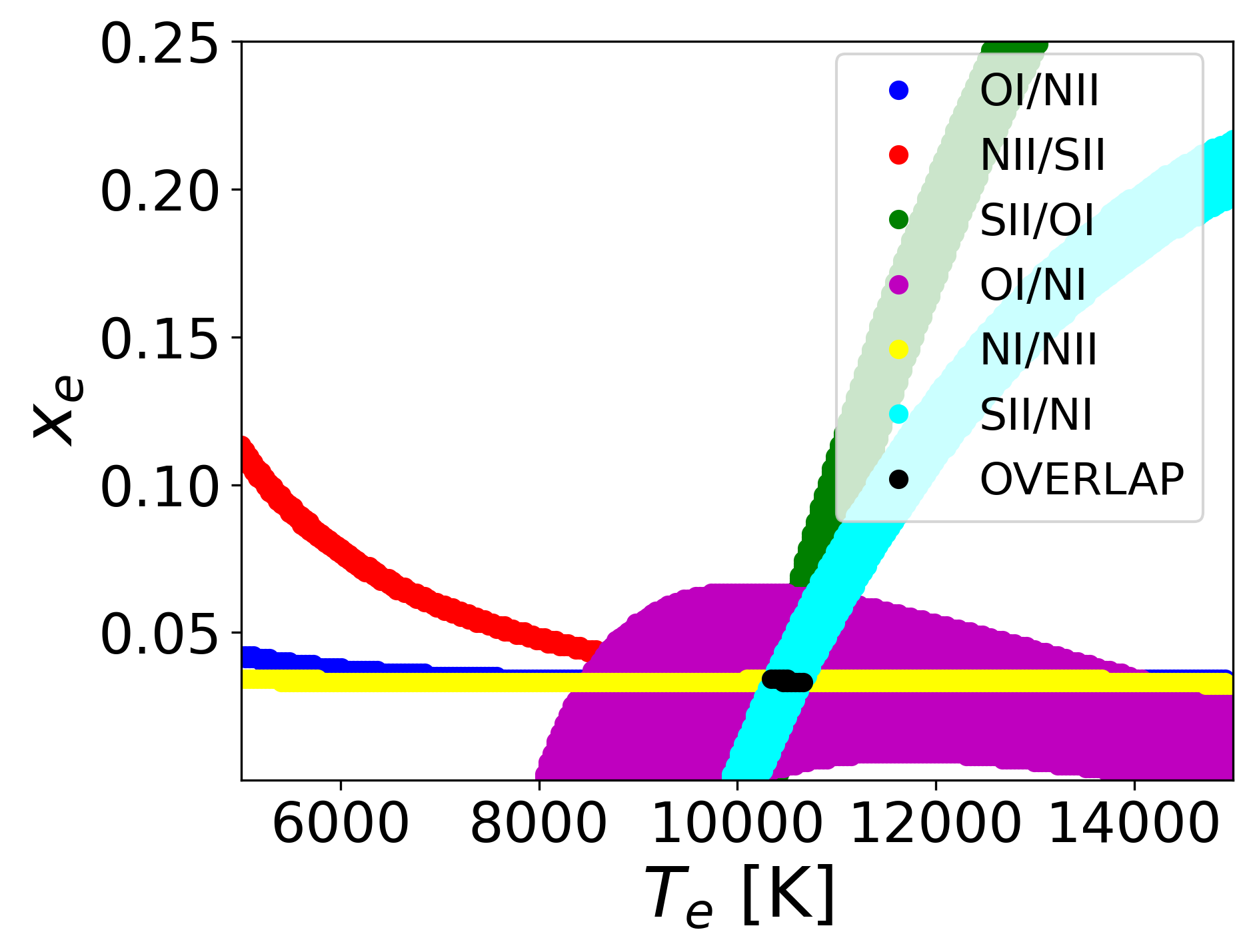}}
\hspace{1em} 
\subfloat[\small{$A_V =0.3$, $t=10^{12}\,\text{s}$}]{\includegraphics[trim=0 0 0 0, clip, width=0.31 \textwidth]{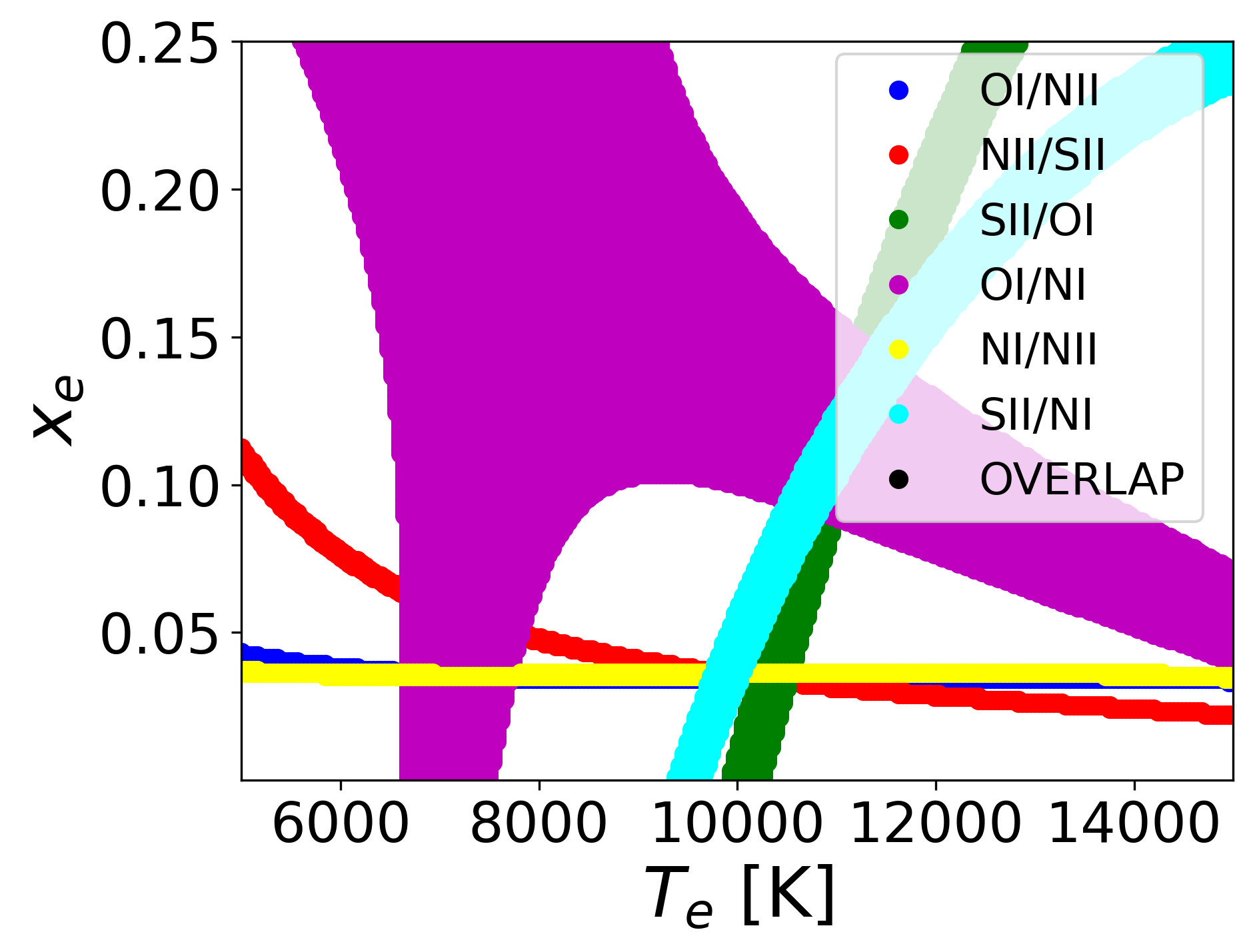}} 
\hspace{1em}   
\subfloat[\small{$A_V =0.6$, $t=10^{12}\,\text{s}$}]{\includegraphics[trim=0 0 0 0, clip, width=0.31 \textwidth]{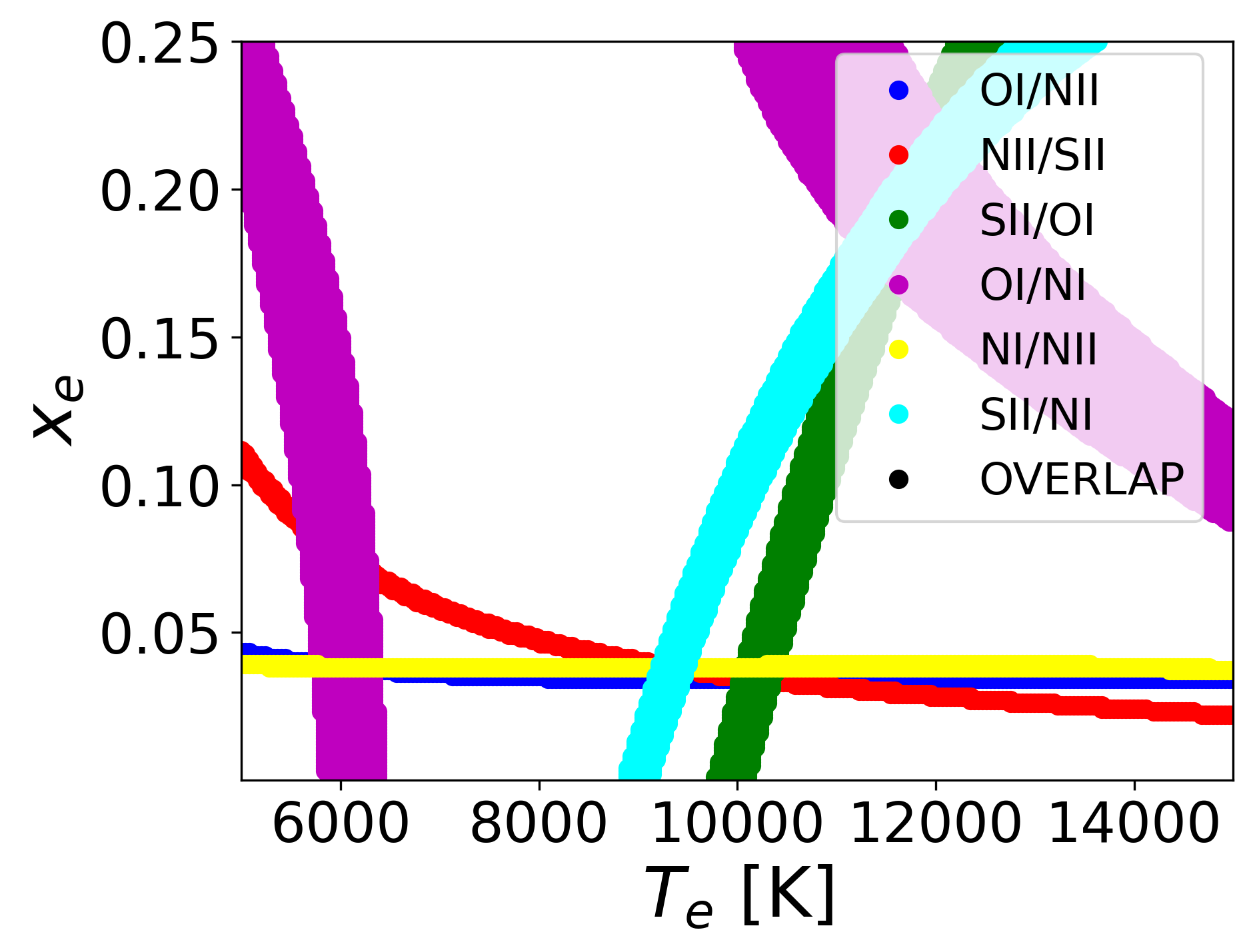}} 
\hspace{1em} 
\caption{\small{Same as Fig.\,\ref{fig:all_BE_diagrams_A} but with the additional line ratio: NI = [N\,I]$\lambda\lambda$5198+5200.}}\label{fig:all_BE_diagrams_B}
\end{figure*} 
 
\clearpage
 
\begin{figure*}[h] 
\centering
\subfloat[\small{$A_V =0.0$, $t=10^6\,\text{s}$}]{\includegraphics[trim=0 0 0 0, clip, width=0.31  \textwidth]{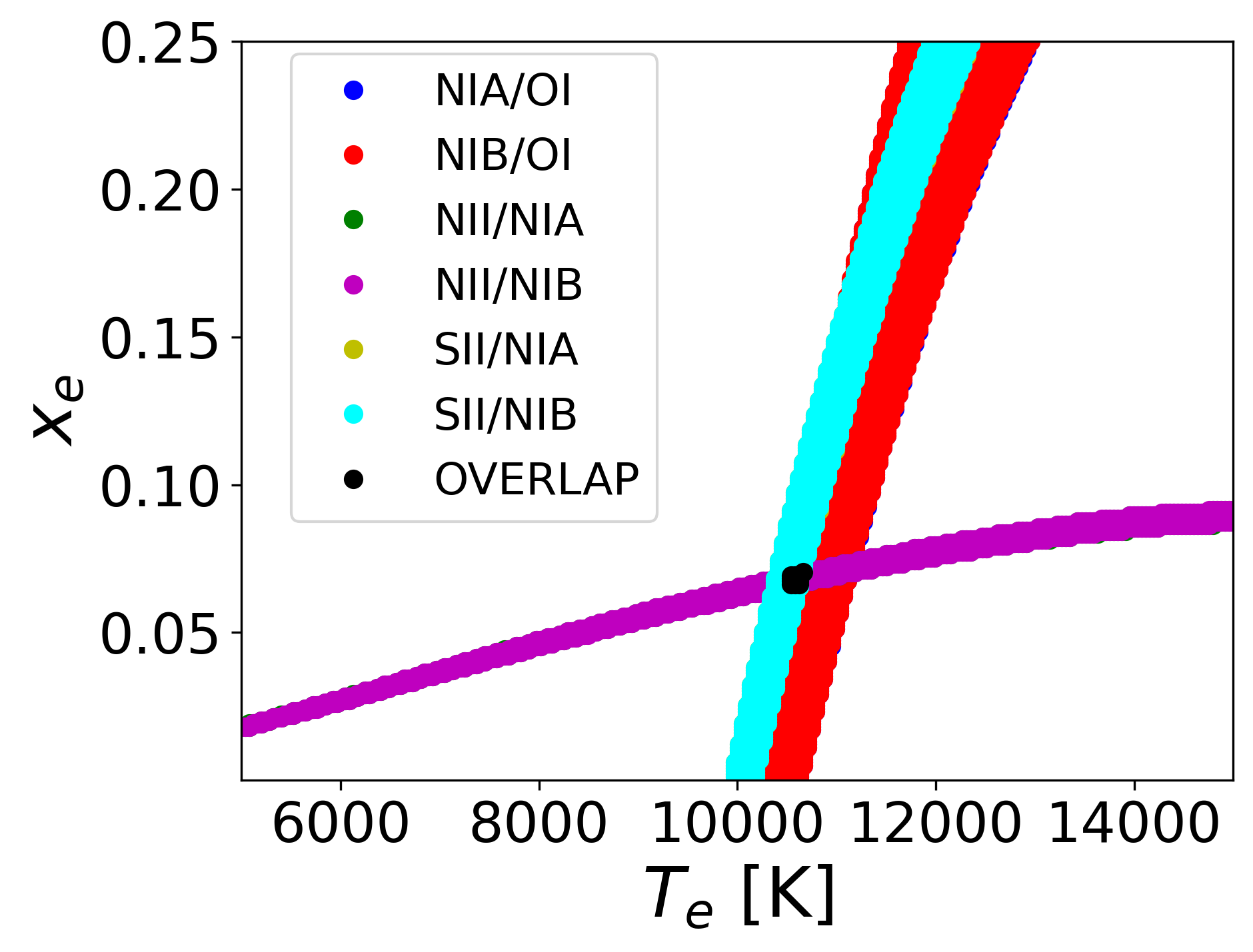}} 
\hspace{1em}   
\subfloat[\small{$A_V =0.3$, $t=10^6\,\text{s}$}]{\includegraphics[trim=0 0 0 0, clip, width=0.31  \textwidth]{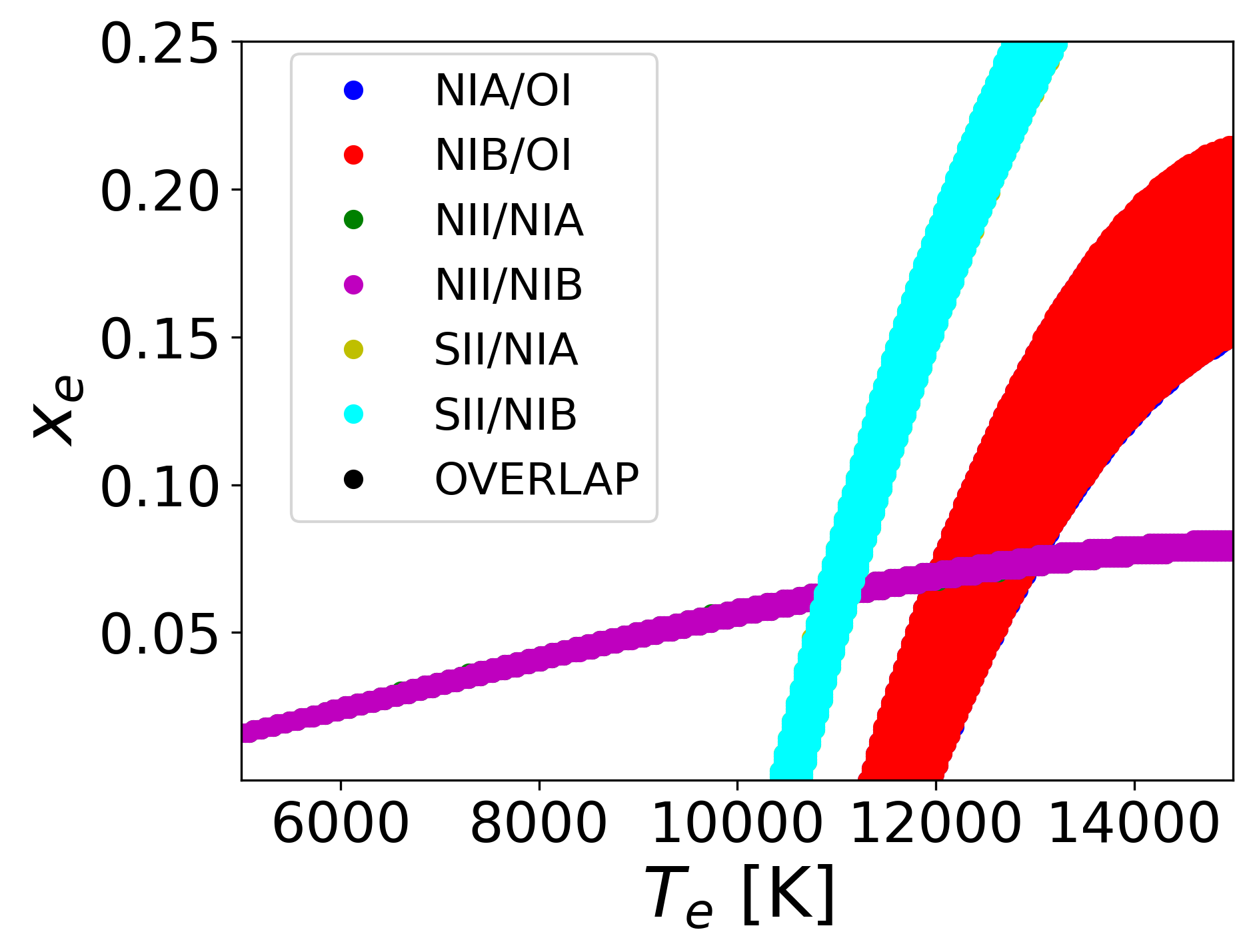}}
\hspace{1em} 
\subfloat[\small{$A_V =0.6$, $t=10^6\,\text{s}$}]{\includegraphics[trim=0 0 0 0, clip, width=0.31  \textwidth]{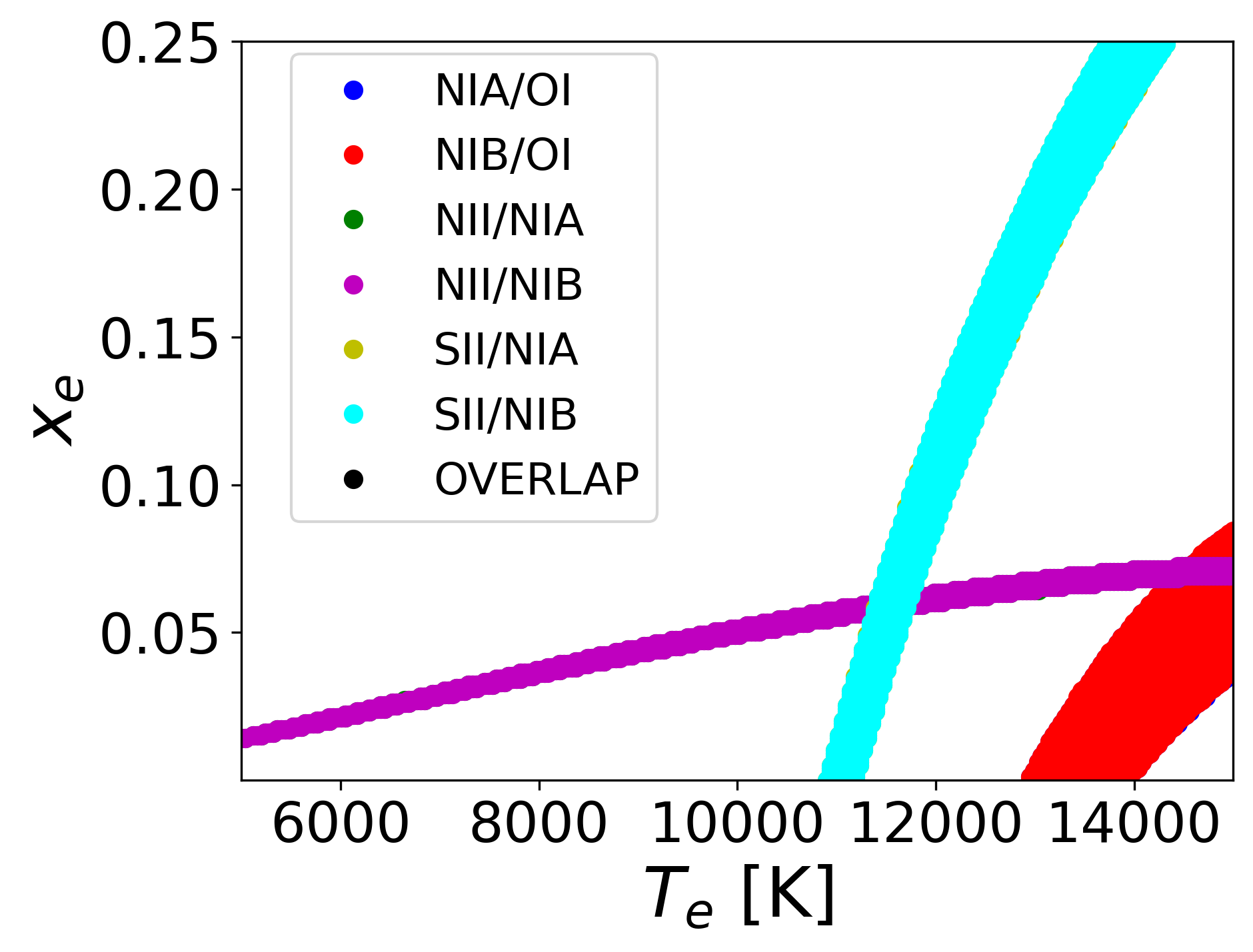}}
\hspace{1em} \\
\subfloat[\small{$A_V =0.0$, $t=10^8\,\text{s}$}]{\includegraphics[trim=0 0 0 0, clip, width=0.31 \textwidth]{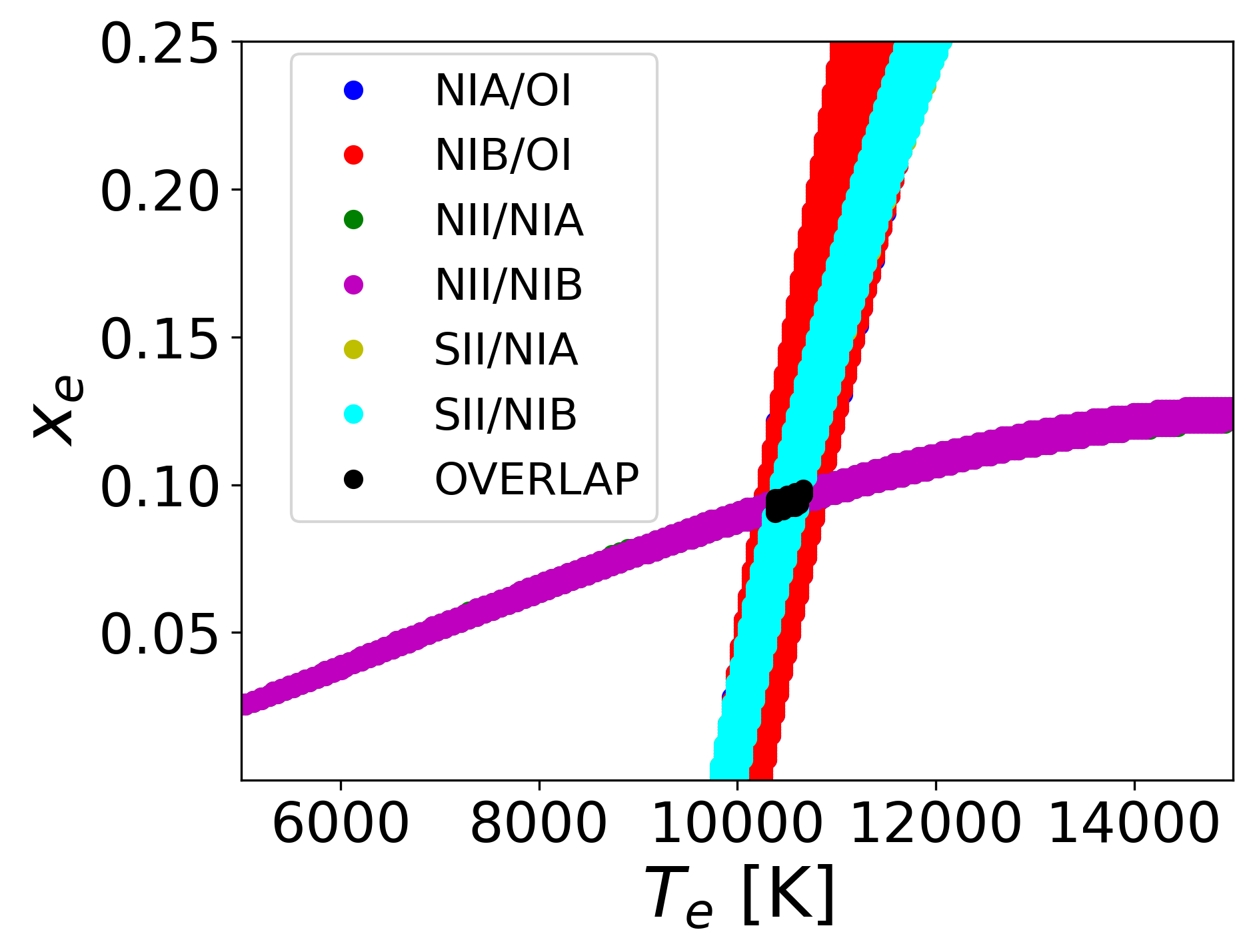}} 
\hspace{1em}   
\subfloat[\small{$A_V =0.3$, $t=10^8\,\text{s}$}]{\includegraphics[trim=0 0 0 0, clip, width=0.31 \textwidth]{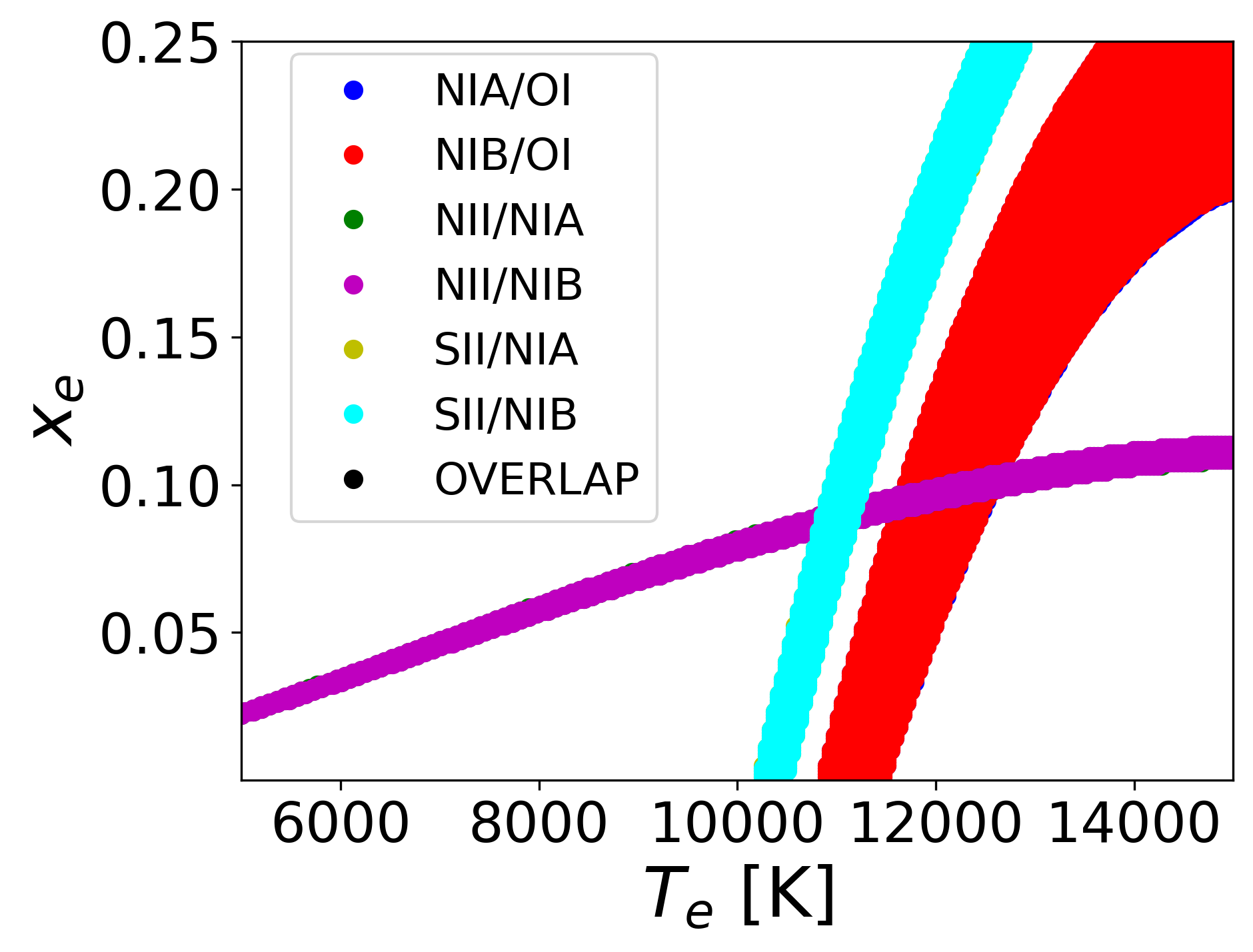}}
\hspace{1em}
\subfloat[\small{$A_V =0.6$, $t=10^8\,\text{s}$}]{\includegraphics[trim=0 0 0 0, clip, width=0.31 \textwidth]{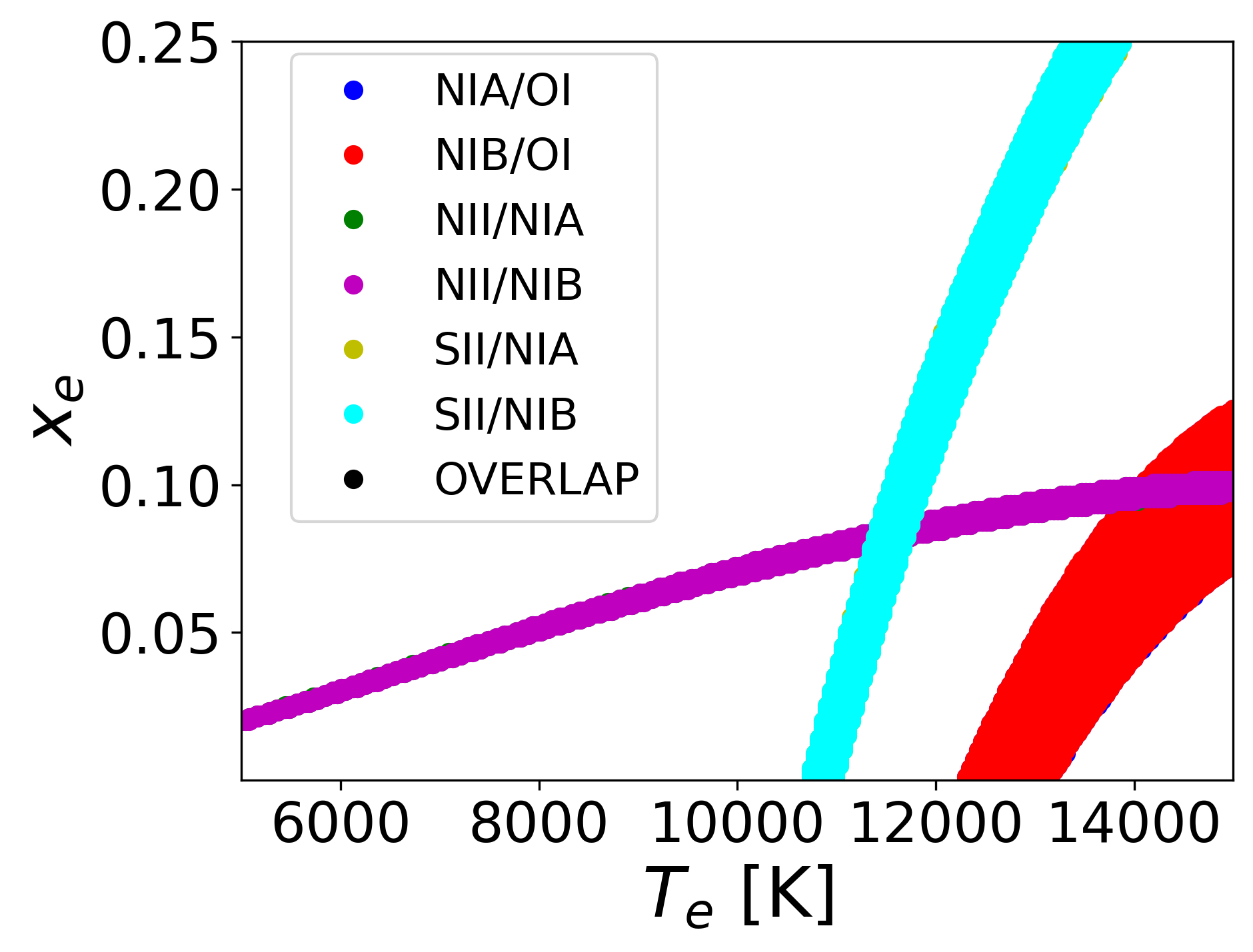}}
\hspace{1em}  \\
\subfloat[\small{$A_V =0.0$, $t=10^{10}\,\text{s}$}]{\includegraphics[trim=0 0 0 0, clip, width=0.31  \textwidth]{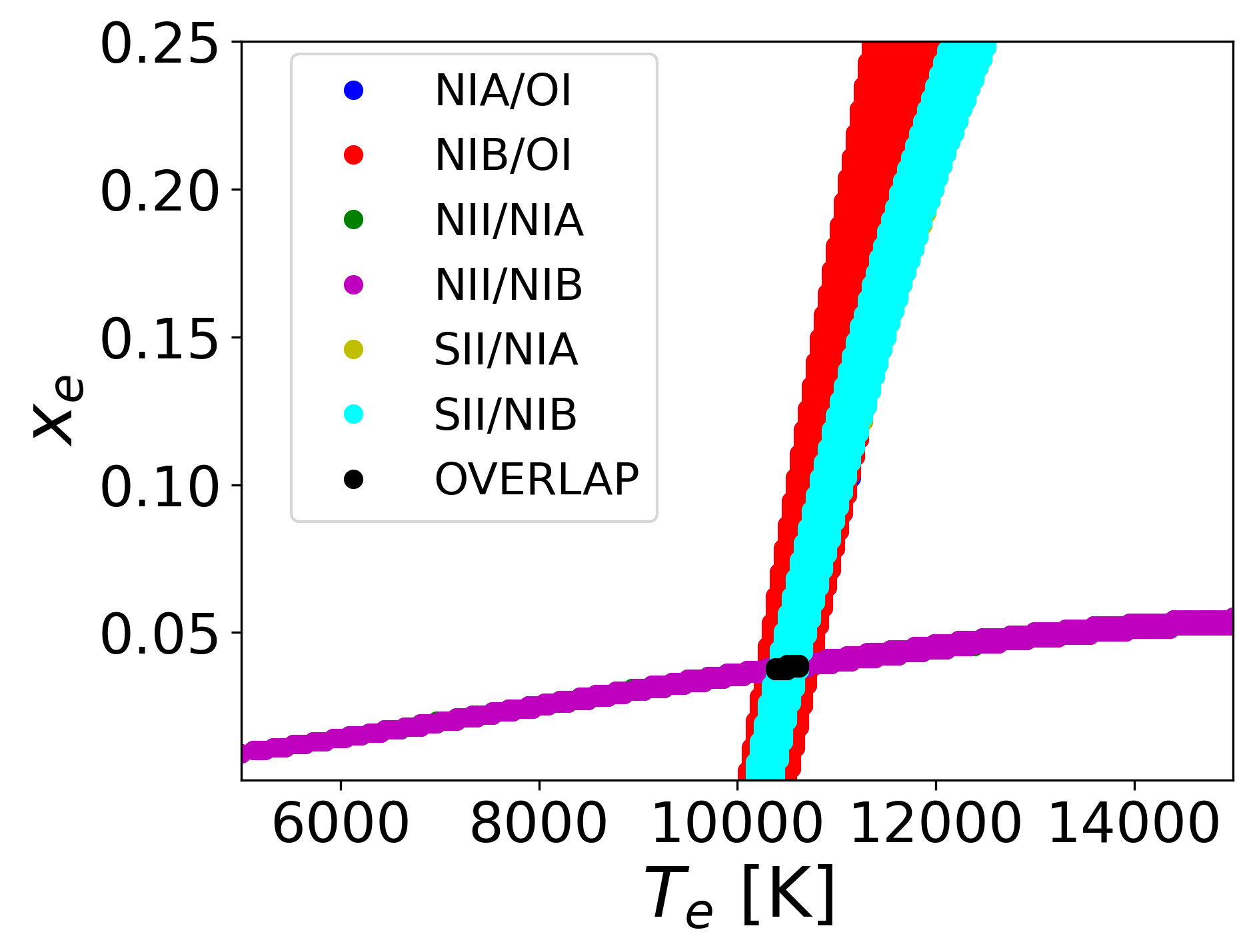}} 
\hspace{1em}   
\subfloat[\small{$A_V =0.3$, $t=10^{10}\,\text{s}$}]{\includegraphics[trim=0 0 0 0, clip, width=0.31 \textwidth]{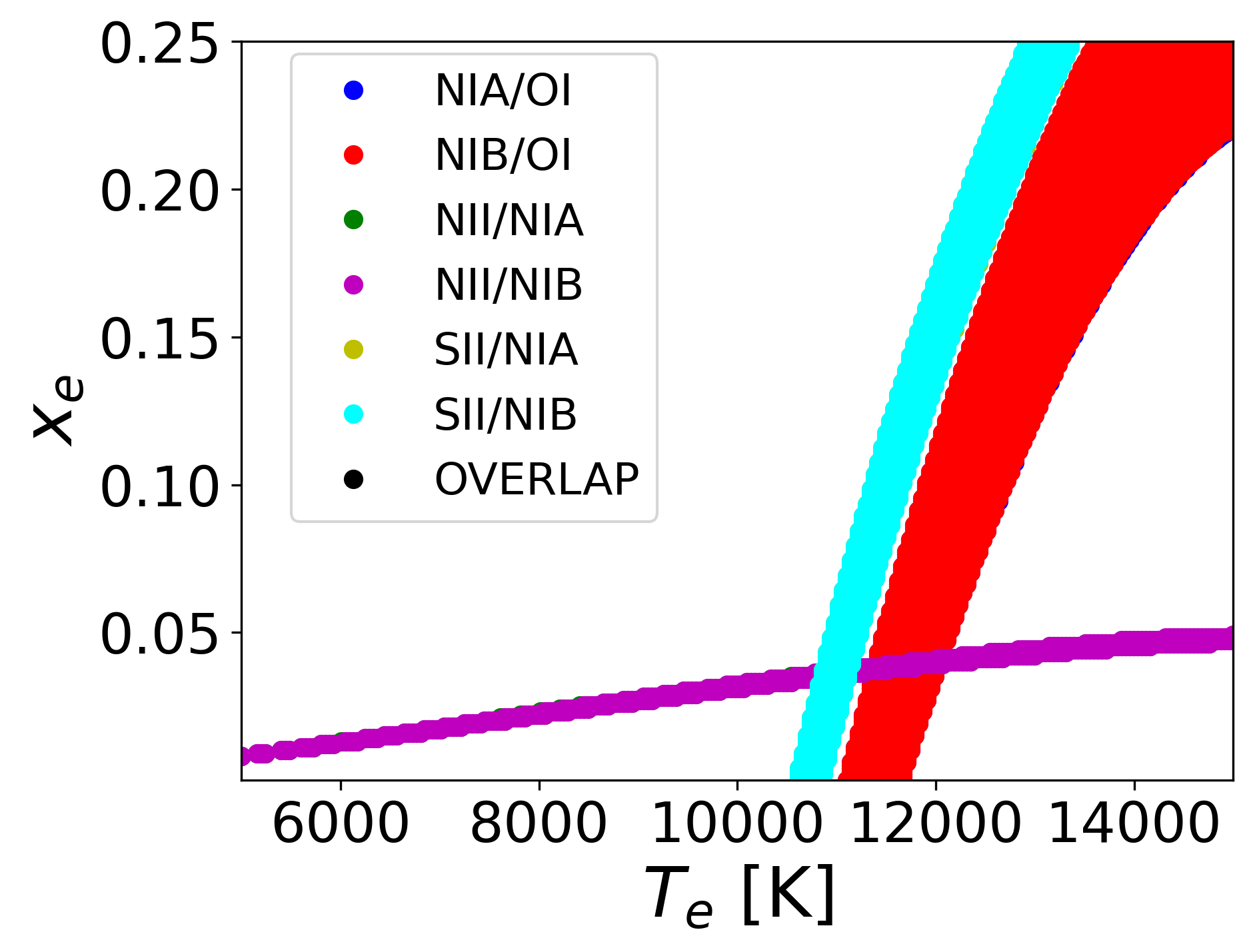}} 
\hspace{1em} 
\subfloat[\small{$A_V =0.6$, $t=10^{10}\,\text{s}$}]{\includegraphics[trim=0 0 0 0, clip, width=0.31  \textwidth]{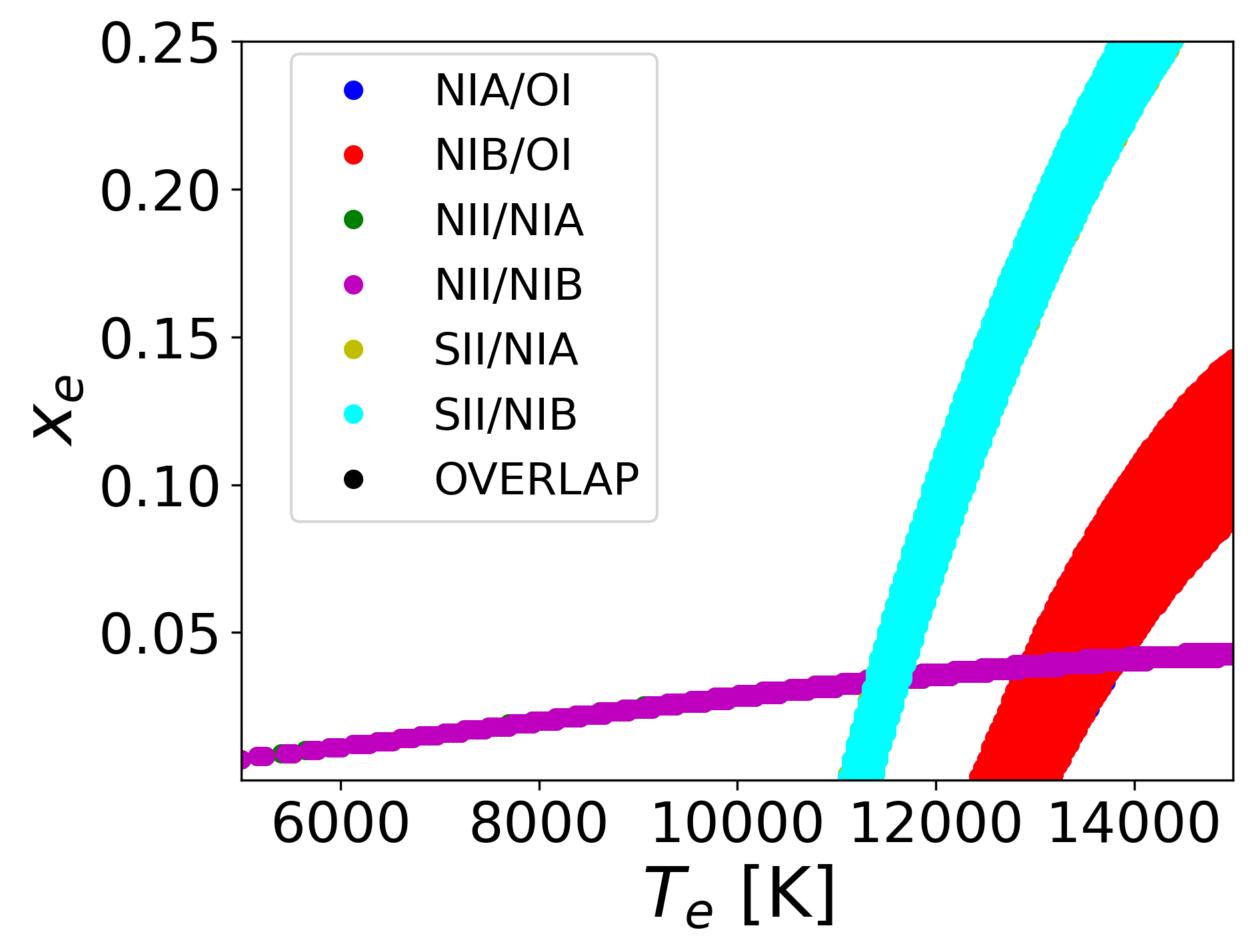}} 
\hspace{1em}    \\
\subfloat[\small{$A_V =0.0$, $t=10^{12}\,\text{s}$}]{\includegraphics[trim=0 0 0 0, clip, width=0.31  \textwidth]{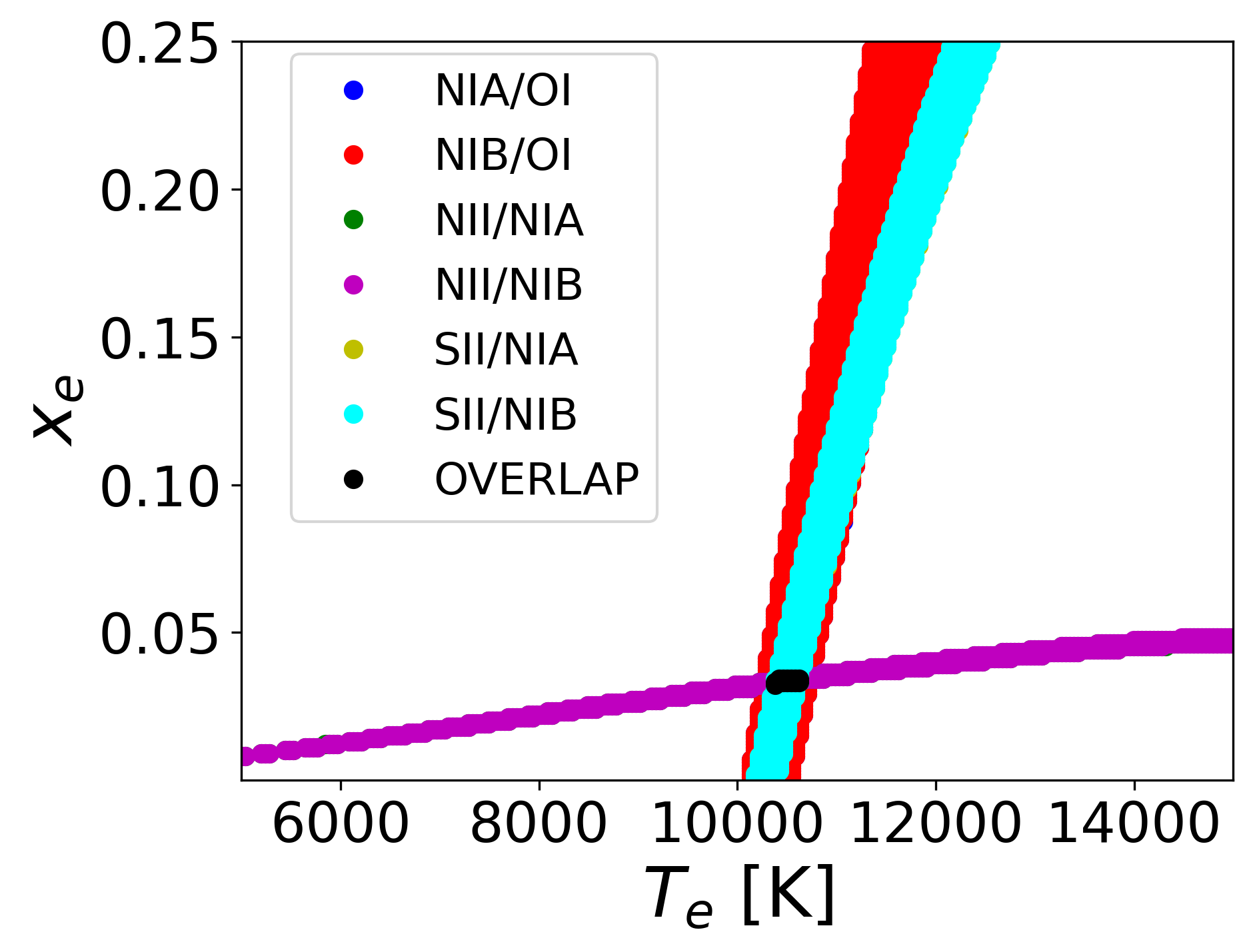}}
\hspace{1em} 
\subfloat[\small{$A_V =0.3$, $t=10^{12}\,\text{s}$}]{\includegraphics[trim=0 0 0 0, clip, width=0.31 \textwidth]{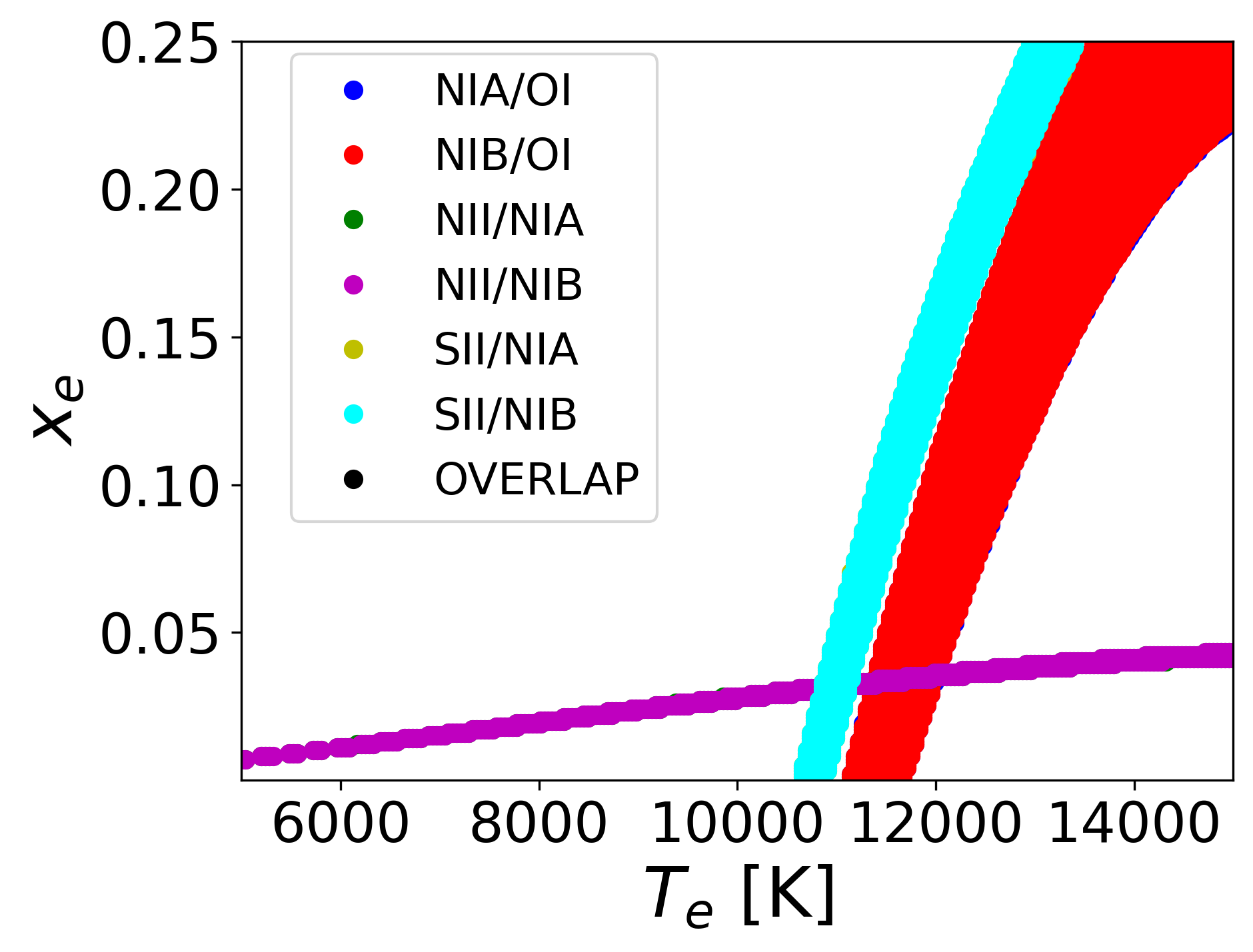}} 
\hspace{1em}   
\subfloat[\small{$A_V =0.6$, $t=10^{12}\,\text{s}$}]{\includegraphics[trim=0 0 0 0, clip, width=0.31\textwidth]{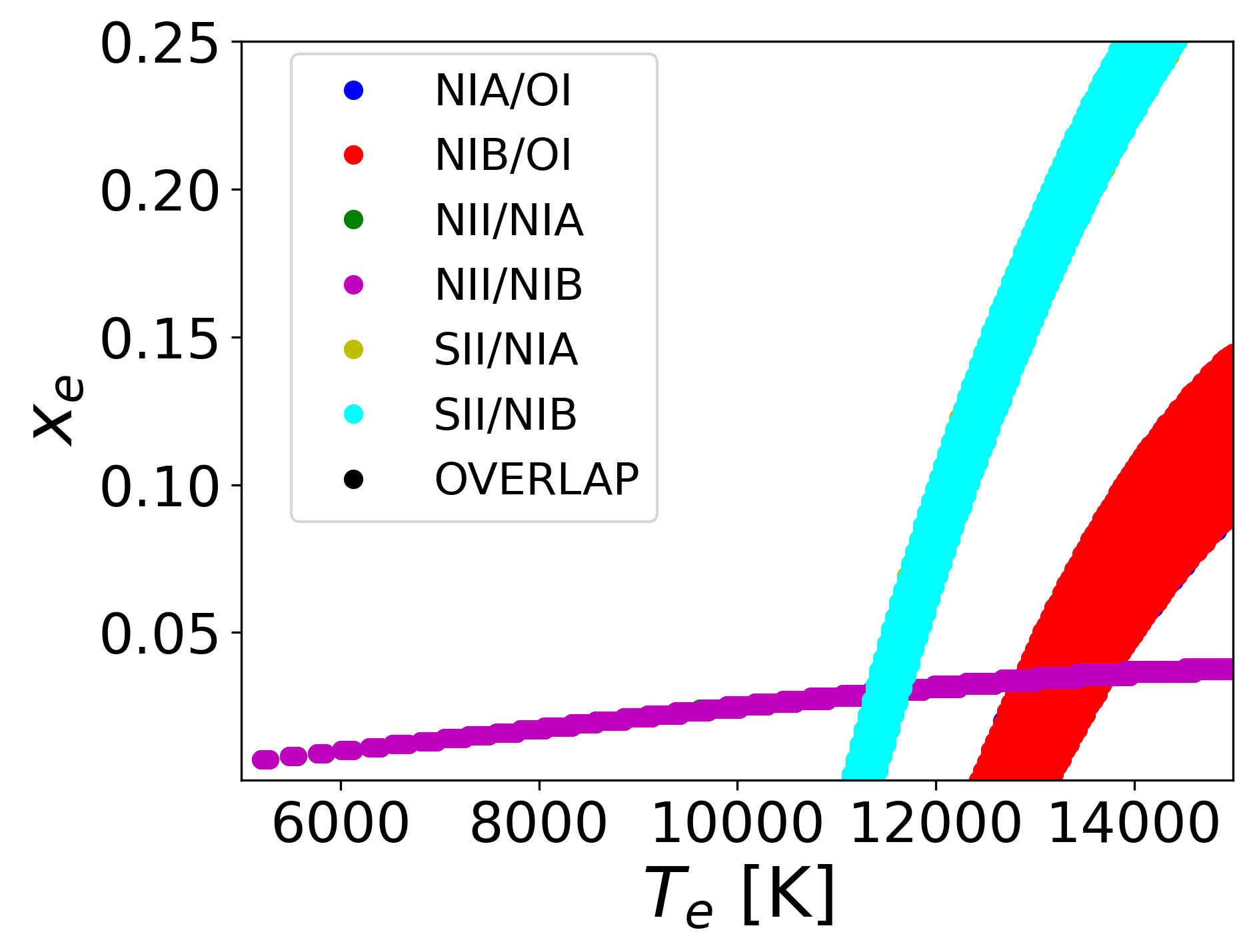}} 
\hspace{1em}  
\caption{\small{Same as Fig.\,\ref{fig:all_BE_diagrams_A} but with the additional line ratio: NIA = [N\,I]$\lambda\lambda$10397.7+10398.1, NIB = [N\,I]$\lambda\lambda$10407.1+10407.5.}}\label{fig:all_BE_diagrams_C}
\end{figure*} 

\clearpage

\begin{figure*}[h] 
\centering
\subfloat[\small{$A_V =0.0$, $t=10^6\,\text{s}$}]{\includegraphics[trim=0 0 0 0, clip, width=0.31  \textwidth]{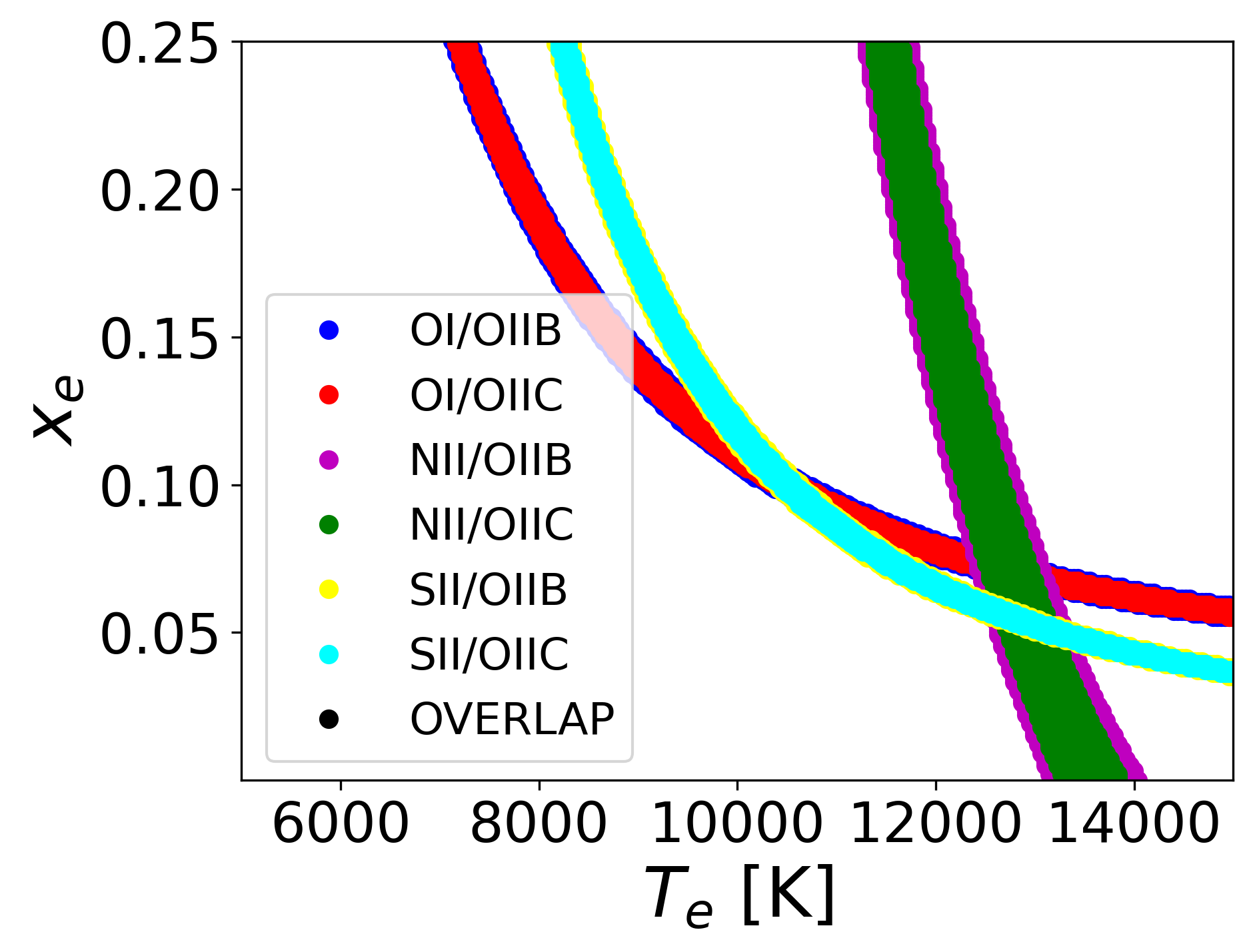}} 
\hspace{1em}   
\subfloat[\small{$A_V =0.3$, $t=10^6\,\text{s}$}]{\includegraphics[trim=0 0 0 0, clip, width=0.31  \textwidth]{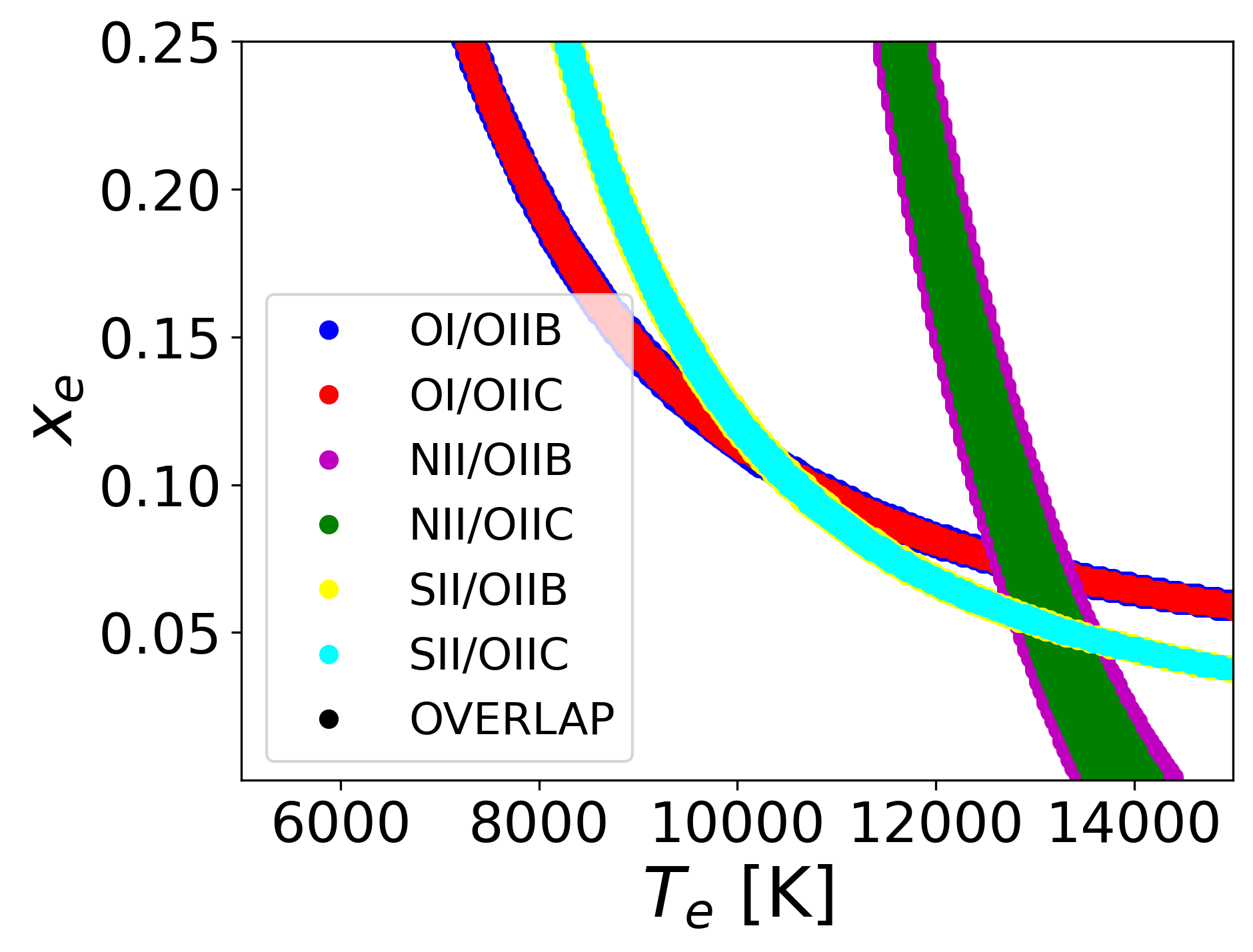}}
\hspace{1em} 
\subfloat[\small{$A_V =0.6$, $t=10^6\,\text{s}$}]{\includegraphics[trim=0 0 0 0, clip, width=0.31 \textwidth]{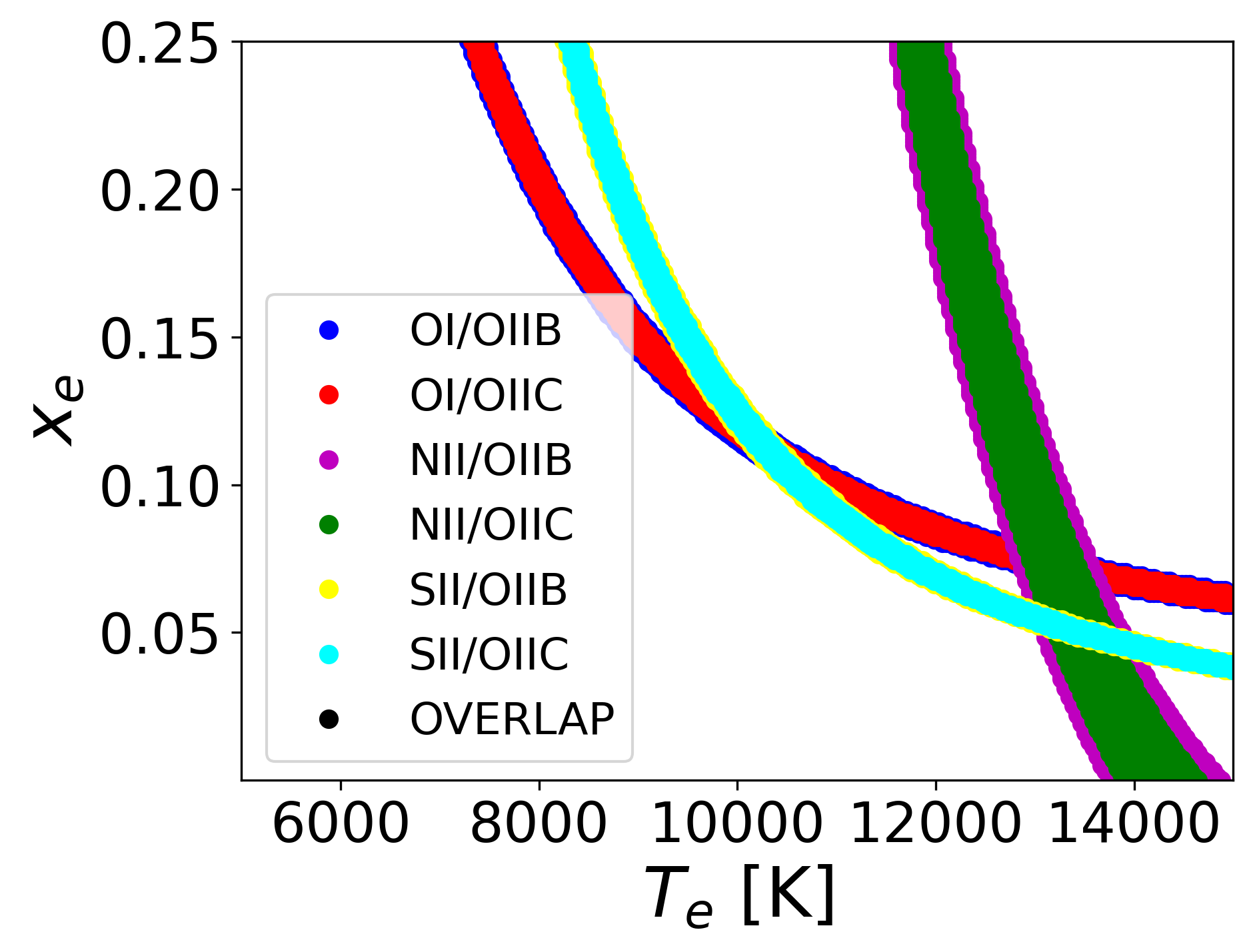}}
\hspace{1em} \\
\subfloat[\small{$A_V =0.0$, $t=10^8\,\text{s}$}]{\includegraphics[trim=0 0 0 0, clip, width=0.31 \textwidth]{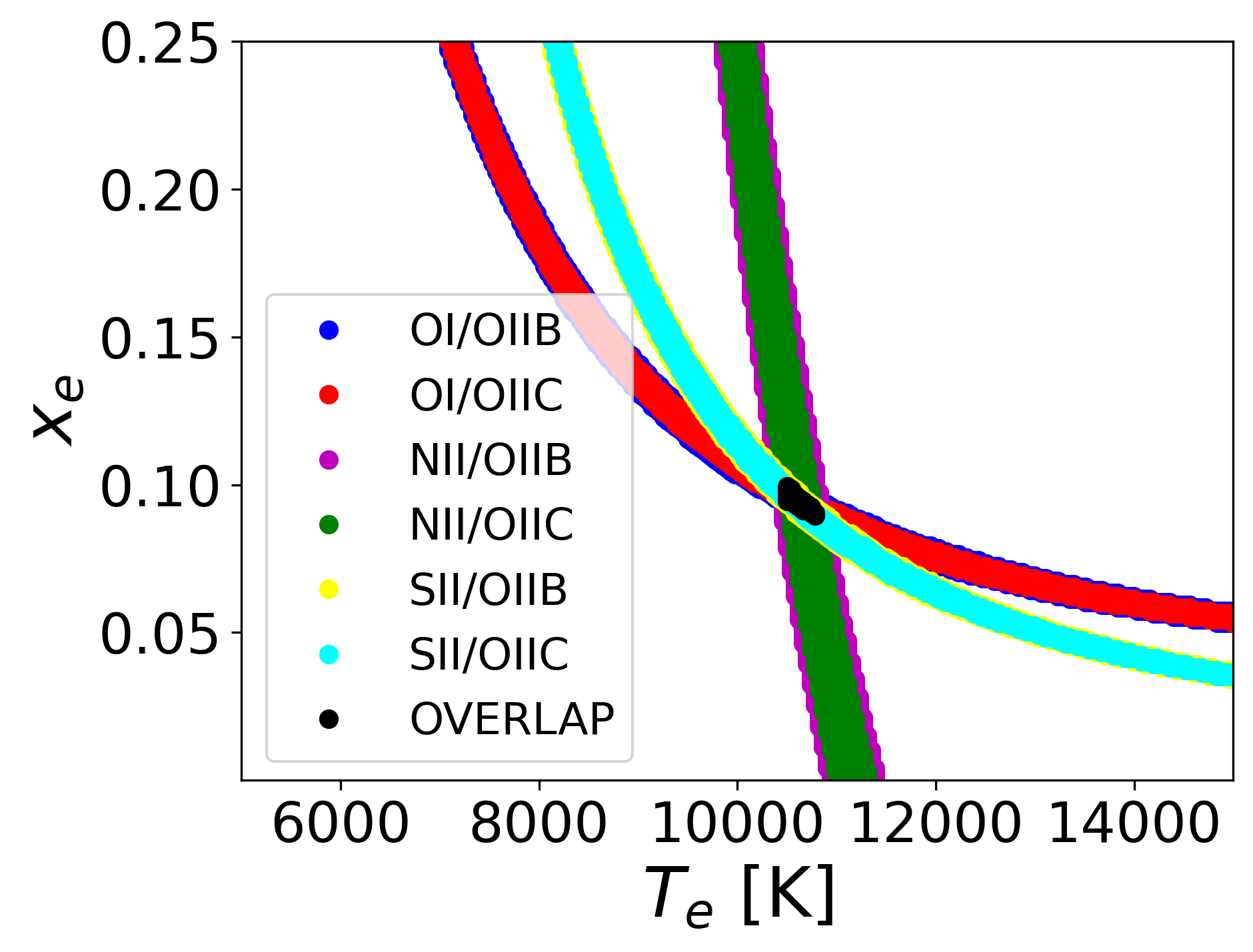}} 
\hspace{1em}   
\subfloat[\small{$A_V =0.3$, $t=10^8\,\text{s}$}]{\includegraphics[trim=0 0 0 0, clip, width=0.31 \textwidth]{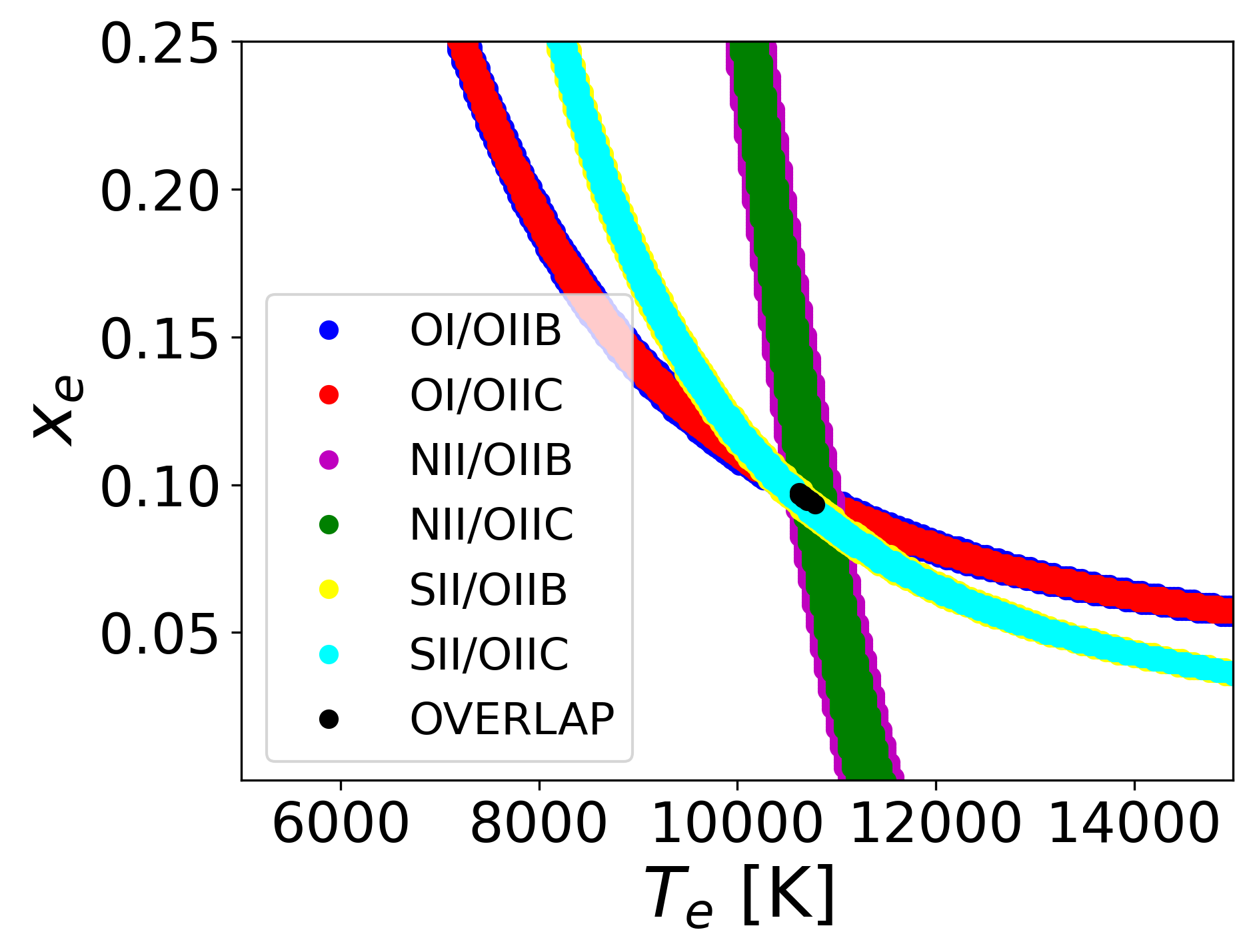}}
\hspace{1em}
\subfloat[\small{$A_V =0.6$, $t=10^8\,\text{s}$}]{\includegraphics[trim=0 0 0 0, clip, width=0.31 \textwidth]{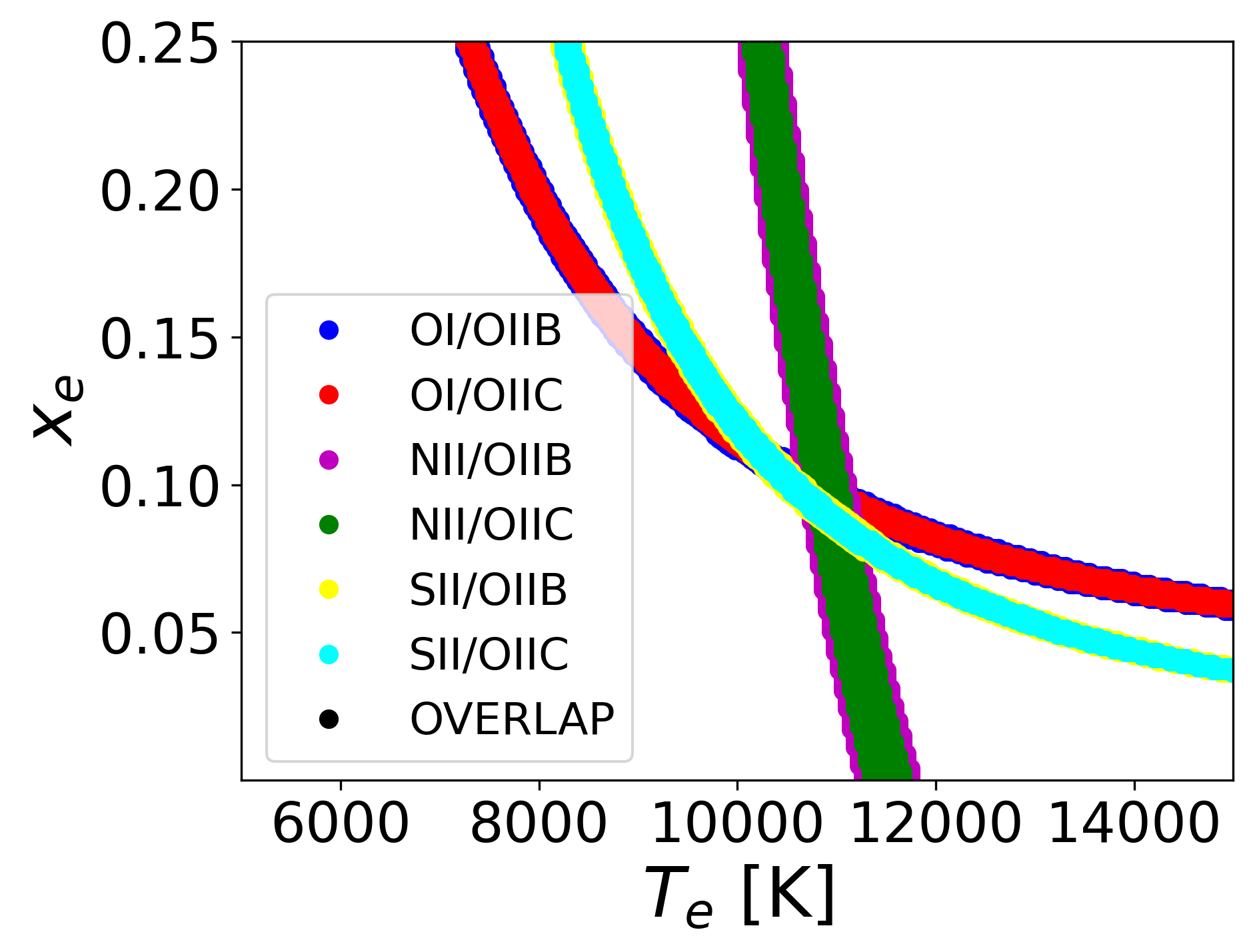}}
\hspace{1em}  \\
\subfloat[\small{$A_V =0.0$, $t=10^{10}\,\text{s}$}]{\includegraphics[trim=0 0 0 0, clip, width=0.31  \textwidth]{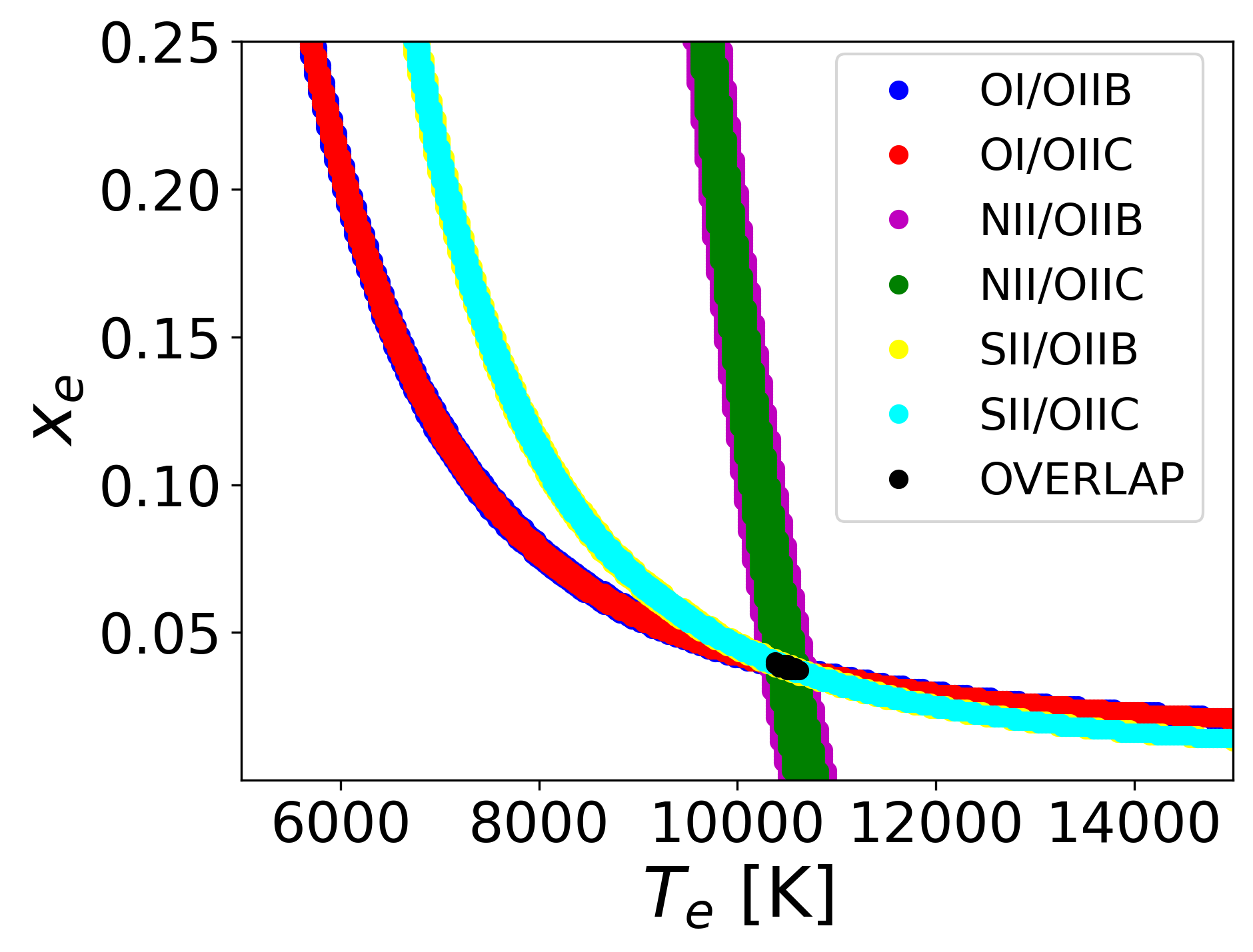}} 
\hspace{1em}   
\subfloat[\small{$A_V =0.3$, $t=10^{10}\,\text{s}$}]{\includegraphics[trim=0 0 0 0, clip, width=0.31  \textwidth]{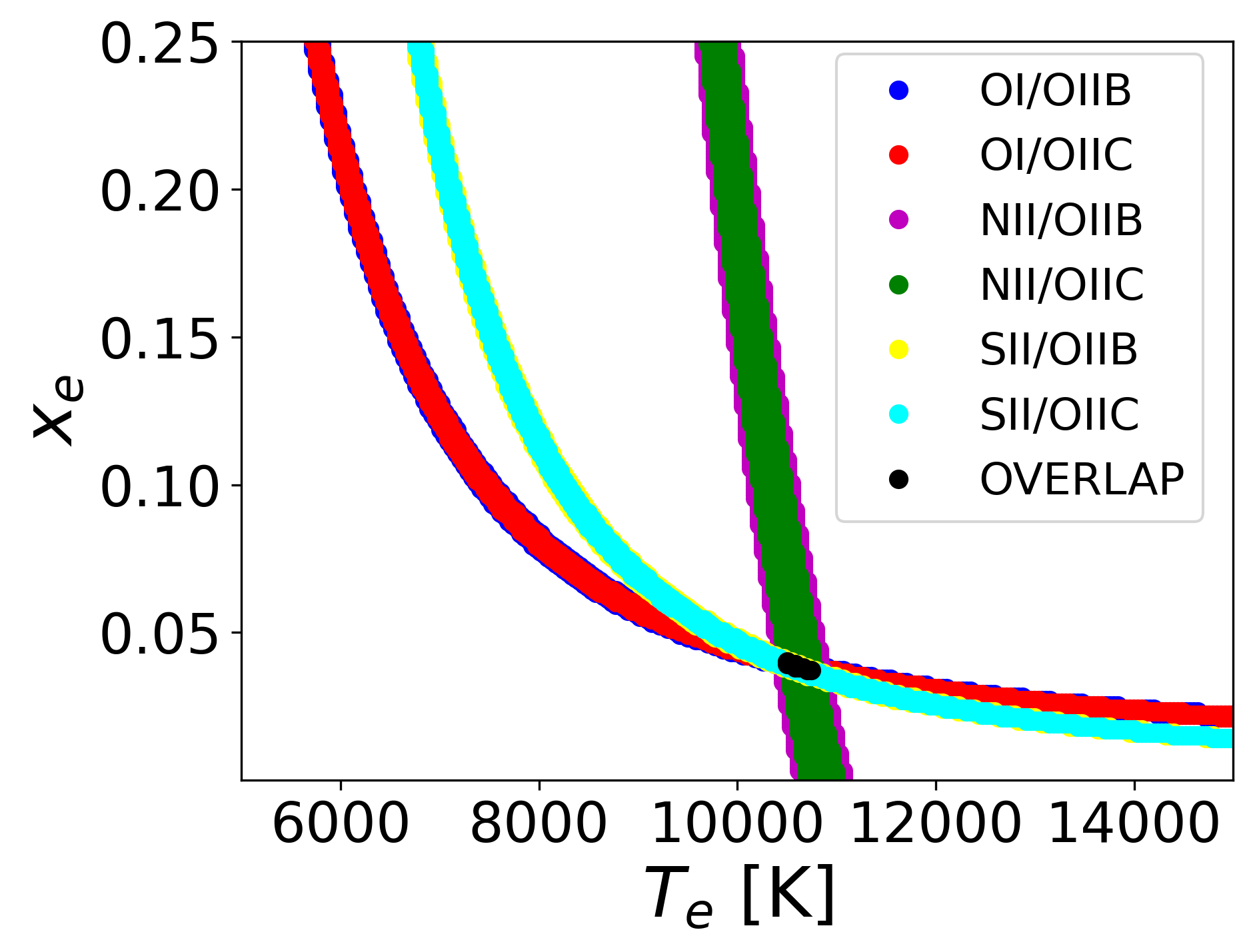}} 
\hspace{1em} 
\subfloat[\small{$A_V =0.6$, $t=10^{10}\,\text{s}$}]{\includegraphics[trim=0 0 0 0, clip, width=0.31  \textwidth]{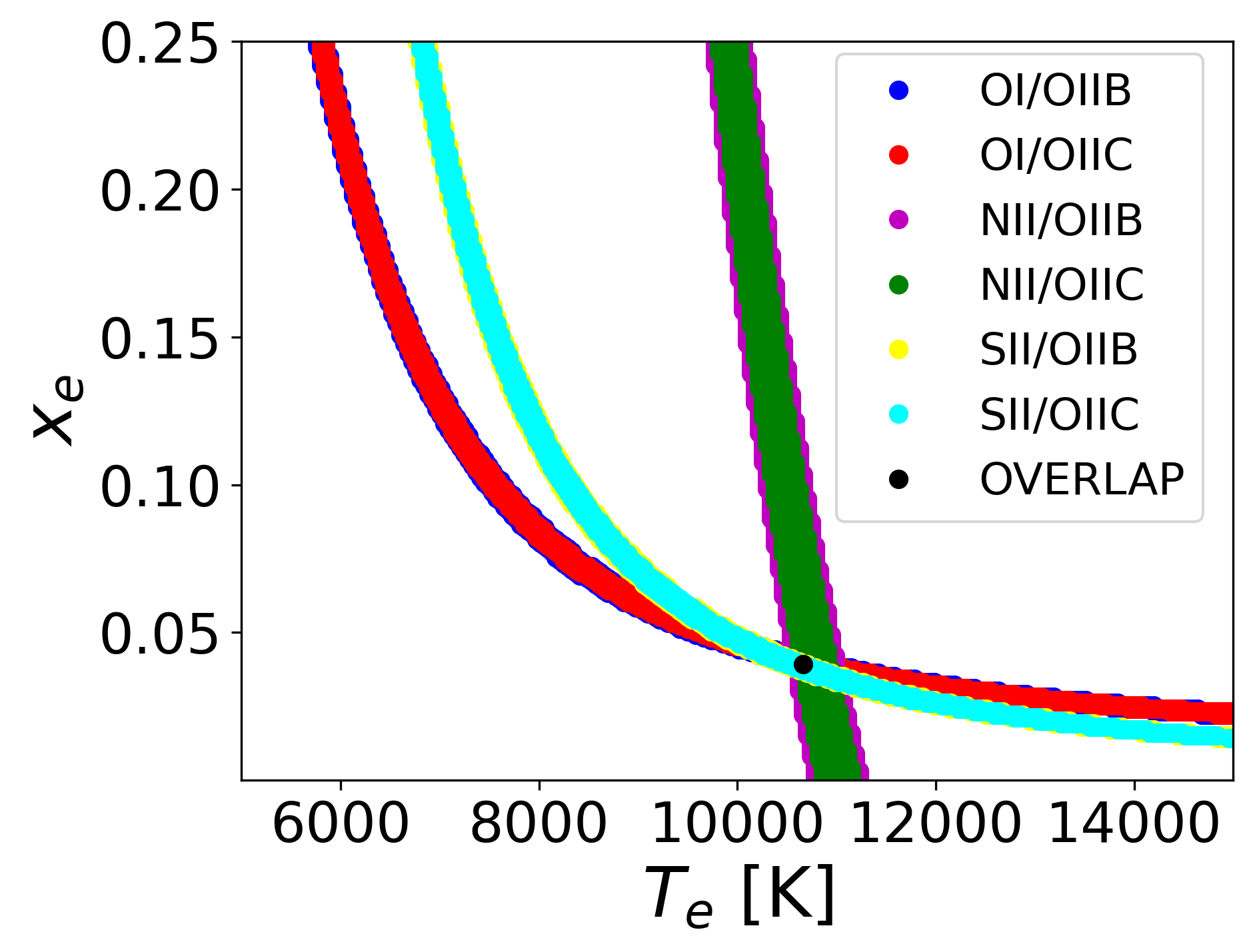}} 
\hspace{1em}    \\
\subfloat[\small{$A_V =0.0$, $t=10^{12}\,\text{s}$}]{\includegraphics[trim=0 0 0 0, clip, width=0.31  \textwidth]{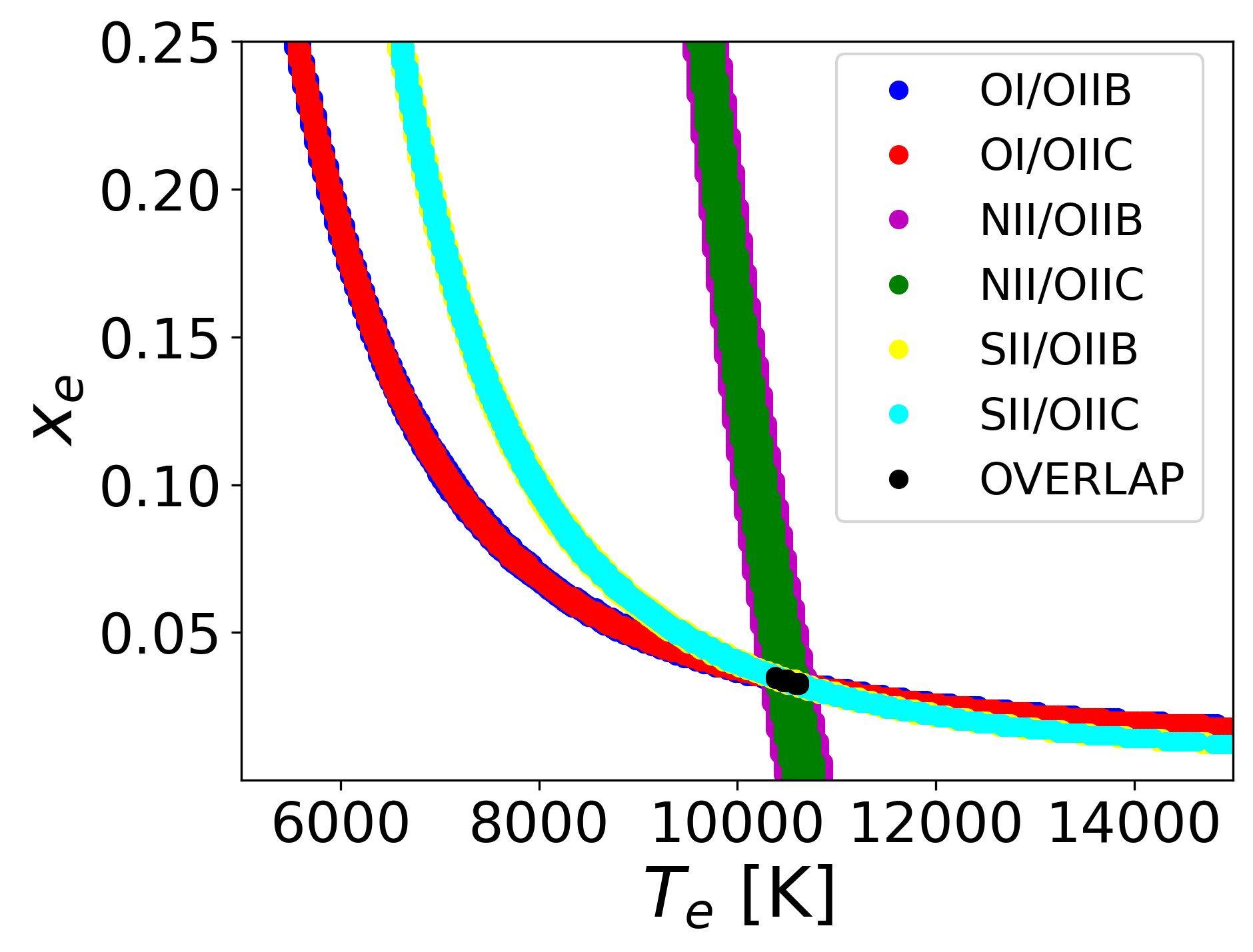}}
\hspace{1em} 
\subfloat[\small{$A_V =0.3$, $t=10^{12}\,\text{s}$}]{\includegraphics[trim=0 0 0 0, clip, width=0.31\textwidth]{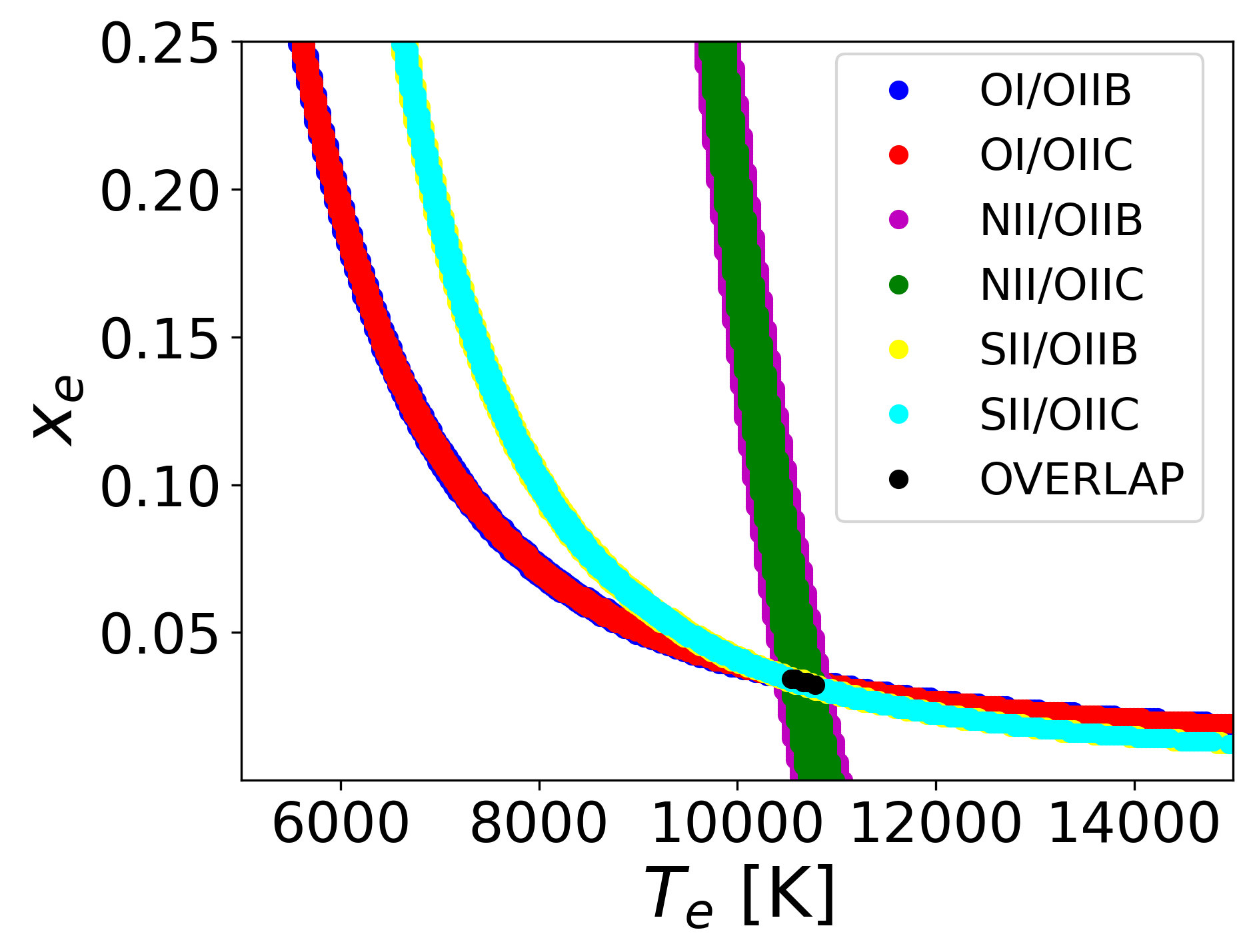}} 
\hspace{1em}   
\subfloat[\small{$A_V =0.6$, $t=10^{12}\,\text{s}$}]{\includegraphics[trim=0 0 0 0, clip, width=0.31\textwidth]{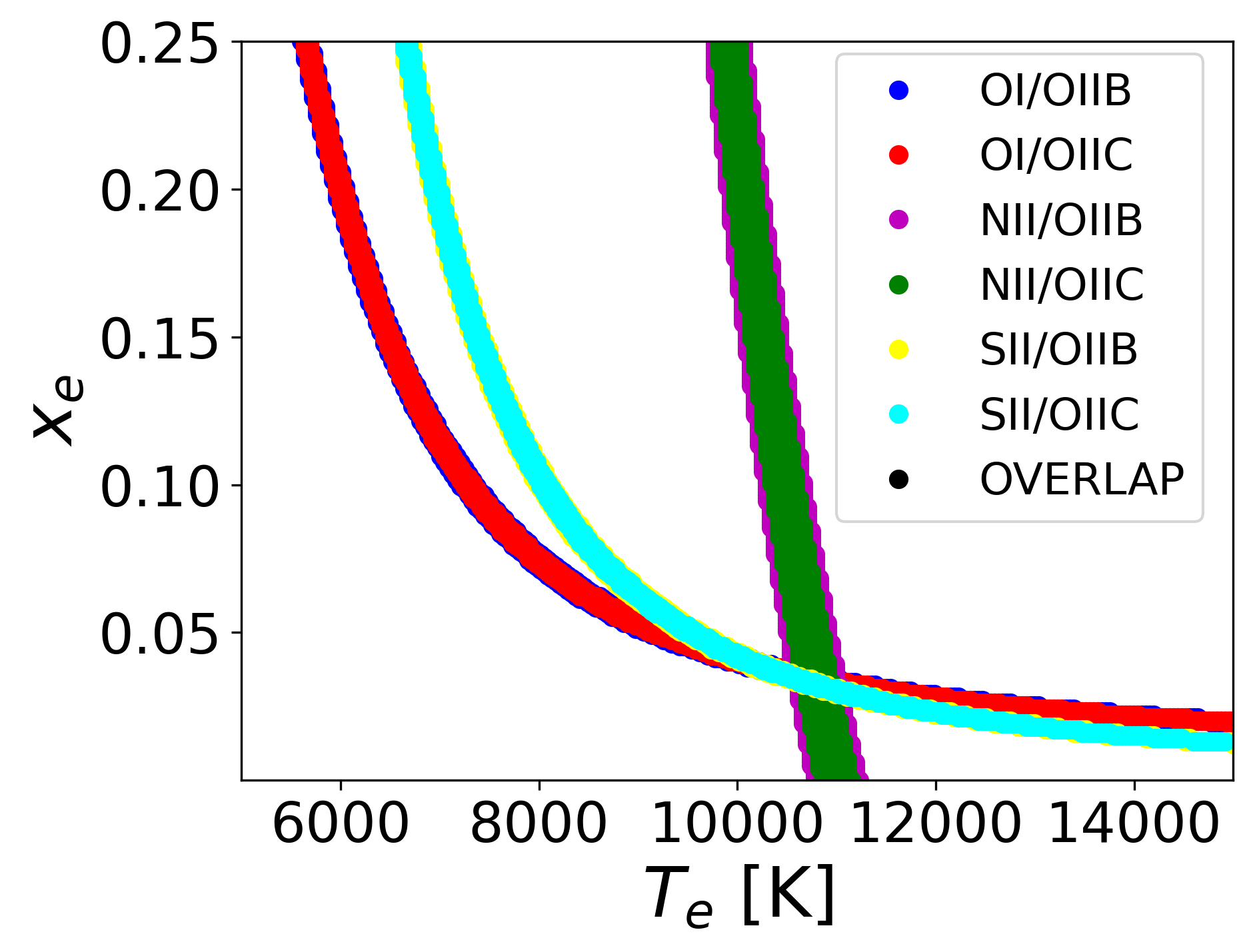}} 
\hspace{1em}  
\caption{\small{Same as Fig.\,\ref{fig:all_BE_diagrams_A} but with the additional line ratio: OIIB = [O\,II]$\lambda\lambda$7319+7320, OIIC = [O\,II]$\lambda\lambda$7329+7331.}}\label{fig:all_BE_diagrams_D}
\end{figure*} 

\clearpage

\section*{Appendix D - PV diagrams of the Par Lup 3-4 outflow}\label{appendix:E}


\begin{figure*}[h]
\centering
\subfloat{\includegraphics[trim=0 0 0 0, clip, width=0.2  \textwidth]{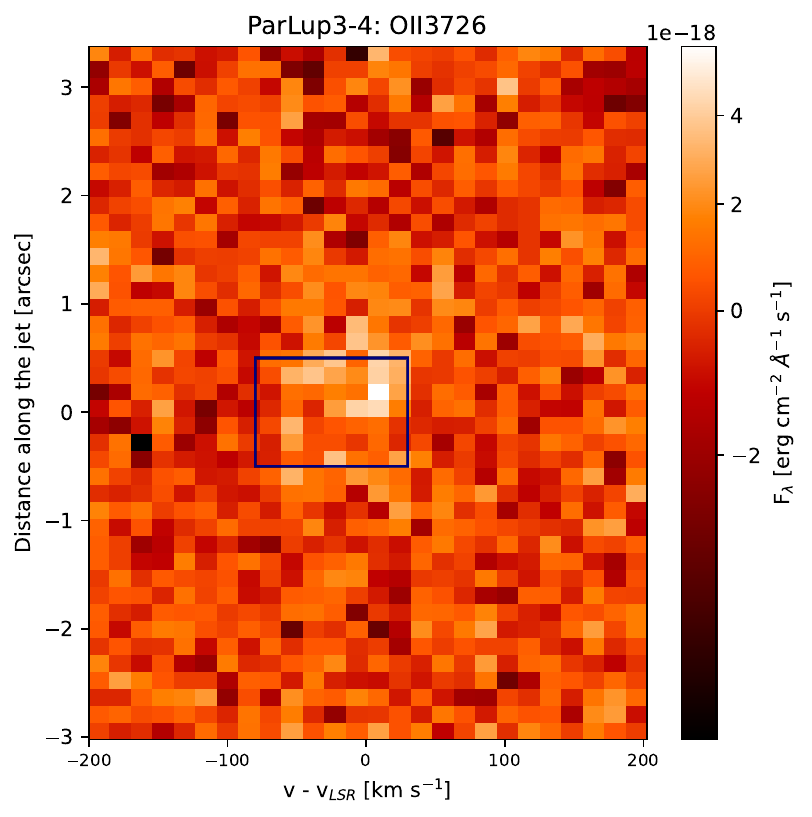}}
\hfill 
\subfloat{\includegraphics[trim=0 0 0 0, clip, width=0.2  \textwidth]{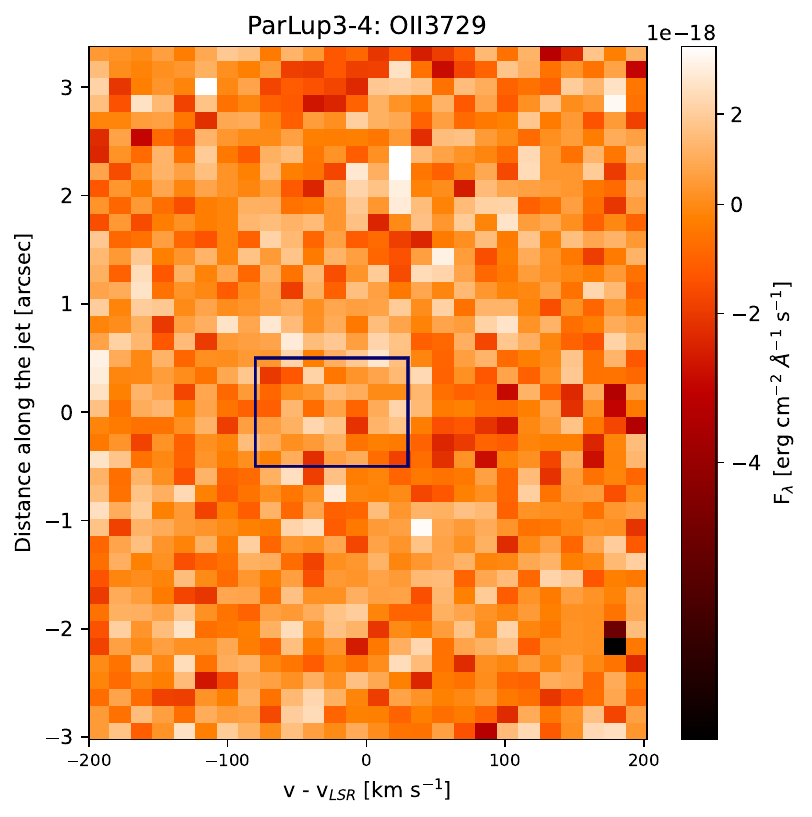}}
\hfill  
\subfloat{\includegraphics[trim=0 0 0 0, clip, width=0.2 \textwidth]{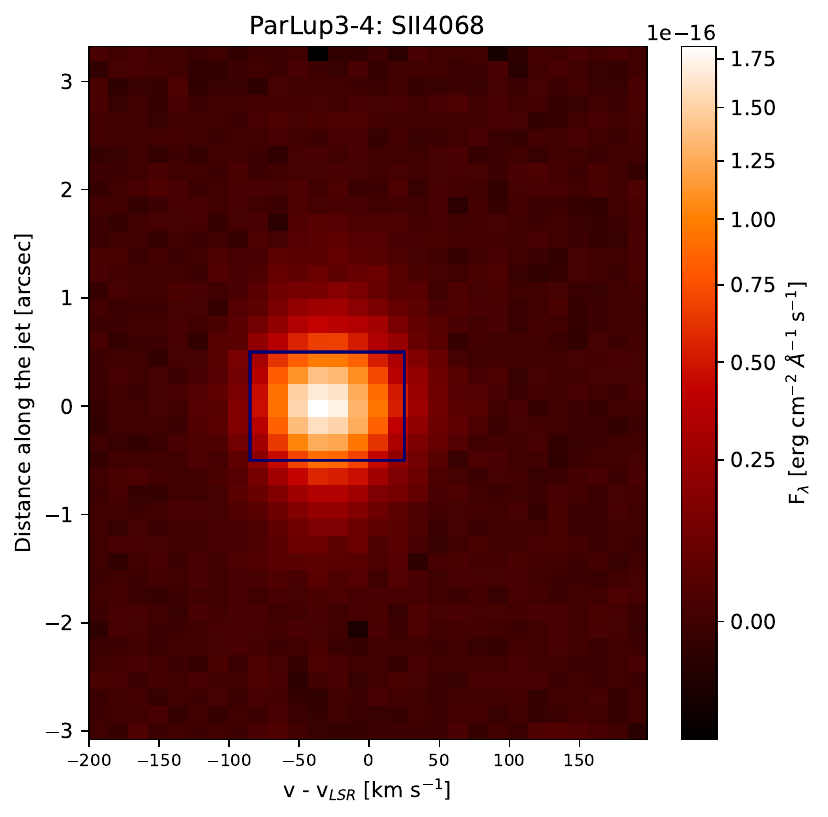}}
\hfill 
\subfloat{\includegraphics[trim=0 0 0 0, clip, width=0.2 \textwidth]{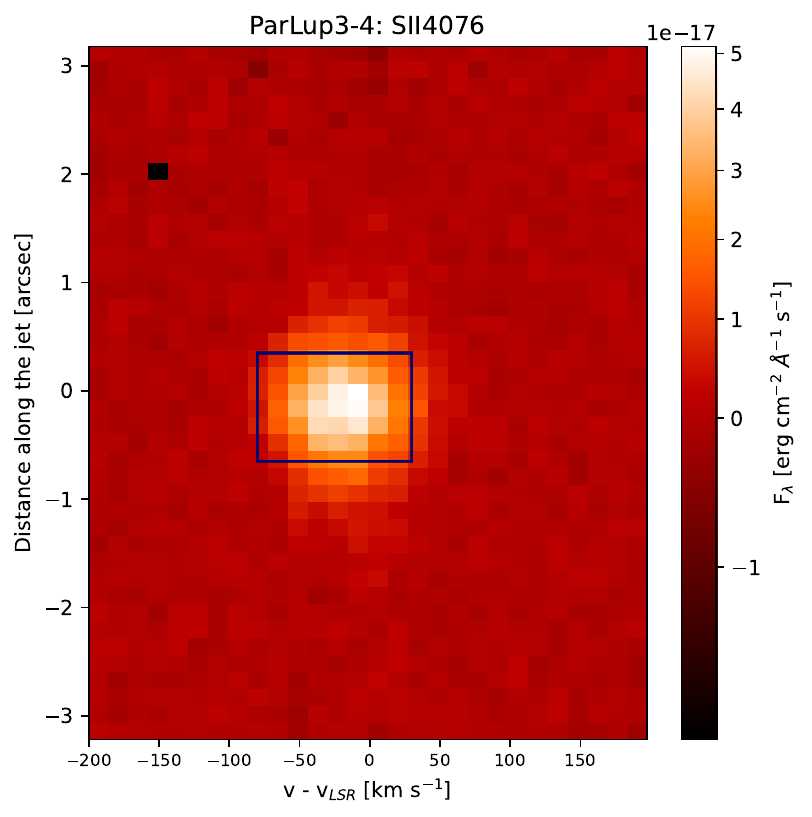}}
\hfill  
\subfloat{\includegraphics[trim=0 0 0 0, clip, width=0.2 \textwidth]{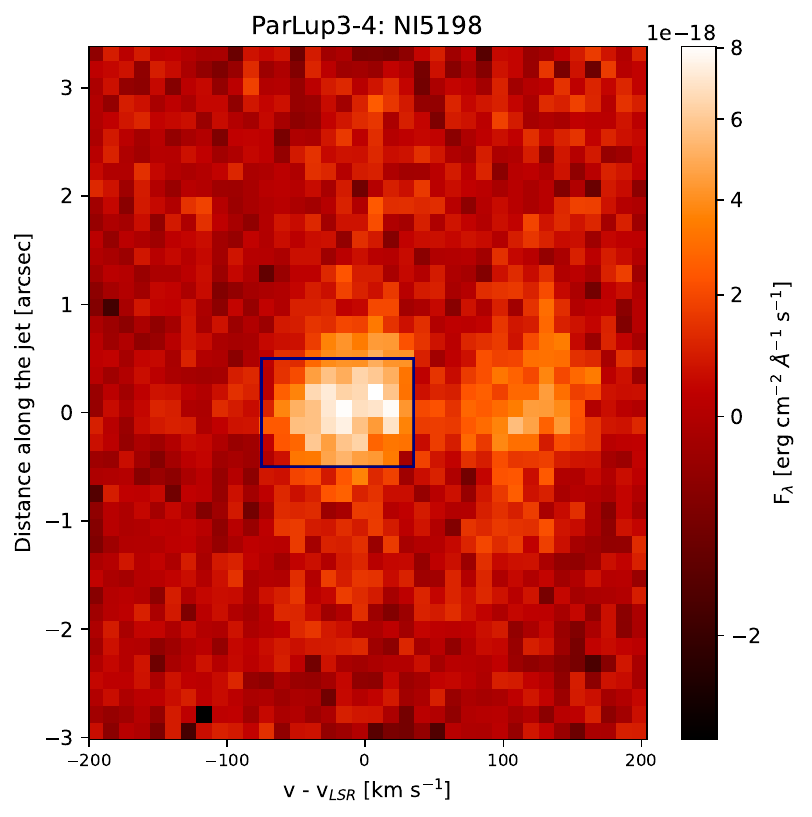}}
\hfill  \\
\subfloat{\includegraphics[trim=0 0 0 0, clip, width=0.2 \textwidth]{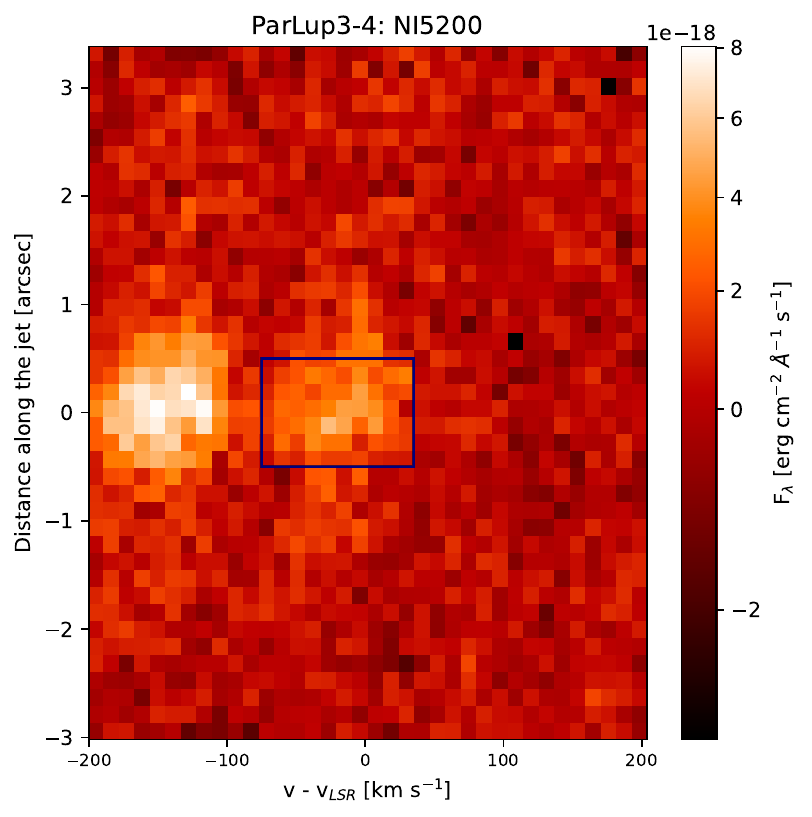}}
\hfill  
\subfloat{\includegraphics[trim=0 0 0 0, clip, width=0.2 \textwidth]{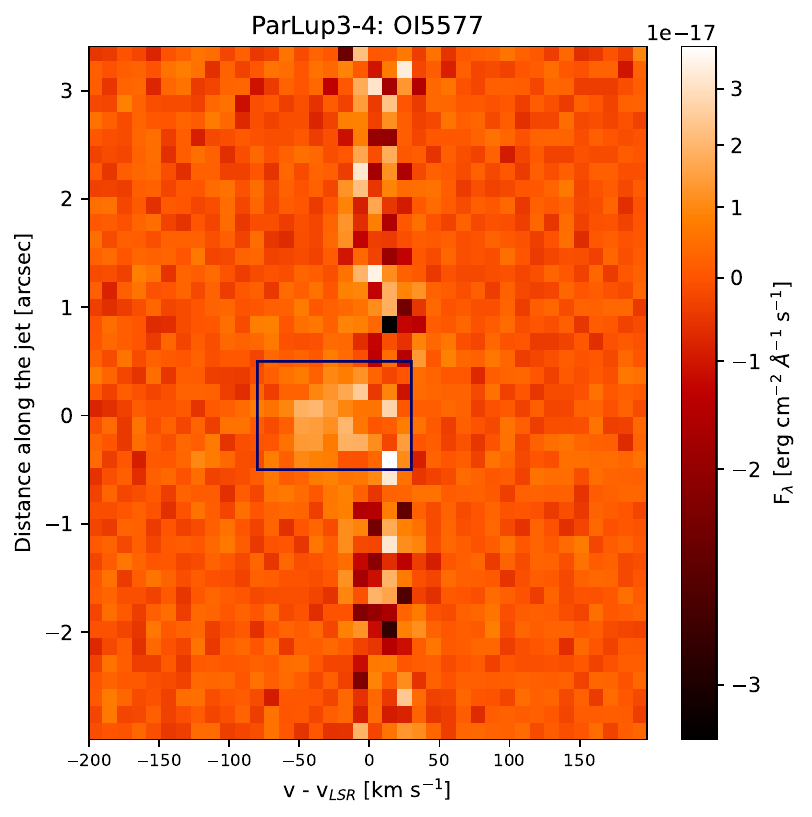}}
\hfill  
\subfloat{\includegraphics[trim=0 0 0 0, clip, width=0.2 \textwidth]{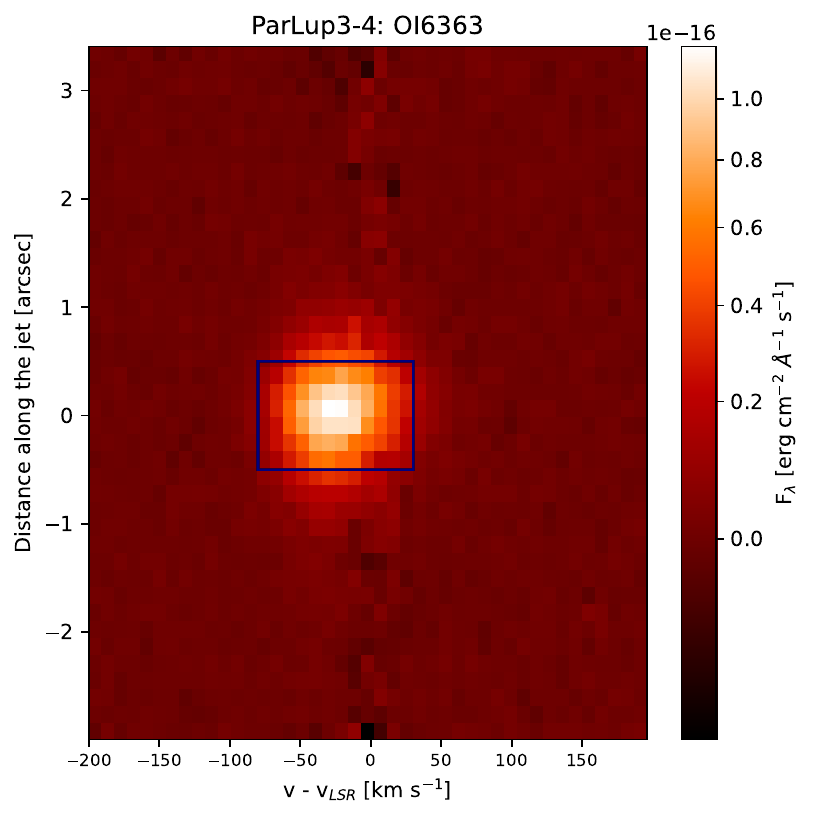}}
\hfill
\subfloat{\includegraphics[trim=0 0 0 0, clip, width=0.2 \textwidth]{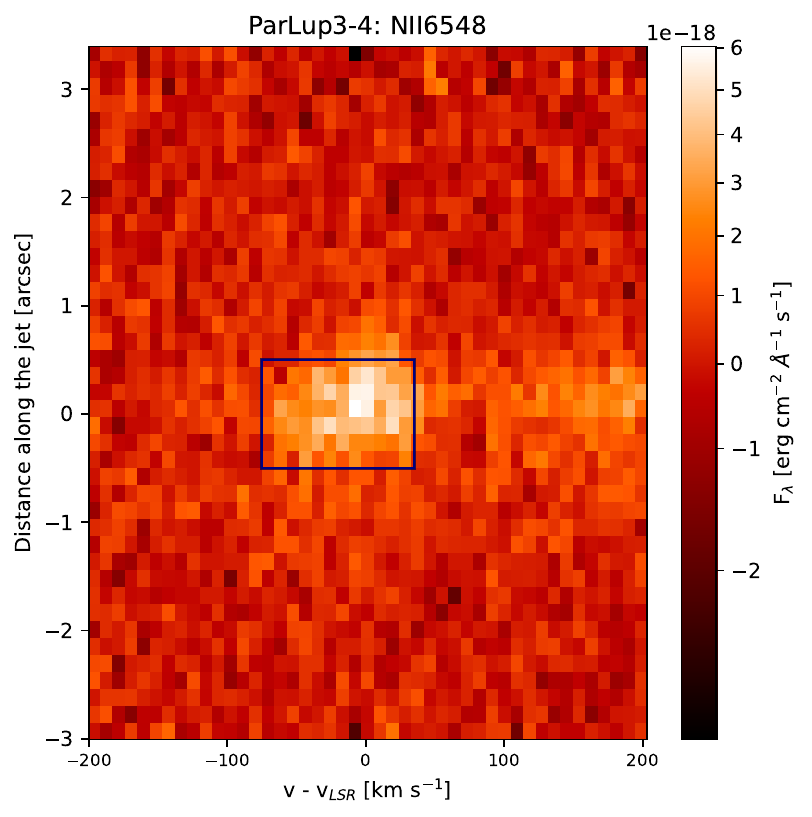}}
\hfill 
\subfloat{\includegraphics[trim=0 0 0 0, clip, width=0.2 \textwidth]{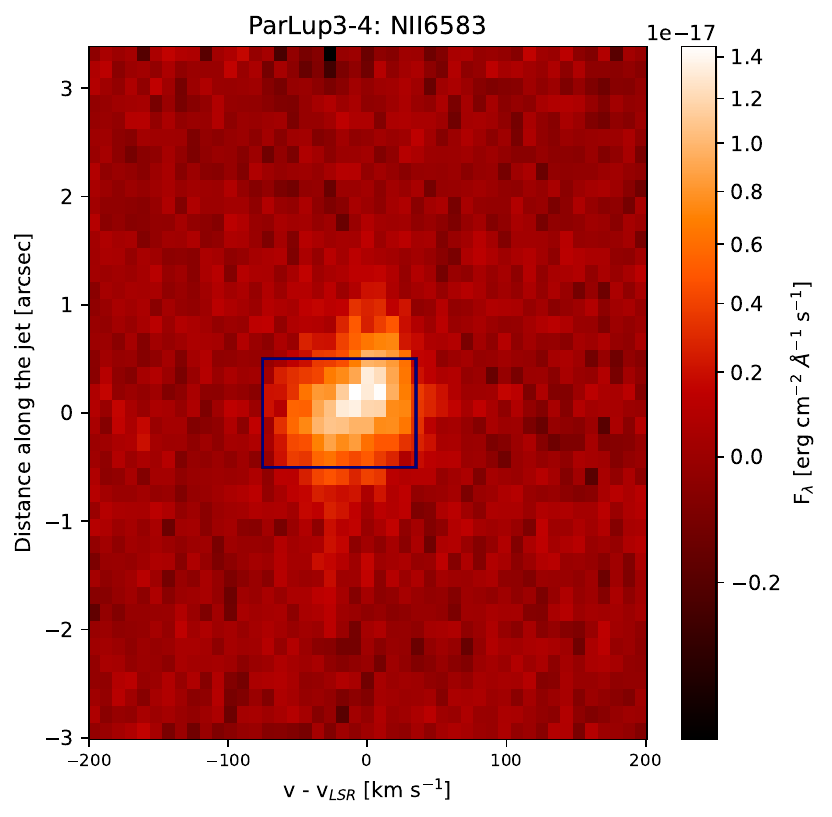}}
\hfill \\   
\subfloat{\includegraphics[trim=0 0 0 0, clip, width=0.2 \textwidth]{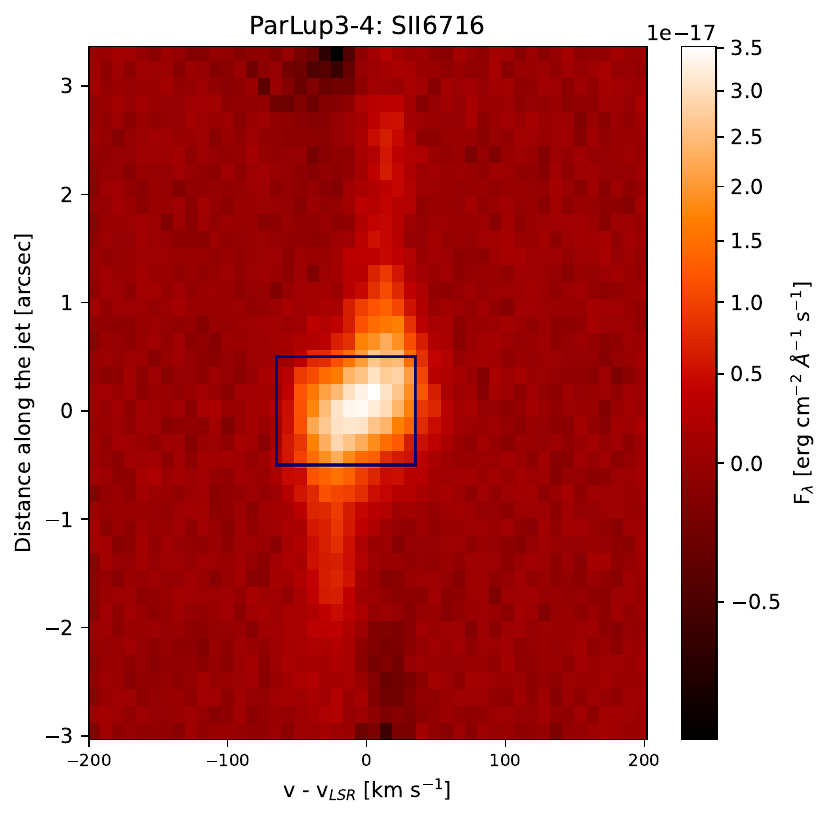}}
\hfill
\subfloat{\includegraphics[trim=0 0 0 0, clip, width=0.2 \textwidth]{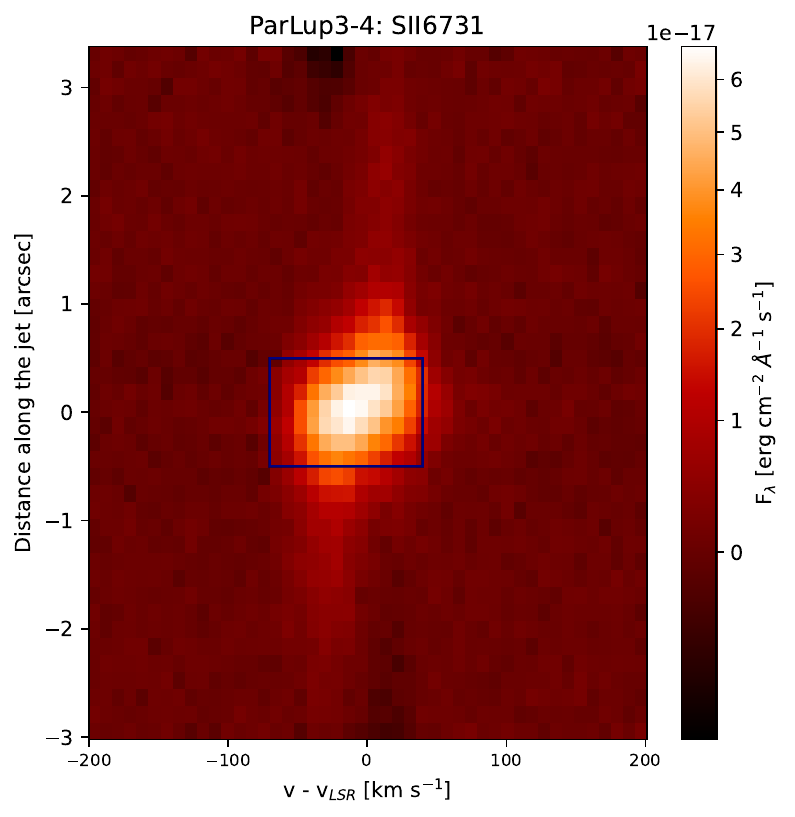}}
\hfill    
\subfloat{\includegraphics[trim=0 0 0 0, clip, width=0.2 \textwidth]{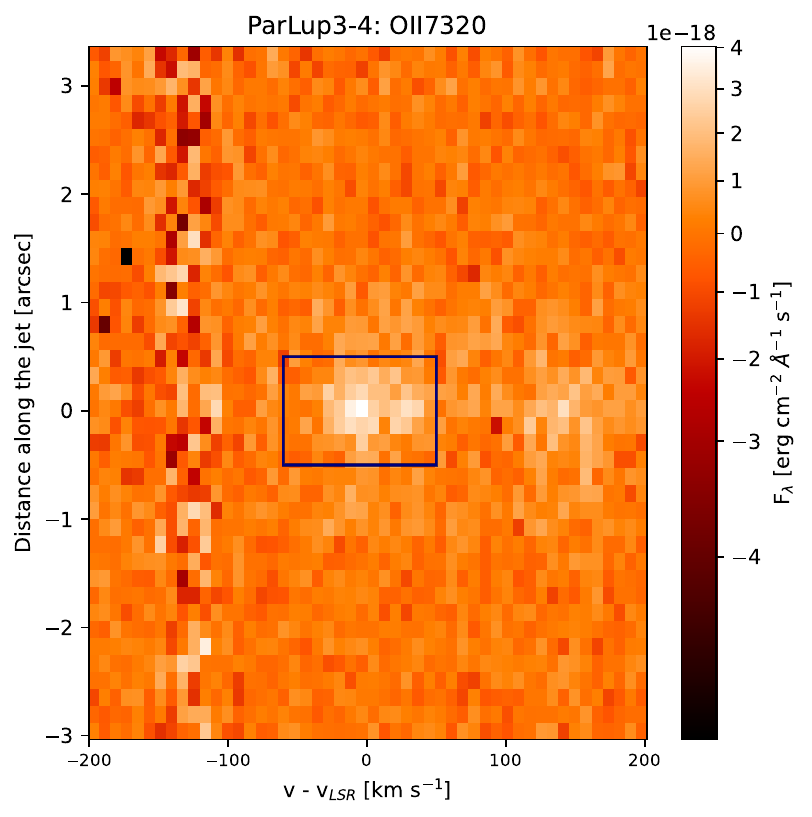}}
\hfill
\subfloat{\includegraphics[trim=0 0 0 0, clip, width=0.2 \textwidth]{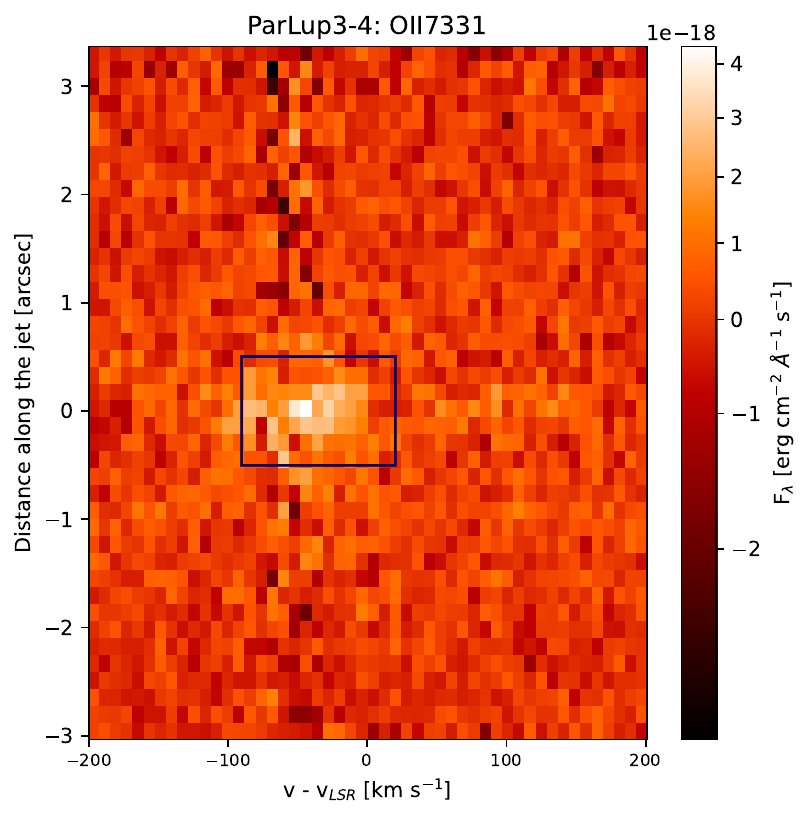}}
\hfill 
\subfloat{\includegraphics[trim=0 0 0 0, clip, width=0.2 \textwidth]{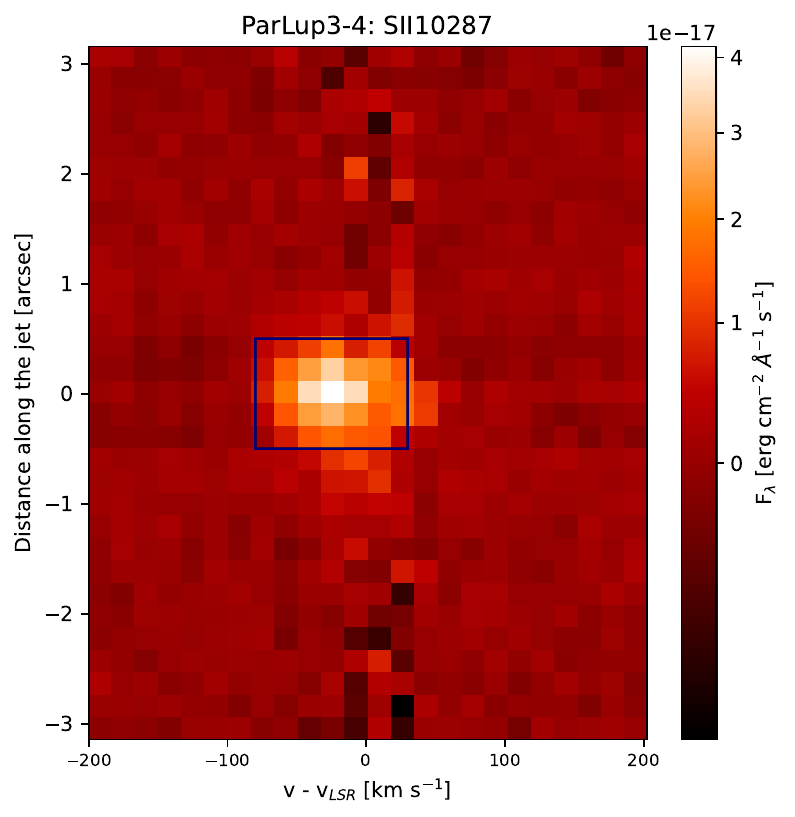}}
\hfill  \\
\subfloat{\includegraphics[trim=0 0 0 0, clip, width=0.2 \textwidth]{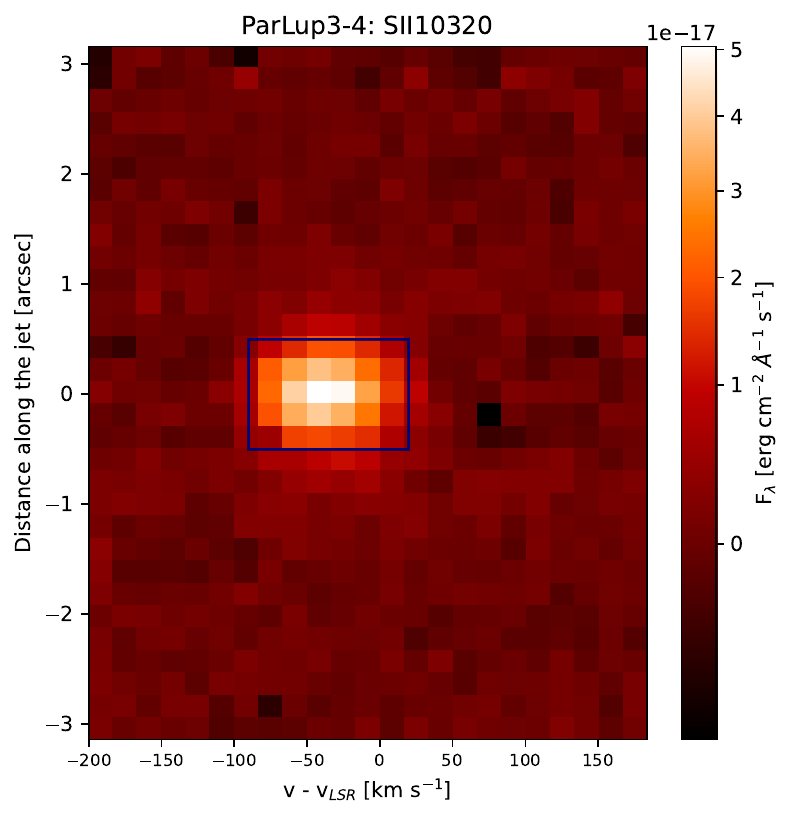}}
\hfill  
\subfloat{\includegraphics[trim=0 0 0 0, clip, width=0.2 \textwidth]{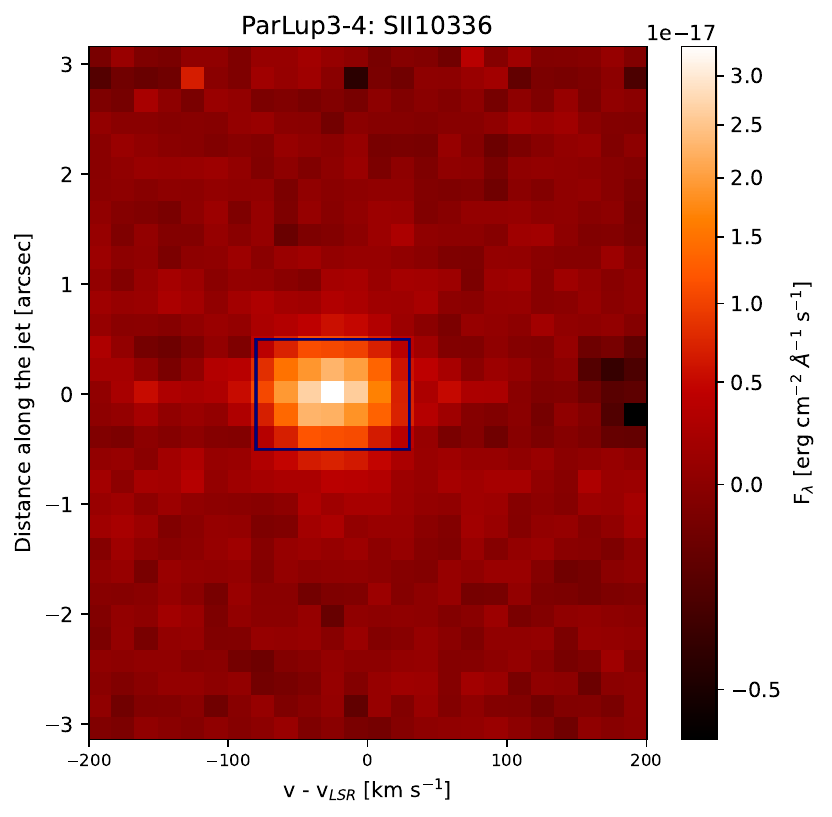}}
\hfill 
\subfloat{\includegraphics[trim=0 0 0 0, clip, width=0.2 \textwidth]{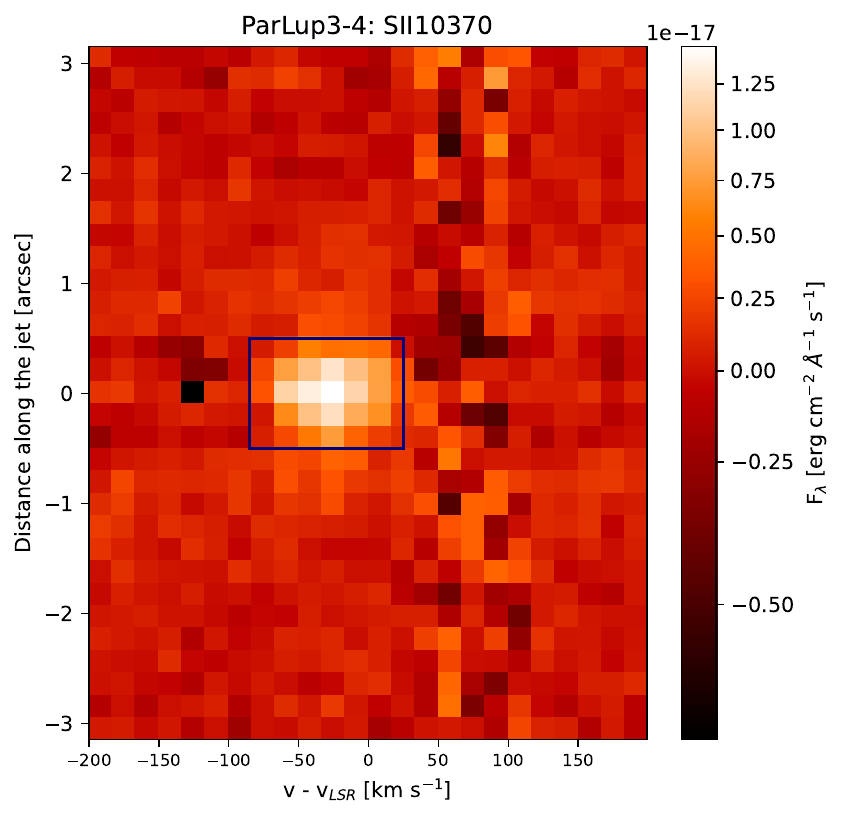}}
\hfill  
\subfloat{\includegraphics[trim=0 0 0 0, clip, width=0.2 \textwidth]{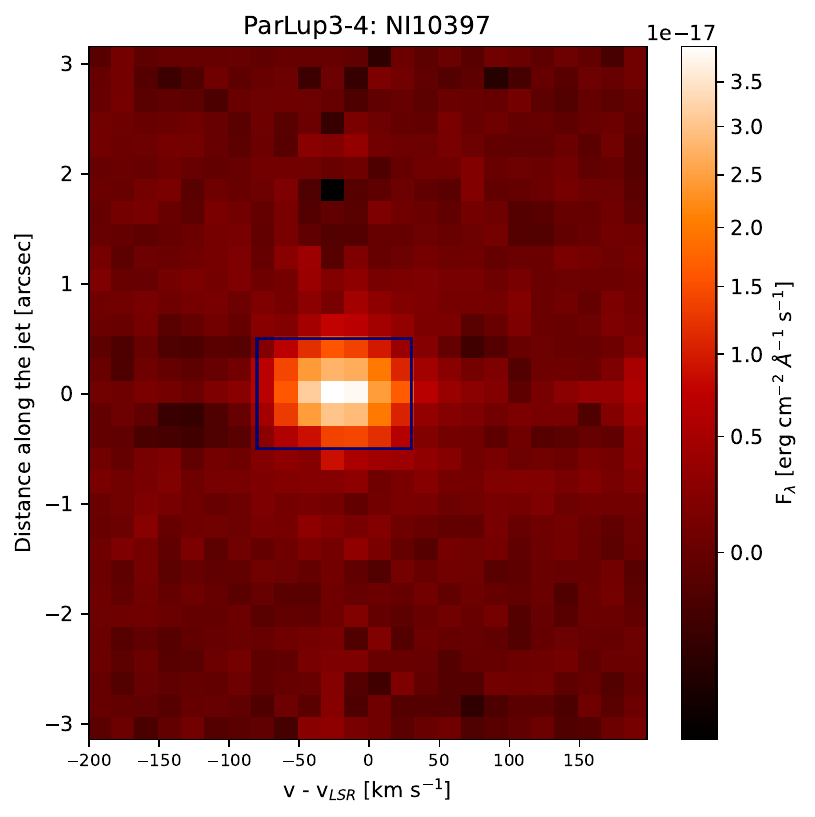}}
\hfill 
\subfloat{\includegraphics[trim=0 0 0 0, clip, width=0.2 \textwidth]{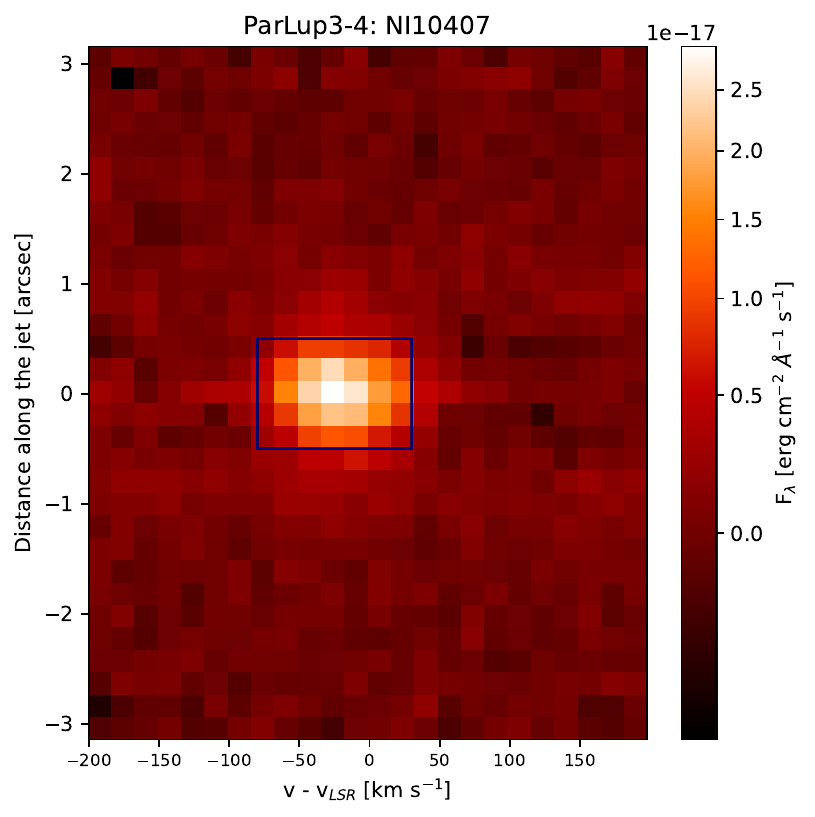}}
\hfill 
\caption{\small{Position-velocity diagrams for the BE99 lines detected towards Par Lup 3-4.}}\label{fig:all_pv_diagrams_boxes} 
\end{figure*}

\clearpage

\twocolumn
 
\section*{Appendix E - Extracted relative flux values}\label{appendix:F}

 {\renewcommand{\arraystretch}{1.2}
\begin{table}[htb]   \tiny
\caption{\small{Observed, normalised line fluxes at the driving source of Par Lup 3-4. We normalised the line fluxes to the bright [O\,I]$\lambda$6300 line by setting its line flux to the fiducial value of 100. For the comparison with W14 we multiplied the NIR fluxes stated in W14 with the factor 0.25 to correct for the overcorrection therein. Blended lines are indicated with (bl).}}\label{table:line_fluxes}
\centering \scalebox{1.0}{
\begin{tabular}{  |c||c|c|  }
\hline\hline
 \textbf{Line ID} &  $F$ (this work) & W14 \\  
 \hline
 $[$O\,II$]\lambda$3726   & $<1$  & not detected\\
 $[$O\,II$]\lambda$3729   & $<1$  & not detected\\
 $[$S\,II$]\lambda$4068   & $39.48 \pm 3.9$  & $39.55 \pm 0.08$ \\
 $[$S\,II$]\lambda$4076   & $ 11.01 \pm 1.1 $ & $ 11.20 \pm 0.08$ \\
 $[$O\,III$]\lambda$4959  & $<1$  & not detected\\
 $[$O\,III$]\lambda$5007  & $<1$  & not detected\\
 $[$N\,I$]\lambda$5198    & $ 2.36 \pm 0.24$ &  $2.46 \pm 0.08$\\
 $[$N\,I$]\lambda$5200    & $ 1.28 \pm 0.13$ &  $1.37 \pm 0.08$ \\  \hline
 $[$O\,I$]\lambda$5577    & $<5.0$ &  $ 4.67\pm 0.12 $  \\
 $[$N\,II$]\lambda$5755   & $<1$  & not detected\\
 $[$O\,I$]\lambda$6300    & $100.00 \pm 2.80$ &  $100 \pm 0.12$  \\
 $[$O\,I$]\lambda$6363    & $33.95\pm 2.00$ &  $32.94\pm 0.12$ \\
 $[$N\,II$]\lambda$6548   & $1.87 \pm 0.58$ & $1.89 \pm 0.08$\\
 $[$N\,II$]\lambda$6583   & $4.32 \pm 0.69$ & $4.23 \pm 0.08$\\ 
 $[$S\,II$]\lambda$6716   & $11.55 \pm 0.85$ & $11.92 \pm 0.08$\\
 $[$S\,II$]\lambda$6731   & $22.91 \pm 1.07$ & $22.67 \pm 0.08$ \\ 
 $[$OvII$]\lambda$7320 (bl.) & $<1.0$ &  $ 1.17\pm 0.08 $  \\
 $[$O\,II$]\lambda$7330 (bl.) & $<1.0$ &  $0.68 \pm 0.08 $  \\
 $[$S\,III$]\lambda$9068  & $<1$  & not detected\\
 $[$SvIII$]\lambda$9530  & $<1$  & not detected\\
   \hline
 $[$S\,II$]\lambda$10287 & $ 4.67 \pm 0.47$ & $3.91 \pm 0.24$ (corr.)   \\
 $[$S\,II$]\lambda$10320 & $6.06 \pm 0.61$  & $4.87 \pm 0.24$ (corr.) \\
 $[$S\,II$]\lambda$10336 &  $3.72 \pm 0.37$ & $3.04 \pm 0.24$ (corr.) \\
 $[$S\,II$]\lambda$10370 & $1.74 \pm 0.17$ & $1.47 \pm 0.24$ (corr.) \\  
 $[$N\,I$]\lambda$10398 (bl.) & $4.63 \pm 0.46$ & $4.03 \pm 0.24$ (corr.) \\  
 $[$N\,I$]\lambda$10407 (bl.) & $3.53 \pm 0.35$ & $3.14 \pm 0.24$ (corr.) \\
  \hline\hline
 \end{tabular} }
\end{table}

\end{document}